\documentclass[
 aip,
 amsmath,
 amssymb,
 numerical,
 pop,
 reprint,
 onecolumn
]{revtex4-1}

\usepackage{graphicx}
\usepackage{dcolumn}
\usepackage{bm}

\usepackage{mathptmx}
\usepackage{etoolbox}

\usepackage{float}

\usepackage{enumitem,amssymb}
\newlist{todolist}{itemize}{2}
\setlist[todolist]{label=$\square$}
\usepackage{pifont}

\usepackage[colorlinks=true,citecolor=purple,linkcolor=purple,urlcolor=purple,filecolor=red]{hyperref}

\usepackage{soul} 
\usepackage{xcolor}

\usepackage{lmodern}

\newcommand{\Ra}{R_{\mathrm{a}}}
\newcommand{\Rp}{R_{\mathrm{p}}}
\newcommand{\mB}{\mathrm{B}}
\newcommand{\md}{\mathrm{d}}
\newcommand{\mD}{\mathrm{D}}
\newcommand{\me}{\mathrm{e}}
\newcommand{\mi}{\mathrm{i}}
\newcommand{\mJ}{\mathrm{J}}
\newcommand{\mRe}{\mathrm{Re}}
\newcommand{\mIm}{\mathrm{Im}}
\newcommand{\bJ}{\mathbf{J}}
\newcommand{\bX}{\mathbf{X}}
\newcommand{\br}{\mathbf{r}}
\newcommand{\bp}{\mathbf{p}}
\newcommand{\bq}{\mathbf{q}}
\newcommand{\bv}{\mathbf{v}}

\newcommand{\bF}{\mathbf{F}}
\newcommand{\bzero}{\mathbf{0}}
\newcommand{\btheta}{\bm{\theta}}
\newcommand{\bn}{\mathbf{n}}
\newcommand{\bL}{\mathbf{L}}
\newcommand{\bA}{\mathbf{A}}
\newcommand{\bOm}{\bm{\Omega}}
\newcommand{\nn}{\nonumber}
\newcommand{\micro}[1]{#1_{\mkern2mu\textsc{m}}}
\newcommand{\p}{\partial}
\newcommand{\half}{\tfrac{1}{2}}
\newcommand{\omegaR}{\omega_{\mathrm{R}}}

\newcommand{\pattern}{\Omega_{\mathrm{p}}}
\newcommand{\tdyn}{t_{\mathrm{dyn}}}
\newcommand{\trelax}{t_{\mathrm{relax}}}
\newcommand{\Etot}{E_{\mathrm{tot}}}
\newcommand{\bmin}{b_{\mathrm{min}}}
\newcommand{\bmax}{b_{\mathrm{max}}}
\newcommand{\tNR}{t_{\mathrm{2BR}}}
\newcommand{\mpc}{\mathrm{pc}}
\newcommand{\mkpc}{\mathrm{kpc}}
\newcommand{\mkm}{\mathrm{km}}
\newcommand{\ms}{\mathrm{s}}

\newcommand{\mesc}{\mathrm{esc}}
\newcommand{\cst}{\mathrm{cst.}}
\newcommand{\msource}{\mathrm{source}}
\newcommand{\mbare}{\mathrm{bare}}
\newcommand{\msA}{\mathsf{A}}
\newcommand{\msB}{\mathsf{B}}
\newcommand{\msD}{\mathsf{D}}
\newcommand{\msE}{\mathsf{E}}
\newcommand{\msI}{\mathsf{I}}
\newcommand{\msM}{\mathsf{M}}
\newcommand{\mszero}{\mathsf{0}}
\newcommand{\mcE}{\mathcal{E}}

\makeatletter
\def\@email#1#2{%
 \endgroup
 \patchcmd{\titleblock@produce}
  {\frontmatter@RRAPformat}
  {\frontmatter@RRAPformat{\produce@RRAP{*#1\href{mailto:#2}{#2}}}\frontmatter@RRAPformat}
  {}{}
}%
\makeatother

\begin{document}

\preprint{AIP/123-QED}

\title[\textcolor{gray}{Hamilton \& Fouvry | \textit{Kinetic Theory of Stellar Systems} | Physics of Plasmas Tutorials | DRAFT: \today}
 ]{\textcolor{purple}{Kinetic Theory of Stellar Systems: A Tutorial} 
}
\author{\textcolor{black}{Chris Hamilton}}
 \email{chamilton@ias.edu}
\affiliation{Institute for Advanced Study, Einstein Drive, Princeton, NJ 08540, USA} 
\author{\textcolor{black}{Jean-Baptiste Fouvry}}
\affiliation{CNRS and Sorbonne Universit\'{e}, UMR 7095, Institut d’Astrophysique de Paris, 98 bis Boulevard Arago, F-75014 Paris, France
}

\date{\today}

\begin{abstract}

Stellar systems --- globular and nuclear star clusters, elliptical and spiral galaxies and their surrounding dark matter haloes, and so on --- are ubiquitous characters in the evolutionary tale of our Universe.
This tutorial article is an introduction to the collective dynamical evolution of the very large numbers of stars and/or other self-gravitating objects that comprise such systems, i.e.\ their \textit{kinetic theory}. 

We begin by introducing the basic phenomenology of stellar systems, and explaining why and when we must develop a kinetic theory that transcends the traditional two-body relaxation picture of Chandrasekhar.
We then study the individual orbits that comprise stellar systems, how those orbits are modified by linear and nonlinear perturbations, how a system responds self-consistently to fluctuations in its own gravitational potential, and how one can predict the long term evolutionary fate of a stellar system in both quasilinear and nonlinear regimes. 
Though our treatment is necessarily mathematical, we develop the formalism only to the extent that it facilitates real calculations. Each section is bolstered with intuitive illustrations
and we give many examples throughout the text of the equations being applied to
topics of major astrophysical importance, such as 
radial migration, spiral instabilities, and dynamical friction on galactic bars.

Furthermore, in the 1960s and 1970s the kinetic theory of stellar systems was a fledgling subject which developed in tandem with the kinetic theory of \textit{plasmas}.
However, the two fields have long since diverged as their practitioners have focused on ever more specialized and technical issues. This tendency, coupled with the famous obscurity of astronomical jargon,
 means that today relatively few plasma physicists  are aware that their knowledge is directly applicable in the beautiful arena of galaxy evolution, and relatively few galactic astronomers know of the plasma-theoretic foundations upon which a portion of their subject is built.
 Yet once one has become fluent in
both Plasmaish and Galacticese, and has a dictionary relating the two,
one can pull ideas directly from one field to solve a problem in the other.
Therefore, another aim of this tutorial article is to provide our plasma colleagues 
with a jargon-light understanding of the key properties of stellar systems, to offer them
the theoretical minimum necessary to engage with the modern stellar dynamics literature, to point out the 
many direct analogies between stellar- and plasma-kinetic calculations, 
and ultimately to convince them that stellar dynamics and plasma kinetics are, in a deep, beautiful and \textit{useful} sense, the same thing.
\end{abstract}

\maketitle

    \section*{}

    \tableofcontents

    \newpage

\section{Introduction}
\label{sec:Introduction}

The evolution of our Universe is a great unfolding drama in which 
galaxies play a fundamental role.
These galaxies grow and die; they pulse and throb; they swallow gas and they belch it back out again;
they merge together and they tear one another apart.
The study of the collective dynamical behavior that occurs inside galaxies --- and of the violent lives they lead, the tumultuous dark matter environments in which they sit,
and the star clusters they cradle --- is best termed the \textit{kinetic theory of stellar systems}.
This is the subject with which we are concerned in this tutorial article.

The kinetic theory of stellar systems rests upon two main theoretical pillars, and our tutorial article aims to 
introduce colleagues to both of them.
The first pillar is the theory of \textit{orbits}.
By this we mean the orbital motion of stars and/or dark matter particles in gravitational fields; or, more broadly, the dynamics of point particles subject to some Hamiltonian flow.
The basic language of this topic is angle-action coordinates.
These are not only the natural variables with which to describe smooth orbits in quiescent gravitational fields, but also to understand how these orbits are modified by \textit{perturbations} of various kinds. These perturbations tend to have the most dramatic effect on orbits near \textit{resonances}, where the frequency of the perturbation matches 
the natural frequency of the orbit.
The first aim of our tutorial article is to introduce colleagues to these concepts. 
We employ as little heavy formalism as possible, assuming that readers already have a passing familiarity with Hamiltonian mechanics, our aim being to help them understand how they apply in the \textit{stellar dynamical context}.

The second pillar upon which our subject is built is the collective dynamics of many bodies,  and here our theoretical foundation is plasma kinetics. 
Why is this the case in a treatise on gravity?
Well, every high school student knows that electromagnetism and gravity are intellectually adjacent subjects:
up to a sign, the Coulombic electrical force and the Newtonian gravitational force are equivalent.
What is less well appreciated is that this adjacency extends much further than simple ${ \pm 1/r^2}$ force laws.
Indeed, the mathematical methods developed to probe the collective behavior of protons and electrons in astrophysical and fusion plasmas are in many cases \textit{precisely the same} as those needed to 
describe the dynamics of stars, spiral arms and dark matter in galaxies like the Milky Way.
Concepts like two-body relaxation,
linear response theory, Landau damping, resonant wave-particle interactions and phase mixing 
all had origins in plasma theory but are now part of
the standard lexicon of galactic dynamics.
But these two fields long ago went their separate ways, and in much of today's literature  are separated by a near-impenetrable wall of jargon. 
In the second part of this tutorial article we develop the tools needed to describe collective behavior in galaxies. We aim to provide colleagues with the theoretical minimum needed to appreciate galactic-dynamical problems. In so doing we aim to convey --- to those on both the stellar-dynamical and plasma-theoretic sides of the aisle --- 
just some of the deep connections between the two fields,
and to encourage much greater cross-talk between these two estranged, yet intimately related, 
subjects.

In contrast with most other treatises on stellar (or `galactic') dynamics and galaxy evolution, we approach our subject from the point of view of theoretical physics rather than practical astronomy. Thus, we spill almost no ink over such crucial concepts as mass-to-light ratios and chemical evolution and dust extinction (though we happily refer the reader to the many excellent texts that do\cite{Binney1998-mz,falcon2012secular,mo2010galaxy}). Our models are necessarily rather simple, in some cases even crude, since our goal here is to provide physical insight rather than to accurately reproduce observational data.
On the other hand, we do \textit{not} simply develop mathematical formalism for its own sake. Almost every formal calculation we present is accompanied by a concrete example of its application.
We hope this approach underscores our key message, which is that the kinetic theory of stellar systems 
is not merely the mathematical fancy of a few specialists. Instead, it can and should aid our understanding of galaxy evolution; and it passes the basic test of all good physics, by telling us things about the real world that we did not already know.

\subsection{A (very) brief astronomy lesson}
\label{sec:Astronomy}

\begin{figure}[htbp!]
    \centering
   \includegraphics[width=0.35\textwidth]{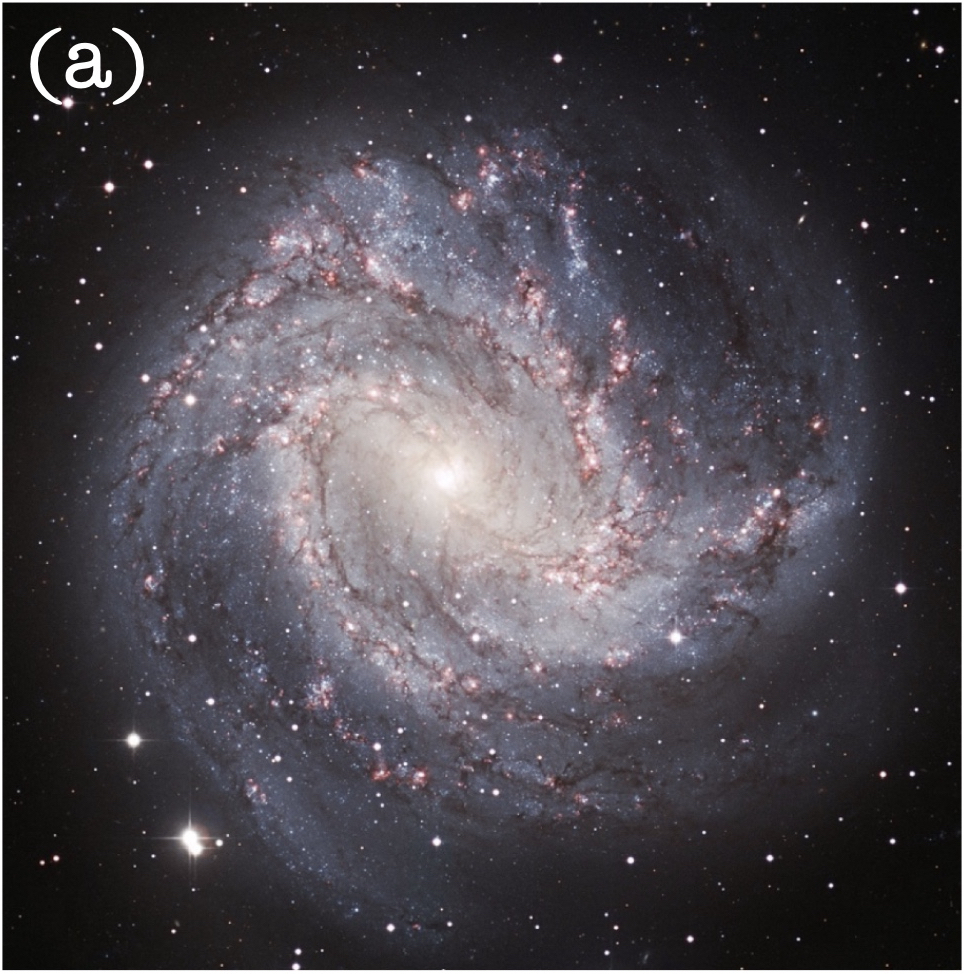}
   \includegraphics[width=0.3545\textwidth]{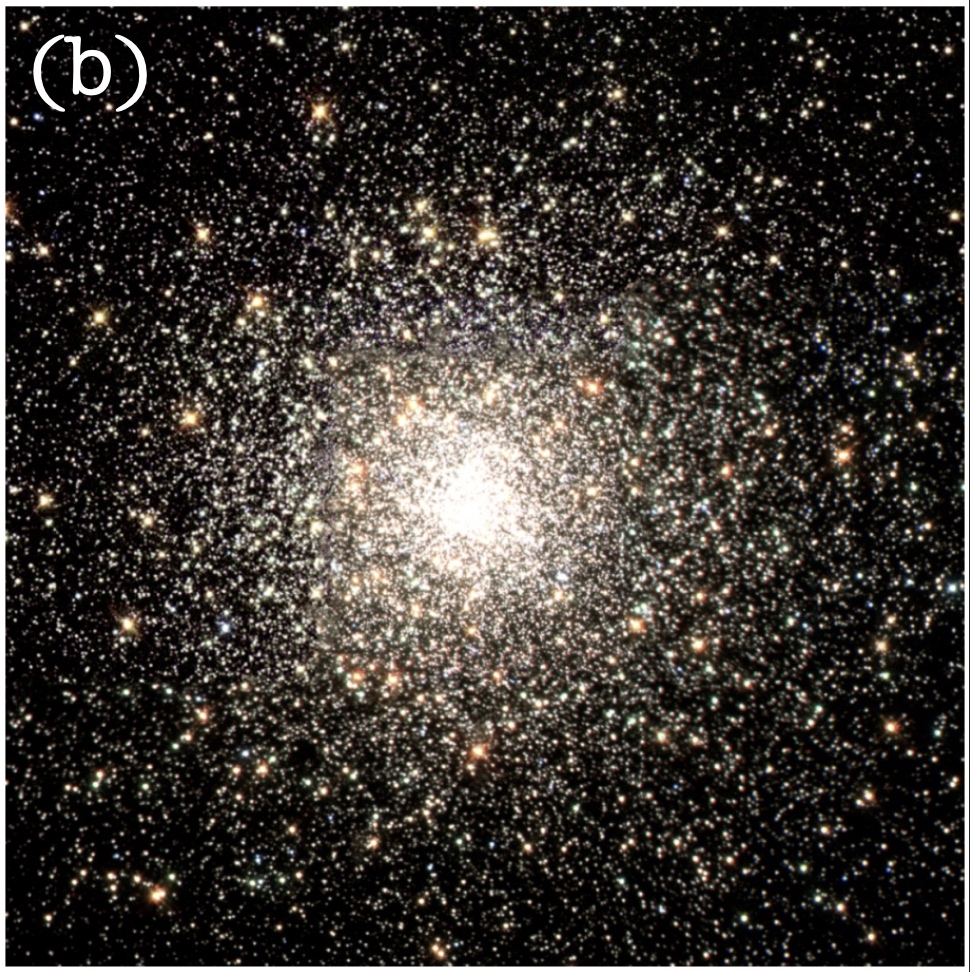}
    \caption{(a) The spiral galaxy M83 (the Southern Pinwheel). Credit: ESO. (b) The globular cluster M80. Credit: NASA, The Hubble Heritage Team, STScI, AURA.
        \label{fig:Spiral_Globular}}
\end{figure}

To set the context for this tutorial article, and to get an idea of the characteristic scales involved,
we invite you to consider Figure~\ref{fig:Spiral_Globular}.
Panel (a) shows an image of a \textit{disk galaxy} called the Southern Pinwheel, 
captured by the la Silla Observatory in the mountains of Chile.
We have chosen this image not because it is visually stunning --- though who could deny that? --- 
but because it exhibits several of the paradigmatic features of disk galaxies, and is in fact thought to resemble our own Galaxy, the Milky Way. Like the Milky Way, the Southern Pinwheel has an elongated or \textit{barred} structure at its center and exhibits beautiful \textit{spiral arms}.
The red knots strung along the arms are H$\alpha$ emission from regions of recent star formation, meaning stars have been born there in the last few tens of Myr (where $1$ Myr  $\equiv10^6$ years),
while the bluer light comes from much older stars, up to ages of $\sim 10$ Gyr (where $1$ Gyr $\equiv 10^9$ years; the Big Bang happened 13.7 Gyr ago, and galaxies formed in the first few Gyr that followed\cite{mo2010galaxy}).  The image is haphazardly obscured by thin, dark \textit{dust lanes}, regions wherein visible starlight has been absorbed.  The whole galaxy is a few tens of kiloparsecs (kpc) across, where $1$ pc $\sim 3\times 10^{16}$m. (The distance from the Sun to the next-closest star, Proxima Centauri, is of order $\sim 1$ pc).
From our perspective, we are seeing the Southern Pinwheel galaxy more or less face-on, but if we were to see it from the side we would find that it was very thin, with a central bulge --- like two fried eggs glued back to back.
It consists of around $N \sim 10^{11}$ stars, most of which are orbiting in this very thin, nearly two-dimensional disk plane, and they mostly move on (quasi-)\textit{circular orbits} around the center of the galaxy, all rotating in the same direction. 
This means that the motion of stars in the galaxy is highly ordered, a.k.a.\ \textit{dynamically cold}. Another way to say this is that the stellar disk, like a plasma beam, has low velocity dispersion.  The key \textit{dynamical timescale}
in this disk is the azimuthal orbital period of one of the stars, which is typically $\tdyn \gtrsim 100$ Myr, and this timescale increases with orbital radius.
Since galaxies are $\sim 10$ Gyr old, we deduce that 
most stars have completed fewer than $100$ orbits in the lifetime of the galaxy. 

Now let us contrast this with panel (b) of  Figure~\ref{fig:Spiral_Globular}, which is an image of the \textit{globular cluster} M80. This is another optical image, taken with the Hubble Space Telescope.
Globular clusters are much smaller than galaxies, with typical radii of $\sim 10 \mpc$ --- in fact, a galaxy like the Southern Pinwheel typically contains many hundreds or even thousands of globular clusters. Despite their relative 
smallness, globular clusters have a much higher density of stars than do galaxies. The cluster M80 contains 
something like $N \sim 10^5$ stars, so the typical distance between stars is $\sim (10^5/\mpc^3)^{-1/3} \sim 0.02 \mpc$, of order $50$ times smaller than the distance from the Sun to Proxima Centauri. Globular clusters typically contain very little gas, and hence exhibit almost no ongoing star formation. 
This is why there is a much narrower range of colors in this image compared to the one beside it.
Moreover, unlike disk galaxies, globular clusters are (nearly) \textit{spherically symmetric}, so the image of M80 would not look very different if we were able to view it from another angle. Also unlike disk galaxies, they are disordered, \textit{dynamically hot} systems, meaning that the motion of their stars has no strongly preferred direction (like the isotropic Maxwellian in an electrostatic plasma). More mathematically, at any given point $\br$ in the cluster, the local phase space distribution function $f(\br, \bv)$ depends primarily on the magnitude, rather than the direction, of the velocity $\bv$.
Though they are of a comparable age to galaxies --- having formed $\sim 10$ Gyr ago --- globular clusters are much more `dynamically old' than galaxies, since the typical orbital period of their stars is only $\tdyn \sim 10^5$yr.  Thus each star in a globular cluster has undergone $\sim 10^5$ orbits in its lifetime. 

To summarize: disk galaxies are \textit{dynamically young and cold}, globular clusters are \textit{dynamically old and hot}.

\subsection{Two mental models}

\begin{figure}[htbp!]
    \centering
   \includegraphics[width=0.7\textwidth]{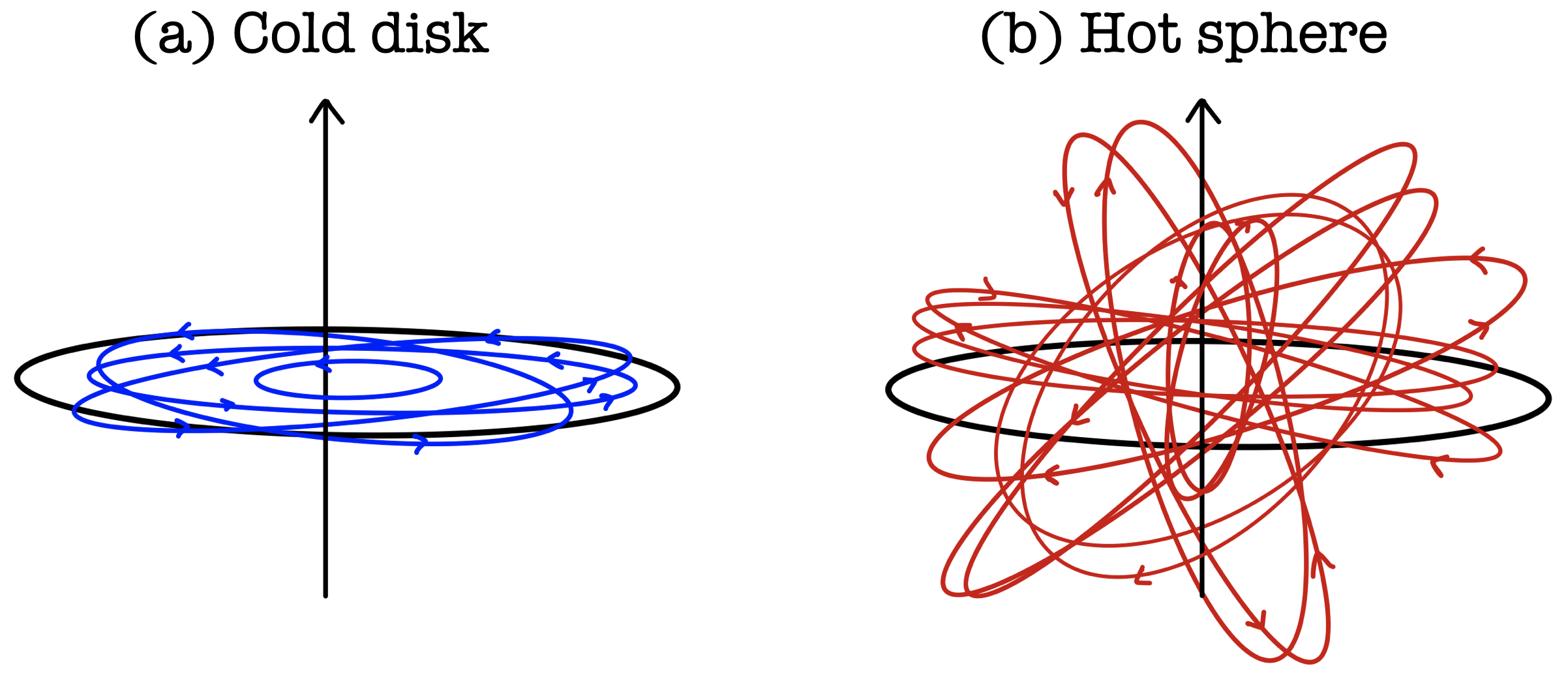}
    \caption{Schematic illustration of typical orbits in (a) cold disks and (b) hot spheres (idealized versions of the archetypical systems shown in Figure~\ref{fig:Spiral_Globular}). In a cold disk most orbits are nearly circular and confined to low inclinations relative to the disk plane. By contrast, in a hot sphere the orbits are near-isotropically distributed and have a broad range of eccentricities.}
            \label{fig:Two_Mental_Models}
\end{figure}

The brief astronomy lesson above motivates us to introduce two simple mental models 
which we will refer to throughout this tutorial article:
 the \textit{cold, rotating axisymmetric disk} and the \textit{hot, non-rotating spherical cluster} --- see Figure~\ref{fig:Two_Mental_Models} for an illustration.
These are really just (over-)simplified versions of disk galaxies and globular clusters respectively, 
and constitute clean examples of stellar systems that exhibit completely different dynamical behaviors.
Some of the characteristic properties of these two prototypical systems are summarized in Table~\ref{table:Two_Mental_Models}.
\begin{table}
\caption{Characteristic scales for the two prototypical systems considered in this article, the disk galaxy and the globular cluster.\label{table:Two_Mental_Models}}
\begin{ruledtabular}
\begin{tabular}{ccc}
 &Disk galaxy &Globular cluster\\
\hline
Mental model & Cold, rotating axisymmetric disk & Hot, nonrotating sphere\\
Size & $ 10$s of kpc &  10 pc \\
Age & $ 10$ Gyr & $ 10$ Gyr\\
Number of stars $N$ & $ 10^{11}$ & $ 10^5$\\
Orbital period $\tdyn $ & $ 100$ Myr & $ 0.1$ Myr
\end{tabular}
\end{ruledtabular}
\end{table}

We will find that the cold disk is analogous to a non-thermal plasma (with collective oscillations/instabilities and wave-particle interactions playing a key role), while the hot sphere is more analogous to a thermal plasma or a classical gas (with a kinetic evolution primarily driven by discrete, local two-body encounters).  But there are many subtleties and differences to be encountered along the way, and gravitational systems are capable of behaviors that have no analogue in gas or plasma kinetics.

It is also worth mentioning that throughout this article we will, for concreteness we often refer to the motion of `stars'.  However, we are also very much interested in the motion
of dark matter particles, which obey the same set of equations as stars, and are just as important a component of galactic dynamics.

\subsection{Plan for the rest of this article}

The rest of this tutorial article is organized as follows.
In \S\ref{sec:Basics} we introduce some of the basic properties of $N$-body self-gravitating systems,
discuss the extent to which self-gravitating systems play by the same thermodynamic rules as gases and plasmas,
and give a qualitative idea of how these systems evolve on long timescales.
Then we begin with the theory of orbits.
In \S\ref{sec:Mean_Field_Dynamics} we approximate a galaxy's gravitational potential by a smooth `mean field' potential and study the orbits of individual stars in this mean field, with the help of angle-action variables.
In \S\ref{sec:P_Theory} we show how individual orbits respond to both linear and nonlinear perturbations and describe the theory of resonant trapping.
Our proper study of kinetic theory begins in \S\ref{sec:secular}, where 
we introduce the Klimontovich equation for an $N$-body system's \textit{phase space distribution function} (DF) and derive a generic set of equations governing its evolution.
In \S\ref{sec:Linear_Response} we study the self-consistent linear response of stellar systems to weak perturbations, and then in \S\ref{sec:Quasilinear_Evolution} we use those linear solutions to construct a quasilinear theory of statistically-averaged evolution, culminating in the Balescu--Lenard equation.
Our final topic, in \S\ref{sec:Nonlinear_Kinetics}, is nonlinear kinetic theory, particularly in the presence of 
orbit trapping and with applications to bar-halo friction and the saturation of spiral instabilities.

Throughout these sections we continually draw direct analogies between gravitational and plasma systems,
as well as giving many concrete examples of the equations working in practice. 
Lastly, in \S\ref{sec:Discussion} we touch on some other areas of galactic dynamics that have plasma-theoretic analogues, and give references for further reading.

We mention here that the standard reference in this subject is the textbook \textit{Galactic Dynamics}\cite{Binney2008-ou} by Binney \& Tremaine (2008), which we will hereafter refer to as BT08.
While our article is far more narrow in scope than BT08, much of what we discuss cannot be found there or in any other textbook or publication. The main exceptions to this rule are the introductory sections  \S\S\ref{sec:Basics}-\ref{sec:Mean_Field_Dynamics}, where by necessity there is significant overlap with BT08. Where helpful, we point the reader to specific sections of BT08 in which more details can be found. 

Finally, this tutorial grew out of notes that we both compiled as lecturers of the Kinetic Theory of Stellar Systems course at Oxford. Lecture materials, problem sets, etc. can be found online\cite{Hamilton.web}.

\section{\texorpdfstring{Basic properties of $N$-body self-gravitating systems}{Basic notions of large-N self-gravitating systems}}
\label{sec:Basics}

The simplest theoretical approach to describing a stellar system is to imagine it as a gas of stars, and to apply the traditional methods of equilibrium thermodynamics and two-body collision theory. In this section we will attempt to do just that, and we will find that while these classical notions can be useful in characterizing stellar systems, they also break down completely in some circumstances ---  and that in the strict sense, in self-gravitating systems 
there is no such thing as equilibrium. The insights we gain from this failure will point us towards the more complete  kinetic description that we will develop in later sections.

\subsection{Virial equilibrium}
\label{sec:virial}

The first thing to mention is that
the speeds of stars in galaxies and star clusters are nearly always of order $10$s or $100$s of km s$^{-1}$, much less than the speed of light. This means that relativistic corrections to Newtonian gravity can almost always be ignored.
 There are exceptions, particularly for stars orbiting around supermassive black holes in galactic centers\cite{Alexander2017-nj},
 but for our purposes Newtonian gravity is an excellent approximation.  Also, while stars do exhibit a spectrum of different masses, for simplicity we will always assume that our system consists of a single mass species, unless we explicitly state otherwise.
 
We therefore consider a system of $N \gg 1$ stars of equal mass $m$ (usually, but not necessarily, taken to be of order a solar mass $M_\odot$), interacting only via Newtonian gravity.
The gravitational potential energy due to the interaction between two stars with positions $\br$ and $\br'$ is
\begin{equation}
        U(\br, \br') = -\frac{Gm^2}{\vert \br - \br'\vert},
\end{equation}
where $G = 4.3\times 10^{-3}$ pc $M_\odot^{-1}$km s$^{-1}$ is Newton's gravitational constant (here $M_\odot = 2.0\times 10^{30}$kg is the mass of the Sun). 
Newton's second law then tells us that the force on particle $i$ is 
\begin{equation}
   m \frac{\md \bv_i}{\md t} = -\frac{\p }{\p \br_i} \sum_{j \neq i}^N U(\br_i, \br_j) = -Gm^2\sum_{j\neq i}^N \frac{\br_i - \br_j}{\vert \br_i - \br_j \vert^3},
   \label{eqn:Newton}
\end{equation}
where $\bv_i \equiv \md \br_i/\md t$ is the $i$th star's velocity.

Gravity is an exclusively attractive force --- two masses always attract and never repel --- which means that unlike in plasma, there is no shielding, nor is there any sensible notion of a `Debye length' over which forces are screened.
Since $N$ is large ($\sim 10^{11}$ in the cold disk, $\sim 10^5$ in the hot sphere), and since every star is attracting every other star, what stops the entire system from collapsing under its own weight and forming a huge black hole?  The answer depends on the context. In the hot sphere, the random motions of the stars 
counterbalance the gravitational attraction (and can be thought of as an effective `pressure').
In the case of a cold disk it is the rotational support (i.e.\ the fact that the disk is spinning) that counterbalances gravity.

To put these notions on a more mathematical basis, we now derive the \textit{virial theorem}.  For an isolated 
$N$-body system we can choose the origin of our coordinates to be the fixed center of mass.  Then the moment of inertia of the $N$-body system with respect to the origin is
\begin{equation}
    I = \sum_{i=1}^N m \br_i^2.
\end{equation}
Let us now take the second derivative of this quantity with respect to time
\begin{equation}
    \frac{\md^2 I}{\md t^2} = 2\left( \sum_{i=1}^N m\bv_i^2 
    +
    \sum_{i=1}^N m\br_i\cdot \frac{\md \bv_i}{\md t} 
    \right).
\end{equation}
Using the equation of motion~\eqref{eqn:Newton}, we can write the second term on the
right hand side as 
\begin{align}
    \sum_{i=1}^N m\br_i\cdot \frac{\md \bv_i}{\md t} = & -Gm^2 \sum_{i=1}^N \sum_{j\neq i}^N \br_i \cdot  \frac{\br_i - \br_j}{\vert \br_i - \br_j \vert^3}
    \nn
    \\
    =& - \half  Gm^2 \sum_{i=1}^N \sum_{j\neq i}^N \frac{(\br_i - \br_j)\cdot (\br_i - \br_j)}{\vert \br_i - \br_j \vert^3}
    \nn
    \\
    =& \sum_{i=1}^N \sum_{j > i}^N U(\br_i, \br_j) .
\end{align}
For an equilibrium system, 
the moment of inertia can fluctuate around some average value taken over a few dynamical times,
but that average value must not itself change. 
Denoting such a time average by $\langle . \rangle$, we have that for a system in \textit{virial equilibrium},
\begin{equation}
    2K + W = 0,
    \label{eqn:virial}
\end{equation}
where 
\begin{equation}
    K = \bigg\langle \half m\sum_{i = 1}^{N} \bv_i^2 \bigg\rangle, \quad\quad W = \bigg\langle \sum_{i<j}^{N}U(\br_i,\br_j) \bigg\rangle,
\end{equation}
are the averaged total kinetic and total potential energies of the system respectively.
In other words, in virial equilibrium, the kinetic and potential parts of the energy of the system balance each other to within an order-unity factor.

A globular cluster (right panel of Figure~\ref{fig:Spiral_Globular}) is an excellent example of a system in virial equilibrium. Its smooth, near-spherical shape reflects the fact that it is not, on average, undergoing any violent bulk motions, despite the stars inside it whizzing around like a swarm of bees on the timescale $\tdyn$.
Note that even though a cluster is in virial equilibrium, that equilibrium \textit{does} still evolve; but the evolution of this `quasi-stationary state' occurs over a timescale much longer than $\tdyn$, as we will see (\S\ref{eqn:Evolution}). 

The virial theorem allows us to make quick order-of-magnitude estimates about typical random velocities of stars, $\sigma$. 
For instance, consider that a globular cluster has a mass $M \approx Nm \sim 10^5M_\odot$ and a radius $R$ of a few parsecs.
Using the virial theorem we can calculate the typical velocity dispersion of stars in this cluster. Roughly, we have $K \approx M\sigma^2/2$ and $W \approx -GM^2/R$, so using \eqref{eqn:virial} we get
\begin{equation}
\sigma \sim \sqrt{\frac{GM}{R}} \approx 10\, \mkm \, \ms^{-1}\left( \frac{M}{10^5M_\odot}\right)^{1/2} \!\! \left( \frac{R}{4 \mpc}\right)^{-1/2} ,
    \label{eqn:cluster_sigma}
\end{equation}
in agreement with what is observed (and much less than $c$, justifying our use of Newtonian gravity).

\begin{figure}[htbp!]
    \centering
       \includegraphics[width=0.35\textwidth]{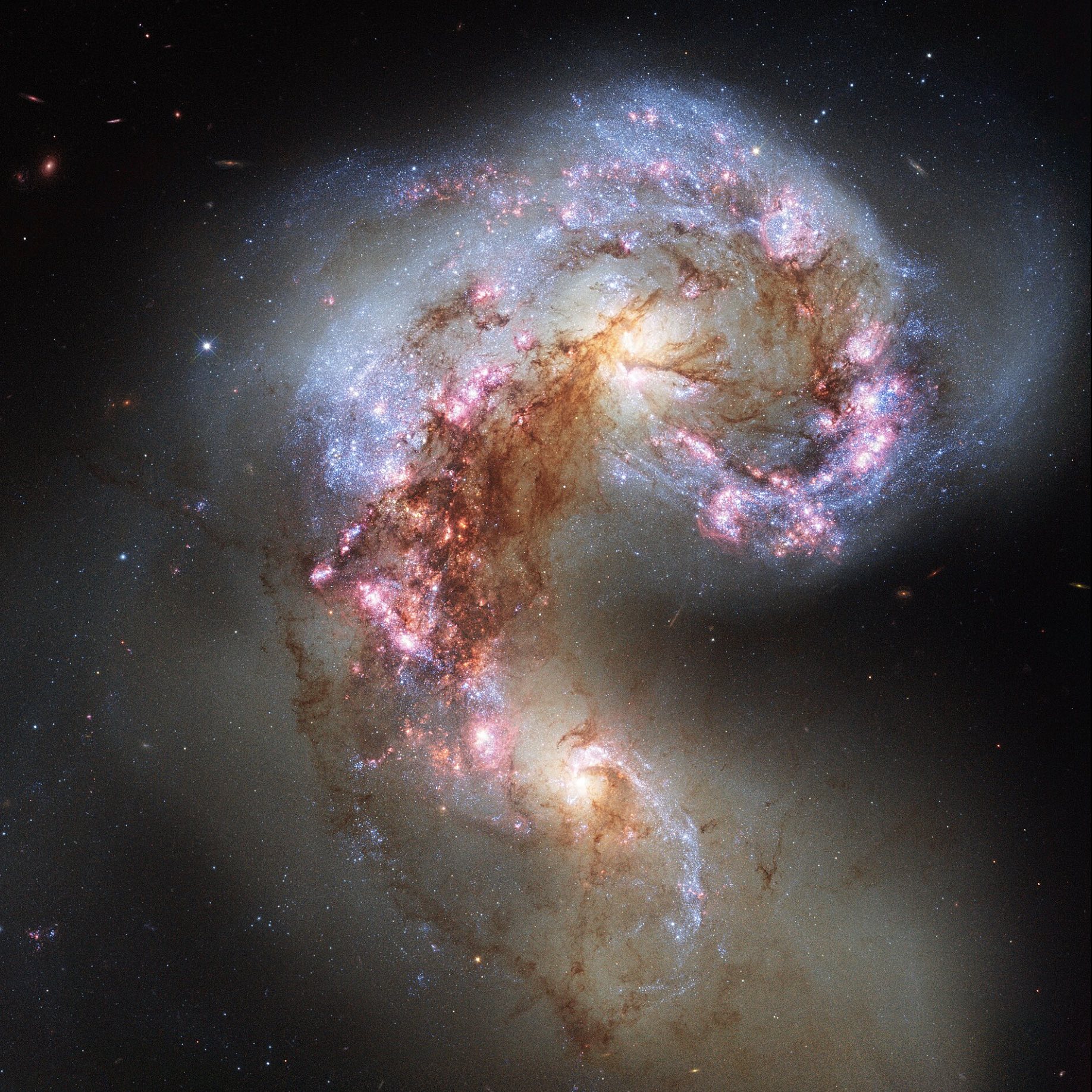}
    \caption{The Antennae galaxies NGC4038 and NGC4039.  These two colliding galaxies are in the process of merging into a single structure.  It is easy to see by eye that this system is going to undergo significant morphological changes on the typical timescale $\sim \tdyn$. Hence, it is not in virial equilibrium.
    Credit: ESA/Hubble.
        \label{fig:Antennae}}
\end{figure}

Finally, let us contrast globular clusters with an example of a system which is very much \textit{not} in virial equilibrium. Figure~\ref{fig:Antennae} shows the Antennae system, which consists of two galaxies that have collided and are in the process of merging.
For this system, equation~\eqref{eqn:virial} does not hold; instead energy is rapidly being transferred between kinetic and potential forms on the timescale $\tdyn$ as the two galaxies perform their cosmic dance.  But we know from observations and numerical simulations of such galaxies that they ultimately \textit{do} virialize, and that this virialization is nearly complete after only a few $\tdyn$ (which in this case still means a billion years or more).
This rather rapid evolution towards a virialized state is called \textit{violent relaxation}\cite{Lynden-Bell1967-er}.
The precise mechanics of violent relaxation, and the question of which particular virial equilibrium state the system will fall into, is a matter of ongoing research not only in stellar dynamics but also in plasma physics\cite{ewart2022collisionless}. 

For the rest of this article we will only consider virialized stellar systems whose bulk properties are evolving on timescales $\gg \tdyn$.  Isolated disk galaxies and globular clusters both fall into this category.

\subsection{Thermal equilibrium?}
\label{sec:Thermal_Equilibrium}

We mentioned above that virial equilibria are not true equilibria of stellar systems --- rather, they are just quasi-equilibria that evolve on timescales much longer than the dynamical time $\tdyn$.
But why is this the case?  After all, this is not what happens in either a gas or an electrostatic plasma:  there, the natural thermal equilibrium distribution of particles' positions and velocities is a homogeneous Maxwellian.
Unfortunately, gravity does not play by those thermodynamic rules.

To see why, suppose we wished to describe a globular cluster as a `stellar gas' in thermal equilibrium.
We could make a natural definition of the cluster's temperature $T$ by
relating it to the total kinetic energy
\begin{equation}
    K = \frac{3}{2}N k_\mB T,
\end{equation}
where $N$ is the number of stars.
Then the total energy of the cluster $\Etot$
is the sum of the kinetic and potential energies
\begin{equation}
    \Etot = K+W.
\end{equation}
But from the virial theorem~\eqref{eqn:virial} we know that $W = -2K$, so $\Etot = -K < 0$.
The fact that $\Etot$ is negative is not unusual: it just means that the system is gravitationally bound.  The heat capacity is then
\begin{equation}
        C = \frac{\p \Etot}{\p T} = -\frac{\p K}{\p T} = -\frac{3}{2}Nk_\mB < 0.
        \label{eqn:Heat_Capacity}
\end{equation}
Notice that $C$ is \textit{negative}.  This \textit{is} strange: if you add heat to the system (increase its total energy $\Etot$), it cools down!
In other words, the more energy is given to the cluster,
the lower the amplitude of the random velocities inside it.

The negativity of the heat capacity implies that it is impossible for a globular cluster ever to reach a thermal equilibrium.
For instance, consider a cluster at temperature $T$ 
immersed in a heat bath of the same temperature.
Suppose there is some infinitesimal negative fluctuation in the cluster's energy, 
$-\vert \delta E\vert $. By~\eqref{eqn:Heat_Capacity}, this will cause the cluster 
to achieve a higher temperature by $\delta T  = (-\vert \delta E\vert )/C > 0$, i.e.\ with higher velocity dispersion.
Since the temperature is now higher, the system loses more heat to the bath,
which then makes its temperature increase even further, and so on without limit.

Nor can one achieve thermal equilibrium in an isolated stellar system (i.e.\ in the absence of any heat bath).  The reason is that any star with a sufficiently large velocity is unbound from the system and hence will escape to infinity, a process known as \textit{evaporation}.
This precludes the system from ever attaining a Maxwellian velocity distribution function (DF), since the cluster would otherwise have a finite phase space density out to $v \to \infty$. 

One might argue at this point that the Maxwellian is only an idealization, and not being able to \textit{exactly} 
realize a Maxwellian DF is a rather pedantic argument for the non-existence of thermal equilibrium. (Indeed, all speeds are ultimately bounded by the speed of light, but this does not stop us from employing a Maxwellian DF when describing plasmas!).  To justify our claim further,
let us calculate the rate at which stars \textit{would} evaporate from a hypothetical Maxwellian stellar system. 
As we will argue in \S\ref{sec:Mean_Field_Dynamics},
the motion of individual stars on the timescale $\sim \tdyn$ 
is very accurately described by a \textit{mean field approximation}, in which we ignore the complexities of the $N$-body system and instead consider the stars to be moving in a \textit{smooth} gravitational potential $\Phi(\br)$,
in precisely the same way as one imagines the unperturbed motion
in a homogeneous electrostatic plasma to be a straight line.
Then the energy per unit mass of any star with velocity $v$ and position $\br$ 
is $v^2/2 + \Phi(\br)$, meaning that any star with $v > v_\mesc (\br) \equiv \sqrt{-2\Phi(\br)}$ will evaporate. Averaging this over the entire system gives us the 
typical escape speed
\begin{equation}
        \sigma_\mesc^2 \approx \frac{1}{M}\int \md \br \, \rho(\br) v_\mesc^2(\br) = -\frac{2}{M}\int \md \br \, \rho(\br) \Phi(\br),
\end{equation}
where $\rho_0(\br)$ is the mass density of stars at position $\br$, and $M$ is the total cluster mass.
This is obviously related to the total potential energy of the system $W$.  Being careful not to over-count the pairwise interactions, and using the virial theorem~\eqref{eqn:virial}, we get
\begin{equation}
    \sigma_\mesc^2 \approx -\frac{4W}{M} = \frac{8K}{M} = 4\sigma^2,
\end{equation}
where $\sigma^2 = N^{-1}\langle\sum_{i=1}^N \bv_i^2 \rangle$ is the cluster's velocity dispersion.
Supposing the cluster did have a Maxwellian DF with the same velocity dispersion,
we can estimate the fraction of unbound stars by
\begin{equation}
    f_\mesc \approx \frac{\int_{2\sigma}^\infty \md v\, 
    v^2 \me^{-3v^2/2\sigma^2}}{\int_0^\infty \md v\, v^2 \me^{-3v^2/2\sigma^2}} \approx \frac{1}{140}.
\end{equation}
In other words, we would expect about $1\%$ of an isolated stellar system 
to evaporate away
on the timescale of refilling of the Maxwellian tail, which is roughly
a relaxation time (see the next subsection for the definition of this term)
The equivalent situation in a collisional plasma would be that $1\%$ of the electrons disappeared every mean-free time!
Thus, even if the cluster `wanted' to reach a Maxwellian steady-state, it could not do so because stars are constantly evaporating from the high-velocity tail.  

Evaporation is a real and important effect in cluster dynamics; indeed there are certain star clusters called \textit{open clusters} with $N\sim 10^3$ (some of which you can see with the naked eye, like the Pleiades) whose lifetime is limited by evaporation to a few hundred Myr. In other words, there were star clusters visible to the dinosaurs that have since evaporated into the vastness of space and are no longer visible to us. But at no point was their DF ever Maxwellian.

With minor modifications, the above arguments also hold for cold, differentially rotating disks, as well as other virialized self-gravitating structures. The bottom line is that \textit{strict thermal equilibrium is impossible in a self-gravitating system}, and there is no universal requirement for stellar systems to `tend towards the Maxwellian' as gases and plasmas do. 
Fundamentally, this is stems from the purely attractive nature of gravity: since the potential energy of a very loosely bound particle decays like $1/r$, there are infinitely many ways to rearrange the phase space coordinates of very distant particles with no energy cost.
In the language of statistical mechanics, this means that in isolated self-gravitating systems there is no microcanonical ensemble, no canonical ensemble, energy is not extensive, and entropy is not bounded (so there exists no maximum entropy state\cite{Binney2008-ou,campa2014physics}).
Concepts like entropy are still useful in stellar dynamics, particularly if the phase space is bounded by, e.g.\@, tidal truncation (meaning the system is not truly isolated) or by the approximate conservation of certain invariants (meaning there is a quasi-maximum entropy state on short enough timescales)\cite{tremaine2015statistical}.  However, in general one has to be very careful about naive application of `standard' statistical mechanical results to gravitational systems.


\subsection{So what does happen to a virialized stellar system?}
\label{eqn:Evolution}

We have argued that there exists no unevolving thermal equilibrium state for stellar systems. 
What, then, does happen to stellar systems after they reach virialization?
Roughly speaking, the answer is that they move slowly from one virialized quasi-equilibrium state to another on a timescale that we call the \textit{relaxation time}, $t_\mathrm{relax}$. 

How do we calculate the relaxation time?  Roughly speaking, it is the timescale over which a typical star `forgets' its initial orbit.  A proper theory of relaxation requires 
an understanding of stellar orbits (\S\ref{sec:Mean_Field_Dynamics}),
but for now the intuition from gas or plasma kinetics is good enough. Hence we simply ask, supposing an unperturbed star was moving in a straight line, 
how long would it take for its velocity $\bv$ to de-correlate from its initial value, 
i.e.\ to accumulate an impulse $\vert \Delta \bv \vert \approx \vert \bv\vert $?

\begin{figure*}
    \centering
       \includegraphics[width=0.65\textwidth]{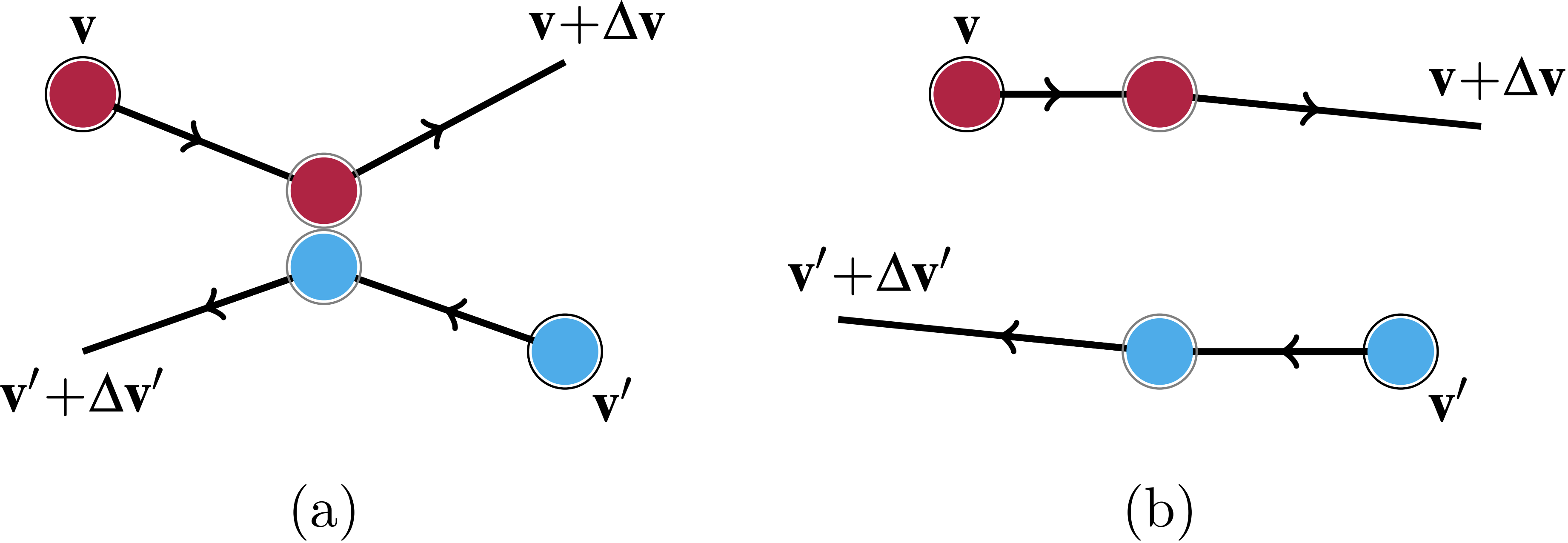}
    \caption{(a) Relaxation in a gas involves physical contact collisions (or at least, very short-range interactions) between molecules, meaning a molecule's velocity can be changed significantly after a single collision. (b) In a stellar system, by contrast, two-body encounters are mostly weak, long-range nudges. Later in the article we will deal with \textit{collective effects}, whereby one must think beyond bare two-body processes and account for the coherent oscillations of very large numbers of stars at once.}
        \label{fig:short_long}
\end{figure*}

In a gas these impulses occur because a given molecule collides with some other molecule roughly once per mean-free time. Hence, after a few such collisions, this molecule has completely forgotten its initial state (see Figure~\ref{fig:short_long}a).
This is not so in a stellar system. Even in a dense system like a globular cluster, the distance between single stars is such that direct physical collisions are very rare\cite{Hills1976-ea}.
Instead, in a cluster the evolution of a star's orbit is mostly driven by a very large number of weak nudges from all the other stars it is interacting with simultaneously (Figure~\ref{fig:short_long}b).
In the coming subsections we will 
follow Chandrasekhar and estimate the relaxation timescale
due to a series of such nudges (\S\ref{sec:Chandra}),
a calculation 
completely analogous to the Spitzer two-body collision calculation from plasma kinetics.
We will also pause to comment on the confusing words `collisional' and `collisionless' as applied to stellar systems (\S\ref{sec:Collisional_Collisionless}).
Then (\S\ref{sec:Where_Now}) we will show that Chandrasekhar's theory is very successful at describing the evolution of hot spheres (in the same way that Spitzer's theory works for describing collisional plasmas) but 
fails completely to describe the evolution of cold disks (as does Spitzer's theory when dealing with collisionless plasma out of equilibrium).
This will be the jumping-off point where stellar dynamics 
ceases to consist of a series of rough estimates; thereafter, things get quantitative.

\subsubsection{Chandrasekhar's theory of two-body relaxation}
\label{sec:Chandra}

Here we give the simplest version of Chandrasekhar's calculation of the relaxation time\cite{Chandrasekhar1943}, based on his theory of two-body relaxation, and primarily following \S1.2.1 of BT08. (For a more complete treatment of Chandrasekhar theory, see Chapter 7 of BT08).  What we will present is a gravitational version of Rutherford scattering.

To begin, we will ignore all the complications associated with real stellar systems and instead
consider a single `test' star moving in a statistically homogeneous box of static `field' stars of mass $m$.
Let the test star travel in a perfectly straight line until it approaches a field star
at impact parameter $b$.  The goal now is to calculate the nudge to the test star's velocity,  $\delta \bv(b)$.

\begin{figure}[htbp!]
    \centering
       \includegraphics[width=0.45\textwidth]{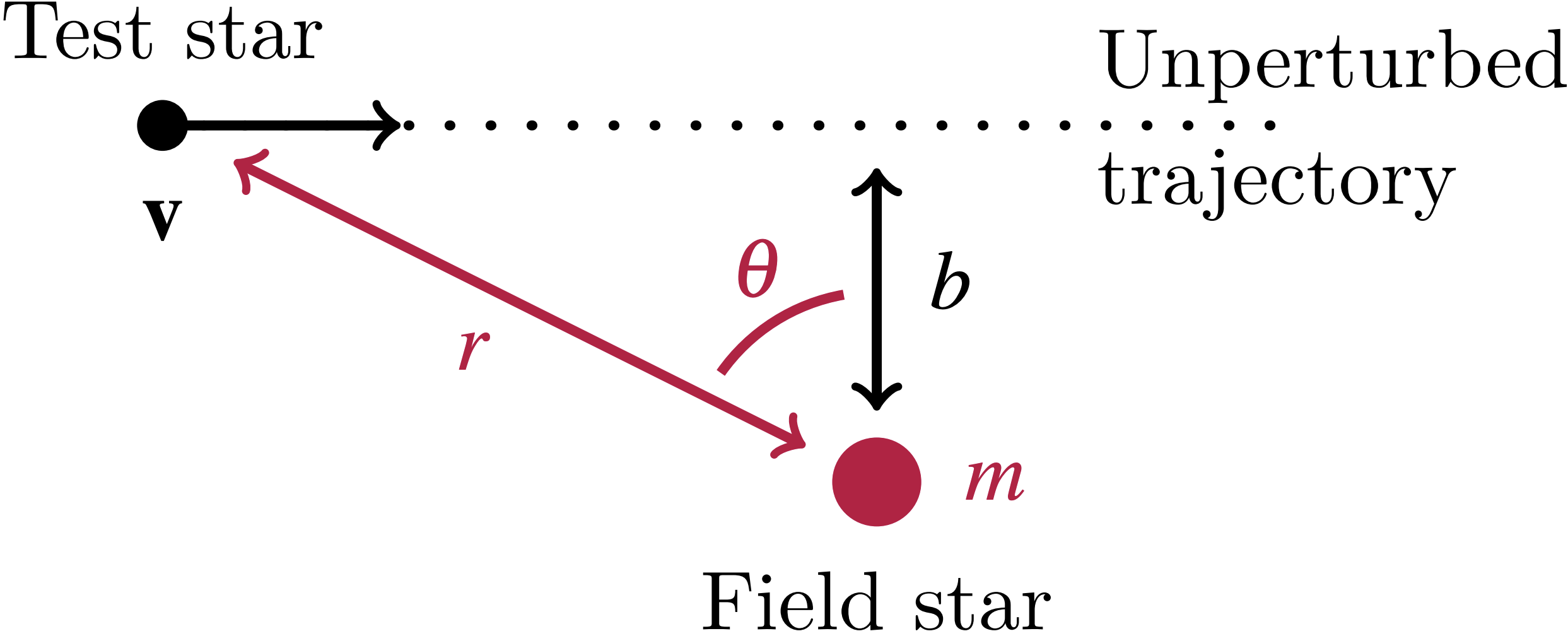}
    \caption{A test star encounters a field star of mass $m$ at impact parameter $b$.  Assuming the encounter to be weak, the lowest-order nudge to the test star's velocity, $\delta \bv$, can be calculated by integrating along the unperturbed trajectory (straight dotted line).
        \label{fig:Chandra}}
\end{figure}

Since we are going to assume that the interaction between the two stars is weak, 
to lowest order we can approximate the trajectory of the test star as a straight line throughout the encounter (see Figure~\ref{fig:Chandra}).
Suppose without loss of generality that the closest approach of the test star to the field star occurs at $t=0$. We may then calculate the total impulse it receives by integrating from $t=-\infty$ to $t=\infty$.
Then the acceleration the test star feels in the direction parallel to $\bv$ while flying towards the field star ($t<0$) is exactly canceled by the retardation it feels as it flies away ($t>0$), meaning that there is no overall velocity change in the parallel direction.
Hence, we need only to calculate the total \textit{perpendicular} velocity nudge $\delta v_\perp$.

The perpendicular acceleration field felt by the test star is (see Figure~\ref{fig:Chandra} for the definition of $r$ and $\theta$):
\begin{equation}
    F_\perp = \frac{Gm}{r^2}\cos \theta = \frac{Gm}{b^2}\left[ 1 + \left( \frac{vt}{b}\right)^{2}\right]^{-3/2}.
\end{equation}
Integrating over all time, we get
\begin{equation}
    \delta v_\perp = \int_{-\infty}^\infty \md t\, F_\perp = \frac{Gm}{bv}\int_{-\infty}^\infty \frac{\md s}{(1+s^2)^{3/2}} = \frac{2Gm}{bv}.
    \label{eqn:deltavperp}
\end{equation}
As one might expect, the more massive the field star, and the smaller the impact parameter $b$, the 
larger the velocity impulse the test star experiences.  Moreover, it is sensible that the impulse decays with $v$, since the larger is $v$ the shorter is the duration of the encounter. 
On the other hand, the formula~\eqref{eqn:deltavperp} also predicts its own failure: if $\delta v_\perp$ is sufficiently large that 
it is comparable to the original speed $v$, then the assumed straight-line trajectory is obviously a poor approximation.
Note that for a fixed $m$ and $v$, this breakdown occurs at a critical impact parameter
$b_{90} = 2Gm/v^2$, so-called because this in fact produces a $90$-degree deflection.
For impact parameters $b \lesssim b_{90}$ one ought really to solve the two-body encounter problem without making any approximations (which is not difficult --- see~\S{3.2} of BT08).
Luckily, such strong encounters are rare and  make little difference to the bulk evolution of the system. We will ignore them from now on.

Chandrasekhar's next step is to assume that relaxation in a stellar system consists of nothing more than an uncorrelated series of these weak two-body encounters.  Under this assumption we
can calculate the relaxation time as follows. Let the number density of stars in the box be $n$. 
Then in a time $T$ the field star passes
a total of $\approx nv\, T\, 2\pi b\,\md b$ field stars at impact  parameter $\in [b, b+\md b]$. 
The mean-square kick to the velocity that the test star receives in time $T$ is therefore 
\begin{equation}
   \sum \delta v_\perp^2 \approx \int_{\bmax}^{\bmin} nv\, T\, 2\pi b\,\md b \times \delta v_\perp^2 = \frac{8\pi G^2m^2n}{v}T \ln \Lambda,
   \label{eqn:Sum_deltavperpsquared}
\end{equation}
where $\ln \Lambda \equiv \ln (\bmax / \bmin)$ is the {Coulomb logarithm}, which is
familiar from the corresponding plasma calculation of charged particle scattering\cite{Lifshitz1981-zf}.

We seem to have a problem here, since taking either $\bmin \to 0$ or $\bmax \to \infty$, or both, causes the total velocity impulse to diverge (albeit in the most undramatic way, i.e.\ logarithmically).  So, 
what values of $\bmin$ and $\bmax$ should we take?
Usually one takes 
$\bmax$ to be comparable to 
the scale of the system $\sim R$, since there cannot be any encounters more distant than this (this is different from plasma, in which one can safely take $\bmax$ of order the Debye screening length\cite{Lifshitz1981-zf}).
Moreover, one typically takes $\bmin \approx b_{90}$ because, as we mentioned above, strong encounters are usually rare and unimportant. In fact, while setting the Coulomb logarithm accurately is an important problem for those modeling globular cluster evolution\cite{heggie2003gravitational}, from a theorist's standpoint the answer is \textit{it does not matter too much because the model is only heuristic anyway}. For instance, encounters with impact parameters $b$ on the order of the size of the system certainly cannot be thought of as straight lines, since they must take into account the curved geometry of the true orbits of the stars (\S\ref{sec:Where_Now}). The standard choice $\Lambda \approx R/b_{90}$
is good enough for obtaining order-of-magnitude estimates, which is all we care about here.

If we now set $\sum \delta v_\perp^2$
equal to the original squared speed $v^2$, we can identify $T=t_\mathrm{2BR}$ as the \textit{two-body relaxation time}
\begin{equation}
   \tNR \approx \frac{v^3}{8\pi G^2m^2 n \ln \Lambda}.
\end{equation}
Let us relate this more clearly to the number of stars in the system $N$.
Let the side-length of the box be $\approx R$ --- which is a proxy for the radius of the star cluster or galaxy in which we are interested.  
Then we know from the virial theorem that typically $v^2 \approx GNm/R$ (see equation~\ref{eqn:cluster_sigma}).
We may also set the typical crossing time to be $\tdyn \approx R/v$ and the typical number density to be $n\approx N/(4\pi R^3/3)$. Then taking $\bmax = R$ and $\bmin = b_{90} = 2Gm/v^2$ we get $\Lambda \approx Rv^2/(Gm) \approx N$, so that 
\begin{equation}
    \tNR \approx \frac{0.1 \, N}{\ln N} \tdyn.
    \label{eqn:2BR}
\end{equation}
To be even more heuristic, we will often drop the factors of $0.1$ and $\ln N$ and simply state that
the two-body relaxation time is on the order of $\tNR \sim N \tdyn$.

Plugging in characteristic numbers from Table~\ref{table:Two_Mental_Models}, we find that for a globular cluster $\tNR \sim 10$ Gyr, comparable to the age of the cluster.
This suggests that the simple physics of uncorrelated two-body encounters is of major importance for determining the evolution of globular clusters.
However, for a disk galaxy this theory 
suggests evolution should only occur on the excruciatingly long timescale $\tNR \sim 10^{10}$ Gyr,  nine orders of magnitude longer than the age of the Universe\footnote{Actually, the argument we have made here does not really apply to the geometry of a thin disk. Accounting for the disky geometry, the estimate on the right hand side of~\eqref{eqn:2BR} 
is reduced by a factor like $\sim 10^4$ (see equation~{5} of~\citet{Sellwood2014-te} and the surrounding discussion). Nevertheless, even this shortened two-body relaxation time is much, much longer than the age of the Universe.}.  And yet we know from observations and simulations that disk galaxies can evolve significantly on a timescale of $\gtrsim$ a few Gyr. The estimate~\eqref{eqn:2BR} tells us that star-star encounters are much too inefficient to have produced this evolution, so a more sophisticated kinetic theory is required (\S\ref{sec:Where_Now}).

Note that equation \eqref{eqn:2BR} is directly analogous to the 
inverse two-body collision time in an electrostatic plasma, 
$t_\mathrm{coll} \sim (\Lambda_\mathrm{p}/\log \Lambda_\mathrm{p}) \omega_\mathrm{p}^{-1}$, where $\Lambda_\mathrm{p}$ is the plasma parameter (the number of particles per Debye sphere), and $\omega_\mathrm{p}$ is the plasma frequency\cite{Diamond2010-mh}.
Thus, there is a direct analogy between the plasma parameter $\Lambda_\mathrm{p}$ and the number of stars $N$ (at least for stellar systems whose evolution is driven purely by finite-$N$ fluctuations, with little collective amplification). In particular, the inverse of these quantities, $1/\Lambda_\mathrm{p}\ll1$ and $1/N \ll 1$, 
are the small numbers with which one can expand the BBGKY hierarchy in weakly-coupled plasmas and stellar systems respectively\cite{Chavanis2013-eg}.  These numbers being small is equivalent to the statement that
the kinetic energy of each particle is large compared to the 
\textit{fluctuations} in the potential energy (recall that the \textit{mean} potential energy in a stellar system is of order the kinetic energy, see \S\ref{sec:virial}).

\subsubsection{`Collisional' and `collisionless'}
\label{sec:Collisional_Collisionless}

It is worth commenting here on the (often extremely confusing) use of the words \textit{collisional} and \textit{collisionless}
when describing stellar systems.  These words were partly inherited from plasma theory, where they have a fairly concrete meaning: roughly, interactions that occur on scales smaller than the Debye scale are `collisions', and if these can be ignored then the system is `collisionless'.  
In a stellar system, however, there is no shielding, no Debye length,
and no natural way to decouple small-scale and large-scale gravitational potential fluctuations
--- every star is interacting gravitationally with every other star, always.

Sometimes in the gravitational context,  the word `collisionless' is used as a shorthand for `a system whose naive two-body relaxation time~\eqref{eqn:2BR} is much longer than its age' (e.g.\@, a disk or an elliptical galaxy) while `collisional' systems are those with $\tNR$ less than or comparable to their age (basically, star clusters).  
Alternatively, the word `collisional' is sometimes used to refer to self-interacting dark matter (SIDM)\cite{moore2000collisional}, a model of dark matter involving a finite cross section for two dark matter particles to annihilate, by contrast with standard `collisionless' dark matter which has no such cross section. We could give many other examples of these words being used for slightly different purposes. It is therefore no surprise that this semantic issue frequently causes confusion among non-experts.

There are two sensible ways to deal with the issue. The first way is to call a system `collisional' if and only if the finite number of particles $N$ matters to its evolution, and to call it `collisionless' otherwise. 
The second way is to avoid using the words `collisional' and `collisionless' altogether.
We advocate the second way, although these words are so ingrained in the literature that sometimes this is impossible.

\subsubsection{Where do we go from here?}
\label{sec:Where_Now}
\begin{figure}[htbp!]
    \centering
       \includegraphics[width=0.38\textwidth]{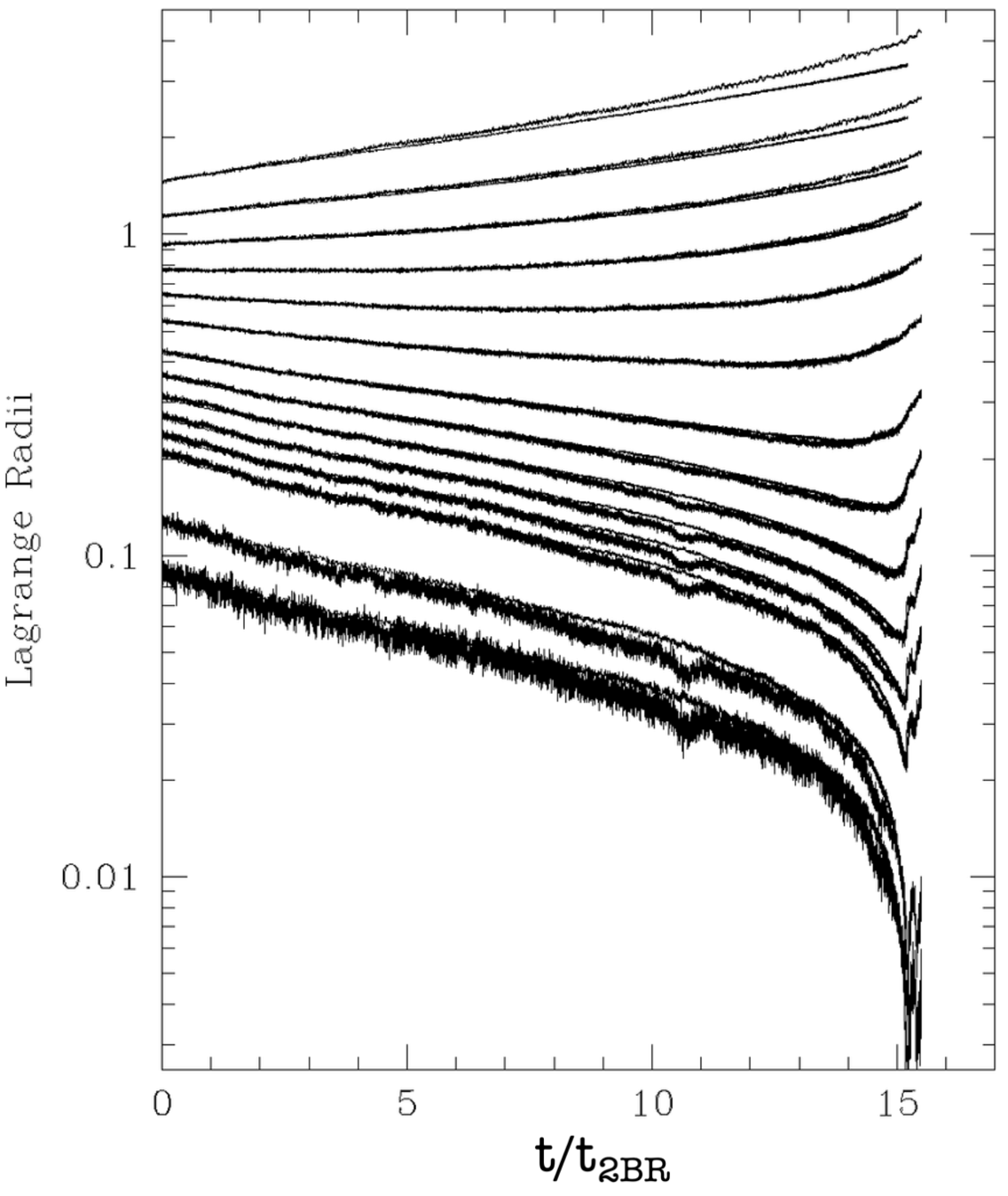}
    \caption{Long term evolution of a globular cluster 
    (adapted from K. J. Joshi, F. A. Rasio, and S. Portegies Zwart, ApJ 540, 969 (2000)\cite{joshi2000monte}).
    The lines show the evolution of the `Lagrange radii' which enclose $0.35$, $1$, $3.5$, $5$,  $7$, $10$, $14$, $20$, $30$, $40$, $50$, $60$, $70$, and $80$ percent of the cluster mass respectively. In each case, both the results of direct $N$-body simulation and of a Monte--Carlo calculation\cite{Henon1971} based on Chandrasekhar's theory are shown.
    For most times and Lagrange radii, the two sets of results are indistinguishable.}
        \label{fig:Joshi}
\end{figure}
\begin{figure}[htbp!]
    \centering
       \includegraphics[width=0.9\textwidth]{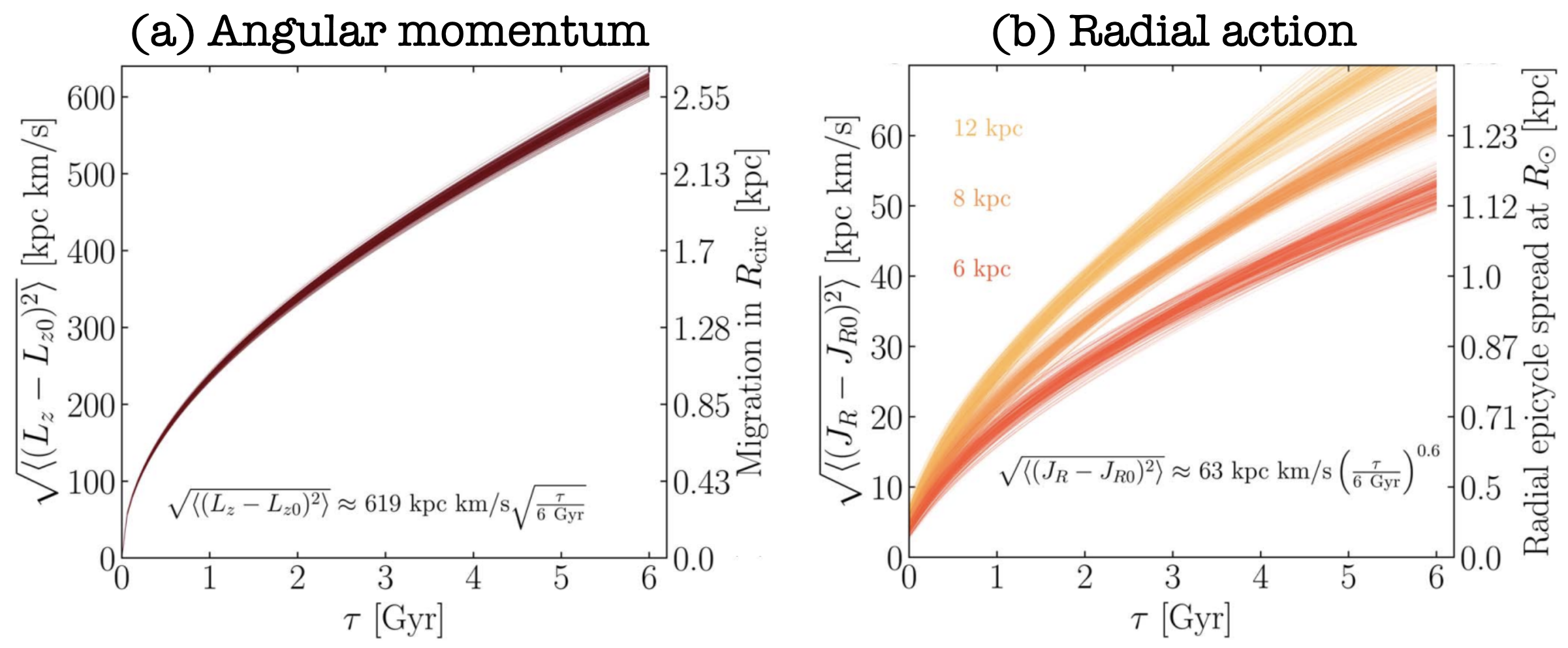}
    \caption{Adapted from N. Frankel, J. Sanders, Y.-S. Ting, and H.-W. Rix, ApJ 896, 15 (2020)\cite{Frankel2020-vy}.
    Each panel shows the evolution of an orbital integral of motion  (namely (a) the component of angular momentum perpendicular to the galactic disk and (b) the radial action, which will be introduced properly in \S\ref{sec:angleaction}) of stars in the Milky Way over $6$ Gyr, as inferred from Gaia data.
    In the right panel, different colors correspond to groups of stars with different initial Galactocentric radii.}
        \label{fig:Frankel}
\end{figure}

It turns out that when applied to the evolution of hot spherical clusters, Chandrasekhar's picture does a remarkably good job of estimating the true relaxation time $t_\mathrm{relax}$ and various other evolutionary properties\cite{heggie2003gravitational}, 
although the value of the Coulomb logarithm has to be tweaked\cite{Henon1975,giersz1994statistics}.
Moreover, one can use Chandrasekhar's ideas to spin up a full kinetic theory --- i.e.\ a kinetic equation of the form $\p f/\p t = C[f]$, with $C$ calculated by assuming relaxation consists of two-body encounters as above --- and this works well in hot spheres too\cite{Fouvry2021-ra}. 
To demonstrate this, we present Figure~\ref{fig:Joshi} (adapted from~\citet{joshi2000monte}), which shows the 
evolution of `Lagrange radii' of a spherical globular cluster in a numerical simulation.
These Lagrange radii are the radii enclosing a fixed percentage of the cluster's mass at any given time. 
First of all, we see that the inner Lagrange radii gradually decrease with time (corresponding to the contraction of the cluster core, leading ultimately to the `gravothermal catastrophe'\cite{LyndenBellWood1968}) and the outer radii slowly increase (corresponding to the expansion of the cluster's envelope). This evolution happens very slowly (the time unit on the horizontal axis is the two-body relaxation time $t_\mathrm{2BR} \gg t_\mathrm{dyn}$), corresponding to the gradual evolution from one virialized state to another.
Looking more closely, one can see that for each Lagrange radius in the panel there are actually two lines plotted.
One of these was calculated from a direct $N$-body simulation of the `exact' equations of motion of all the stars, whereas the other employed a Monte--Carlo calculation based upon Chandrasekhar's theory of two-body encounters.  The fact that across much of the figure these two lines are indistinguishable is testament to the success of Chandrasekhar theory in describing hot spherical clusters.

On the other hand, as mentioned at the end of \S\ref{sec:Chandra}, Chandrasekhar's theory fails completely 
to describe the evolution
of cold disk galaxies.
First of all, 
disk galaxies develop long-range collective structures  like the spiral arms in Figure~\ref{fig:Spiral_Globular}a, and
these have no place in Chandrasekhar's picture. 
Also, Chandrasekhar's theory drastically overestimates the timescale on which these systems relax.
As an example of this, consider Figure~\ref{fig:Frankel} (adapted from~\citet{Frankel2020-vy}).
This figure was generated by fitting a simple diffusion model to the orbital actions (to be introduced in \S\ref{sec:angleaction}) of stars in the Milky Way using data from Gaia, a space observatory 
which has been mapping the kinematics of stars in our Galaxy since 2014.
Panel (a) shows the diffusion of the stellar angular momenta 
(essentially the guiding radius of near-circular orbits around the galaxy), while 
panel (b) shows the evolution of those same stars' mean radial actions (essentially their orbital `eccentricities').
In both panels, we see that significant evolution can occur on the timescale of a few Gyr.
To get an estimate for the true relaxation time in this system, note that the orbital angular momenta of stars on circular orbits (speed $v_\phi \approx 250$ km s$^{-1}$) near the Solar radius in the Milky Way ($R \approx 8$ kpc) is $L = Rv_\varphi \approx 2000$ kpc km s$^{-1}$.
The relaxation time we infer from Figure~\ref{fig:Frankel}a
is therefore of order $t_\mathrm{relax} \approx (2000/619)^2 \times 6$ Gyr $\approx 60$ Gyr, many orders of magnitude shorter than would be suggested by the Chandrasekhar estimate~\eqref{eqn:2BR}. And even though this relaxation time is still several times
longer than the age of the Galaxy ($\sim 10$ Gyr), it can still have a major effect on galaxy evolution, as we will see in \S\ref{sec:Dynamical_friction}.

Chandrasekhar's theory fails because it is based on a number of false assumptions, for instance:
\begin{itemize}[itemsep=0pt, topsep=4pt]
\item Stellar systems are inherently \textit{inhomogeneous} (see Figure~\ref{fig:Spiral_Globular}!) but the theory assumed homogeneity.
\item Because of the inhomogeneity, unperturbed stellar orbits are not endless straight lines, but instead are \textit{quasiperiodic} (\S\ref{sec:Mean_Field_Dynamics}). This leads to \textit{resonant} interactions.
\item Stars do not behave in an independent, uncorrelated way but instead tend to oscillate \textit{collectively}. Thus, one cannot replace all interactions between stars with a superposition of uncorrelated two-body encounters. 
\end{itemize}
These statements apply to both globular clusters and disk galaxies, which raises the question: why does Chandrasekhar's theory work to describe the former but not the latter?
The answer lies in the fact that globular clusters are dynamically \textit{hot}
whereas disk galaxies are dynamically \textit{cold}.
The hotness of clusters means that they act like thermal plasmas: collective effects are weak and successive fluctuations  are mostly uncorrelated, as different parts of the system have no way to efficiently communicate with one another. On the other hand, the highly-ordered orbital structure of disk galaxies means that great numbers of stars
can have long-lived coherent dynamical interactions. 
The result is that disk galaxies, like non-equilibrium plasmas, are replete with large-scale collective oscillations and possible instabilities.
Such effects are able to redistribute energy and angular momentum much more efficiently --- and in a fundamentally different way --- than the mechanism of incoherent two-body relaxation.

It follows that to describe the behavior of disk galaxies, we need to leave Chandrasekhar's theory behind and develop a new kinetic theory, one capable of dealing with the complications listed above.
That is one of the aims for the rest of this tutorial article.  But before we can do kinetics, we must learn how to 
describe the orbits of stars in inhomogeneous, but still fixed and smooth, mean field gravitational potentials (\S\ref{sec:Mean_Field_Dynamics}), and then figure out what happens to those orbits when they are perturbed (\S\ref{sec:P_Theory}).

\section{Orbits in mean field potentials}
\label{sec:Mean_Field_Dynamics}

As we mentioned in \S\ref{sec:Thermal_Equilibrium}, on the timescale $\sim t_\mathrm{dyn}$ the orbits of stars in a virialized $N$-body system can be very well described by ignoring the discrete nature of system and instead treating individual stars as test particles orbiting in a smooth potential.  This is a good approximation for two reasons: (i) the gravitational force between two objects is long range, decreasing as the square of the distance between them, and (ii) the number of stars in the system, $N$, is very large.  The consequence of (i) is that the force acting on any individual star is typically not dominated by its nearest neighbors, but rather by the collective gravity of simultaneous distant interactions. The consequence of (ii) is that the Poisson fluctuations in the number of stars in a given sub-volume tends to be small, so the collective gravitational force due to stars in this sub-volume fluctuates only weakly on the timescale $\tdyn$.
The average value of this collective force, combining all the sub-volumes, is then typically very smooth
(see also \S1.2 of BT08).

Another key feature of virialized stellar systems is that they have rather simple bulk geometry: they are normally approximately spherical (in the case of globular clusters, some dwarf galaxies, etc.), or thin and axisymmetric (like a disk galaxy),
or possibly ellipsoidal/triaxial (like some massive elliptical galaxies).
This simple geometry leads to (mostly) simple, regular orbits, and not many chaotic orbits.  
In this section, we will describe how stars
orbit within a simple class of smooth mean field potentials.  


\subsection{Mean field orbits}

In elementary courses on plasma physics there is no great heed paid to 
`mean field dynamics' because those dynamics are trivial: in a homogeneous electrostatic plasma, mean field `orbits' are just straight lines: $\br(t) = \bv t + \br(0)$ and $\bv = \bv(0)=$ cst.
On the contrary, stellar systems are inherently inhomogeneous, which means that 
the unperturbed mean field orbits of stars are not straight lines. 
This is something plasma physicists also have to deal with when considering orbits in systems with nontrivial geometry and especially with magnetic fields, such as in toroidal tokamak fusion plasmas or the Earth's magnetosphere\cite{boozer2004physics,zonca2015nonlinear,chen2016physics}. 

If the gravitational potential is $\Phi(\br)$, 
the equations of motion of a test star are 
\begin{subequations}
\begin{align}
        \frac{\md \br}{\md t} & = \bv, \quad\quad
        \label{eqn:EOM_r}
        \\
        \frac{\md \bv}{\md t} & = - \frac{\p \Phi}{\p \br}.
        \label{eqn:EOM_v}
\end{align}
\label{eqn:EOM_rv}%
\end{subequations}
In practice,  we cannot solve these 
equations of motion explicitly except in a few very simple potentials (e.g.\@, the Kepler potential $\Phi(\br) \propto -1/\vert \br \vert$, and the harmonic potential $\Phi(\br) \propto \vert \br \vert^2$).  But 
we can solve the equations via numerical integration --- this is called `integrating the orbit'. Modern numerical packages like \texttt{Galpy}\cite{Bovy2015-it}, \texttt{Gala}\cite{price-whelan2015-gala} and \texttt{AGAMA}\cite{Vasiliev2019-agama}
allow one to integrate and analyze mean field orbits in pre-defined potentials with ease.

\begin{figure}[htbp!]
    \centering
   \includegraphics[width=0.37\textwidth]{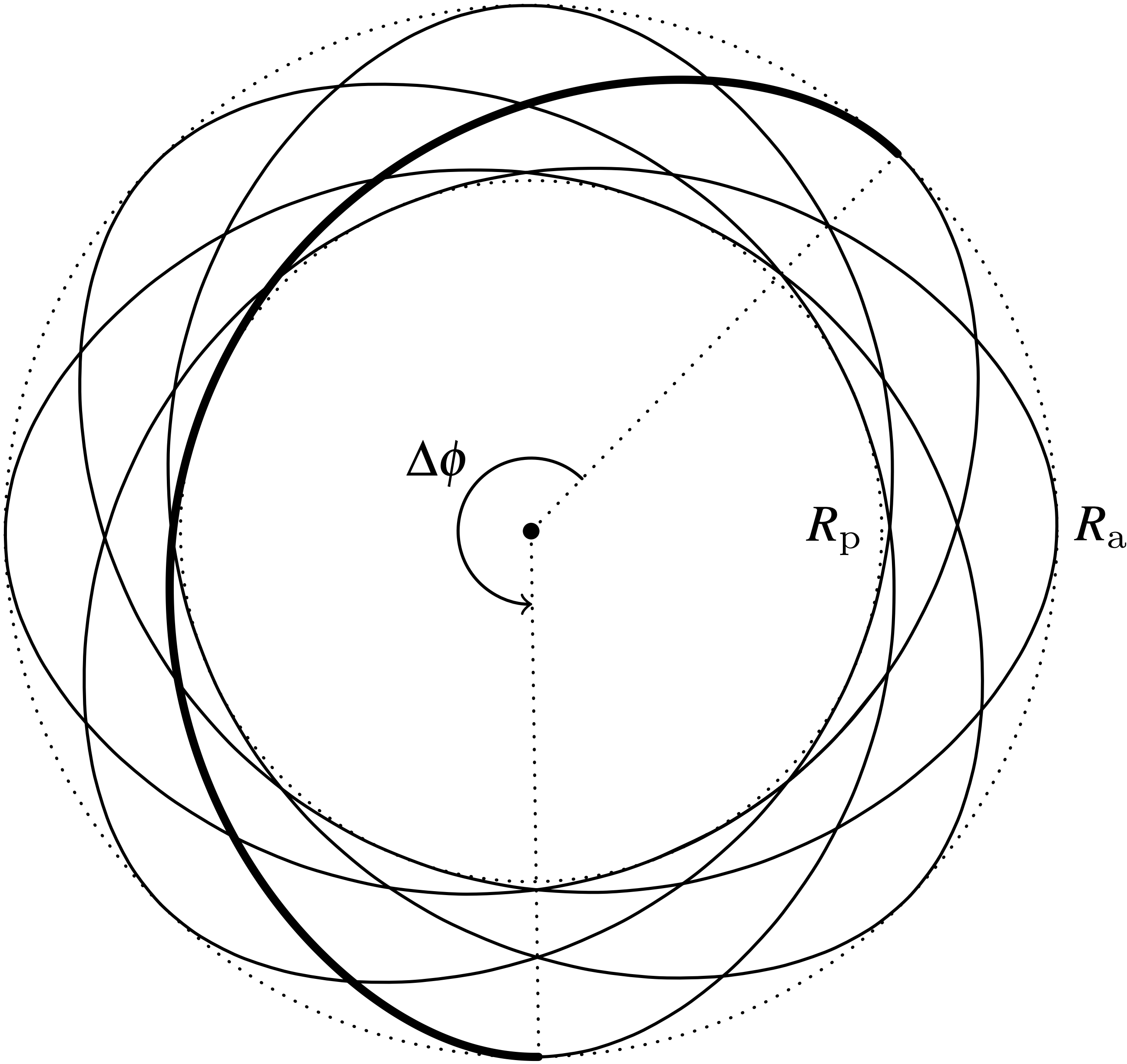}
    \caption{Nearly all orbits in central potentials $\Phi(r)$ look like this: a two-dimensional planar `rosette'.
Using in-plane polar coordinates $(\phi, R)$, each 
orbit is a combination of an azimuthal oscillation in the $\phi$ direction 
    and a radial oscillation between $\Rp$ and $\Ra$.
    The thick black line shows the trajectory over one radial period, from $\Ra$ to $\Rp$ and back again.}
         \label{fig:rosette}
\end{figure}

In general, numerical orbit integration reveals a wide array of possible orbit families --- see Chapter 3 of BT08 for an overview.  However, for 
simplicity, throughout this article we will only consider one simple orbit family, which is the planar rosette orbit shown in Figure~\ref{fig:rosette}.
This is (nearly) the only type of orbit that is possible in \textit{central} potentials $\Phi(r)$, where $r \equiv \vert \br \vert$. It is therefore the 
 overwhelmingly dominant orbit family in both our mental models, the hot sphere and the cold axisymmetric disk (Figure~\ref{fig:Two_Mental_Models}). By symmetry, motion in central potentials conserves each component of the specific orbital angular momentum vector, $\bL = \br \times \bp$, and so each such orbit is confined to a two-dimensional plane (which we can take to be the plane $Z=0$ without loss of generality).
From now on we describe orbits in this plane with a standard $(\phi, R)$ polar coordinate system centered at $r=0$.

\subsubsection{Orbits in central potentials}
\label{sec:Central}

We will now show how to describe the rosette-type orbit of Figure~\ref{fig:rosette} quantitatively.
We see from Figure~\ref{fig:rosette} that the orbit consists of 
a combination of an azimuthal circulation in the positive-$\phi$ direction, 
    and a radial oscillation in $R$ between the fixed \textit{pericentre} $\Rp$ and \textit{apocentre} $\Ra$.
    We can calculate $\Rp$ and $\Ra$ in terms of the two conserved quantities of the star's motion, namely its specific energy and specific angular momentum (we will drop the word `specific' from now on)
\begin{subequations}
\begin{align}
E & = \half \left(\frac{\md R}{\md t}\right)^2 + \half \frac{L^2}{R^2} + \Phi(R), 
\label{eqn:def_E}
\\
L & = R^2 \frac{\md \phi}{\md t}.
\label{eqn:def_L}
\end{align}
\label{eqn:E_L}%
\end{subequations}
The peri/apocentre are the radial turning points of the orbit; that is, they are solutions to $v_R(\Rp) = v_R(\Ra) = 0$, where $v_R = \md R/\md t$ is the radial velocity
    \begin{equation}
        v_R(R) \equiv \sqrt{2[E - \Phi(R)] - L^2/R^2}.
        \label{eqn:def_vr}
\end{equation}
In central potentials,
    there is then a one-to-one correspondence between the integrals of motion $(E,L)$ and the peri/apocentric distances $(\Rp, \Ra)$:
\begin{subequations}
\begin{align}
E(\Rp, \Ra) &= \frac{\Ra^2\Phi(\Ra) - \Rp^2\Phi(\Rp)}{\Ra^2 - \Rp^2},
\label{def_E_from_rpra}
\\
L(\Rp, \Ra) &= \sqrt{\frac{2[\Phi(\Ra) - \Phi(\Rp)]}{\Rp^{-2} - \Ra^{-2}}}.
\label{def_L_from_rpra}
\end{align}
\label{def_EL_fromrpra}%
\end{subequations}

   We can also calculate the frequencies of the radial and azimuthal oscillations as follows.
   First, the radial frequency is 
   \begin{equation}
\Omega_R = \frac{2\pi}{T_R},
   \label{eqn:Radial_Frequency}    
   \end{equation}
   where $T_R$ is the radial period
   \begin{equation}
       T_{R}(\Rp, \Ra) = 2 \int_{\Rp}^{\Ra} \frac{\md R}{v_R(R)}.
       \label{eqn:Tr}
   \end{equation}
The azimuthal frequency is 
   \begin{equation}
\Omega_\phi = \Omega_R \, \frac{\vert \Delta \phi \vert}{2\pi},
   \label{eqn:Azimuthal_Frequency}    
   \end{equation}
where $\vert \Delta \phi\vert $
is the increment in azimuthal angle over one radial period reading
\begin{equation}
    \vert \Delta \phi(\Rp, \Ra) \vert = 2L \int_{\Rp}^{\Ra} \frac{\md R}{R^2 v_R(R)}.
    \label{eqn}
\end{equation}
(In all sensible spherical potentials, the increment $\vert \Delta \phi(\Rp, \Ra) \vert$ lies between $\pi$ and $2\pi$). In other words, all the characteristic quantities describing the orbit can be calculated
given the two numbers $(\Rp, \Ra)$, or equivalently the two constants of motion $(E, L)$.
This is in contrast to a homogeneous plasma, where one can label an orbit just by its velocity $\bv$.

To further illustrate what these rosette orbits look like, 
consider motion in the spherical \textit{Plummer potential}, which is often used as a simple model for 
the gravitational potentials in globular clusters:
\begin{equation}
    \Phi(r) = - \frac{GM}{\sqrt{r^2 + b^2 }}, \quad\quad \mathrm{(Plummer\,\, potential).}
    \label{eqn:Plummer}
\end{equation}
Here $b$ is the scale radius of the potential (for a globular cluster $b\sim 1$pc --- see Table~\ref{table:Two_Mental_Models}).
The Plummer potential has very simple asymptotic forms. For $r\gg b$ it approximates the Keplerian potential of a point mass, $ \Phi \approx -GM/r$.  In this potential all orbits are closed ellipses with focus at the origin, i.e. they have $ \vert \Delta \phi(\Rp, \Ra) \vert = 2\pi$. 
For $r \ll b$ the (non-constant part of the) Plummer potential is quadratic $ \Phi \propto r^2$, 
which corresponds to motion in a spherical `harmonic core' of constant density (see BT08, \S3.1).
All orbits in the harmonic core are also closed ellipses, but this time they are \textit{centered} on the origin, and have 
$ \vert \Delta \phi(\Rp, \Ra) \vert = \pi$.

\begin{figure*}
    \centering
   \includegraphics[width=0.85\textwidth]{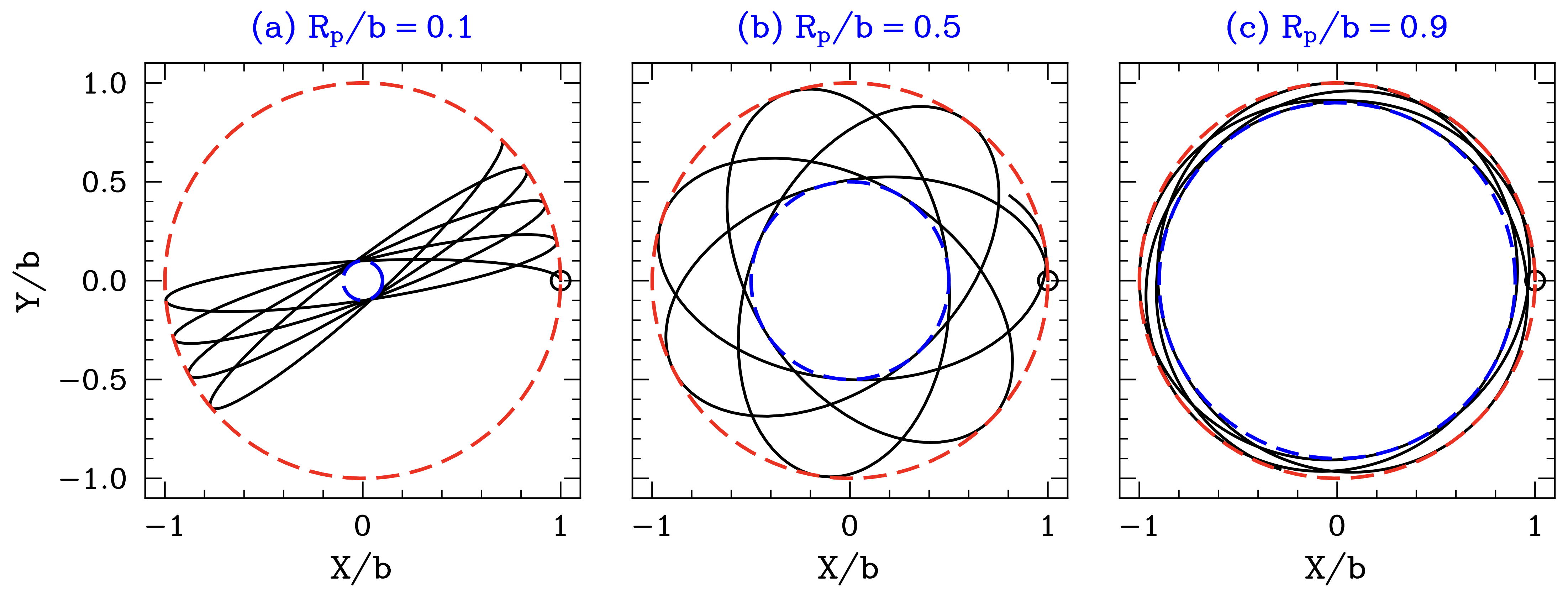}
    \caption{Three orbits in the Plummer potential~\eqref{eqn:Plummer}. In each case the orbit starts at its apocentric distance at the location $(X, Y) = (b,0)$. Each orbit has the same apocenter $\Ra= b$ (red dashed circles) but different pericenters $\Rp$ (blue dashed circles), as indicated above each panel.
    As a result, orbit (a) is almost radial ($J_R\gg L$), orbit (c) is almost circular ($J_R\ll L$), and orbit (b) is somewhere in-between ($J_R\sim L$) --- see~\S\ref{sec:Examples_AA} for the definition of the radial action $J_{R}$.
        \label{fig:Plummer}}
\end{figure*}

In Figure~\ref{fig:Plummer} we show three orbits integrated in the potential~\eqref{eqn:Plummer} in the $(\phi, R)$ plane, each with the same $R_a = b$ but different $\Rp$, as labeled above each panel.
Orbit (a) is nearly radial, with $\Rp \ll \Ra$, whereas orbit (c) is nearly circular, with $\Rp \approx \Ra$, and orbit (b) is somewhere in-between.  

The orbit labels  $(\Rp, \Ra)$, or equivalently $(E, L)$, contain no information about the orbit's instantaneous radial or azimuthal \textit{phase},
just as labeling an orbit by $\bv$ in a plasma 
does not tell us its instantaneous position in the box $\br$.  
But it turns out that the phase of the orbits also exhibit rather simple behavior, as we investigate next.

\subsection{Quasiperiodicity}
\label{sec:quasiperiodicity}

A crucial feature of orbits in many smooth, regular galactic potentials revealed by
numerical orbit integration is that they are \textit{quasiperiodic}. By this we mean that, very often, one can Fourier-analyze a numerically integrated orbit and 
write it as
\begin{equation}
    \br(t) = \sum_{{\mathbf{n}}} \br_{{\mathbf{n}}} \, \me^{ \mi {\mathbf{n}} \cdot \bOm t}.
\end{equation}
Here the sum is over all $d$-dimensional vectors comprised of integers, where $d$ is the number of dimensions, e.g.\@, $(0,0,1)$, $(0,1,1)$, ... etc. in the case of $d=3$. The vector $\bOm$ is made up of $d$ frequencies that are characteristic of the orbit in question.
Quasiperiodicity is also a common feature of particle orbits in tokamaks, magnetospheres, and so on.


That the orbits in Figure~\ref{fig:Plummer} are quasiperiodic might have been guessed by eye, but this fact also extends to a much broader range of orbits in realistic galactic potentials\cite{Binney1982,Binney1984}.
Mathematically, that an orbit is  quasiperiodic can be proven in cases where the number of known independent isolating integrals of motion is (at least) equal to the number of dimensions (which is always true for orbits in spherically symmetric potentials). However, it is a non-trivial empirical fact that quasiperiodicity also extends to orbits in many realistic \textit{axisymmetric} potentials ${\Phi}(R, Z)$  for which the orbit is fully three-dimensional and there are only two obvious independent constants of motion ($E$ and the component of angular momentum along the symmetry axis, $L_Z$)\cite{Binney1982,Binney1984}. Normally in a system with more dimensions than there are independent integrals of motion, one would expect some degree of chaos; instead, it is believed that orbits in axisymmetric potentials almost always possess a third independent integral of motion, whose analytic form we do not know but which makes the orbits regular and hence quasiperiodic (though this is   not generically guaranteed\cite{HenonHeiles1964}).

While triaxial potentials (like ellipsoids) can also harbor families of quasiperiodic orbits, they also tend to have many more chaotic orbits too, which arise in the parts of phase space between the regions of distinct quasiperiodic families. As such, in the rest of this tutorial article we will suppose that the exact potential of the system is close to, or exactly, axisymmetric, and that in the limit of pure axisymmetry there is only one regular (quasiperiodic) family of orbits we need to worry about, and no chaos.
The central potentials $\Phi(r)$ are of this nature.




\subsection{Angle-action variables}
\label{sec:angleaction}

The upshot of quasiperiodicity is that stellar orbits can be thought of as \textit{a superposition of oscillators} with fixed frequencies $\bOm$.  
Here we will show how this fact allows us to construct a coordinate system in which this mean field orbital motion is trivial, namely angle-action coordinates.  An alternative and more detailed introduction to these coordinates and how they are used
can be found in \S3.5 of BT08.

Let $(\bq, \bp)$ be a set of canonical coordinates and momenta on phase space (a reader who is not familiar with canonical coordinates can consult \S D.4 of BT08). Now consider a star moving according to the generic Hamiltonian $H(\bq, \bp, t)$.
The equations of motion for that star are
\begin{subequations}
\begin{align}
\frac{\md \bq}{\md t} & = \frac{\p H}{\p \bp},
\label{eqn:Hamilton_q}
\\
\frac{\md \bp}{\md t} & = -\frac{\p H}{\p \bq}.
\label{eqn:Hamilton_p}
\end{align}
\label{eqn:Hamilton}%
\end{subequations}
We can also use these canonical coordinates to introduce a bilinear antisymmetric functional on phase space
called the \textit{Poisson bracket}.
For any two functions $g(\bq, \bp)$ and $h(\bq, \bp)$ we define
\begin{equation}
    [g, h] = \frac{\p g}{\p \bq} \cdot \frac{\p h}{ \p \bp} -  \frac{\p g}{\p \bp} \cdot \frac{\p h}{\p \bq}.
    \label{eqn:poisson_bracket}
\end{equation}
This will be useful later.

A perfectly valid choice for the canonical coordinates $(\bq, \bp)$ is the position and velocity $(\br, \bv)$.
However, the lack of straight line orbits in stellar systems 
means that $(\br, \bv)$ are awkward variables
with which to formulate a kinetic theory.
We would like instead to work as much as possible with quantities that are invariant along orbits.
An \textit{integral of motion} $I(\br, \bv)$ is any function of phase space coordinates that is constant along an orbit.
As argued in \S\ref{sec:quasiperiodicity}, for all the mean field orbits we care about in this tutorial article (that is, orbits that are quasiperiodic), 
there exist at least as many independent integrals of motion as there are degrees of freedom in the system (normally 2 or 3).
One can construct as many mutually \textit{dependent} integrals of motion as one wishes, since any function of integrals  of motion is itself an integral of motion.

The obvious next question is: given that these integrals of motion exist, can we use them as coordinates for our phase space?
The answer is yes, but if we want to preserve the Hamiltonian structure of our system --- i.e.\ if we want the new coordinates to be \textit{canonical} --- then we have to be careful about which integrals to use.
It turns out that there is a special set of integrals called \textit{actions} $\bJ$,
which \textit{are} canonical momenta on phase space\cite{Goldstein1950}, and can be complemented by canonical coordinates.
Specifically, these actions have the special property that the
Hamiltonian of the underlying quasiperiodic system $H(\bq, \bp)$
can be written exclusively in terms of them, ${ H \!=\! H (\bJ) }$, and 
they can be complemented by canonical coordinates or \textit{angles} $\btheta$, 
which are $2\pi$-periodic.
In these special angle-action coordinates $(\bq, \bp) = (\btheta,\bJ)$,
Hamilton's equations~\eqref{eqn:Hamilton} take the simple form
\begin{subequations}
\begin{align}
\frac{\md \btheta}{\md t} & = \frac{\p H}{\p \bJ} = \bOm (\bJ),
\label{eqn:Hamilton_theta}
\\
\frac{\md \mathbf{J}}{\md t} & = - \frac{\p H}{\p \btheta} = 0 ,
\label{eqn:Hamilton_J}
\end{align}
\label{HamiltonsA}%
\end{subequations}
where we introduced the orbital frequencies ${ \bOm (\mathbf{J}) \!=\! \p H / \p \mathbf{J} }$.
In these coordinates, the orbits are straight, horizontal lines, as one has
\begin{subequations}
\begin{align}
\btheta (t) & = \btheta({0}) + \bOm (\mathbf{J}) \, t \,\,\,\,\,\,\,\,(\mathrm{mod}\,  \, 2\pi),
\label{eqn:motion_theta}
\\
\mathbf{J} (t) & = \bJ(0) = \cst
\label{eqn:motion_J}
\end{align}
\label{motionA}%
\end{subequations}

Some further properties of the angle-action coordinates are as follows.
Since they are canonical,
they leave the Poisson bracket~\eqref{eqn:poisson_bracket} invariant, so that one generically has
\begin{equation}
\big[ g,h] = \frac{\p g}{\p \btheta} \cdot \frac{\p h}{\p \mathbf{J}} - \frac{\p g}{\p \mathbf{J}} \cdot \frac{\p h}{\p \btheta} .  
\label{Poisson_invariant}
\end{equation}
As a result, angle-action coordinates also leave the infinitesimal phase space volumes invariant
so that 
\begin{equation} \md \bq \, \md \bp \!=\! \md \br \, \md \bv \!=\! \md \btheta \,  \md \mathbf{J}.
\end{equation}
Furthermore, the periodicity of the angles in $d$ dimensions means that 
any phase space function can be Fourier expanded as
\begin{equation}
G (\btheta , \mathbf{J}) = \sum_{{\mathbf{n}}} G_{{\mathbf{n}}} (\mathbf{J}) \, \me^{ \mi {\mathbf{n}} \cdot \btheta} ,
\label{eqn:Fourier_any}
\end{equation}
with inverse
\begin{equation}
G_{{\mathbf{n}}} (\mathbf{J}) = \!\! \int \!\! \frac{\md \btheta}{(2 \pi)^{d}} \, G (\btheta , \mathbf{J}) \, \me^{- \mi {\mathbf{n}} \cdot \btheta} .
\label{Fourier_expansion}
\end{equation}
Actions are also \textit{adiabatic invariants}:
roughly speaking, if $H$ evolves on a timescale longer than the dynamical time, ${ \tdyn \!\sim\! 1/\vert \bOm \vert }$,
an orbit governed by $H$ will evolve in such a way that ${ \mathbf{J} \!=\! \cst }$ (but see~\citet{weinberg1994adiabaticI,weinberg1994adiabaticII,weinberg1994adiabaticIII} for a more precise statement).

Note that we have not \textit{proven} the existence of the angle-action variables, nor have we explained how to generate them from scratch (via the process of writing down a suitable canonical generating function).  All this is standard material which is explained rather neatly (if heuristically) in \S3.5 of BT08, or much more rigorously in the books by~\citet{Lichtenberg2013-pw} and~\citet{Arnold1989-ui}.
Throughout the rest of this tutorial article we will simply assume that the angle-action coordinates can be constructed for the problem at hand.

Also, it is important to note that angles and actions are just coordinates, and that once we have constructed them, we can use them to describe the motion of particles in Hamiltonians other than $H(\bJ)$.  In particular, we often care about systems in which the motion is nearly, but not perfectly, quasiperiodic (such as a spherical stellar system  perturbed  by an infalling satellite). In such a case we can construct the angle-action variables $(\btheta, \bJ)$ corresponding to the corresponding perfectly periodic system, then describe the real (perturbed) system by expressing its Hamiltonian in these coordinates, $H(\btheta, \bJ)$, deriving the corresponding equation of motion $\md \btheta/\md t = \p H/\p\bJ$ and $\md \bJ/\md t = - \p H/\p \btheta$, etc. We will see how this construction works in practice starting in \S\ref{sec:mean_field_fluctuations}.

\subsubsection{Example: Angle-action variables for a multiperiodic homogeneous system}
\label{sec:Homogeneous}

As a first example, we consider a fictitious \textit{homogeneous stellar system},
which provides a direct analogy with homogeneous plasmas.
To this end, let us assume that our stellar system is placed within a periodic three-dimensional box
of side-length $L$. We also assume that the mean potential vanishes, i.e.\ $\Phi_{0} = 0$,
so that unperturbed trajectories are straight lines.
In this case, the system's angle-action coordinates and the associated
orbital frequencies are given by
\begin{equation}
    \btheta = \frac{2\pi }{L}\br, \quad\quad  \mathbf{J} = \frac{L}{2\pi} \bv, \quad\quad \bOm = \frac{2\pi}{L}\bv.
    \label{eq:AA_homogeneous}
\end{equation}
This makes clear a crucial correspondence between (fictitious) homogeneous stellar systems and homogeneous electrostatic plasmas.
In both systems, mean field orbits are labelled
by their velocity $\bv$, and, up to a prefactor, these velocities
are also the orbit's intrinsic frequencies $\bOm$. This means that (i) the gravitational/electrostatic potential $\Phi(\br)$ is a function only of angles $\btheta$ and not actions $\mathbf{J}$, and (ii) the orbital frequencies $\bOm(\mathbf{J})$ are trivial (linear) functions of the actions.  These two features are special to homogeneous systems, and greatly simplify the kinetic description (see e.g.\ \S\ref{sec:Response_Example_Homogeneous}).

\subsubsection{Example: Angle-action variables for the 1D harmonic oscillator}
\label{sec:1D_Oscillator}

\begin{figure}[htbp!]
  \centering
\includegraphics[width=0.55\textwidth]{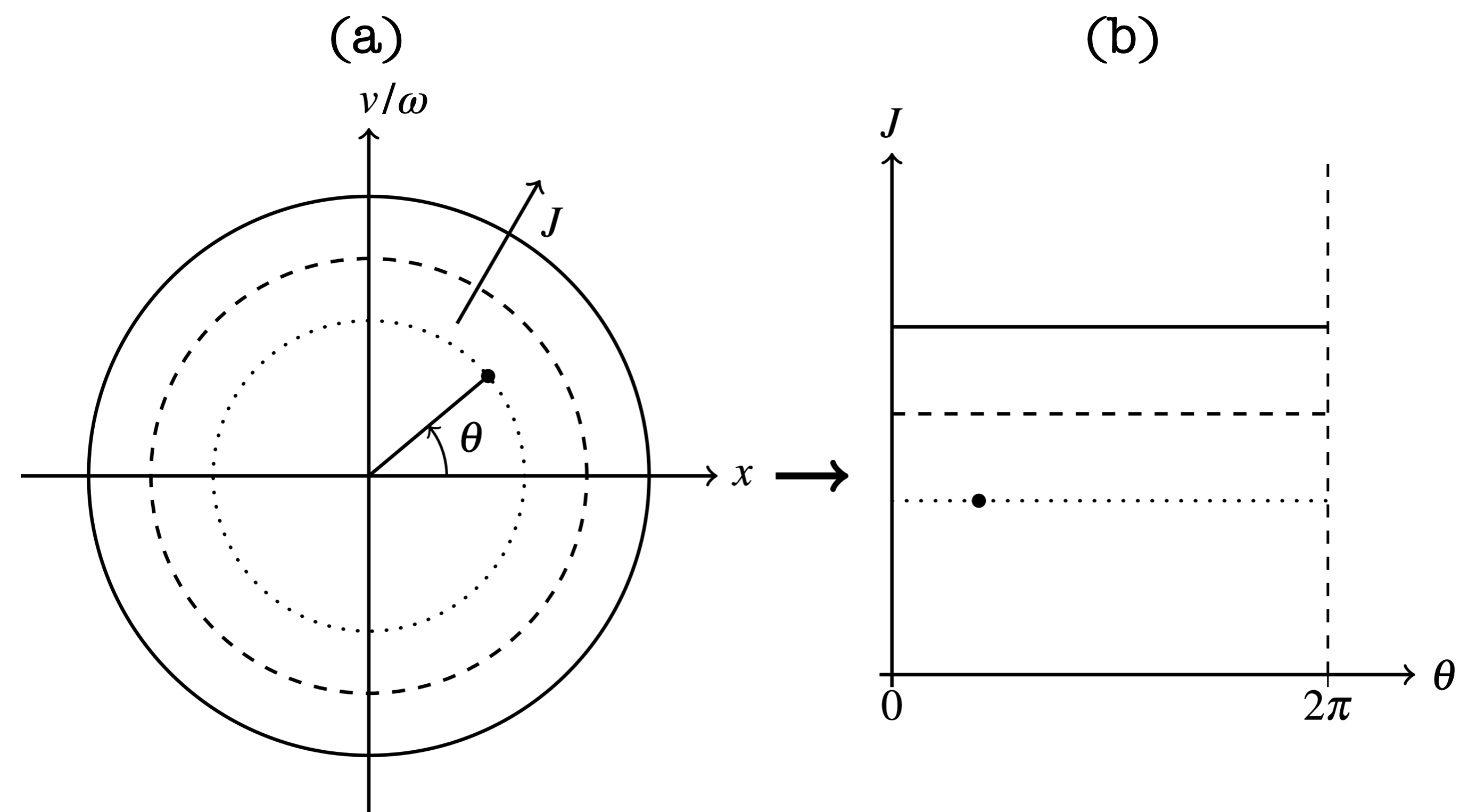}
    \caption{Phase space diagram of a harmonic oscillator, $H=(v^2+ \omega^2 x^2)/2$.
    Panel (a) shows particle trajectories in the physical phase space ${ (x, v/\omega) }$,
    which take the form of concentric circles.
    Here, the action $J$ should therefore be seen as a label for the circle associated with the orbit,
    and the angle $\theta$ should be seen as the position along the circle.
    Panel (b) shows the same trajectories in angle-action space ${ (\theta , J) }$.
    In these coordinates, the mean field motions are straight lines.
    The action $J$ is conserved, while the angle $\theta$ evolves linearly in time.
        \label{fig:IllustrationAngleAction}}
\end{figure}

Let us now consider the case of a one-dimensional harmonic oscillator whose Hamiltonian is 
$H=(v^2 + \omega^2 x^2)/2$.
As illustrated in Figure~\ref{fig:IllustrationAngleAction}, in this case,
orbits take the form of concentric circles in the phase space $(x , v/\omega)$.

We define the action via ${ H = J \omega }$, along with the angle $\theta$
so that
\begin{equation}
x = \sqrt{\frac{2 J}{ \omega}} \, \sin \theta , \quad\quad v = \sqrt{2 J \omega} \, \cos \theta .
\label{eq:AA_oscillator}
\end{equation}
Following~\eqref{Poisson_invariant}, one can check that $[\theta , J] = 1$; thus, $ (\theta , J) $ are canonical coordinates.
Finally, orbits which followed circles in $(x, v/\omega)$ space now traverse straight horizontal lines in $(\theta, J)$ space.

\subsubsection{Example: Angle-action variables for orbits in a central potential} 
\label{sec:Examples_AA}

For central potentials, $\Phi(R)$, the two-dimensional rosette motion
can be described using action variables $J_1=L$ and $J_2 = J_R$, where 
$L$ is the usual angular momentum (see equation~\ref{eqn:E_L}) and 
$J_{R}$ is the \textit{radial action}
\begin{equation}
J_{R} = \frac{1}{\pi} \!\! \int_{\Rp}^{\Ra} \!\! \md R \, v_{R} (R) ,
\label{def_Jr}
\end{equation}
with the radial velocity, $v_R$, given by equation~\eqref{eqn:def_vr}.
This is directly analogous to the longitudinal adiabatic invariant $J=m \oint v_\parallel \md s$ for particle motion in a magnetic mirror\cite{bittencourt2013fundamentals}.
The conjugate angles are the azimuthal and radial angles
\begin{subequations}
\begin{align}
\theta_1 {} & = \theta_\phi = \phi + 
\int_{\Rp}^{R} \md R' \, \frac{\Omega_\phi - L/R'^2 }{v_R(R')},
\label{eqn:azimuthal_angle}
\\
\theta_2 {} & = \theta_R = \Omega_R \int_{\Rp}^{R} \frac{\md R'}{v_R(R')}.
\label{eqn:radial_angle}
\end{align}
\label{eqn:def_angles}%
\end{subequations}
These angles evolve
at the rate of the azimuthal frequency $\Omega_{\phi}$
and radial frequency $\Omega_R$ respectively.

\begin{figure*}
  \centering
\includegraphics[width=0.605\textwidth]{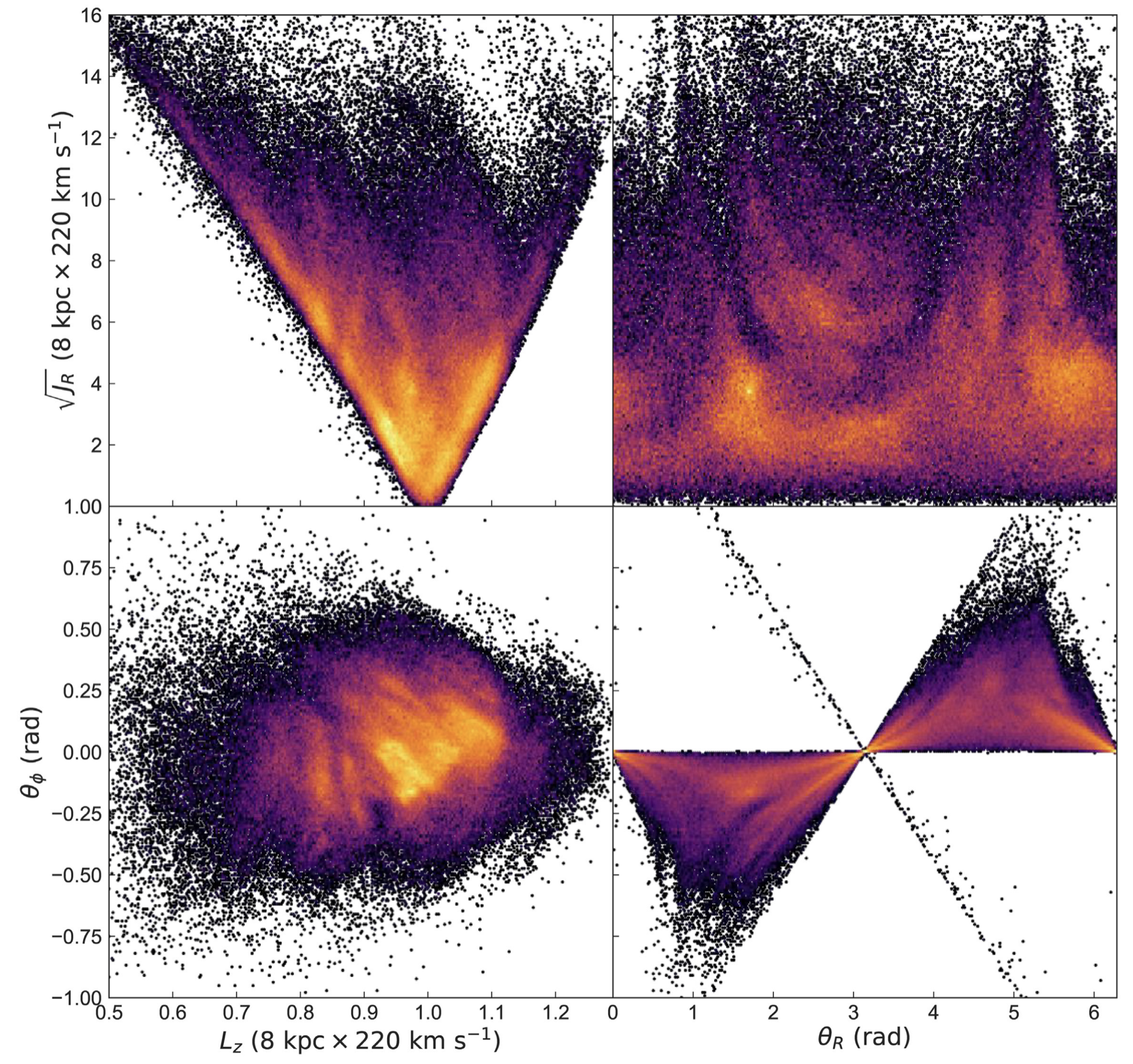}
    \caption{The distribution of stars' orbits in our local patch of the Milky Way disk (a.k.a.\ the Solar neighborhood),
    plotted
    in angle-action space, using data from the Gaia satellite.
    This figure was taken from J. A. S. Hunt, M. W. Bub, J. Bovy, J. T. Mackereth, W. H. Trick, and D. Kawata, MNRAS 490, 1026 (2019)\cite{Hunt2019-qo}.
    Another detailed guide to features in the action space (top left panel) can be found in~\citet{Trick2019-qp}.
        \label{fig:hunt}}
\end{figure*}

Figure~\ref{fig:hunt} (taken from~\citet{Hunt2019-qo}) presents recent data from the spacecraft Gaia,
showing the distribution of millions of stellar orbits in our local patch of the Milky Way disk, 
plotted in various slices through angle-action space.
Each panel is full of structure, some of which is due to selection effects (since we are unable to access data across the whole Galaxy) but much of which is due to dynamical interactions between stars, spiral arms, the Galaxy's central bar, and so on. Though we will not pursue any details here,
understanding structures like those in Figure~\ref{fig:hunt}
is a major goal of modern Galactic astronomy.

\subsubsection{Example: Angle-action variables for near-circular orbits}
\label{sec:Example_Near_Circular}

We have introduced angle-action coordinates as the natural variables with which to describe quasiperiodic orbits, since those orbits are superpositions of (generally anharmonic) oscillations at discrete frequencies $\bm{\Omega}(\bJ)$.  However, 
a disadvantage of angle-action coordinates is that we can rarely make explicit the mapping $(\mathbf{r},\mathbf{v}) \leftrightarrow (\btheta, \mathbf{J})$.
 On the other hand we \textit{can} write down an explicit mapping if we restrict ourselves to near-circular orbits (which is a highly relevant regime  for describing cold disks). There, we can employ the
 \textit{epicyclic approximation}.
 In this limit, 
 each orbit consists of circular motion at the \textit{guiding center} radius $R_\mathrm{g}$, plus a superposition of \textit{harmonic} oscillations
 in the radial and azimuthal coordinates.  In Appendix \ref{sec:epicyclic_approximation}
we show how to construct explicit angle-action coordinates in the epicyclic approximation.

\subsection{Summary and outlook}
\label{sec:Sum_Out}

For the remainder of this tutorial article we assume that we can always construct angle-action variables $(\btheta, \mathbf{J})$, such that the mean field Hamiltonian depends only on $\mathbf{J}$  (which is itself a nontrivial technical problem\cite{sanders2016review}).
It is then tempting to think of actions as the `velocity' variables in our new phase space and the angles as our new `positions'.
And indeed, many  formulae from plasma kinetics will more-or-less hold in stellar dynamics just by making the substitutions $\bv \to \mathbf{J}$ and $\br \to \btheta$ (see especially Table \ref{table:BL_Comparison}). 
However,  there are two key mathematical differences that make stellar kinetics 
much harder than plasma kinetics.
The first is that unlike in a homogeneous plasma, the canonical coordinates change at rates that are in general nonlinear functions of the canonical momenta; i.e.\ $\bOm(\mathbf{J})$ is typically a complicated (nonlinear) function of $\mathbf{J}$, whereas in plasma we have $\bOm \propto \bJ \propto \bv$.
The other key difference is that the fluctuating gravitational potential in these variables (which we will introduce in \S\ref{sec:P_Theory}) depends on \textit{both} the canonical coordinates and momenta, i.e.\ $\delta \Phi(\br, t) = \delta \Phi(\btheta, \mathbf{J}, t)$.
As we will see, these differences make the physics of stellar systems in some ways richer dynamically than that of electrostatic plasmas, 
but also make explicit calculation more difficult.

\section{Perturbed orbits}
\label{sec:P_Theory}


Now that we know how to describe stellar orbits in smooth central potentials,
 we are ready to consider what happens if we perturb those orbits.
In particular, in this section we
discuss how stellar orbits in an axisymmetric, razor-thin (i.e.\ two-dimensional) disk are modified when we introduce 
a rigidly-rotating non-axisymmetric perturbation (e.g.\ a spiral arm pattern).
The discussion will be familiar to plasma physicists who have studied, for example, the interaction of ions with toroidal Alfven waves in tokamaks\cite{boozer2004physics,chen2016physics}.

\subsection{Mean field and perturbations}
\label{sec:MF_and_Pert}

In order to construct angle-action variables we need a system that is integrable.
Since we know that orbits in axisymmetric potentials are nearly always integrable, we will first consider an axisymmetrized version of our (in general non-axisymmetric) system, and construct the angle-action variables associated with motion in that axisymmetrized system. We then treat non-axisymmetries as perturbations.

To begin our construction, at a given time let the \textit{exact}, in general non-axisymmetric, gravitational potential of the whole two-dimensional system (including external perturbations, if they are present) be $\phi(\br) = \phi(\varphi, R)$. Then the exact Hamiltonian of a test star orbiting in this system is 
\begin{equation}
    \label{eq:Exact_Ham}
    H(\br, \mathbf{v}) = \frac{1}{2} \mathbf{v}^2 + \Phi(\br).
\end{equation}
The azimuthal average of $H$ is 
\begin{equation}
    \label{eq:Axi_Ham}
    \overline{H}(R, \mathbf{v}) \equiv \frac{1}{2} \mathbf{v}^{2} + \overline{\Phi}(R),
\end{equation}
where for any function $q(\br)$ we define
\begin{equation}
    \overline{q}(R) \equiv \frac{1}{2\pi}\int \md \varphi \, q(\br).
\end{equation}
We know that we can construct angle-action variables $(\btheta, \bJ)$ such that $\overline{H}$ only depends on $\bJ$. 
Given these angle-action variables, we can express any function on phase space in terms of them via $\br = \br(\btheta, \bJ)$ and $\mathbf{v} = \mathbf{v}(\btheta, \bJ)$. 
In particular, we can return to our original exact Hamiltonian $H$ and split it into a $\btheta$-independent `mean-field' part and a $\btheta$-dependent `fluctuation':
\begin{equation}
    \label{eq:H1}
    H [ \br( \btheta, \bJ), \mathbf{v}(\btheta, \bJ)]  = H_0(\bJ) + \delta H(\btheta, \bJ),
\end{equation}
where the $\btheta$-average of some quantity $q(\btheta, \bJ)$ is
\begin{equation}
    \label{eq:Definiton_Angle_Average}
    q_0(\bJ) \equiv \int \frac{\md \btheta}{(2\pi)^2} \, q(\btheta, \bJ).
\end{equation}
Here, the `$0$' subscript reminds us that $q_0$ is equivalent to the zeroth Fourier mode of $q$ when expanded in angles (see equation \eqref{eqn:Fourier_any} above). 

One can show that the `mean-field Hamiltonian' $H_0$ and the `axisymmetric Hamiltonian' $\overline{H}$ are equivalent (even though this is \textit{not} generally true of the potential alone, i.e. $\Phi_0 \neq \overline{\Phi}$).
Further, we have that 
\begin{equation}
    \label{eqn:Non_Axi_Phi}
    \delta H(\btheta, \bJ) = \Phi(\br) - \overline{\Phi}(R) \equiv \delta \Phi(\btheta, \bJ),
\end{equation}
i.e., the angle-dependent part of the Hamiltonian is just the non-axisymmetric part of the potential. Hence, from now on we write
\begin{equation}
    \label{eqn:H}
    H[ \br(\btheta, \bJ), \mathbf{v}(\btheta, \bJ)] = H_0(\bJ) + \delta \Phi(\btheta, \bJ),
\end{equation}
and refer to $H_0$ and $\delta \Phi$ as the mean-field Hamiltonian and the potential perturbation respectively.

\subsection{Rotating potential perturbation}

We now assume that the non-axisymmetric potential perturbation $\delta \Phi$ rotates in the $\phi$ direction with fixed pattern speed $\pattern$:
\begin{equation}
\delta \Phi =    \delta\Phi(\phi - \pattern t, R).
\label{eqn:Rigidly_Rotating_Perturbation}
\end{equation}
Given that the perturbation is periodic in $\phi - \pattern t$,
using equation~\eqref{eqn:azimuthal_angle} and the fact that $R$ is a function only of $\theta_R$ and $\mathbf{J}$ (through the implicit relation~\ref{eqn:radial_angle}),
we can always expand $\delta \Phi$ as a Fourier series (c.f.\ equation~\ref{eqn:Fourier_any})
\begin{equation}
    \delta \Phi(\bm{\theta},\mathbf{J},t) = \sum_{\mathbf{n}\neq \bm{0}} \delta \Phi_{\mathbf{n}}(\mathbf{J}) \, \me^{\mi (\mathbf{n}\cdot\bm{\theta} - n_\phi \pattern t)},
    \label{eqn:delta_Phi_Fourier}
    \end{equation}
where $\mathbf{n} = (n_\phi, n_R)$.  For a perturbation with fixed azimuthal harmonic number $m>0$ (e.g.\ an $m$-armed spiral with a sinusoidal azimuthal profile), the only terms that survive in the expansion~\eqref{eqn:delta_Phi_Fourier} are those with $n_\phi = \pm m$.
More generally, bar/spiral perturbations can usually be built as a superposition of a small number of $m$-harmonic components.

\subsection{Resonances}

A star's orbit evolves according to Hamilton's equations~\eqref{HamiltonsA}.
Thus, using the Fourier expansion~\eqref{eqn:delta_Phi_Fourier},
its action evolves as 
$\mathbf{J} = \mathbf{J}(0) + \delta \mathbf{J}(t)$,
where
\begin{equation}
    \frac{\md \delta \mathbf{J}}{\md t} = -\mi \sum_{\mathbf{n}} \mathbf{n} \,\delta \Phi_{\mathbf{n}}(\mathbf{J}) \, \me^{\mi (\mathbf{n}\cdot\bm{\theta} - n_\phi \pattern t)}.
    \label{eqn:ddeltaJ_dt}
\end{equation}
Given this equation, it will not surprise the reader that the majority of interesting orbital modifications by rigidly rotating potential perturbations occur at \textit{resonances}, i.e.\ locations $\bJ$ in phase space where
\begin{equation}
     \mathbf{N}\cdot \bm{\Omega}(\mathbf{J}) = N_\phi \pattern.
     \label{eqn:resonance_condition}
 \end{equation}
 for some vector of integers $\mathbf{N}=(N_\phi, N_R)$.  That is because these are locations where, substituting the unperturbed trajectories \eqref{eqn:motion_theta}-\eqref{eqn:motion_J}, the argument of the exponential on the right hand side of \eqref{eqn:ddeltaJ_dt} is constant.

For concreteness, suppose the perturbation in question has azimuthal harmonic number $m$.  Then
the most important values of $\mathbf{N}$ --- which we are sometimes referred to as the perturbation's \textit{major resonances} --- are
\begin{subequations}
\begin{alignat}{2}
 \mathbf{N}_\mathrm{CR} & = (m, 0)
&& \quad \text{--} \quad \text{corotation resonance (CR)} ,
 \label{eqn:N_CR}
 \\
     \mathbf{N}_\mathrm{ILR} & = (m, -1)
   && \quad \text{--} \quad \text{inner Lindblad resonance (ILR)} ,
        \label{eqn:N_ILR}
\\
    \mathbf{N}_\mathrm{OLR} & = (m, 1)
  && \quad \text{--} \quad \text{outer Lindblad resonance (OLR)} .
        \label{eqn:N_OLR}        
\end{alignat}
\label{eqn:def_N_res}%
\end{subequations}
These resonances tend to dominate the in-plane evolution of disk galaxies 
driven by large-scale perturbations like spirals and bars. Resonances with $\vert N_R\vert \geq 2$ also exist, but are usually subdominant because bars and spirals have Fourier spectra which are dominated by the lowest few $n_R$. On the other hand, resonances with $\vert N_R\vert \geq 2$ are very important for understanding, e.g.\@, the dynamical friction felt by an orbiting point mass\cite{Weinberg1989-cj,Kaur2018-wp,Banik2021-ug,Banik2022-rj,Kaur2022-tb}.

The resonances \eqref{eqn:def_N_res} have a simple physical interpretation in the important case of near-circular orbits (see~\S\ref{sec:Example_Near_Circular} and Appendix~\ref{sec:epicyclic_approximation}).
In that case, the corotation resonance is located at the orbital radius
$R_\mathrm{CR}$ such that an orbit with guiding radius $R_\mathrm{g} =R_\mathrm{CR}$ advances in azimuth at exactly the same rate as the perturbation, i.e.\ $\Omega_\mathrm{c}(R_\mathrm{CR}) = \pattern$,
with $\Omega_{\mathrm{c}}$ the azimuthal frequency for circular orbits.
If we now imagine keeping that star's orbit near-circular but increasing
$R_\mathrm{g}$ slightly, it will begin to circulate
at a frequency slightly smaller than $\pattern$.  As we continue this process, eventually $R_\mathrm{g}$ will be sufficiently large that the difference $\pattern - \Omega_\mathrm{c}$ is equal to the frequency of radial epicycles $\kappa$ (equation~\ref{eqn:Radial_Epicyclic_Frequency}); this is the outer Lindblad resonance.
Had we performed a similar  experiment but instead gradually made the circular orbit \textit{smaller} than $R_\mathrm{g}$,
the star would begin to orbit faster than $\pattern$, until $\pattern - \Omega_\mathrm{c}$ coincided with $-\kappa$; this is the inner Lindblad resonance.  In summary, for epicyclic orbits we can write the resonance condition as
\begin{equation}
    m \, (\pattern - \Omega_{\mathrm{c}}) = n \kappa,
    \label{eqn:CR_Lindblad_Condition}
\end{equation}
with $n = 0, 1, -1$ for CR, OLR and ILR respectively.
For an illustration of these resonant orbits, see e.g.\ Figure 6.10 of BT08.


\subsubsection{Example: Orbits perturbed by a spiral potential}
\label{sec:Example_Perturbed_Orbits}

\begin{figure}[htbp!]
\centering
\includegraphics[width=0.45\linewidth]{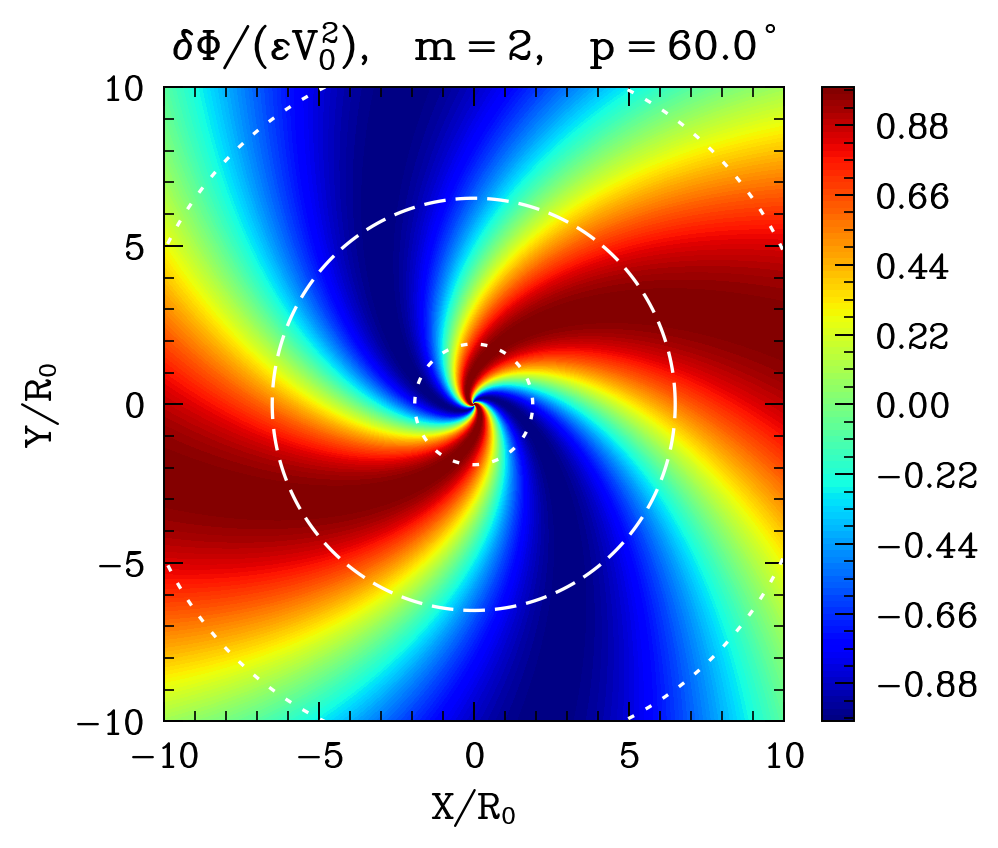}
\caption{Contours of the spiral potential, equation~\eqref{eqn:Phi_Spiral}, for $m=2$ and $p=60^\circ$ (the azimuthal phase is arbitrary).
The dotted and dashed white curves show the radial locations of the ILR, CR and OLR respectively assuming $\pattern = V_0/(6.5R_0)$, 
but this is only meant to guide the eye --- the logarithmic spiral~\eqref{eqn:Phi_Spiral} and the logarithmic halo~\eqref{eqn:Phi_Logarithmic} are both scale-free, so nothing about the figure would change if we modified $\pattern$
apart from the radii of these white circles.}
\label{fig:Spiral_Potential_Contours_m_2_p_60.0}
\end{figure}

To illustrate the morphology of perturbed orbits and the importance of resonances, consider the motion of test stars in the potential $\Phi = \overline{\Phi} + \delta\Phi$,
where $\overline{\Phi}$ is the logarithmic halo potential 
\begin{equation}
    \overline{\Phi}(R) = V_0^2 \ln (R/R_0),
        \label{eqn:Phi_Logarithmic}
\end{equation}
and $R_0$ is an arbitrary scale length.
This potential gives rise to a \textit{flat rotation curve}, meaning that all stars on circular orbits ($J_R=0$) have the same orbital speed: $v_\varphi^2 = R \, \partial \overline{\Phi}/\partial R = V_0^2 = \cst$
This is in fact a reasonably realistic model for many galactic disks\cite{mo2010galaxy}.  
For $\delta \Phi$, we choose the potential of a steadily-rotating \textit{logarithmic spiral}
\begin{equation}
    \delta \Phi(\phi, R, t) = - \varepsilon  V_0^2 \cos[\alpha\ln (R/R_0) - m(\phi-\pattern t-\phi_\mathrm{p})], \\\,\,\,\,\,\,\,\,\,\,\,\,\,\,\,\,\,\,\,\,\,  \alpha \equiv m \cot p.
    \label{eqn:Phi_Spiral}
\end{equation}
This corresponds to a spiral with $m$ arms, pattern speed $\pattern$,
and pitch angle $p$ (so $p\to 0$ is a very tightly wrapped trailing spiral, which gets less and less tightly wound as $p \to 90^\circ$ --- for much more detail, see~\S{6.1} of BT08).
The rotation period of this spiral is simply, $T_{\mathrm{p}} = 2 \pi / \Omega_{\mathrm{p}}$.
In Figure~\ref{fig:Spiral_Potential_Contours_m_2_p_60.0}
we plot contours of $\delta \Phi/(\varepsilon V_0^2)$ for $m=2$ and $p=60^\circ$ (a rather loosely wrapped arm).

We now fix the pattern speed to be $\pattern = 0.154 \, V_0/R_0$, so that the corotation radius of circular orbits is $R_\mathrm{CR} = 6.5R_0$. We take the spiral strength to be $\varepsilon = 0.01$, the pitch angle $p=60^\circ$, and the initial phase $\phi_\mathrm{p} = 0$. 
We initialized the orbits of nine
test stars in this potential, putting each star on a circular orbit with speed $V_0$ and
azimuthal location $\phi(t = 0) = 0$. 
The initial conditions of the stars therefore differed only in their initial orbital radii $R(0)$, 
which we sampled linearly between $R_\mathrm{ILR}$
and $R_\mathrm{OLR}$ inclusive. 
We integrated each orbit for a total time $20 T_\mathrm{p}$.

\begin{figure}[htbp!]
\centering
\includegraphics[width=0.75\linewidth]{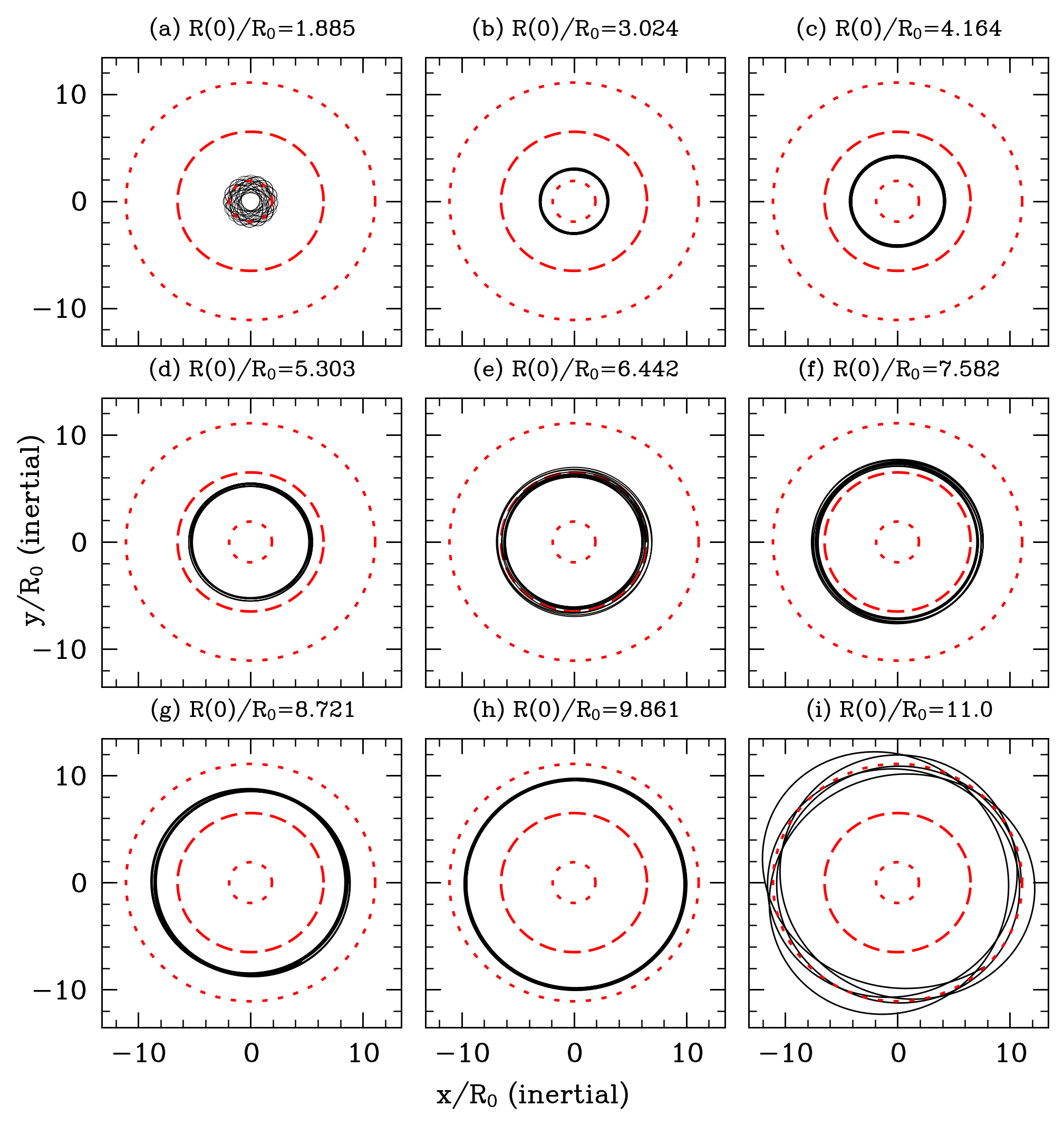}
\caption{Orbits of stars in the combined logarithmic halo~\eqref{eqn:Phi_Logarithmic} and spiral~\eqref{eqn:Phi_Spiral} potentials. All stars are initially on circular orbits with speed $V_0$ but differ in their initial orbital radii $R(0)$. The dotted and dashed lines show the Lindblad and corotation radii, as in Figure~\ref{fig:Spiral_Potential_Contours_m_2_p_60.0}. In particular, in panels (a), (e) and (i) the orbit is trapped at the ILR, CR and OLR respectively (see Figure \ref{fig:3x3_rotating}).}
\label{fig:3x3_inertial}
\end{figure}

\begin{figure}[htbp!]
\centering
\includegraphics[width=0.75\linewidth]{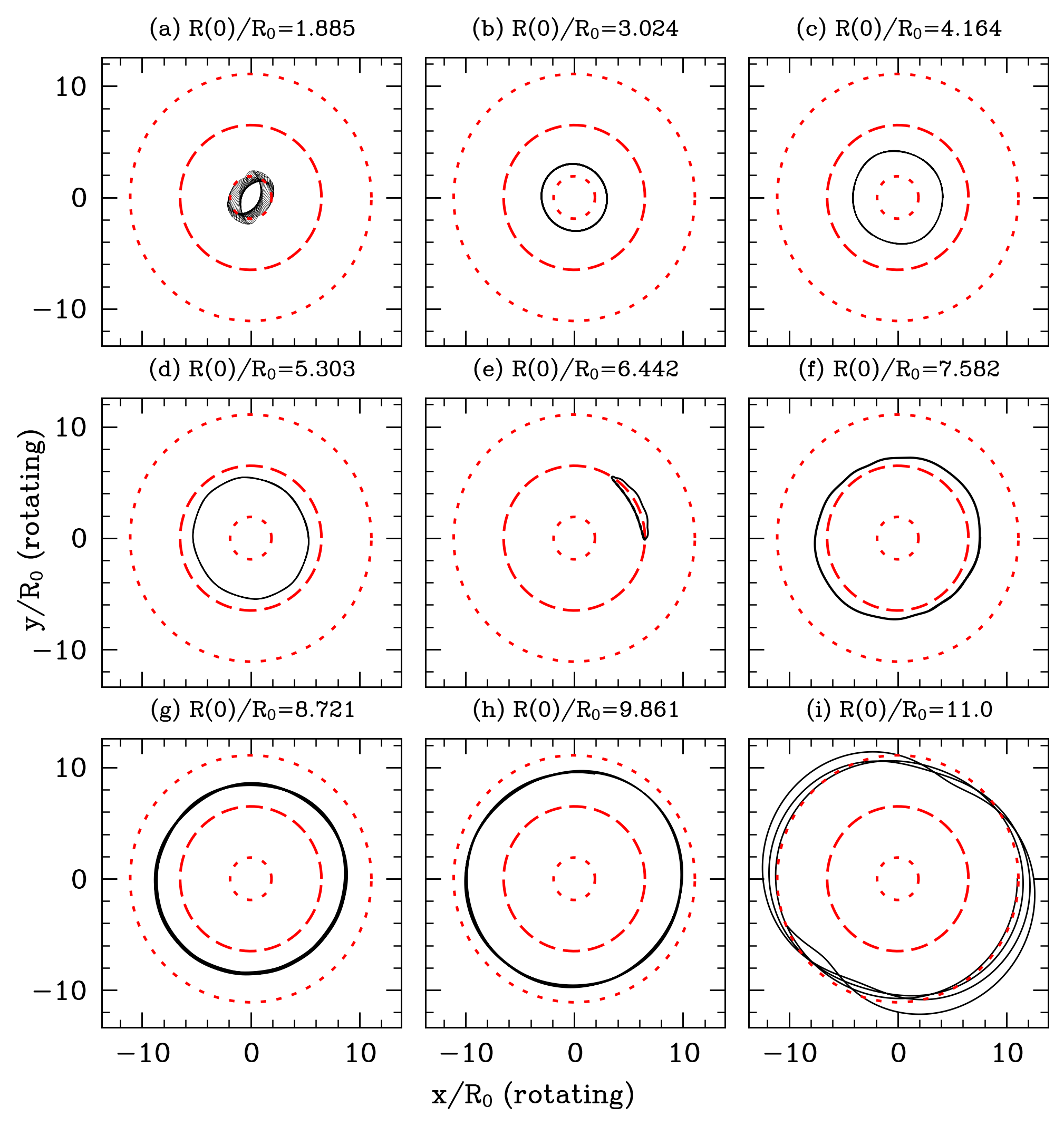}
\caption{As in Figure \ref{fig:3x3_inertial} except in a frame that rotates in the $\phi$ direction at rate $\pattern$ (so the spiral pattern is stationary).}
\label{fig:3x3_rotating}
\end{figure}

In Figures~\ref{fig:3x3_inertial} and \ref{fig:3x3_rotating} we show the resulting orbits.
In each panel, the black line shows the trajectory of a star whose initial radius $R(0)$ is given at the top of the panel. we also show the corotation radius of near-circular orbits 
$R_\mathrm{CR}$ with a dashed red line,  and the corresponding 
Lindblad radii $R_\mathrm{ILR/OLR}$ with dotted red lines.
In Figure \ref{fig:3x3_inertial} we plot the trajectories in the inertial 
frame of the galaxy, while in Figure \ref{fig:3x3_rotating} we plot them in a frame that rotates azimuthally at the rate $\pattern$ (so the spiral potential~\eqref{eqn:Phi_Spiral} is fixed).

These figures reveal that initially circular orbits are not dramatically distorted if the star sits far from a resonance.  In this case, the orbits are well described with linear perturbation theory which we introduce in \S\ref{sec:Linearly_Perturbed_Orbits}.
On the other hand, the orbits are changed qualitatively if they are located near a resonance.
In particular, panel (a) shows an orbit that is `trapped' at the ILR, which is the key resonance responsible for e.g.\ the formation of stellar bars. Meanwhile, panel (e) shows an  orbit that is trapped at corotation. When viewed in the rotating frame these orbits do not complete any azimuthal loops, but are rather `stuck' on one side of the spiral, and librate back and forth in the trough of the spiral potential.
Trapped orbits at corotation are especially important for determining the slowdown rate of spinning stellar bars (\S\ref{sec:Dynamical_friction}), the 
saturation of spiral instabilities (\S\ref{sec:saturation}), and so on. (They are also very familiar to plasma physicists, who are used to particles `bouncing' in the troughs of waves if the wave's phase velocity matches the particle velocity).
Finally, panel (i) shows an orbit trapped at the OLR.
\\
\\
The rest of this section is devoted to a mathematical description of how
stellar orbits respond to a perturbation of the form 
\eqref{eqn:delta_Phi_Fourier}.
We start by discussing a quantity that is rigorously conserved for such orbits (\S\ref{sec:Jacobi}).
We then describe how one can approximate the evolution of 
linearly perturbed orbits (\S\ref{sec:Linearly_Perturbed_Orbits}).  Finally, we show how the nonlinearly distorted orbits that can develop
near resonances can be treated analytically in the pendulum approximation (\S\ref{sec:Pendulum_Approximation}).


\subsection{Jacobi integral}
\label{sec:Jacobi}

One thing we can immediately say about test star motion in the Hamiltonian~\eqref{eqn:H} is that it preserves neither angular momentum $L$ nor energy $E$ (since $\delta \Phi$ is both non-axisymmetric and time-dependent).
Yet, it \textit{does} conserve the Jacobi integral (see~\S{3.3.2} of BT08):
\begin{equation}
    E_\mathrm{J} \equiv E - \pattern L.
\end{equation}
This follows because $H_\mathrm{J} \equiv H-\pattern L$ is just the Hamiltonian in the frame co-rotating with the perturbation, and in that frame the Hamiltonian is time-independent and hence conserved (see \S3.3.2 of BT08 for proof).
Since $E_\mathrm{J}=\cst$, any interaction of a test star with a constant amplitude, rigidly rotating potential perturbation 
must produce changes to the star's energy $\Delta E$ and angular momentum $\Delta L$ that are related by 
\begin{equation}
    \Delta E = \pattern \, \Delta L.
\end{equation}
If $\Delta E$ is sufficiently small, we can approximate it by $\Delta E \approx (\partial E/\partial L)\Delta L + (\partial E/\partial J_R)\Delta J_R = \Omega_\phi \Delta L + \Omega_R \Delta J_R$, following the generic definition of orbital frequencies in~\eqref{eqn:Hamilton_theta}.
This allows us to relate the change in radial action $\Delta J_R$ to the change in angular momentum $\Delta L$:
\begin{equation}
    \Delta J_R \approx \frac{\pattern - \Omega_\phi}{\Omega_R} \Delta L.
    \label{eqn:Sellwood_Binney}
\end{equation}
For instance, stars that corotate with the perturbation, $\Omega_\phi = \pattern$, 
do not experience changes to their radial actions to first order.



\subsection{Linear perturbations}
\label{sec:Linearly_Perturbed_Orbits}


Suppose the perturbation strength $\vert \delta \Phi \vert/\vert H_0\vert \sim \varepsilon \ll 1$ (just as in equation~\ref{eqn:Phi_Spiral}). Then to
zeroth order in $\varepsilon$ we have $\delta \mathbf{J}^{(0)} = \bm{0}$, i.e.\ the star's trajectory is just a straight line through angle-action space,  $\mathbf{J} = \mathbf{J}_0$ and $\btheta = \btheta_0 + \bOm t$. Plugging this back in to the right hand side of~\eqref{eqn:ddeltaJ_dt} and integrating forward in time, we get 
\begin{equation}
    \delta \mathbf{J}^{(1)}(t) = - \mi \sum_{\mathbf{n}} \mathbf{n} \, \delta \Phi_{\mathbf{n}}(\mathbf{J}_0) \, \me^{\mi \mathbf{n}\cdot\bm{\theta}_0}  \frac{\me^{\mi (\mathbf{n} \cdot \bOm(\mathbf{J}_0)- n_\phi \pattern) t}-1}{\mi (\mathbf{n} \cdot \bOm(\mathbf{J}_0)- n_\phi \pattern)},
    \label{eqn:deltaJ_1}
\end{equation}
Thus to first order in $\varepsilon$ the
perturbation~\eqref{eqn:delta_Phi_Fourier} 
nudges the star's orbital action by an amount $\delta \mathbf{J}^{(1)}$
which is proportional to $\varepsilon$ in magnitude and oscillatory in time, \textit{except} in the case where the star's mean field orbit is in resonance with the perturber, namely when there exists some pair of integers
$\mathbf{n} = \mathbf{N} = (N_\phi, N_R)$ such that~\eqref{eqn:resonance_condition} holds for $\bJ=\bJ_0$.
In the resonant case, \eqref{eqn:deltaJ_1} predicts that
$\delta \mathbf{J}$ will grow linearly in time, since the $\mathbf{N}^{\mathrm{th}}$ Fourier  component of the forcing~\eqref{eqn:delta_Phi_Fourier} 
is constant in the frame moving with the unperturbed stellar orbit.

In reality, it is precisely near these resonant locations that linear perturbation theory will eventually break down. In that case, one should instead turn to an alternative, \textit{nonlinear} description of the dynamics.

\subsection{The pendulum approximation}
\label{sec:Pendulum_Approximation}

To describe the near-resonant orbits analytically we turn to the `pendulum approximation'.
The main idea of the pendulum approximation is 
that near resonances, the linear approximations used in \S\ref{sec:Linearly_Perturbed_Orbits} only truly break down along certain special `resonant directions' in angle-action space, namely those directions
where the argument of the exponential in~\eqref{eqn:delta_Phi_Fourier}
varies slowly for some resonance ${\mathbf{n}} = \mathbf{N}$.
This allows us to average out the oscillatory motion in the `nonresonant' directions, and thereby reduce the complicated resonant dynamics to the integrable problem of a one-dimensional pendulum.

 For definiteness, let us set the number of dimensions to $d=2$.  Then $\btheta=(\theta_\phi, \theta_R)$ and $\mathbf{J}=(J_\phi, J_R)$.
 We now
 make a canonical transformation (see \S D.4 of BT08) to a new set of coordinates, which are again angle-action coordinates of the unperturbed problem\cite{Lichtenberg2013-pw,Binney2020-mw}.
Precisely, we map
$(\btheta, \mathbf{J}) \to (\btheta',\mathbf{J}')$, where $\btheta' = (\theta_\mathrm{f}, \theta_{\mathrm{s}})$ consists of the `fast' and `slow' angles 
\begin{subequations}
\begin{align}
\theta_{\mathrm{f}} & \equiv \theta_R,
\label{eqn:def_theta_f}
\\
\theta_s &\equiv \mathbf{N}\cdot \boldsymbol{\theta} - N_\phi \pattern t,
\label{eqn:def_theta_s}
\end{align}
\label{eqn:slow_angle}%
\end{subequations}
and $\mathbf{J}' = (J_{\mathrm{f}}, J_\mathrm{s})$
consists of the corresponding fast and slow actions
\begin{subequations}
\begin{align}
 J_{\mathrm{f}} & \equiv J_R - \frac{N_R}{N_\phi} L, 
 \label{eqn:def_Jf}
 \\
J_\mathrm{s} & \equiv \frac{L}{N_\phi}.
\label{end:def_Js}
\end{align}
\label{eqn:slow_action}%
\end{subequations}
(In plasma, the analogue of, e.g.\@, equation \eqref{eqn:slow_angle} would be a coordinate transform $x' = x - v_\mathrm{ph}t$, where $v_\mathrm{ph}$ is the phase velocity of a wave moving along the $x$ axis).
Having made this transformation, we may rewrite the Hamiltonian $H$ (equations \eqref{eqn:H}-\eqref{eqn:Rigidly_Rotating_Perturbation}) in terms of the new coordinates:
\begin{equation}
    H(\btheta',\mathbf{J}') = H_0(\mathbf{J}')  - N_\phi \pattern J_\mathrm{s} 
   + \sum_{\mathbf{k}\neq \bm{0}} \Psi_{\mathbf{k}}(\mathbf{J}') \me^{\mi \mathbf{k}\cdot \bm{\theta}'},
    \label{eqn:Hamiltonian_slowfastangles}
\end{equation}
where $\mathbf{k} = (k_\mathrm{f}, k_\mathrm{s})$ is a vector of integers and 
we have expanded $\delta \Phi$ as a Fourier series in the new angles $\btheta'$, i.e.
written $\delta \Phi(\mathbf{r}, t) = \sum_{\mathbf{k}} \Psi_{\mathbf{k}}(\mathbf{J}') \me^{\mi \mathbf{k}\cdot \btheta'}$.  The coefficients $\Psi_{\mathbf{k}}$ are easily computed for a simple bar or spiral model\cite{Tremaine1984-wt,Chiba2022-qt}.
The special thing about the form~\eqref{eqn:Hamiltonian_slowfastangles} of the Hamiltonian is that it has no explicit time dependence (or rather, the time-dependence has been absorbed into the definition of the angle $\theta_\mathrm{s}$; see equation~\ref{eqn:slow_angle}).

The fast angles $\theta_\mathrm{f}$ evolve on the orbital timescale whereas $\theta_\mathrm{s}$ evolves on the much longer timescale ${ \sim (\mathbf{N}\cdot \bm{\Omega} \!-\! N_\phi \pattern)^{-1} }$.  Thus we choose to average $H$ over the unimportant fast motion, and work instead with the simpler Hamiltonian $h \equiv (2\pi)^{-1}\int \md \theta_\mathrm{f} H$ at fixed $J_\mathrm{f}$.
We also expand $h$ around the resonance.
To do this, we let the resonant action be $\mathbf{J}_\mathrm{res}$ (meaning that equation~\eqref{eqn:resonance_condition} is satisfied for $\bJ = \mathbf{J}_\mathrm{res}$).
In the transformed coordinates this is $\mathbf{J}'_\mathrm{res} = ({J}_\mathrm{f}, J_\mathrm{s, res})$, and so we can define
\begin{equation}
    I \equiv J_\mathrm{s} - J_{\mathrm{s, res}} , \,\,\,\,\,\,\,\,\,\,\, 
    \varphi_k = \theta_\mathrm{s} + \arg(\Psi_k) / k,
    \label{eqn:I_phi_definitions}
\end{equation}
with $\varphi_k \in [-\pi, \pi]$ (and we used the shorthand $\Psi_{(0,k_\mathrm{s})} = \Psi_k$). Expanding $h$ for small $I$, and assuming that a single Fourier component $k$ dominates the perturbation, we find that at each fixed ${J}_\mathrm{f}$
the dynamics reduces to motion in the `slow plane'
  $(\theta_\mathrm{s}, J_\mathrm{s})$ dictated by  the pendulum Hamiltonian
\begin{equation}
    h (\varphi, I) = \frac{1}{2}GI^2 - F\cos k\varphi,
    \label{eqn:resonant_Hamiltonian}
\end{equation}
where $\varphi = \varphi_k$ and 
\begin{equation}
        G({J}_\mathrm{f}) \equiv \frac{\partial^2 H_0}{\partial J_\mathrm{s}^2}\bigg\vert_{J_{\mathrm{s,res}}}, \quad
F({J}_\mathrm{f}) \equiv -2\vert \Psi_k \vert_{J_{\mathrm{s,res}}} .
    \label{eqn:F_and_G}
\end{equation}
Here, $J_{\mathrm{s, res}}({J}_\mathrm{f})$ is the resonant value of the slow action at fixed fast action, $\vert \Psi_k\vert_{J_{\mathrm{s,res}}}$ is the strength of the bar perturbation on resonance, and $\varphi \in [-\pi, \pi]$ is just a phase-shifted slow angle.
Note that in contrast with linear perturbation theory (\S\ref{sec:Linearly_Perturbed_Orbits}),
in this case if the perturbation strength is 
$\vert \delta \Phi \vert/\vert H_0\vert \sim \varepsilon \ll 1$, 
then the `perturbing' part of the Hamiltonian \eqref{eqn:resonant_Hamiltonian} has relative strength $\vert F/h\vert \sim   \mathcal{O}(\varepsilon^{1/2})$, and the timescale for evolution due to this perturbation is therefore $t_\mathrm{lib}\sim \varepsilon^{-1/2}t_\mathrm{dyn}$.
Appreciating this difference is crucial when it comes to developing an appropriate kinetic theory (see \S\ref{sec:Nonlin_Nonpert}).

\begin{figure}[hbtp!]
\centering
\includegraphics[width=0.425\linewidth]{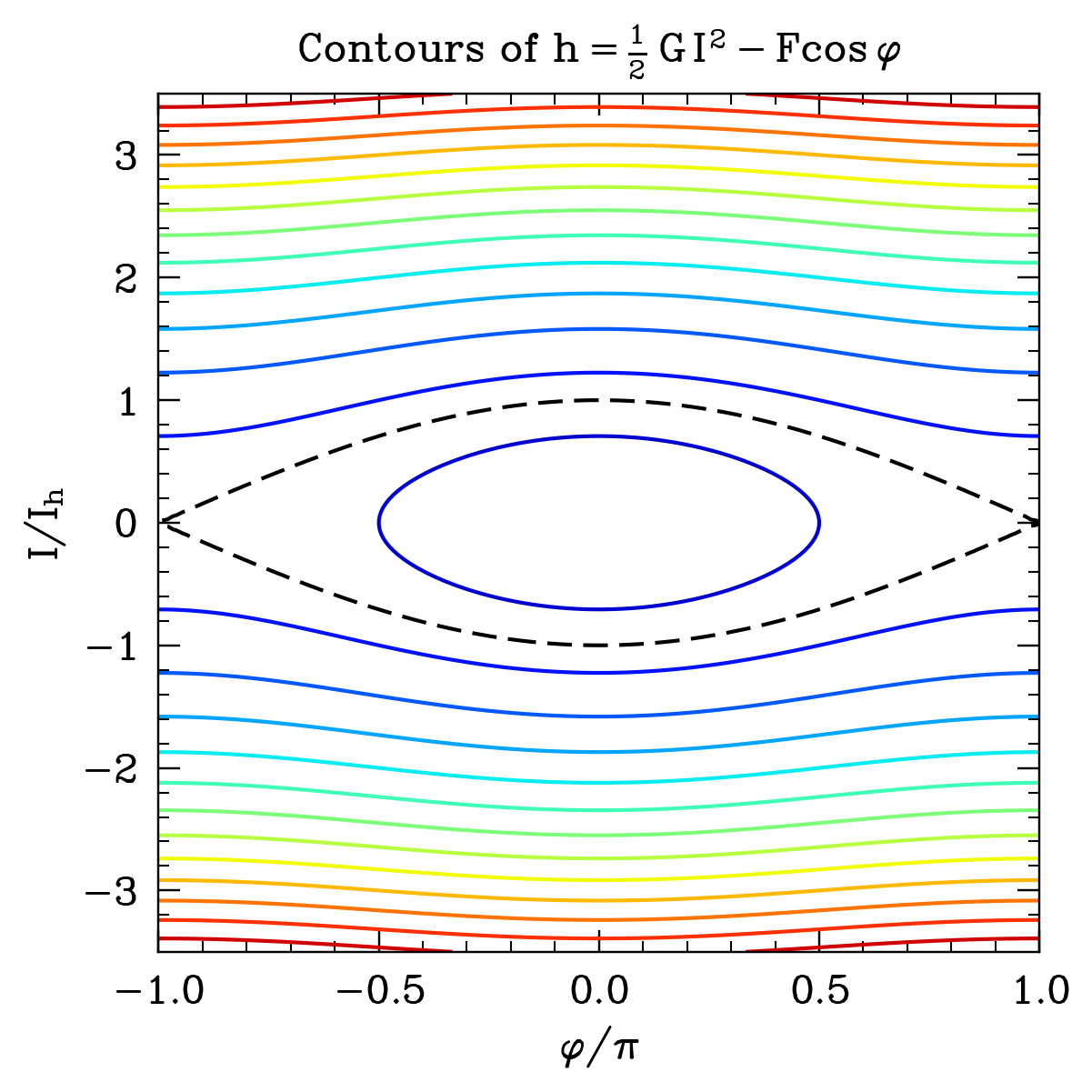}
\caption{Pendulum phase space. Colors show contours of constant $h(\varphi, I)$ (see equation~\ref{eqn:resonant_Hamiltonian}) for $k=1$. The black dashed line shows the separatrix between librating (i.e.\ trapped) and circulating (i.e.\ untrapped) orbit families.}
\label{fig:Phase}
\end{figure}

The variables $(\varphi, I)$ are canonical variables for the pendulum Hamiltonian~\eqref{eqn:resonant_Hamiltonian}.
The pendulum moves at constant `energy' $h$,  either on an untrapped `circulating' orbit with $h > F$ or a trapped `librating'  orbit with $h < F$ (the separatrix between these two families is $h = F$).
The \textit{libration period} for oscillations around $(\varphi, I) = (0,0)$ is
\begin{equation}
    t_\mathrm{lib} \equiv 2\pi / \omega_\mathrm{lib} \,\,\,\,\,\,\,\,\,\,\mathrm{where}\,\,\,\,\,\,\,\,\,\,\,\omega_\mathrm{lib}({J}_\mathrm{f}) \equiv \sqrt{kF G}.
    \label{eqn:t_libration}
\end{equation}
The maximum width in $I$ of the librating `island' is at $\varphi = 0$, where it spans $I \in [-I_\mathrm{h}, I_\mathrm{h}]$ and $I_\mathrm{h}$ is the \textit{island half-width}:
\begin{equation}
    I_\mathrm{h}({J}_\mathrm{f}) \equiv 2\sqrt{F / G},
    \label{eqn:half_width}
\end{equation}
which is independent of $k$.
The contours of this Hamiltonian are plotted in Figure~\ref{fig:Phase} for the case with $k=1$.
The separatrix between librating and circulating orbit families is shown with a black dashed line.

\subsubsection{Example: Corotation resonance, flat rotation curve}
\label{sec:CCF}

Let us illustrate the pendulum dynamics with a concrete example.  This requires stipulation of the resonance $\mathbf{N}$ under consideration.
Let us consider the corotation resonance (CR),
as defined in equation~\eqref{eqn:N_CR},
with the resonance location at the angular momentum value $L = L_\mathrm{CR}(J_R)$
for a given radial action $J_R$.
Then we have
\begin{subequations}
\begin{align}
&N_\phi = m, 
\,\,\,\,\,\,\,\,\,\,
   N_R = 0, 
   \,\,\,\,\,\,\,\,\,\,
k=1,
\label{eqn:para_trap_1}
\\
 &  J_\mathrm{s} = L/m,
   \,\,\,\,\,\,\,\,\,\,
   J_\mathrm{f} = J_R,
\,\,\,\,\,\,\,\,\,\,
   I = (L-L_\mathrm{CR})/m,
   \label{eqn:para_trap_2}
\\
  & \varphi = m(\theta_\phi - \pattern t),
   \,\,\,\,\,\,\,\,\,\,
 \vert \Psi_1 \vert = \vert \delta\Phi_{(m,0)}\vert.
 \label{eqn:para_trap_3}
 \end{align}
 \label{eqn:def_para_trap}%
\end{subequations}
We must also stipulate a background potential; the simplest choice is the logarithmic halo potential
\eqref{eqn:Phi_Logarithmic}, which gives rise to a flat rotation curve with circular speed $V_0$.
With this choice, for initially circular orbits ($J_\mathrm{f} = J_R=0$) we get 
\begin{subequations}
\begin{align}
& L = Rv_\phi = RV_0,
\label{eqn:para_pendulum_1}
\\
& \Omega_\mathrm{c} (L) = V_0/R = V_0^2/L,
\label{eqn:para_pendulum_2}
\\
& G = -N_\phi^2 V_0^2/L_{\mathrm{CR}}^2 = -N_\phi^2/R_\mathrm{CR}^2.
\label{eqn:para_pendulum_3}
\end{align}
\label{eqn:para_pendulum}%
\end{subequations}
We now combine these two choices, i.e.\ we  consider circular, corotating orbits in a logarithmic halo.
In this case the slow action $J_\mathrm{s}$ is just proportional to the guiding radius $R$, with the  resonant value $J_\mathrm{s,res}$ occurring at radius $R_\mathrm{CR}$,  and the slow angle is (up to a phase shift) equivalent to $m[\phi - \int_0^t \md t' \pattern(t')]$.
Thus, in this case, one can think of the transformation into slow and fast angle-action variables simply as `moving into the rotating frame' (or in the plasma context, shifting to a frame in which the wave's phase velocity is zero).

With these choices, the libration time~\eqref{eqn:t_libration} can be estimated as 
\begin{subequations}
\begin{align}
    t_\mathrm{lib} & \approx \frac{t_\mathrm{dyn}}{m \, \varepsilon^{1/2}} = \frac{2\pi}{m \, \varepsilon^{1/2}\,\pattern}
    \label{eqn:t_lib_CR}
    \\
& \approx 1 \, \mathrm{Gyr} \times \bigg( \frac{m}{2} \bigg)^{-1}
\bigg( \frac{\varepsilon}{0.01} \bigg)^{-1/2}
\bigg( \frac{t_\mathrm{dyn}}{200\, \mathrm{Myr}} \bigg),
\label{eqn:t_lib_CR_unit}
\end{align}
\label{eqn:t_lib_full}%
\end{subequations}
where $t_\mathrm{dyn} = 2\pi R/V_0$ is the circular orbital period at radius $R$,
and all quantities must be evaluated at the resonance location $R=R_\mathrm{CR}$.
Clearly, $t_\mathrm{lib}$ is always several times longer than the orbital period $t_\mathrm{dyn}$.
    Moreover, equation~\eqref{eqn:half_width} lets us estimate the radial width of the resonant island for these orbits:
\begin{subequations}
\begin{align}
    R_\mathrm{h} &= \frac{mI_\mathrm{h}}{V_0} \approx 2 \varepsilon^{1/2} R_\mathrm{CR}
    \label{eqn:half-radius}
        \\
&\approx 1.3 \, \mathrm{kpc} \times
\bigg( \frac{\varepsilon}{0.01} \bigg)^{1/2}
\bigg( \frac{R_\mathrm{CR}}{6.5\, \mathrm{kpc}} \bigg),
\label{eqn:half-radius_units}
\end{align}
\label{eqn:half_radius_full}%
\end{subequations}
Realistic spirals like those in the Milky Way today have $\varepsilon \sim$ several percent or more\cite{Eilers2020-na}.
Thus, the resonant island at corotation can span one or more kiloparsecs in radius.  This must be borne in mind whenever one makes a `narrow resonance' approximation, since $R_\mathrm{h}$ can also be comparable to the radial scale over which the galaxy's underlying density profile changes significantly ($\approx 3$ kpc for the stellar disk of the Milky Way).

\begin{figure}[htbp!]
\centering
\includegraphics[width=0.45\linewidth]{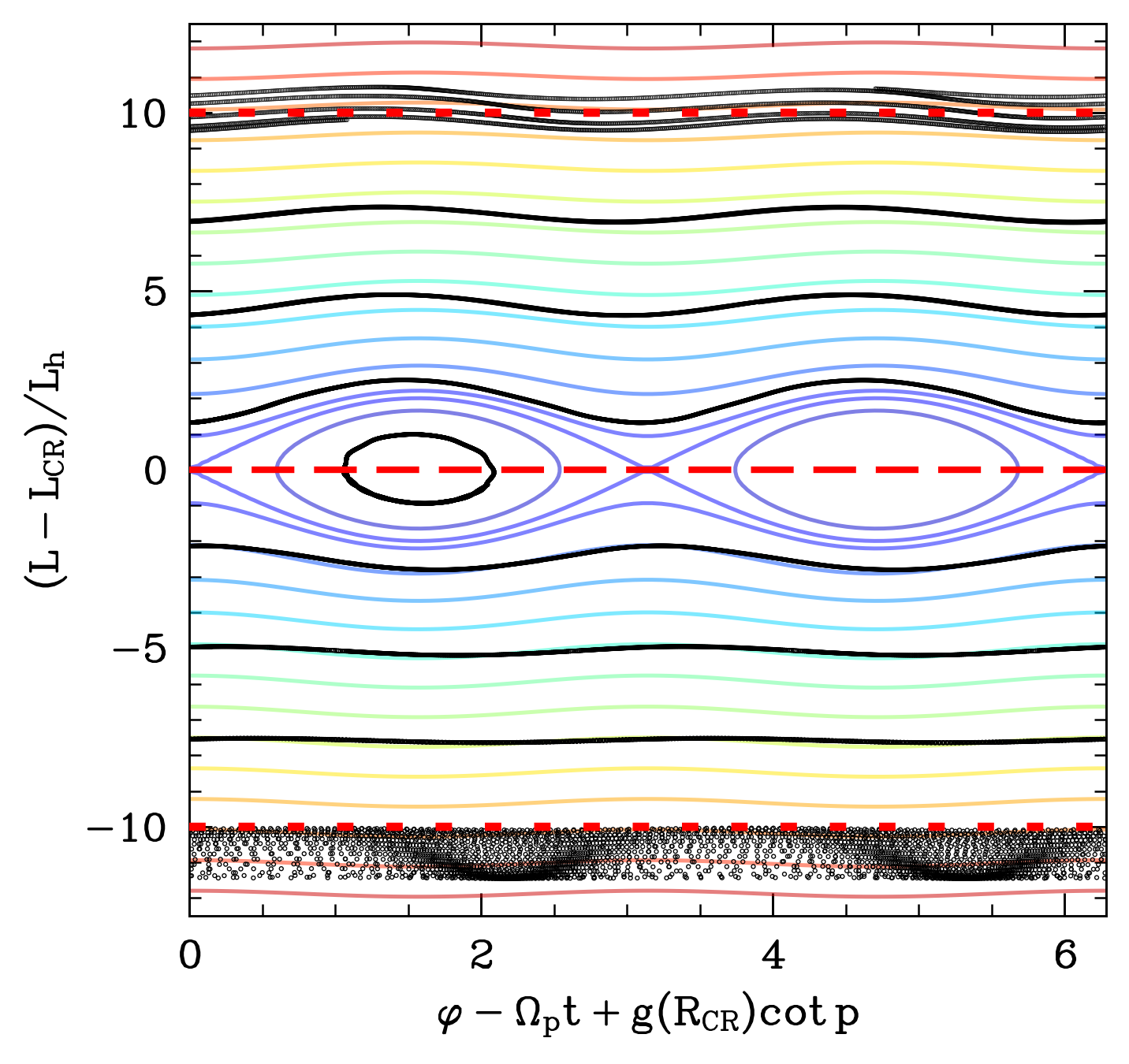}
\caption{The same orbits as in Figure~\ref{fig:3x3_rotating}, but plotted in the space of azimuthal angle (in the rotating frame) and angular momentum (relative to the corotation resonance value). Up to constants, these are equivalent to the slow angle and slow action at corotation (equations~\ref{eqn:slow_angle} and~\ref{eqn:slow_action}).
The background colored lines are contours of the associated pendulum Hamiltonian (c.f. Figure~\ref{fig:Phase}).
The dashed and dotted red lines have the same meaning as in Figure~\ref{fig:3x3_rotating}.}
\label{fig:Pendulum_m_2_p_60.0}
\end{figure}

As a more quantitative example, let us apply our pendulum formalism to the nine orbits shown in Figure~\ref{fig:3x3_rotating}. 
In Figure~\ref{fig:Pendulum_m_2_p_60.0} we 
show these nine trajectories through the space of azimuthal 
angle $\phi - \pattern t + \ln(R_\mathrm{CR}/R_0) \cot p$
and angular momentum $L$.
Up to constants, these are equivalent to the slow angle-action pair $(\varphi, I)$ near corotation.
With the colored lines we plot contours of the pendulum Hamiltonian (c.f.\ Figure~\ref{fig:Phase})
 using the parameters for the corotation resonance (equations \ref{eqn:def_para_trap}-\ref{eqn:para_pendulum}).
 The dashed and dotted red lines again show the ILR, CR and OLR from bottom to top respectively.
We see that the pendulum approximation does a good job of predicting the orbit shapes near corotation. The slight mismatch in the orbit shapes is mostly because the pendulum theory ignores all but one Fourier harmonic $k$ (as such, the agreement with theory gets better (resp.\@ worse) if we use more open (resp.\@ tight) spirals, i.e.\ take $p> 60^\circ$ (resp.\@ $p < 60^\circ$)).
The pendulum approximation can even work quite well \textit{far away} from the CR.  Mostly this is because away from corotation, orbits are not perturbed much in angular momentum so the approximation $L=$ const. is a good one.
However, we see that the pendulum approximation begins to break down close to the ILR and OLR, where resonant effects not associated with corotation start to play an important role in the dynamics. Near those resonances, angular momentum is clearly not conserved, and to describe the dynamics accurately one must instead shift to the proper slow angle-action coordinates appropriate to each resonance.

\subsubsection{Example: Radial migration}
\label{sec:Radial_Migration}

\textit{Radial migration} is the name given to any dynamical process that takes a star on a circular or nearly-circular orbit in a galactic disk and alters its angular momentum (or equivalently its guiding radius) while producing little or no change in its radial action.
The ubiquity of radial migration in galactic disks, and the importance of this especially for the evolution of chemical gradients, was first appreciated by~\citet{Sellwood2002-lv}.
Here we perform two numerical experiments similar to the ones shown in their paper,
in order to demonstrate the difference between linear (\S\ref{sec:Linearly_Perturbed_Orbits}) and nonlinear (\S\ref{sec:Pendulum_Approximation}) perturbations.

We consider a similar setup to the one we described in \S\ref{sec:Example_Perturbed_Orbits}
by integrating an array of test particles in the logarithmic halo potential $\overline{\Phi}$ with an accompanying spiral potential perturbation $\delta \Phi$,
except this time we modify the potential perturbation~\eqref{eqn:Phi_Spiral} by multiplying it by a Gaussian in time:
\begin{equation}
    \delta \Phi(\phi, R, t) = - \varepsilon  V_0^2 \cos[\alpha\ln (R/R_0) - m(\phi-\pattern t-\phi_\mathrm{p})] \times \me^{-(t-t_\mathrm{peak})^2/(2\,\tau^2)}.
    \label{eqn:Phi_Spiral_Transient}
\end{equation}
This corresponds to a transient spiral which grows and decays on the timescale $\approx \tau$, and whose amplitude reaches its peak at $t=t_\mathrm{peak}$.
All other parameters are set to the same values as in Figure~\ref{fig:3x3_inertial}.
With these parameters we can use equation~\eqref{eqn:t_lib_CR} to estimate the libration time at peak spiral amplitude to be $t_\mathrm{lib} \approx 5T_\mathrm{p}$.
For the initial conditions of the stars 
we use circular orbits with radii randomly sampled from the range $R(0)/R_0 \in [4, 9]$, which is centered on the corotation radius but does \textit{not} reach as far as the ILR/OLR.
We choose the initial azimuthal angles $\phi(0)$ randomly in $[0,2\pi]$.
We  integrate for a total time of  $10T_\mathrm{p}$, and always set the time of peak amplitude to be half-way through the integration, $t_\mathrm{peak}=5T_\mathrm{p}$.

\begin{figure}[htbp!]
    \centering
    \includegraphics[width=0.95\textwidth]{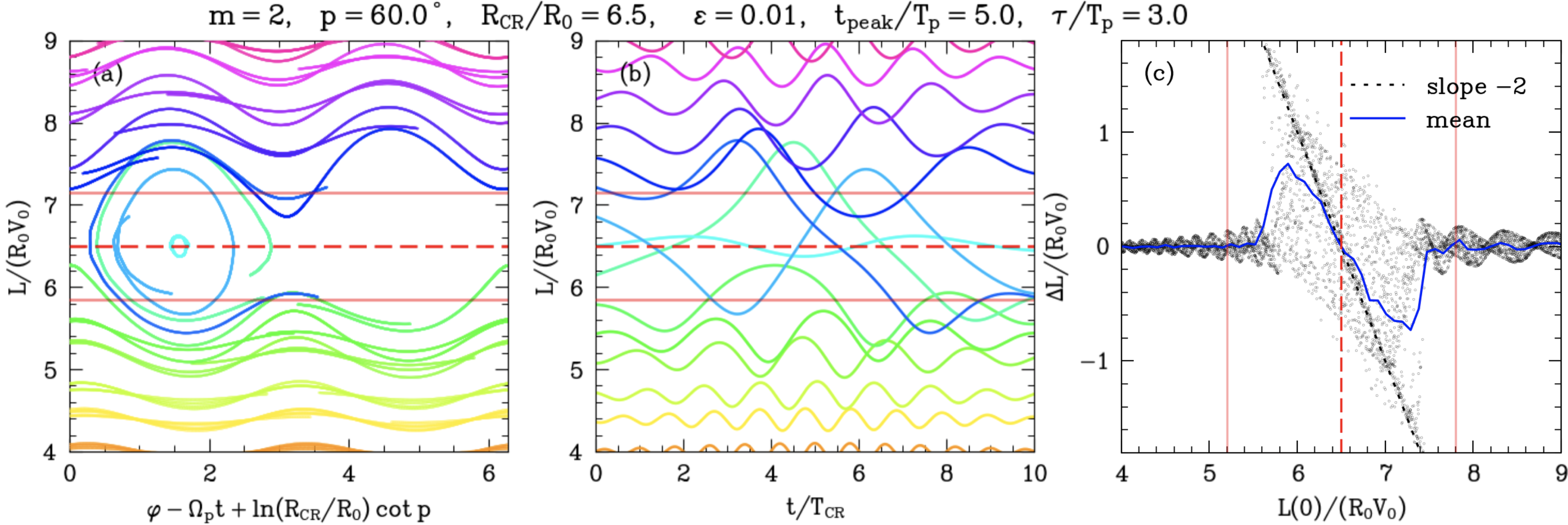}
    \caption{Radial migration in a galactic disk.  We integrate the orbits of 5\,000 stars that are initially on circular orbits with random azimuths $\phi(0)$ and radii chosen randomly in the range $R(0)/R_0 \in [4, 9]$.
    The stars are perturbed by a transient spiral potential~\eqref{eqn:Phi_Spiral_Transient} with lifetime $\approx \tau = 3T_\mathrm{p}$. Panels (a) and (b) show the evolution of the angular momenta of a handful of these stars as functions of azimuthal angle in the rotating frame and time respectively.  Panel (c) shows the 
    change in angular momentum experienced by each particle by the end of the simulation as a function of its initial angular momentum $L(0)$.
    In each panel the red dashed line corresponds to corotation,  and the solid pink lines are offset from corotation by approximately the resonance half-width $L_\mathrm{h} =m I_\mathrm{h}$ (equation \ref{eqn:half_width}) at peak spiral amplitude.}
    \label{fig:RadMig1}
\end{figure}
\begin{figure}[htbp!]
    \centering
    \includegraphics[width=0.965\textwidth]{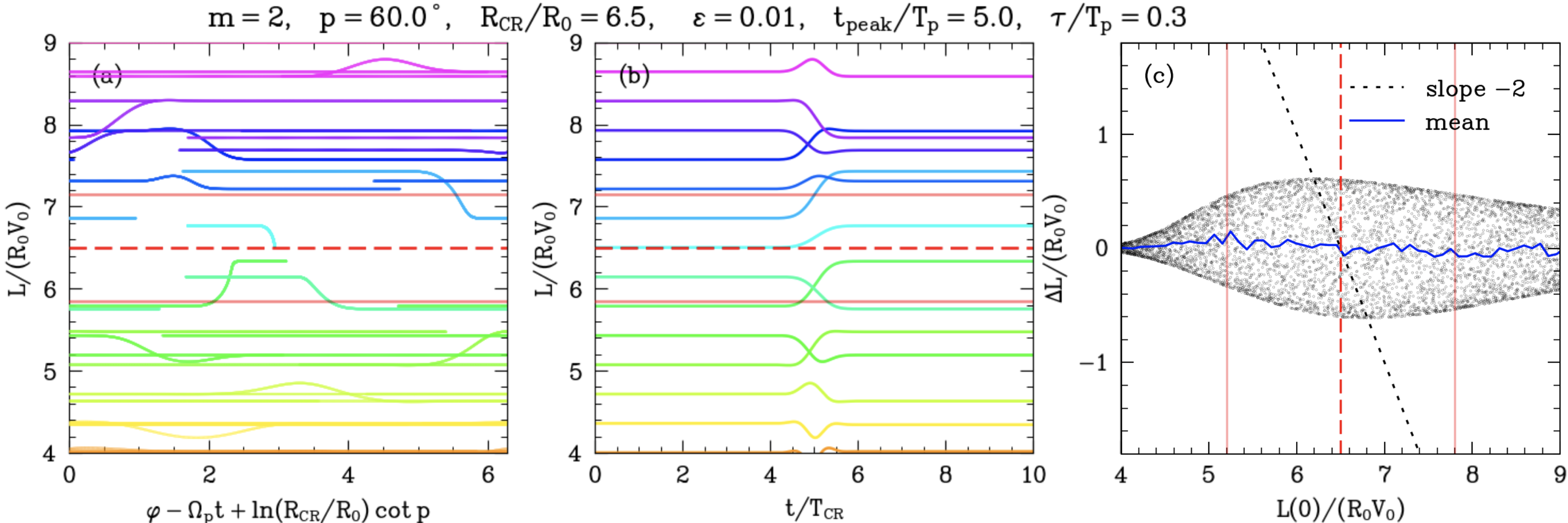}
    \caption{As in Figure~\ref{fig:RadMig1}, except the spiral grows and decays much more quickly, $\tau=0.3T_\mathrm{p}$.}
    \label{fig:RadMig2}
\end{figure}
First, we integrated the equations of motion of 5\,000 test stars in this spiral potential,
choosing the growth/decay time to be $\tau = 3T_\mathrm{p}$.
In panels (a) and (b) of Figure~\ref{fig:RadMig1} we show the resulting evolution of the angular momenta of a randomly selected subset of these stars, shown with different colors, as a function of azimuthal angle and time.
In panel (c) we show a scatter plot of the change in angular momentum, $\Delta L\equiv L(10T_\mathrm{p}) - L(0)$, for every star, as well as the mean $\langle \Delta L \rangle$ at each fixed $L(0)$.
The red dashed line in these panels shows the corotation resonance while the solid pink lines are displaced from corotation by $\approx L_\mathrm{h}$ (equation~\ref{eqn:half_width}),
the approximate resonance width at peak spiral amplitude.
We see that close to the resonance, the spiral traps stars and causes them to undergo large shifts in their angular momenta (up to the resonance width $\approx L_\mathrm{h}$). As the spiral's amplitude decays it `drops off' these stars at angular momenta which are often rather different from their initial values.
The blue line in panel (c) makes it clear that there is a preference for stars initially inside corotation ($L(0) < L_\mathrm{CR}$) to receive positive $\Delta L$, and vice versa. 
To understand this, note that in the limit where the libration time at peak amplitude is very short, a star which was originally located \textit{inside} corotation will librate back and forth across $L=L_\mathrm{CR}$ several times before being deposited at some new $L = L(0)+\Delta L$.
This new $L$ could be either inside or outside $L_\mathrm{CR}$ with roughly equal probability, so on average the star will have moved \textit{out} ($\Delta L > 0$).
Meanwhile, far from the resonance the stars respond roughly adiabatically to the perturbation and so $\Delta L$ is clustered around zero.
We also mention that in panel (c), many stars gather on the line with slope $-2$ that passes through $\Delta L = 0$ at corotation.  This line corresponds to stars that are initially located at $L(0) = L_\mathrm{CR}-\ell$ being deposited at $L_\mathrm{CR}+\ell$\cite{petit1986satellite,hasegawa1990distant}.

Second, in Figure~\ref{fig:RadMig2} we rerun the same experiment except with a much shorter-lived spiral,
$\tau = 0.3T_\mathrm{p}$.
We see from panel (a) that in this case, no orbits manage to undergo a full libration inside the trapping region.
As a result, there is no preferential direction of the transport of stellar angular momenta across the resonance, so the
the mean $\Delta L$ in panel (c) is approximately zero for all initial $L(0)$. In fact, since the perturbation lasts for so short a time, the resonance location is not so significant, meaning the distribution of $\Delta L$ values does not depend particularly strongly on $L(0$).
In fact,  the envelope of $\Delta L$ values is significantly broader far away from corotation than it was in Figure~\ref{fig:RadMig1} (so the variance $\langle (\Delta L)^2 \rangle$ is relatively large even though $\langle \Delta L \rangle$ is close to zero).

From the results of these numerical experiments we can extract two key physical principles, which apply 
ubiquitously in the kinetic theory of both stellar systems and plasmas:
\begin{itemize}[itemsep=0pt, topsep=4pt]
\item The efficiency with which different perturbations shuffle the integrals of motion of a distribution of particles depends on the spatio-temporal correlation spectrum of those perturbations (just as in, e.g.\@, the classic stochastic acceleration study of~\citet{Sturrock1966}). It is therefore imperative to develop a theory which can account for the \textit{statistical} evolution of a distribution of particles when exposed to such a perturbation spectrum, be it externally or self-consistently generated. This is what we will do in the rest of this tutorial article, starting in \S\ref{sec:secular}.
\item  If a single coherent perturbation dominates the potential $\delta \Phi$ over a timescale $\gtrsim t_\mathrm{lib}$ then it will trap resonant orbits nonlinearly, and this may drive the system down a substantially different evolutionary path compared to the case with no trapping. But if some process is able to interrupt the trapping process on a timescale $\ll t_\mathrm{lib}$, then nonlinear trapping can be ignored and one can apply linear perturbation theory (\S\ref{sec:Linearly_Perturbed_Orbits}) with confidence. This `interruption' can be due to the rapid decay of the perturbation as in the example above, or because of a time-dependent $\pattern$
and the accompanying movement of the resonance location through phase space\cite{Tremaine1984-wt,chiba2023dynamical},
or it can be due to some collisional/diffusive process which kicks particles in and out of the resonance region. We address the last of these possibilities in \S\ref{sec:Resonance_Diffusion}.
We also mention that the interruption of nonlinear trapping by time-dependent wave amplitudes, collisionality, and so on is 
very commonly studied in plasma theory\cite{johnston1971dominant,Auerbach1977-xe,Duarte2019-mi,Tolman2021-ly}.
\end{itemize}

\section{Fundamentals of kinetic theory}
\label{sec:secular}

In \S\ref{sec:Basics}, we introduced the idea
that an arbitrary gravitationally-bound ensemble of stars reaches virial equilibrium
after a few dynamical timescales $\tdyn$.
It then moves through a series of 
such equilibria on the relaxation timescale $\trelax \gg \tdyn$.
Then, in \S\S\ref{sec:Mean_Field_Dynamics}-\ref{sec:P_Theory},
we saw how to describe the quasiperiodic orbits of stars in smooth, regular, time-independent potentials $\overline{\Phi}$, became familiar with the machinery of angle-action variables, and discussed the  modification of orbits by some perturbation $\delta\Phi$.   Here we will set up the remainder of the article
by bringing these two sets of ideas together. 
Our aim is to describe a quasi-equilibrium stellar system as a collection of quasiperiodic orbits, 
and then to develop a quantitative theory of how fluctuations in the true gravitational potential cause
that quasi-equilibrium to change slowly over time.

\subsection{Klimontovich description}

We start with the microscopic \textit{Klimontovich} (DF),
or empirical DF,
\begin{equation}
  \label{eq:microscopic-distribution}
  \micro{f}(\br,\bv,t)
  = m \sum_{i=1}^N\delta\bigl(\br - \br_i(t)\bigr)
  \delta\bigl(\bv - \bv_i(t)\bigr).
\end{equation}
(Here and in what follows, the subscript `M' stands for `microscopic').
This function encodes the exact position $\br_i(t)$ and velocity $\bv_i(t)$ 
of every star 
$i$ at time $t$, which are themselves the exact solutions to the Newtonian equations of motion
~\eqref{eqn:Newton}.
Therefore, for indistinguishable particles,
the Klimontovich function~\eqref{eq:microscopic-distribution}
contains complete information about the system.
But at this stage it is totally useless because if we 
knew all this information we would not need kinetic theory.  Nevertheless, 
describing the system in this way is what will allow us to most conveniently formulate that kinetic theory.

Note that $f$ is normalized such that ${ \int \md \br\md \bv f = Nm = M }$, the total mass of the stars.
Thus, if there is no external forcing, the exact microscopic gravitational potential is
\begin{equation}
  \label{eq:microscopic-potential}
  \micro{\Phi}(\br,t)
  = \int\md{\br}' \md \bv' \, \psi(\br,\br')
  \micro{f}(\br',\bv',t),
\end{equation}
where $\psi(\br, \br') =-G/|\br-\br'|$.
This is simply the formal solution of the
Poisson equation 
\begin{equation}
    \nabla^2 \micro{\Phi} = 4 \pi G \micro{\rho},
    \end{equation}
    where 
the microscopic mass density is
${ \micro{\rho}=\int\md \bv\,\micro{f} }$.
It follows that every particle $i$ satisfies Hamilton equations~\eqref{eqn:Hamilton},
namely\footnote{Note the subtlety associated with self-interactions, i.e.\ contributions from $\psi (\br_{i}, \br_{i})$. As long as one assumes that there are no self-forces, this is fine.}
\begin{subequations}
\begin{align}
\frac{\md \br_i}{\md t} & = \frac{\p \micro{H}(\br_i, \bv_i, t)}{\p \bv_i},
\label{eqn:Hamilton_i_r}
    \\
     \frac{\md \bv_i}{\md t} & = -\frac{\p \micro{H}(\br_i, \bv_i, t)}{\p \br_i},
     \label{eqn:Hamilton_i_v}
     \end{align}
\label{eqn:Hamilton_i}%
\end{subequations}
where $\micro{H}$ is the microscopic one-particle Hamiltonian
\begin{equation}
  \label{eq:self-consistent-hamiltonian}
  \micro{H}(\br, \bv, t) = \half \bv^{2} + \micro{\Phi}(\br, t) .
\end{equation}
Using equations~\eqref{eq:microscopic-distribution}-\eqref{eq:self-consistent-hamiltonian} 
it is now easy to check that $\micro{f}(\br, \bv, t)$ exactly satisfies
the \textit{Klimontovich equation}
\begin{equation}
\frac{\md \micro{f}}{\md t} =   \frac{\p \micro{f}}{\p t} + \bv\cdot\frac{\p \micro{f}}{\p \br}
  - \frac{\p \micro{\Phi}}{\p \br} \cdot \frac{\p \micro{f}}{\p \bv} = 0,
\end{equation}
which we notice can also
be written in a pleasingly coordinate-free form as
\begin{equation}
  \label{eq:klimontovich}
\frac{\p \micro{f}}{\p t} + [{\micro{f}},{\micro{H}}] = 0. 
\end{equation}
This form of the equation is particularly convenient when we wish to perform a canonical coordinate transform, e.g.\ from position and velocity to angle-action variables.
Physically, equation~\eqref{eq:klimontovich} just says that each infinitesimal `pixel' of phase-space fluid is invariant
if one follows it along the trajectory prescribed by the exact Hamiltonian $\micro{H}$ (Liouville's theorem).

\subsection{Angle-averaged and angle-dependent quantities}
\label{sec:mean_field_fluctuations}

We now take the formal step of 
splitting all quantities into an angle-averaged
part and an angle-dependent `fluctuation' (which at this stage is not necessarily small).
Thus we write the microscopic DF as
\begin{equation}
  \label{eq:mean-plus-fluctuations-f}
  \micro{f} = f_0 + \delta f,
\end{equation}
where
\begin{equation}
    f_0(\bJ, t) \equiv \int \frac{\md \btheta}{(2\pi)^d} \micro{f} (\btheta, \bJ, t).
\end{equation}
Obviously if we were to convert back to real space coordinates, then $f_0$, for a galactic disc, would be axisymmetric (independent of $\phi$), 
while all non-axisymmetric fluctuations would be contained in $\delta f$.
The equivalent in a homogeneous plasma would be a DF $f_0$ that depended only on velocity, with all position-dependent fluctuations carried in $\delta f$.
Similarly, the scalar potential $\micro{\Phi}$ as defined in
\eqref{eq:microscopic-potential} is a linear functional of the DF $\micro{f}$, as is the microscopic Hamiltonian $\micro{H}$ defined in~\eqref{eq:self-consistent-hamiltonian}. Thus, in
correspondence with~\eqref{eq:mean-plus-fluctuations-f} we may
write
\begin{equation}
  \label{eq:mean-plus-fluctuations-H}
  \micro{H} = H_0 + \delta\Phi.
\end{equation}
By construction, $H_0$ is the part of the Hamiltonian that only depends on actions $\bJ$, while any $\btheta$-dependence (and hence any nonaxisymmetry of the potential) is contained in $\delta \Phi$ (see \S\ref{sec:MF_and_Pert}).
In a homogeneous system, $H_0 = v^2/2$.
Since the transformation to angle-action coordinates (\S\ref{sec:angleaction}) is canonical, it conserves the infinitesimal volumes ${ \md \br \md \bv = \md \btheta \md \mathbf{J} }$ and so
the Klimontovich DF
\eqref{eq:microscopic-distribution} can be expressed as
\begin{equation}
  \micro{f}(\btheta,\mathbf{J},t)
  = m\sum_{i=1}^N\delta\bigl(\btheta - \btheta_i(t)\bigr)
  \delta\bigl(\mathbf{J} - \mathbf{J}_i(t)\bigr).
\end{equation}

We note the subtlety that even though all non-axisymmetry of the DF $\micro{f}$ is contained in $\delta f$, the quantity $\delta f$ may also contain an axisymmetric piece.  Related to this is the fact that
$f_0$ is \textit{not} equivalent to the axisymmetric part of $\micro{f}$.
This means that, for example, one cannot in general construct the axisymmetric part of the exact surface density distribution $\micro{\overline{\Sigma}} (R) = (2\pi)^{-1} \int \md \phi\int\md \bv \micro{f}$ of a disk from $f_0(\bJ)$ alone: one also needs to know the $\theta_R$-dependent part of the DF.

Let us briefly consider what happens if there are no non-axisymmetric fluctuations (which in reality would require $N \to \infty$).
Setting $\delta f = \delta \Phi = 0$ in~\eqref{eq:mean-plus-fluctuations-f}-\eqref{eq:mean-plus-fluctuations-H}
and plugging the results into~\eqref{eq:klimontovich}, we get\footnote{Equation~\eqref{eqn:CollBol} is usually called the Collisionless Boltzmann Equation in stellar dynamics, or the Vlasov equation in plasma kinetics. But in research papers one is equally likely to see either of those names given to the Klimontovich equation~\eqref{eq:klimontovich}, or to something else entirely, so caution is required.}
\begin{equation}
    \frac{\p f_0}{\p t} + [f_0, H_0] = 0.
    \label{eqn:CollBol}
\end{equation}
The left hand side of this equation can be thought of as a `mean field operator' which advects particles along their mean field trajectories; the equation therefore tells us that 
in the absence of fluctuations, the phase space density along mean field trajectories is conserved, as we would expect. 
In fact, since $f_0$ and $H_0$ only depend on $\bJ$, the Poisson bracket is zero (see equation \eqref{Poisson_invariant}): 
\begin{equation}
    [f_0, H_0] = 0.
    \label{eqn:CollBol_implication}
\end{equation}
This means that $\partial f_0/\partial t=0$, i.e.\ $f_0$ 
cannot change with time even at a fixed phase space location.
Put differently, any $f_0$ that is only a function of $\bJ$
is an equilibrium distribution in the absence of fluctuations, or what is called a \textit{collisionless equilibrium}, or equivalently a quasi-stationary distribution --- a results known as \textit{Jeans' theorem}. In a homogeneous stellar system or electrostatic plasma, 
equation \eqref{eqn:CollBol_implication} is equivalent to the requirement that $[f_0, \tfrac{1}{2} \mathbf{v}^2]=0$,
which is trivially true of any $f_0$ which only depends on velocity. In this case and in the case of inhomogeneous stellar systems, the fact that $f_0$ satisfies \eqref{eqn:CollBol_implication} does not mean that it is a \textit{stable} equilibrium; in general, determining whether some $f_0$ 
is linearly stable requires the linear response theory that we will develop in \S\ref{sec:Linear_Response}.

Let us now reinstate the fluctuations, in particular allowing $f_0$ to be time-dependent.
Plugging the expansions~\eqref{eq:mean-plus-fluctuations-f}, \eqref{eq:mean-plus-fluctuations-H} into~\eqref{eq:klimontovich}
and writing the Poisson bracket out explicitly as in~\eqref{Poisson_invariant},
we get, without any approximations,
\begin{subequations}
\begin{align}
    {} & \frac{\partial f_0}{\partial t}  = \frac{\partial}{\partial \bJ} \cdot \bigg( \int \frac{\md \btheta}{(2\pi)^d} \, \delta f \frac{\partial \delta \Phi}{\partial \btheta} \bigg) ,
    \label{eqn:Axisymmetric_Evolution}
\\
    {} & \frac{\partial \delta f}{\partial t} + \bOm \cdot \frac{\partial \delta f}{\partial \btheta} - \frac{\partial \delta \Phi}{\partial \btheta}\cdot \frac{\partial f_0}{\partial \bJ} = -\frac{\partial}{\partial \btheta} \cdot \left( \delta f \frac{\partial \delta \Phi}{\partial \bJ} \right) + \frac{\partial}{\partial \bJ} \cdot \left( \delta f \frac{\partial \delta \Phi}{\partial \btheta} - \int \frac{\md \btheta}{(2\pi)^d} \, \delta f \frac{\partial \delta \Phi}{\partial \btheta}\right),
    \label{eqn:Fluctuation_Evolution}
    \end{align}
    \label{eqn:Fluctuation_Together}%
\end{subequations}
where  $\bOm(\bJ) \equiv \partial H_0/\partial \bJ$.
Equation~\eqref{eqn:Axisymmetric_Evolution} tells us how the (nonlinear) coupling of nonaxisymmetric `fluctuations' $\delta f$ and $\delta \Phi$ drives evolution of the angle-averaged part of the DF $f_0$.
Note that this equation  takes the form of a continuity equation in action space, as it must, since we are not allowing stars to be created or destroyed. Meanwhile, equation~\eqref{eqn:Fluctuation_Evolution} tells us how the fluctuating part of the DF evolves over time.

If we are doing the problem self-consistently then we must also couple these two equations to the Poisson equation for the fluctuations, namely
\begin{equation}
  \label{eq:fluctuating-potential-angle-action}
  \delta \Phi(\btheta,\mathbf{J},t)
  = \int \md{\btheta}' \md{\mathbf{J}}' \,
  \psi(\btheta,\mathbf{J},\btheta',\mathbf{J}')
 \delta f(\btheta',\mathbf{J}',t).
\end{equation}
In plasma, the interaction kernel $\psi$ has no dependence on the `actions' (velocities), which makes the Poisson equation much easier to solve in practice.

\subsection{Fourier expansion}
\label{sec:Fourier}

At this point it is convenient to introduce a Fourier representation for our angle-dependent quantities.  That is, we use the periodicity of the angle variables $\btheta$ to express $\delta f$ and $\delta \Phi$ as 
Fourier series, like in equation~\eqref{eqn:Fourier_any}, namely
\begin{subequations}
\begin{align}
    \delta f(\btheta, \mathbf{J}, t) = & \sum_{{\mathbf{n}}} \delta f_{{\mathbf{n}}}(\mathbf{J}, t) \, \me^{\mi {\mathbf{n}}\cdot\btheta},
    \label{eqn:Fourier_deltaf}
    \\
        \delta \Phi(\btheta, \mathbf{J}, t) = & \sum_{{\mathbf{n}}} \delta \Phi_{{\mathbf{n}}}(\mathbf{J}, t) \, \me^{\mi {\mathbf{n}}\cdot\btheta},
        \label{eqn:Fourier_deltaPhi}
\end{align}
\label{eqn:Fourier_deltaF_deltaPhi}%
\end{subequations}
Let us now rewrite~\eqref{eqn:Axisymmetric_Evolution} and~\eqref{eqn:Fluctuation_Evolution} in Fourier form so that 
\begin{subequations}
\begin{align}
      & \frac{\p f_0}{\p t} = -\frac{\p }{\p \mathbf{J}} \cdot \sum_{{\mathbf{n}}} \mi {\mathbf{n}}\,  \delta \Phi^*_{{\mathbf{n}}}(\mathbf{J},t)
        \delta f_{{\mathbf{n}}}(\mathbf{J}, t) ,
        \label{eqn:Axisymmetric_Evolution_Fourier}
\\
& \frac{\p \delta f_{{\mathbf{n}}}}{\p t} + \mi {\mathbf{n}}\cdot \bOm  \delta f_{{\mathbf{n}}} -  \mi {\mathbf{n}}\cdot \frac{\p  f_0}{\p \mathbf{J}} \delta \Phi_{{\mathbf{n}}} = \mi \sum_{\bn'}\bn' \cdot \left[ \frac{\partial \delta f_{\bn-\bn'}}{\partial \bJ} \delta 
    \Phi_{\bn'}
    -
    \frac{\partial \delta \Phi_{\bn-\bn'}}{\partial \bJ}
    \delta f_{\bn'}\right].
\label{eqn:Fluctuation_Evolution_Fourier}
\end{align}
\label{eqn:Evolution_Fourier}%
\end{subequations}
Meanwhile the Fourier-transformed version of~\eqref{eq:fluctuating-potential-angle-action} reads
\begin{equation}
  \label{eq:fluctuating-potential-angle-action-Fourier}
\delta \Phi_{{\mathbf{n}}}(\mathbf{J},t)
  = (2\pi)^d \sum_{{\mathbf{n}}'} \int\md{\mathbf{J}}'\,
  \psi_{{\mathbf{n}} {\mathbf{n}}'}(\mathbf{J},\mathbf{J}') \, 
 \delta f_{{\mathbf{n}}'}(\mathbf{J}',t).
\end{equation}
The (bare) coupling coefficients
$\psi_{{\mathbf{n}}{\mathbf{n}}'} (\mathbf{J} , \mathbf{J}')$
encode, in essence, Poisson's equation in angle-action space, through the decomposition
\begin{equation}
\psi (\btheta , \bJ , \btheta^{\prime}, \bJ^{\prime}) = \sum_{\bn , \bn^{\prime}} \psi_{\bn \bn^{\prime}} (\bJ , \bJ^{\prime}) \, \me^{\mi (\bn \cdot \btheta - \bn^{\prime} \cdot \btheta^{\prime})} .
\label{eq:decomp_psi}
\end{equation}
In the limit of an homogeneous multi-periodic plasma
(\S\ref{sec:Homogeneous}), these coefficients are greatly
simplified and become
$ \psi_{{\mathbf{n}}{\mathbf{n}}'} \!\propto\! \delta_{{\mathbf{n}}}^{{\mathbf{n}}'} / |{\mathbf{n}}|^{2} $.
In that limit, note that (i) the only `resonances' permitted are for ${\mathbf{n}} = {\mathbf{n}}'$; and (ii) the coupling between particles does not depend on the actions $(\mathbf{J} , \mathbf{J}')$.  The fact that neither (i) or (ii) hold in general for realistic stellar systems makes linear response calculations much harder than in homogeneous plasmas (\S\ref{sec:Linear_Response}).

\subsection{Linear and quasilinear approximations}
\label{sec:Linear_and_Quasilinear}

The only assumption we have made in deriving equations~\eqref{eqn:Axisymmetric_Evolution_Fourier}-\eqref{eq:fluctuating-potential-angle-action-Fourier}
is that $H_0$ is integrable, i.e.\ that there exists a set of global angle-action coordinates $(\btheta, \bJ)$ such that $H_0=H_0(\bJ)$.
If this is true then equations~\eqref{eqn:Axisymmetric_Evolution_Fourier}-\eqref{eq:fluctuating-potential-angle-action-Fourier} are exact.
Sadly, they are also very difficult to solve.

To render them solvable, we will often make an additional assumption that \textit{angle-dependent fluctuations are small}, i.e.\ $\delta f \ll f_0$ and $\delta \Phi \ll H_0$.
With this, it follows naturally from
equations~\eqref{eqn:Axisymmetric_Evolution_Fourier}-\eqref{eqn:Fluctuation_Evolution_Fourier} that the fluctuation $\delta f$ evolves on the dynamical timescale $t_\mathrm{dyn} \sim 1/\vert \bOm \vert$, whereas the angle-averaged DF $f_0$ --- whose rate of change is \textit{quadratic} in the small fluctuations --- evolves on some much longer timescale $\trelax \gg \tdyn$. This timescale separation is what allows the system to move through a series of virial equilibria, each of which approximately satisfies equation~\eqref{eqn:CollBol}, as we described qualitatively in \S\ref{sec:Where_Now}. For self-consistent problems, $H_0[f_0]$ will also be changing on the timescale $\trelax$, and in practice every time $H_0$ changes significantly, we have to recompute the mapping $(\br , \bv) \to (\btheta, \mathbf{J})$.

Our assumption of weak fluctuations means that we now have a well-defined programme for solving the system 
\eqref{eqn:Axisymmetric_Evolution_Fourier}-\eqref{eq:fluctuating-potential-angle-action-Fourier}.
On the timescale $\sim t_\mathrm{dyn}$, we may \textit{linearize} equation~\eqref{eqn:Fluctuation_Evolution_Fourier} by throwing away the small terms on the right hand side:
\begin{align}
\frac{\p \delta f_{{\mathbf{n}}}}{\p t} + \mi {\mathbf{n}}\cdot \bOm  \delta f_{{\mathbf{n}}} -  \mi {\mathbf{n}}\cdot \frac{\p  f_0}{\p \mathbf{J}} \delta \Phi_{{\mathbf{n}}} = 0,
\label{eqn:Fluctuation_Evolution_Fourier_Linearized}
\end{align}
and then solve it
under the assumption that $f_0$ is fixed.  
If we are doing a self-consistent problem we must couple this linearized equation with the Poisson equation~\eqref{eq:fluctuating-potential-angle-action-Fourier}.
The simultaneous solution of these two linear equations is a nontrivial problem,
and is the subject of \S\ref{sec:Linear_Response}.
Supposing we can do this successfully,
we can then plug the solutions $\delta f$, $\delta \Phi$ into the right hand side of~\eqref{eqn:Axisymmetric_Evolution_Fourier} --- or, if appropriate, its ensemble-averaged cousin~\eqref{eqn:Axisymmetric_Evolution_Fourier_Ensemble} --- and solve for the evolution of $f_0$ on the much longer timescale $t_\mathrm{relax}$.  Such evolution of $f_0$ is called \textit{quasilinear}, since it involves the multiplication of two small quantities, each of which is determined using linear theory.

\begin{figure}[htbp!]
\centering
\includegraphics[width=0.95\linewidth]{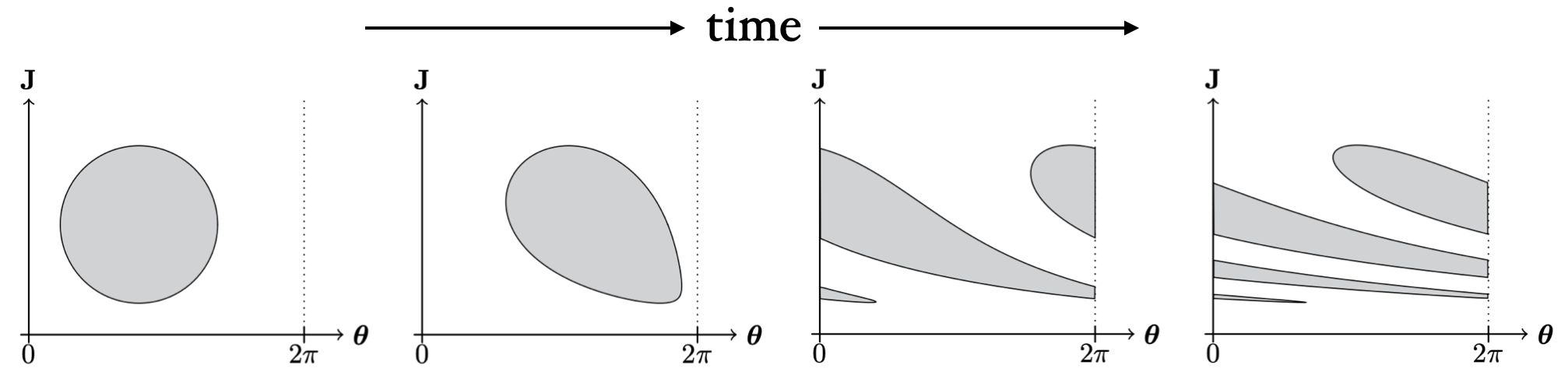}
\caption{Schematic illustration of phase mixing in angle-action space as a function of time. The gray blob represents some phase space DF fluctuation $\delta f$ evolving in the absence of external forces (first term on the right hand side of equation~\ref{eqn:linear_Vlasov_formal_solution}). Unperturbed trajectories follow horizontal lines in angle-action space (Figure \ref{fig:IllustrationAngleAction}b). Since the orbital frequencies $\bOm(\bJ)$ depend on $\bJ$,
particles of different actions 
dephase. This leads to the appearance of ever finer structures in phase space.
\label{fig:PhaseMixing}}
\end{figure}

We emphasize that the linear equation~\eqref{eqn:Fluctuation_Evolution_Fourier_Linearized} is just the kinetic version of the linear perturbation theory that we wrote down for the orbits of individual stars in \S\ref{sec:Linearly_Perturbed_Orbits}.
To see this, note that the formal solution to~\eqref{eqn:Fluctuation_Evolution_Fourier_Linearized} is just
\begin{align}
    \delta f_{{\mathbf{n}}}(\mathbf{J}, t) =  \, \delta f_{{\mathbf{n}}}(\mathbf{J}, 0) \, \me^{- \mi {\mathbf{n}}\cdot\bOm t}
    + \, \mi {\mathbf{n}}\cdot \frac{\p f_0}{\p \mathbf{J}}\int_0^t \md t' \, \me^{- \mi {\mathbf{n}}\cdot \bOm (t-t')} \delta\Phi_{{\mathbf{n}}}(\mathbf{J}, t').
    \label{eqn:linear_Vlasov_formal_solution}
\end{align}
The first term on the right hand side represents the ballistic evolution (i.e. the mean-field motion, along straight lines in angle-action space) of some initial fluctuation $\delta f(\btheta, \mathbf{J}, 0)$ as governed by the mean field Hamiltonian $H_0$, i.e.\ the \textit{phase mixing of unperturbed orbits} $\bJ = $ cst, $\btheta = \bOm t + $ cst. Phase mixing causes the initial fluctuation to oscillate rapidly in action (or, more properly, frequency) space for times $t\gg 1/\vert \bOm \vert$ --- see Figure \ref{fig:PhaseMixing} for illustration.
(Analogously in plasma, phase mixing consists of particles moving along $\bv = $ cst, $\br = \bv t + $ cst and therefore $\delta f(\br, \bv, 0)$ winds up to ever smaller scales in \textit{velocity} space).
The second term on the right hand side of \eqref{eqn:linear_Vlasov_formal_solution} tells us how the DF is modified by the potential perturbation $\delta\Phi$ \textit{as evaluated along the unperturbed trajectory};
in other words it is the result of applying an equation like~\eqref{eqn:deltaJ_1} to every star. Put differently, in any linear or quasilinear kinetic theory, the only particle trajectories that are accounted for are the ballistic unperturbed trajectories set by $H_0$ (straight lines in angle-action space) plus the small (sinusoidal) wiggles instigated by $\delta \Phi$. 
Quasilinear transport is then a consequence of resonances between 
the oscillatory forcing $\delta \Phi$ and these first-order wiggles --- hence enters perturbation theory at second order.

\subsection{Nonlinear and nonperturbative effects}
\label{sec:Nonlin_Nonpert}

One can imagine iterating the above procedure to higher orders in perturbation theory by plugging~\eqref{eqn:linear_Vlasov_formal_solution} back into the right hand side of~\eqref{eqn:Fluctuation_Evolution_Fourier}, calculating $\delta f_{\bn}$ to second order in the perturbation strength, and so on\cite{PichonAubert2006}.
At the level of particle orbits this would correspond to adding `sinusoidal disturbances on top of sinusoidal disturbances' over and over with ever smaller amplitude.
In the case where the perturbations $\delta \Phi$ are stochastic, this approach is usually given the name \textit{weak turbulence theory}. Weak turbulence is the subject of many a plasma monograph 
such as the excellent one by~\citet{kadomtsev1965plasma}, and underlies the resonance broadening theories pioneered by~\citet{Dupree1966-dl} and others\cite{Krommes2002-ka,Diamond2010-mh,Krommes2015-bd}. 
On the other hand, in the case where $\delta \Phi$ is deterministic, the second order calculation leads e.g.\ to the classic
\textit{plasma echo} phenomenon first pointed out by~\citet{Gould1967-dp}.

However, no matter how many iterations of this perturbative scheme one performs, it cannot account for fundamentally nonperturbative effects such as particle trapping by strong potential perturbations (\S\ref{sec:Pendulum_Approximation}). One
way to see this is to note that the perturbative iteration scheme accounts for effects that are of order $\varepsilon$, $\varepsilon^2$, $\varepsilon^3$, etc., where $\varepsilon \sim \vert \delta \Phi/H_0\vert$ measures the strength of the potential perturbations. Nonlinear trapping, though, enters the dynamics at order $\varepsilon^{1/2}$ (\S\ref{sec:Pendulum_Approximation}), and so cannot be recovered by summing any finite number of perturbative terms. Another way to arrive at the same conclusion is to note that trapped orbits (like the one in Figure~\ref{fig:3x3_rotating}e) 
are topologically distinct from untrapped ones (c.f.\ Figure~\ref{fig:3x3_rotating}f), and no amount of iterative sinusoidal deformations of the latter can ever reproduce the former.

These fundamentally nonperturbative effects can be very important for the evolution of the system as a whole, but a kinetic theory that accounts for them cannot start from equation~\eqref{eqn:Fluctuation_Evolution_Fourier_Linearized}. 
In the case where $\delta \Phi$ is stochastic, this regime is termed \textit{strong turbulence}, and formulating an accurate kinetic theory turns out to be very difficult even in the simplest plasmas\cite{Krommes2002-ka}. 
Luckily, in galaxies, $\delta \Phi$ is often dominated by a single coherent perturbation, such as a strong bar or spiral arm.  In this case a simple kinetic theory incorporating trapping \textit{can} be developed using the pendulum approximation (\S\ref{sec:Pendulum_Approximation}), and this will be the subject of \S\ref{sec:Nonlinear_Kinetics}. Moreover, just as we discussed at the end of \S\ref{sec:Radial_Migration},
if nonlinear effects are somehow quenched or interrupted on a short
enough timescale then the problem is \textit{relinearized} and
\eqref{eqn:linear_Vlasov_formal_solution} is rendered (approximately) valid.
We will see a concrete example of this in \S\ref{sec:Resonance_Diffusion}.

\section{Self-consistent linear response theory}
\label{sec:Linear_Response}

In this section we study the self-consistent response of a system to weak perturbations, using the \textit{linear} approximation (\S\ref{sec:Linear_and_Quasilinear}).
We assume throughout that the angle-averaged DF $f_0$
is time-independent, which means that we need to solve equations~\eqref{eq:fluctuating-potential-angle-action-Fourier} and~\eqref{eqn:Fluctuation_Evolution_Fourier_Linearized}
simultaneously, but we can ignore equation~\eqref{eqn:Axisymmetric_Evolution_Fourier}.

For simplicity we will develop the linear theory in the absence of any possible external perturbations, i.e.\ we will assume the only potential fluctuations are derived self-consistently internally from the angle-dependent part of the DF $\delta f$.
However, the case where external perturbations are present is not essentially different (see e.g.\ Chapter 5 of BT08), and we will show a relevant example in \S\ref{sec:Example_Magorrian}.

\subsection{Laplace transforming the linearized Vlasov--Poisson system}
\label{sec:Linear_response_Fourier_Laplace}

Equations~\eqref{eq:fluctuating-potential-angle-action-Fourier} and~\eqref{eqn:Fluctuation_Evolution_Fourier_Linearized} constitute an initial value problem, and are naturally solved via Laplace transform\cite{riley2006mathematical}.  
Let us therefore define the Laplace transform of a function $h(t)$, and its inverse via
\begin{subequations}
\begin{align}
  &  \widetilde{h}(\omega) \equiv \int_0^\infty \md t \, h(t) \, \me^{ \mi \omega t} , 
  \label{eqn:Laplace_Transform}
  \\
  & h(t) = \frac{1}{2\pi}\int_{\mi \sigma - \infty}^{\mi \sigma+\infty} \md \omega \, \widetilde{h}(\omega) \, \me^{ -\mi \omega t} .
    \label{eqn:Inverse_Laplace_Transform}
\end{align}
\label{eqn:def_Laplace}%
\end{subequations}
Recall that the Laplace transform~\eqref{eqn:Laplace_Transform} is defined for complex $\omega$ in the upper-half plane, or more precisely for $\mIm \,\omega > \sigma$ where $\sigma$ is chosen large enough for the integral to converge at $t\to \infty$.  In the inverse transform~\eqref{eqn:Inverse_Laplace_Transform}, the `Bromwich contour'
runs along the line Im$\,\omega = \sigma$ from $\mRe\, \omega = -\infty$
to $\mRe \, \omega = +\infty$, but
can be deformed at will,
provided it never crosses over any singularities of $\widetilde{h}(\omega)$.

Laplace transforming equation~\eqref{eqn:Fluctuation_Evolution_Fourier_Linearized}, we get
\begin{equation}
  - \delta f_{{\mathbf{n}}}(\mathbf{J}, 0)  - \mi (\omega - {\mathbf{n}} \cdot \bOm) \delta \widetilde{f}_{\mathbf{n}} (\mathbf{J}, \omega) =
  \mi {\mathbf{n}} \cdot \frac{\p f_0}{\p \mathbf{J}}
  \delta \widetilde{\Phi}_{{\mathbf{n}}}(\mathbf{J}, \omega) ,
\end{equation}
with $\delta f_{\mathbf{n}} (\bJ , 0)$ is
the initial fluctuation in the DF\@.
The solution to this equation is
\begin{equation}
  \delta \widetilde{f}_{\mathbf{n}} (\mathbf{J}, \omega) = -\frac{{\mathbf{n}} \cdot \p f_0/\p \mathbf{J}}{\omega - {\mathbf{n}}\cdot \bOm} \delta \widetilde{\Phi}_{\mathbf{n}}(\mathbf{J}, \omega) - \frac{\delta f_{{\mathbf{n}}}(\mathbf{J}, 0)}{\mi (\omega - {\mathbf{n}}\cdot \bOm)}.
    \label{eqn:linear-solution-Fourier-Laplace}
\end{equation}
(Note that by taking the inverse Laplace transform of this we can recover the formal solution~\eqref{eqn:linear_Vlasov_formal_solution}).
Meanwhile, the Laplace-transformed version of equation~\eqref{eq:fluctuating-potential-angle-action-Fourier} is
\begin{equation}
  \label{eq:fluctuating-potential-angle-action-Fourier-Laplace}
 \delta \widetilde{\Phi}_{\mathbf{n}}(\mathbf{J},\omega)
  = (2\pi)^d \sum_{{\mathbf{n}}'} \int\md{\mathbf{J}}'\,
\psi_{\mathbf{n} \mathbf{n}'}(\mathbf{J},\mathbf{J}')
 \delta \widetilde{f}_{\mathbf{n}'}(\mathbf{J}',\omega).
\end{equation}
Let us now put equations~\eqref{eqn:linear-solution-Fourier-Laplace} and~\eqref{eq:fluctuating-potential-angle-action-Fourier-Laplace} together in order to eliminate  $\delta \widetilde{f}_{\mathbf{n}} (\bJ , \omega)$.  The result is the following integral equation for the potential fluctuation $\delta \widetilde{\Phi}_{{\mathbf{n}}}(\mathbf{J}, \omega)$:
\begin{align}
& \delta \widetilde{\Phi}_{{\mathbf{n}}}(\mathbf{J}, \omega)  =  -(2\pi)^d\sum_{{\mathbf{n}}'}\int \md \mathbf{J}' \frac{{\mathbf{n}}' \cdot \p f_0/\p \mathbf{J}'}{\omega - {\mathbf{n}}'\cdot \bOm'} \, \psi_{{\mathbf{n}}{\mathbf{n}}'}(\mathbf{J},\mathbf{J}') \,
      \delta \widetilde{\Phi}_{{\mathbf{n}}'}(\mathbf{J}', \omega) - (2\pi)^d
      \sum_{{\mathbf{n}}'}\int \md \mathbf{J}'\frac{\delta f_{{\mathbf{n}}'}(\mathbf{J}', 0)}{\mi (\omega - {\mathbf{n}}'\cdot \bOm')}
      \psi_{{\mathbf{n}}{\mathbf{n}}'}(\mathbf{J},\mathbf{J}'),
      \label{eqn:delta_Phi_integral_equation}
\end{align}
where $\bOm' \equiv \partial H_0(\mathbf{J}')/\partial \bJ'$.

\subsection{Dressed wakes and Landau modes}

Equation~\eqref{eqn:delta_Phi_integral_equation} is a linear operator equation of the form
\begin{equation}
    \vert \delta \widetilde{\Phi} (\omega) \rangle = \widehat{\msM} (\omega)  \vert \delta \widetilde{\Phi} (\omega) \rangle + \vert \delta \widetilde{\Phi}_\msource (\omega) \rangle,
    \label{eqn:Dyson}
\end{equation}
where we have written the `state' of the potential fluctuations at frequency $\omega$ as 
$\vert \delta \widetilde{\Phi} (\omega) \rangle$ and introduced the abstract linear operator 
$\widehat{\msM} (\omega)$.  Note that~\eqref{eqn:Dyson} is not merely a schematic representation of~\eqref{eqn:delta_Phi_integral_equation};
instead, it is what is known in quantum statistical mechanics as a \textit{Dyson equation}\cite{al1965kinetics,chen2016physics}.
One can in fact develop the entire response theory of inhomogeneous systems on the basis of the abstract Dyson equation
and operator theory\cite{Layzer1963-ml,Balescu1963-ye,kalnajs1971dynamics,kato2013perturbation} without reference to the explicit angle-action representation. We will not go 
into the details here, but instead just discuss the physics that can be gleaned from equation~\eqref{eqn:Dyson}.

On the right hand side of~\eqref{eqn:Dyson} we have the source term $\vert \delta \widetilde{\Phi}_\msource(\omega) \rangle$,
which corresponds to the second term on the right hand side of~\eqref{eqn:delta_Phi_integral_equation}.
This is the part of the potential fluctuations that arises purely from the inevitable, finite-$N$ discreteness noise $\delta f_{{\mathbf{n}}}(\mathbf{J}, 0)$, propagated ballistically along mean field trajectories. (Had we included an external potential perturbation in our calculation, it would just be added to this source term). The other term involves the operator $\widehat{\msM}(\omega)$ acting on the potential fluctuation $\vert \delta \widetilde{\Phi} (\omega)\rangle$. It tells us the extent to which
potential fluctuations arise because of perturbations in the DF \textit{which were themselves induced by potential fluctuations}.  Thus we are discussing \textit{collective effects}: perturbations to the DF that seed their own gravitational fields.
Glancing back at~\eqref{eqn:delta_Phi_integral_equation}, we expect these collective effects to be significant if there are fluctuations at frequencies $\omega$ which are close to some of the available stellar frequencies ${\mathbf{n}}' \cdot \bOm'$.
These fluctuations will be the all more efficient provided they come from stars at locations in action space $\mathbf{J}'$ where the mean field DF has a large gradient $\p f_0/\p \mathbf{J}'$.

Another way to understand \eqref{eqn:Dyson} is as follows.
We start with an initial fluctuation
$\vert \delta \widetilde{\Phi}_\msource (\omega) \rangle$ and then ask when we ask what is the self-consistent response of the system is to itself.
First, the fluctuation $\vert \delta \widetilde{\Phi}_\msource (\omega) \rangle$ provokes a response $ \widehat{\msM} (\omega) \vert \delta \widetilde{\Phi}_\msource (\omega) \rangle$. The response to this response is $\widehat{\msM} (\omega) \widehat{\msM} (\omega) \vert \delta \widetilde{\Phi}_\msource (\omega) \rangle$, and so on.
Equating this string of self-consistent responses to the overall potential, we have 
$\vert \delta \widetilde{\Phi} (\omega) \rangle = [1+\widehat{\msM} (\omega) + \widehat{\msM} (\omega)\widehat{\msM} (\omega) + ... ]\vert \delta \widetilde{\Phi}_\msource (\omega) \rangle = [1-\widehat{\msM} (\omega)]^{-1}\vert \delta \widetilde{\Phi}_\msource (\omega) \rangle$, which we can multiply by $[1-\widehat{\msM} (\omega)]$ to arrive at \eqref{eqn:Dyson}.

How do we solve equation~\eqref{eqn:Dyson}? In homogeneous plasma theory it is easy: in that case, the operator notation we have employed here is unnecessary because $\widehat{\msM}(\omega)$ is a scalar (up to proportionality constants, it is equal to $1$ minus the dielectric function $\epsilon_{{\mathbf{k}}}(\omega)$) which `acts' on the potential fluctuation by simple multiplication.
In inhomogeneous systems we are not so lucky. Ultimately this is because,
as we warned in \S\ref{sec:Sum_Out}, the potential fluctuation 
$\delta \Phi(\btheta, \mathbf{J})$ depends on both phase space coordinates \textit{and} momenta, and so $\widehat{\msM}(\omega)$
acts upon these fluctuations in a non-trivial long-range fashion. 
In practice this means we must solve the operator equation by casting
$[\msI - \widehat{\msM}(\omega)]$ in a matrix representation.

The usual technique for doing this is the so-called \textit{biorthogonal basis method}
originally developed by~\citet{Kalnajs1976-gg}. 
The idea is that we can 
expand an arbitrary potential fluctuation in real space in terms of some set of basis functions $\Phi^{(p)}(\br)$,
and similarly for a density fluctuation using some $\rho^{(p)}(\br)$. 
For self-consistent problems, it is intelligent to choose these functions in pairs such that they
solve the Poisson equation,
$ \nabla^2 \Phi^{(p)}=4\pi G\rho^{(p)}$, and are orthogonal, i.e.\ $\int \!\! \md \br \, \Phi^{(p) *}(\br) \, \rho^{(q)} (\br) = 0$ unless $p=q$.
Many such sets are known, and the precise set one should use depends on the problem at hand (but they usually consist of familiar objects like spherical harmonics, Bessel functions, etc.).
Physically, each discrete pair $(\Phi^{(p)}, \rho^{(p)})$ corresponds to a way in which the stellar system can oscillate self-consistently, and the full response of the system can be built up by superposing these discrete oscillations. 
This is reminiscent of elementary quantum mechanics, in which 
one builds the wavefunction of an electron in the hydrogen atom 
as a superposition of orthogonal eigenfunctions.
Mathematically, once we have constructed such a basis, we can project our linear response equations onto it, thereby translating a problem of 
coupled differential equations into one of linear algebra.

The details of the biorthogonal basis method are rather technical, and unimportant for most of what follows, so we relegate them to Appendix~\ref{sec:BasisMethod}, but 
the explicit matrix form of the operator $\widehat{\msM}(\omega)$ that results is
\begin{equation}
\msM_{pq} (\omega) = \frac{(2 \pi)^{d}}{\mcE} \sum_{{\mathbf{n}}} \!\! \int \!\! \md \mathbf{J} \, \frac{{\mathbf{n}} \cdot \p f_{0} / \p \mathbf{J}}{\omega - {\mathbf{n}} \cdot \bOm } \, \Phi^{(p) *}_{{\mathbf{n}}} (\mathbf{J}) \, \Phi^{(q)}_{{\mathbf{n}}} (\mathbf{J}) ,
\label{Fourier_M}
\end{equation}
where $\Phi^{(p)}_{{\mathbf{n}}} (\mathbf{J})$ is the Fourier transform of $\Phi^{(p)}(\br)$ in angle space (see~\ref{eqn:Fourier_any}),
and $\mcE$ is an arbitrary constant with units of energy.
In equation~\eqref{Fourier_M}, ${ \msM (\omega) }$ is usually called the \textit{response matrix}.
There is a striking similarity between equation~\eqref{Fourier_M} and the corresponding quantity in plasma theory, and we explore this connection in \S\ref{sec:Response_Example_Homogeneous}.
Moreover, just as in plasma theory, although we derived~\eqref{Fourier_M} for Im $\omega>0$ we can analytically continue this function to the lower-half $\omega$ plane by taking the action space integral along the Landau contour. In practice, for any given $\bn$, stellar systems typically sustain only a finite range of resonance frequencies,
$\bn \cdot \bOm$. This leads to branch cuts in the response matrix\cite{Barre2011,Barre2013},
as is also the case, for example, in magnetized relativistic plasmas~\cite{Melrose1986}, in the toroidal ion temperature gradient mode\cite{Kim1994,Kuroda1998}, or truncated Maxwellians\cite{Baalrud2013}.
We will not pursue any details here, but refer the reader to Chapter 5 of BT08 and to~\citet{Fouvry2021-jy} for more information.

Having expressed $\widehat{\msM} (\omega)$ as a matrix ${\msM} (\omega)$, the nature of the solution to~\eqref{eqn:Dyson} depends
crucially on whether $\det [\msI - \msM (\omega)]= 0$ or $\det [\msI - \msM (\omega)]\neq 0$:
\begin{itemize}[itemsep=0pt, topsep=4pt]
\item If $\omega$ is such that $\det [\msI - \msM (\omega)]\neq 0$ then the inverse matrix
$[\msI - \msM (\omega)]^{-1}$ exists, and the solution to~\eqref{eqn:Dyson} is simply
\begin{equation}
     \vert \delta \widetilde{\Phi} (\omega) \rangle = [\msI -\msM (\omega)]^{-1} \vert \delta \widetilde{\Phi}_\msource(\omega) \rangle.
     \label{eqn:dressed_noise_schematic}
\end{equation}
This is true for the great majority of $\omega$ values.  Equation~\eqref{eqn:dressed_noise_schematic}
equation tells us that at these frequencies, 
the noise source $\vert \delta \widetilde{\Phi}_\mathrm{source} (\omega) \rangle $ gets amplified by collective effects.  Sometimes this is referred to as the noise being \textit{dressed} by a gravitational \textit{wake}.
In the equivalent electrostatic plasma calculation, $\mathsf{I}-\mathsf{M}$ is just the dielectric function $\epsilon_{\mathbf{k}}(\omega)$; for near-equilibrium plasmas the solution~\eqref{eqn:dressed_noise_schematic} tells us that 
each `bare' charge is locally screened by the Debye cloud of opposite charges that it gathers round itself.
But in the gravitational context there is no screening, and collective effects will often make gravitational perturbations on large scales \textit{stronger}, not weaker. Loosely speaking, whereas Debye clouds tend to \textit{reduce the effective charge}, gravitational wakes tend to \textit{increase the effective mass} hence typically shortening the relaxation time (see~\S\ref{sec:Example_Magorrian} for an example).
This behavior was illustrated beautifully in the context of a local shearing portion of a galactic disk by~\citet{toomre1991spiral}.
\item Alternatively, for $\omega$ such that $\det [\msI - \msM (\omega)] = 0$ the inverse of  $[\msI - \msM (\omega)]$
does \textit{not} exist, which tells us that $\vert \delta \widetilde{\Phi}(\omega) \rangle $ can be non-zero even in the absence of a source term on the right hand side of~\eqref{eqn:Dyson}. These are the so-called \textit{Landau modes} of the system\footnote{Note that in this article we use the nomenclature `Landau mode' regardless of whether the growth rate of the mode is positive, negative or zero.}, and they tend to occur for a discrete set of (generally) complex frequencies $\{ \omega_\mathrm{m} \}$.
 The system is then able to support a discrete set of collective oscillations which, just as electromagnetic waves in vacuum, do not need to be continually driven by some noise source.
Moreover, under fairly weak assumptions one can show that the solution of equation~\eqref{eqn:Dyson} at this frequency will have time-dependence $\propto \me^{\mi \omega_\mathrm{m} t}$ (see~\citet{kalnajs1971dynamics} for a precise statement).
Thus, a system is linearly unstable if there exists a Landau mode whose imaginary part $ \mIm \, \omega_\mathrm{m} > 0$.
Otherwise, the system is stable.
\end{itemize}
This confluence of a `dressed continuum' of fluctuations superposed with a discrete set of 
`Landau modes' is characteristic of both plasmas and stellar systems, although the typical responses of these systems to perturbations can be very different, as we will illustrate with some examples below.

Note that although we use the term \textit{Landau mode} above, 
these are in general the \textit{not} the normal modes of the linearized system, 
because $\omega$ is in general not real. An alternative approach to linear response theory is to use purely real frequencies, in which case the fundamental objects in the theory are (singular) \textit{van Kampen modes}\citep{Lau2021-ae}. One can show that Landau modes are equivalent to a superposition of a continuum of these van Kampen modes.
We will not refer to van Kampen modes again in this article --- we only note that they offer an alternative way to formulate  everything we present in this section.

\subsubsection{Example: Jeans instability in a homogeneous periodic cube}
\label{sec:Response_Example_Homogeneous}

First, let us address directly the analogy between stellar systems and homogeneous plasmas, by 
calculating the response matrix $\msM (\omega)$ for a fictitious three-dimensional homogeneous stellar system placed inside a periodic cube.
As noted in \S\ref{sec:Homogeneous},
in that case, orbits are simple (multi-periodic) straight lines,
and we can take $\btheta \propto \br$ and  $\bJ \propto \bOm \propto \mathbf{v}$ (see equation~\ref{eq:AA_homogeneous}).
The system’s instantaneous potential and DF are linked by 
\begin{equation}
    \Phi(\br) = \int \md \br' \md \bv' f(\br', \bv') \psi(\br, \br'),
\end{equation}
where $\psi(\br,\br')=-G/\vert \br-\br'\vert$ is the gravitational pairwise interaction.
Making use of the periodicity of this system one has
\begin{equation}
    \psi(\btheta, \btheta') = -\frac{G}{L\pi} \sum_{\bp\neq \bzero} \frac{\me^{\mi\bp\cdot (\btheta-\btheta')} }{\vert \bp \vert^2} .
    \label{eq:exp_psi_hom}
\end{equation}
Glancing at~\eqref{eq:quasi_sep}, we find that, in homogeneous systems,
the \textit{natural} basis elements are ${ \Phi^{(\bp)} \propto \me^{\mi \bp \cdot \btheta} / |\bp| }$,
namely plane waves. This is unsurprising, since the 
system is translationally invariant.
In equation~\eqref{eq:exp_psi_hom}, the factor ${ 1/|\bp|^2 }$ is naturally reminiscent
of the ${ 1/|{\mathbf{k}}|^2 }$ factor that arises when Fourier transforming
the electrostatic Poisson equation for plasmas.

With this, the response matrix from~\eqref{Fourier_M} becomes
\begin{equation}
    \msM_{\bp\bq}(\omega) = \frac{GL^2}{\pi} \delta_{\bp}^{\bq} \frac{1}{\vert \bp\vert^2} \int \md \bv
    \, \frac{\bp\cdot \p f_0/\p \bv}{\overline{\omega} - \bp\cdot\bv},
    \label{eqn:M_homogeneous}
\end{equation}
where $\overline{\omega} \equiv (2\pi)^{-1}L\omega$.
Assuming\footnote{Again, the Maxwellian DF is not physically realizable in real astrophysical systems --- see~\S\ref{sec:Thermal_Equilibrium} --- but is well-defined in the periodic cube we are discussing here.} 
that the system’s mean DF follows the Maxwellian distribution
\begin{equation}
    f_0(\bv) = \frac{\rho_0}{(2\pi\sigma^2)^{3/2}}\me^{-\vert \bv \vert^2/(2\sigma^2)},
\end{equation}
where $\rho_0$ is the system’s mean density, and $\sigma$ is the velocity dispersion, we
find from~\eqref{eqn:M_homogeneous} that 
\begin{equation}
    \msM_{\bp\bq}(\omega) =  \delta_{\bp}^{\bq} \frac{1}{\vert \bp\vert^2} \left( \frac{L}{L_\mJ} \right)^2 [1 + \zeta Z(\zeta)],
    \label{M_Maxwellian}
\end{equation}
where $\zeta = \overline{\omega} / (\sqrt{2} |\mathbf{p}| \sigma)$ is a dimensionless frequency, 
$Z$ is the usual plasma dispersion function\cite{FriedConte1961}
\begin{equation}
Z(\zeta) = \frac{1}{\sqrt{\pi}} \int_{-\infty}^\infty \md u \frac{\me^{-u^2}}{u-\zeta},
\end{equation}
and $L_\mJ$ is the so-called \textit{Jeans length}
\begin{equation}
    L_\mJ = \sqrt{\frac{\pi \sigma^2}{G \rho_0}} \approx 10 \, \mkpc \left(\frac{\sigma}{100\, \mathrm{km}\,\mathrm{s}^{-1}} \right)\left( \frac{\rho_0}{10^{8} M_\odot \, \mkpc^{-3}} \right)^{-1/2}.
    \label{eqn:Jeans_Length}
\end{equation}
The main difference between the expression~\eqref{M_Maxwellian} for a self-gravitating system, and the analogous one for
an electrostatic plasma, is the term $1+\zeta Z(\zeta)$ which in a plasma is replaced by $1-\zeta Z(\zeta)$.

\begin{figure}[htbp!]
\centering
\includegraphics[width=0.35\linewidth]{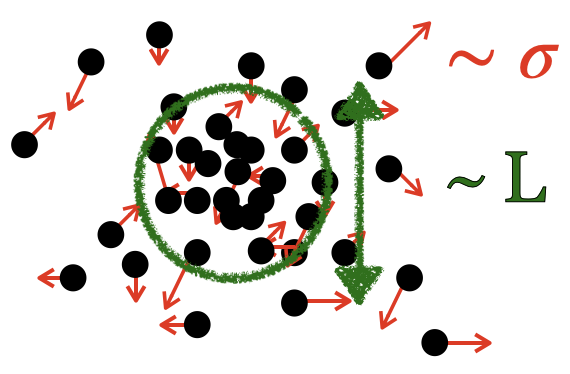}
\caption{Jeans instability as a competition between velocity dispersion and gravitational attraction.
A density fluctuation $\rho_0$ over a length scale $L$ will collapse
under its own gravity if the collapse timescale $\sim 1/\sqrt{G\rho_0}$
is shorter than the crossing time $L/\sigma$ where $\sigma$ is the velocity dispersion.
}
\label{fig:Jeans}
\end{figure}

The dispersion relation for Landau modes is then $\delta_{\bp}^{\bq} - \msM_{\bp\bq}(\omega_\mathrm{m}) = 0$.
By looking for solutions with Re $\omega_\mathrm{m} = 0$ one can show that instability occurs (Im $\omega_\mathrm{m} > 0$) if the system is larger than a critical scale, $L>L_\mJ$ (see~\S{5.2.4} of BT08 for details). Thus, unlike in a plasma, a homogeneous Maxwellian distribution of stars is not necessarily stable to linear perturbations, but instead can be made unstable e.g.\ by making the system dynamically colder (reducing $\sigma$) or more dense (increasing $\rho_0$).
The physics underlying this \textit{Jeans instability} is the competition between gravity and velocity dispersion,
which we sketch in Figure~\ref{fig:Jeans}.
In a stellar system without any velocity dispersion, a density fluctuation $\rho_0$ over a length scale $L$ would collapse
under its own gravity on the timescale $\sim 1/\sqrt{G\rho_0}$. 
However, if the stars have some velocity dispersion $\sigma$, then they can escape the region of size $L$ on the timescale $\sim L/\sigma$.
If the latter timescale is longer than the former, the system cannot avoid collapse: this happens if $L\gtrsim \sigma/\sqrt{G\rho_0}$, which, up to a prefactor,
is the Jeans length~\eqref{eqn:Jeans_Length}.
The Jeans instability is the basic mechanism underlying large-scale structure formation in the Universe\cite{mo2010galaxy}.

\subsubsection{Example: Spiral Landau modes in cold stellar disks}
\label{sec:Response_Example_Imhomogeneous}

Another important example of a Landau mode is a spiral mode in a disk galaxy. In truth, it is still not completely clear what gives rise to the spiral structure that we see in real galaxies like that in the top panel of Figure~\ref{fig:Spiral_Globular}.
 \begin{figure}[htbp!]
  \centering
\includegraphics[width=0.8\textwidth]{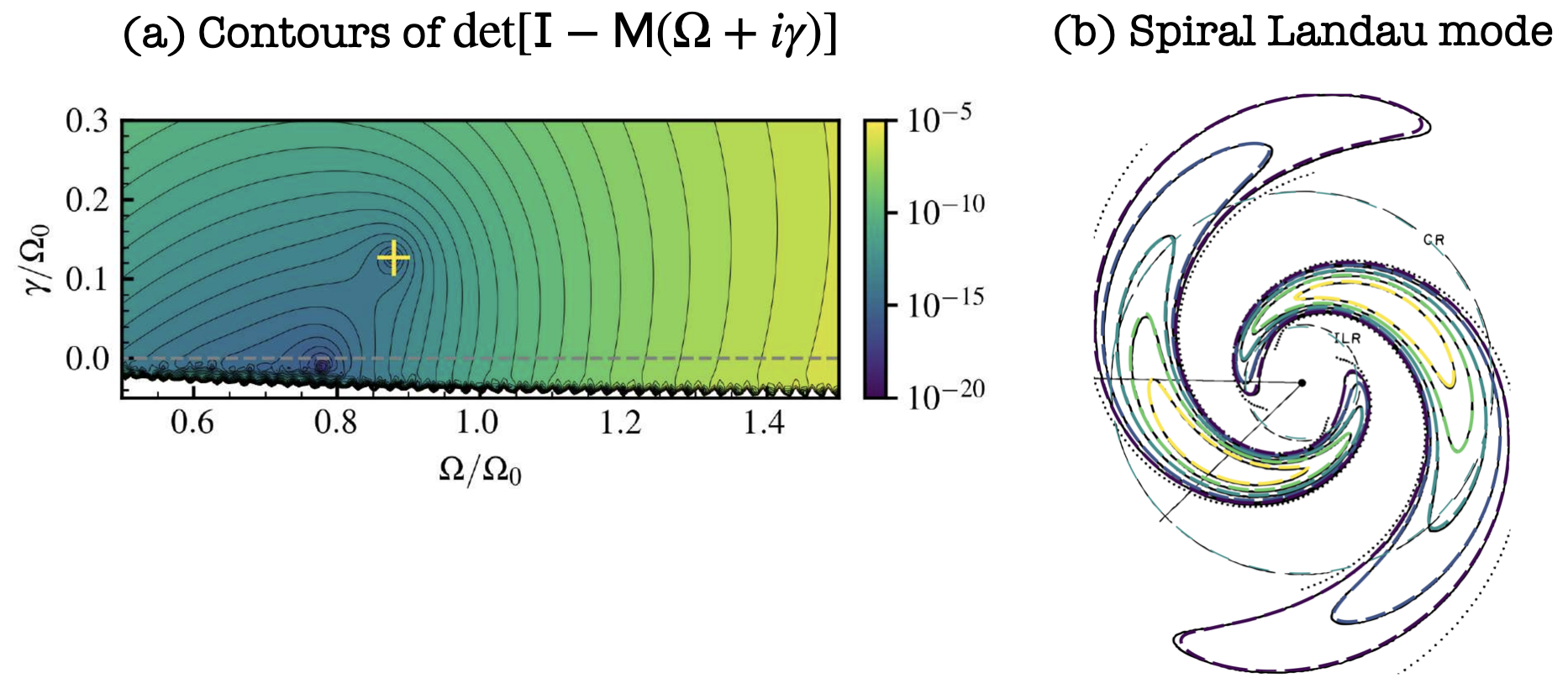}
\caption{Determination of an unstable spiral Landau mode in a truncated Mestel disk following~\citet{zang1976stability}. 
Panel (a) shows the determinant of ${ \mathsf{I}-\mathsf{M} (\omega) }$ in the complex frequency plane; the yellow cross shows where this function passes through zero. Panel (b) shows contours of the corresponding spiral Landau mode.
Figure adapted from M. S. Petersen, M. Roule, J.-B. Fouvry, C. Pichon, and K. Tep, arXiv, 2311.10630 (2023)\cite{petersen2023predicting}.
}
\label{fig:Zang}
\end{figure}
Rather than attempt a review of this fascinating subject here\cite{Kormendy2004,Sellwood2021-pb}, let us restrict ourselves to considering a more artificial system, namely an \textit{isolated, two-dimensional} stellar disk consisting of \textit{equal-mass} stars and \textit{no gas}.  In this case, spirals can arise manifestly as linear instabilities (i.e.\ Landau modes with Im $\omega_\mathrm{m} > 0$) of an underlying DF $f_0(L, J_R)$.

A classic example of such a spiral mode was discovered by~\citet{zang1976stability}, which we reproduce in Figure~\ref{fig:Zang} (adapted from~\citet{petersen2023predicting}).
The model Zang employed was a `Mestel disk' --- a simple, scale-free, two-dimensional disk embedded in a rigid dark matter halo --- to which he added inner and outer radial truncations.
Just as with the Jeans instability (\S\ref{sec:Response_Example_Homogeneous}), one can always make a disk unstable by making it sufficiently dynamically cold and/or dense.
In Zang's case, this was achieved by modifying the 
 \textit{active fraction} of the disk, essentially the ratio between the self-gravitating mass in the disk and that in the unresponsive halo.
Panel (a) of Figure~\ref{fig:Zang} shows the contours of $\det [\msI - \msM (\omega)$] (equation~\ref{Fourier_M}) in the complex frequency plane.
The yellow cross shows the location where this determinant equals zero, which gives the frequency $\omega_\mathrm{m}$ of an unstable spiral Landau mode.
Precisely, the mode rotates with pattern speed $0.88 \, \Omega_0$
and growth rate $ 0.12 \, \Omega_0$, where $\Omega_0$ is a typical dynamical frequency comparable to $\sim 2\pi/t_\mathrm{dyn}$.
The spatial structure of the spiral mode is shown in panel (b);
this is calculated by finding the eigenvector $\bX_\mathrm{m}$ with components $X_\mathrm{m}^p$, such that $\msM (\omega_\mathrm{m}) \cdot \bX_\mathrm{m} = 0$ and then writing
\begin{equation}
    \delta \Phi_\mathrm{m}(\br, t) \propto \mRe \bigg[\sum_p X_\mathrm{m}^p \, \Phi^{(p)}(\br) \, \me^{- \mi \omega_\mathrm{m} t} \bigg].
\end{equation}
(Since this is a linear eigenmode, the overall normalisation is arbitrary).

\begin{figure}[htbp!]
  \centering
\includegraphics[width=0.64\textwidth]{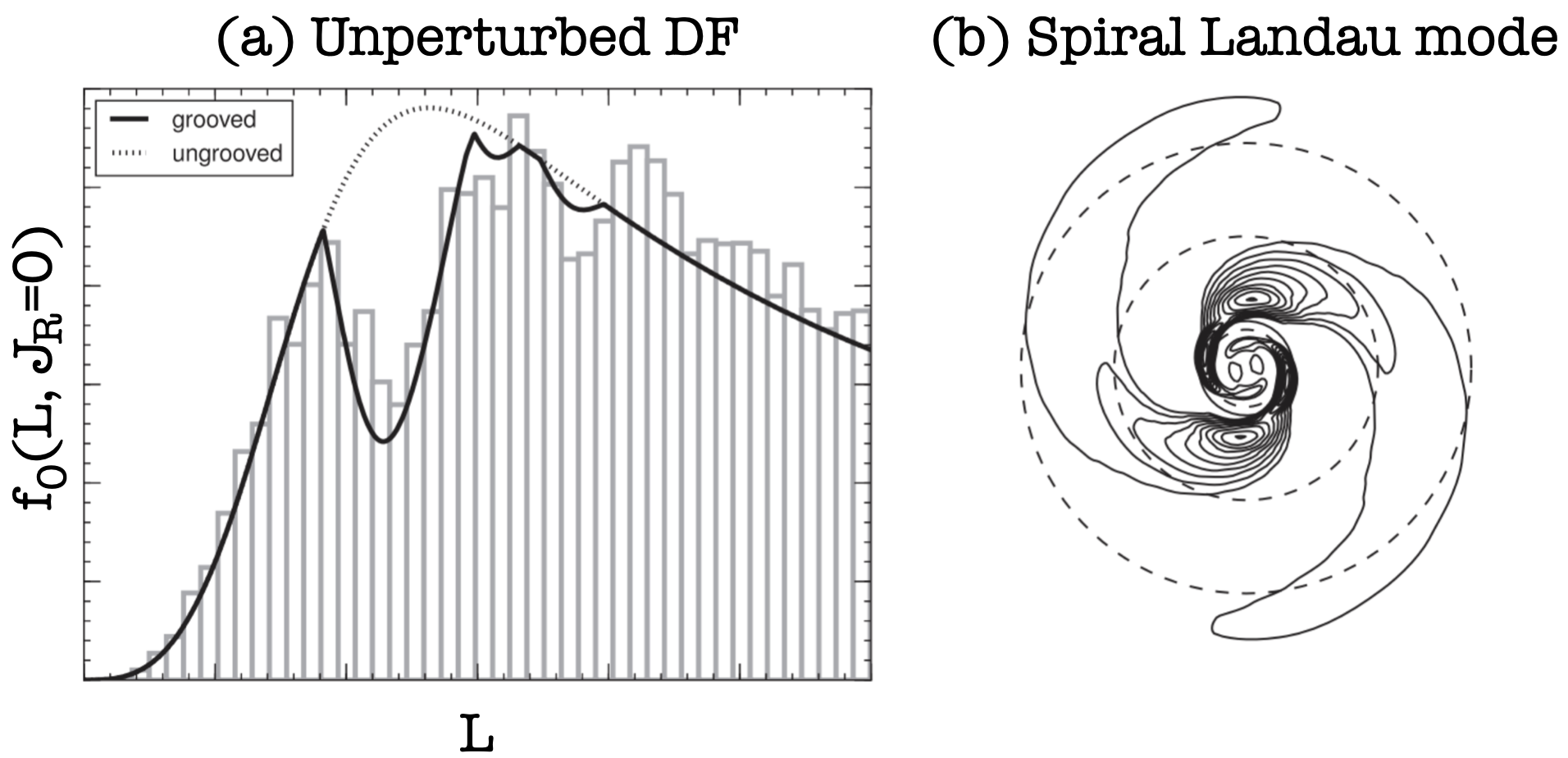}
\caption{Groove-induced instability in a Mestel disk.
Panel (a) shows the $J_R = 0$ cut through the
DF in the stable (ungrooved) and unstable (grooved) case. Panel (b) shows the unstable spiral mode as calculated from linear stability analysis of the grooved disk. The solid circle shows the location of the groove for circular orbits, while the three dashed circles show the ILR, CR, and OLR resonance locations of the resulting spiral mode.
Figure adapted from S. De Rijcke, J.-B. Fouvry, and C. Pichon, MNRAS 484, 3198 (2019)\cite{De_Rijcke2019-uo}.
}
\label{fig:groove}
\end{figure}

The two examples  we have given so far both involved smooth mass distributions that were  made sufficiently cold and/or dense that they became unstable.
However, there are other ways of producing instabilities which are perhaps more familiar to plasma physicists used to the bump-on-tail paradigm\cite{Diamond2010-mh}.
For instance,
\citet{De_Rijcke2019-uo} recently performed a linear stability analysis, like the one we have outlined in this section, for a \textit{grooved} Mestel disk.
That is, they took a Mestel disk with a stable DF --- the $J_R=0$ slice through which is shown with a dotted line in Figure~\ref{fig:groove}a --- and then carved a `groove' in it by removing a portion of the stars in a given range of angular momentum (solid line in Figure~\ref{fig:groove}a).
This grooved disk is also unstable to spiral Landau modes, and 
Figure~\ref{fig:groove}b shows the shape of the dominant mode.

Simulations show that the Landau mode spectrum of stellar disks is often dominated by one, or at most a few, isolated frequencies\cite{Sellwood2014-te}.
This means that spirals in unstable disks are perhaps more analogous to, e.g.,\ Alfvenic eigenmodes in tokamak plasmas than to classic bump-on-tail instabilities in homogeneous plasmas, since the latter tend to harbor a whole spectrum of unstable modes (at least one at every unstable wavenumber $\mathbf{k}$).
Because of this, the evolution of unstable stellar disks might be more aptly described by something like the single-wave kinetics of~\citet{ONeil1965-uy,Berk1990-ic,Berk1996-bv}, etc. than by the typical many-wave quasilinear theory of~\citet{drummond1961non,Kaufman1972-vv}. See~\S\ref{sec:Nonlinear_Kinetics} for more.


\subsection{Explicit wake solution}
\label{sec:Wakes}

We now suppose that our system is stable, meaning that all Landau modes are damped.
An astrophysical example of such a (weakly) damped mode
is the (dipole) ${ \ell = 1 }$ mode of globular clusters\cite{Weinberg1994},
which represent long-lasting sloshing oscillations\cite{Heggie2020}.
We further assume that we may wait long enough that these modes no longer contribute to the response (i.e.\ we wait for $t\gg \vert \mathrm{Im}\,\omega_\mathrm{m}\vert^{-1}$, where $\omega_\mathrm{m}$ is the system's least strongly-damped Landau mode).
In this regime, we can solve~\eqref{eqn:delta_Phi_integral_equation} directly under the assumption that $\omega$ is \textit{not} associated with a Landau mode. Having constructed the solution we will then perform an inverse Laplace transform to get the dressed wake part of the response in the time domain. 

The easiest --- and most physically enlightening --- way to solve~\eqref{eqn:delta_Phi_integral_equation} is to guess what form the solution should have, and then work out the exact details post-hoc.
So, let us look at~\eqref{eqn:delta_Phi_integral_equation}. 
Supposing for a moment that we were to ignore collective effects ($\widehat{\mathsf{M}} = \mathsf{0}$), we would find
\begin{equation}
    [\delta \widetilde{\Phi}_{{\mathbf{n}}}(\mathbf{J}, \omega)]_\mbare = 
-(2\pi)^d
      \sum_{{\mathbf{n}}'}\int \md \mathbf{J}'\frac{\delta f_{{\mathbf{n}}'}(\mathbf{J}', 0)}{\mi (\omega - {\mathbf{n}}'\cdot \bOm')}
      \psi_{{\mathbf{n}}{\mathbf{n}}'}(\mathbf{J},\mathbf{J}'),
            \label{eqn:bare-potential}
\end{equation}
which we call the `bare' solution.
Inverse-Laplace transforming this according to \eqref{eqn:Inverse_Laplace_Transform}, deforming the Bromwich contour so that it encircles the pole at $\omega = {\mathbf{n}'} \cdot \bOm'$ (see Fig f.1 of Fouvry \& Bar-Or\cite{Fouvry2018-gi} for illustration) and closing the contour with a large semicircle in the lower-half plane, we would find
\begin{align}
    [\delta \Phi_{{\mathbf{n}}}(\mathbf{J}, t)]_\mathrm{bare} = (2\pi)^d \sum_{{\mathbf{n}}'}\int & \md \mathbf{J}' \, \delta f_{{\mathbf{n}}'}(\mathbf{J}', 0) \me^{ - \mi {\mathbf{n}}' \cdot \bOm't} 
    \psi_{{\mathbf{n}}{\mathbf{n}}'}(\mathbf{J}, \mathbf{J}').
    \label{eqn:bare_potential_fluctuations}
\end{align}
The physical interpretation of this result is very clear: 
it is the (Fourier transform of the) potential at location $\br(\btheta, \bJ)$ that is created by summing the contributions from every star with initial locations $\br'(0) = \br'(\btheta'(0), \bJ'(0))$, 
assuming those stars simply travel along mean field trajectories
$\btheta' = \bOm' t + \btheta'(0)$ and $\bJ' = \bJ'(0)$.

The ansatz we now make is that the solution 
 to the full equation~\eqref{eqn:delta_Phi_integral_equation} is exactly of this functional form,
except with the bare interaction potential $ \psi_{{\mathbf{n}}{\mathbf{n}}'}(\mathbf{J},\mathbf{J}')$ replaced with some \textit{dressed} (and frequency-dependent) interaction potential $\psi^\md_{{\mathbf{n}}{\mathbf{n}}'}(\mathbf{J},\mathbf{J}',\omega)$, whose form we do not yet know.  That is, we assume the solution has the form
\begin{equation}
    \delta \widetilde{\Phi}_{{\mathbf{n}}}(\mathbf{J}, \omega) = 
-(2\pi)^d
      \sum_{{\mathbf{n}}'}\int \md \mathbf{J}'\frac{\delta f_{{\mathbf{n}}'}(\mathbf{J}', 0)}{ \mi (\omega - {\mathbf{n}}'\cdot \bOm')}
      \psi^\md_{{\mathbf{n}}{\mathbf{n}}'}(\mathbf{J},\mathbf{J}', \omega).
      \label{eqn:ansatz}
\end{equation}
This ansatz is actually very natural if one is acquainted with \textit{Rostoker's principle} in plasma kinetics,
which essentially says that one can consider a hot plasma as a set of completely uncorrelated particles, provided one replaces the bare Coulombic potential of each particle with its dressed, or \textit{screened}, version\cite{Diamond2010-mh}.
(We will use this idea to derive the Balescu--Lenard kinetic equation for stable stellar systems in \S\ref{sec:Rostoker}).
Plugging the ansatz~\eqref{eqn:ansatz} into~\eqref{eqn:delta_Phi_integral_equation} we can read off an implicit 
equation for $ \psi^\md$:
\begin{align}
&\psi^\md_{{\mathbf{n}}{\mathbf{n}}'} (\mathbf{J} , \mathbf{J}' , \omega) = \psi_{{\mathbf{n}}{\mathbf{n}}'} (\mathbf{J} , \mathbf{J}')
\label{self_psid}  - (2 \pi)^{d} \sum_{{\mathbf{n}}^{\prime\prime}} \!\! \int \!\! \md \mathbf{J}^{\prime\prime}
\frac{{\mathbf{n}}^{\prime\prime} \cdot \p f_0 / \p \mathbf{J}^{\prime\prime}}{\omega - {\mathbf{n}}^{\prime\prime} 
\cdot \bOm^{\prime\prime}} \,
\psi_{{\mathbf{n}}{\mathbf{n}}^{\prime\prime}} (\mathbf{J} , \mathbf{J}^{\prime\prime}) \, \psi^\md_{{\mathbf{n}}^{\prime\prime}{\mathbf{n}}'} (\mathbf{J}^{\prime\prime} , \mathbf{J}' , \omega),
\end{align}
where $\bOm'' \equiv \partial H_0(\mathbf{J}'')/\partial \bJ''$.
Thus, we will have constructed a valid solution to equation~\eqref{eqn:delta_Phi_integral_equation} if we can solve equation~\eqref{self_psid} for $\psi^\md$.  While at first sight it may not seem like we have made much progress with this rewriting, it turns out that~\eqref{self_psid} is actually quite easy to solve if we project the various terms onto the biorthogonal basis elements $\Phi_{\bn}^{(p)}(\bJ)$, just as in~\eqref{Fourier_M}.
We leave the details to Appendix~\ref{sec:BasisMethod}, and here just write down the final answer:
\begin{equation}
\psi^\md_{{\mathbf{n}}{\mathbf{n}}'} (\mathbf{J} , \mathbf{J}' , \omega) = - \frac{1}{\mcE} \sum_{p , q} \Phi^{(p)}_{{\mathbf{n}}} (\mathbf{J}) \, [\msI - \msM (\omega)]_{pq}^{-1} \, \Phi^{(q) *}_{{\mathbf{n}}'} (\mathbf{J}') .
\label{dressed_basis}
\end{equation}
We can usefully compare this to the bare interaction, which as we show in Appendix~\ref{sec:BasisMethod}
can be expanded as
\begin{equation}
\psi_{{\mathbf{n}}{\mathbf{n}}'} (\mathbf{J} , \mathbf{J}') = - \frac{1}{\mcE}\sum_{p} \Phi^{(p)}_{{\mathbf{n}}} (\mathbf{J}) \, \Phi^{(p) *}_{{\mathbf{n}}'} (\mathbf{J}').
\label{bare_basis}
\end{equation}
Thus we see that in passing from $\psi \to \psi^\md$ we have simply `divided' the bare interaction by $[\msI - \msM (\omega)]$.
This is analogous to dividing by the dielectric function $\epsilon_{\mathbf{k}}(\omega)$ in electrostatic plasma theory.

We can now inverse Laplace transform equation~\eqref{eqn:ansatz}.
The assumption that Landau modes can be ignored (i.e.\ $t \gg \vert \mathrm{Im} \, \omega_\mathrm{m} \vert^{-1}$) means that when we deform the Bromwich contour we need only 
pick up contributions from the poles at $\omega = {\mathbf{n}}' \cdot \bOm'$, 
and can ignore any singularities of $\psi^\mathrm{d}$ (i.e.\ zeroes of $[\mathsf{I}-\mathsf{M}(\omega)]$, see equation~\ref{dressed_basis}).
The result is
\begin{align}
    \delta \Phi_{{\mathbf{n}}}(\mathbf{J}, t) = (2\pi)^d \sum_{{\mathbf{n}}'}\int & \md \mathbf{J}' \, \delta f_{{\mathbf{n}}'}(\mathbf{J}', 0) \, \me^{ - \mi {\mathbf{n}}' \cdot \bOm't}
    \psi^\md_{{\mathbf{n}}{\mathbf{n}}'}(\mathbf{J}, \mathbf{J}', {\mathbf{n}}'\cdot \bOm'),
    \label{eqn:dressed_potential_fluctuations}
\end{align}
which is the same as~\eqref{eqn:bare_potential_fluctuations} with $\psi \to \psi^\mathrm{d}$.
Thus, we have realized Rostoker's idea: the dressed potential fluctuation is equivalent to 
that carried by the initial fluctuation ($ \delta f_{{\mathbf{n}}'}(\mathbf{J}' , 0)$), transported along the mean field ($\propto \me^{- \mi {\mathbf{n}}' \cdot \bOm' t}$), but with the Newtonian potential of each star replaced with the dressed potential ($ \psi^\md$) evaluated at the mean field frequency (${\mathbf{n}}'\cdot\bOm'$),
with contributions summed over all stars ($\sum_{{\mathbf{n}}'} \int \md \mathbf{J}'$). 
Furthermore,~\eqref{eqn:dressed_potential_fluctuations} can be plugged into the right hand side of equation~\eqref{eqn:linear_Vlasov_formal_solution}
to give an explicit solution for the fluctuations in the DF $\delta f_{{\mathbf{n}}}(\mathbf{J}, t)$.
Note that the basis functions $\Phi^{(p)}$ play explicit no role in equation~\eqref{eqn:dressed_potential_fluctuations};
they are merely a calculational tool which is \textit{only} used to construct the dressed interaction $\psi^\md_{{\mathbf{n}} {\mathbf{n}}'}(\mathbf{J}, \mathbf{J}', \omega)$ for a given mean field $f_0(\mathbf{J})$. 

\subsubsection{Example: Response of a stellar system to a perturbing mass}
\label{sec:Example_Magorrian}

So far in this section we have concentrated on the response of a stellar system to internal noise $\delta f(t=0)$.
However, an entirely analogous calculation shows that even externally imposed potential fluctuations will be `dressed' by the system's self-gravity --- just as an externally imposed point charge will gather its own Debye cloud in an electrostatic plasma.
Here we will not pursue any quantitative details, but instead show a few figures that capture the characteristic features of such a response.

\begin{figure}[htbp!]
\centering
\includegraphics[width=0.6\linewidth]{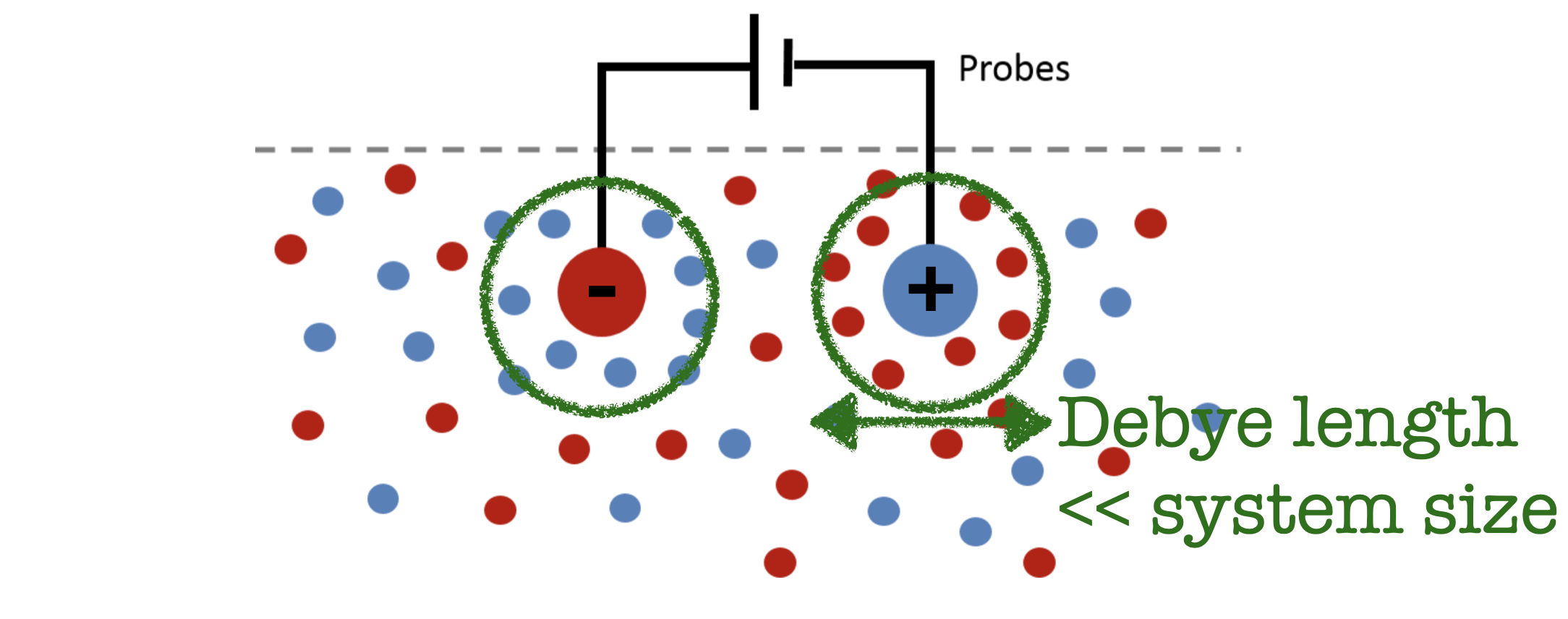}
\caption{Cartoon
(adapted from P. Gibbon, arXiv, 2007.04783 (2020)\cite{gibbon2020introduction})
of the Debye shielding that occurs when electrical probes are placed inside a quasineutral plasma.}
\label{fig:Debye}
\end{figure}

First we present Figure~\ref{fig:Debye}, which is adapted from~\citet{gibbon2020introduction}.
This is a cartoon of a quasineutral electrostatic plasma made of blue positive charges and red negative charges.
When a pair of electrical probes is inserted into the plasma, they 
gather around them Debye clouds consisting of particles of the opposite charge.
In this way, the electrostatic potential of the probes is screened over a Debye length.
This cloud is typically much smaller than the scale of the whole system.

\begin{figure}[htbp!]
  \centering
\includegraphics[width=0.75\textwidth]{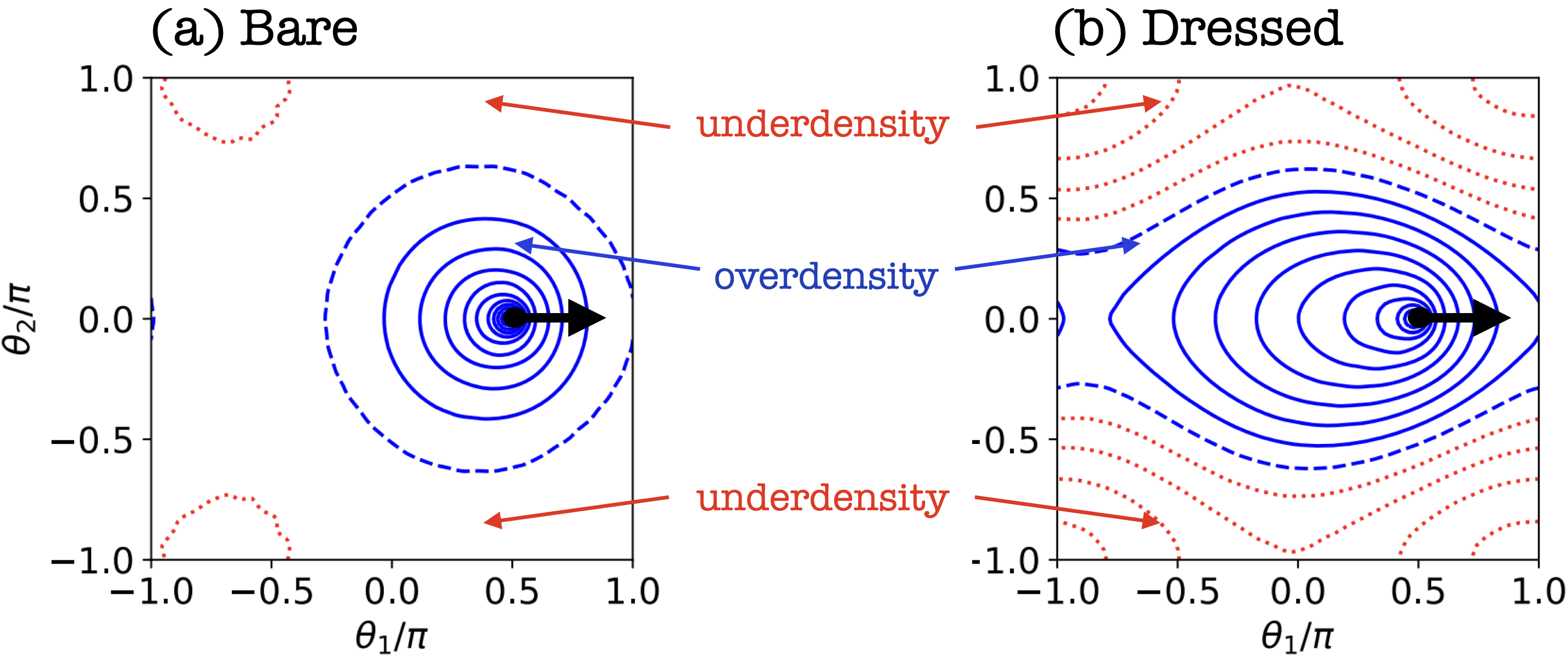}
    \caption{Response of a Maxwellian box of stars to a point mass perturber,
   adapted from J. Magorrian, MNRAS 507, 4840 (2021)\cite{magorrian2021stellar}. 
The point mass (black dot) is driven through the box of size $\approx 0.9L_\mJ$ at constant speed,
and the wake response is computed according to linear theory.
Solid blue contours denote positive overdensity, and red
dotted contours denote negative overdensity, while the blue dashed curve corresponds to zero overdensity.
In panel (a) the stars only feel the gravitational potential of the perturber but not of each other, while in panel (b) they interact with both the perturber and each other.
The same linearly spaced contour levels are
used in both panels.}
\label{fig:Magorrian}
\end{figure}

Next, we turn to Figure~\ref{fig:Magorrian}, which is adapted from~\citet{magorrian2021stellar}.
This is an analogous stellar-dynamical scenario, in which a heavy point mass is driven on a straight line orbit at constant speed through
a homogeneous stellar system with Maxwellian $f_0$ (\S\ref{sec:Response_Example_Homogeneous}).
The size of the box in this case has size $L \approx 0.9L_\mJ$, so the system is (weakly) stable (recall that instability occurs if $L > L_\mathrm{J}$).
The contours in panel (a) show the density response of the system according to linear theory in the case that stars only respond to the point-mass perturber but not to each other.  This we call the `bare' response.
By contrast, panel (b) shows the response density in the `dressed' case in which stars interact both with the perturber and with each other (the contour levels are the same as in panel (a)).
Let us make two simple observations which contrast the behavior of the stellar system shown in these panels with the plasma system sketched in Figure~\ref{fig:Debye}.
(i) The overdense wake in the gravitational system acts to reinforce the perturbation rather than shield it, since there are no opposite charges. Heuristically therefore, one can think of gravitational polarization as a kind of `anti Debye-shielding'.
 (ii) The scale of the induced wake, especially in the dressed case, can be comparable to the scale of the system itself, in  contrast with an electrostatic plasma where the Debye length is usually much smaller than the system.

\begin{figure}[htbp!]
\centering
\includegraphics[width=0.55\textwidth]{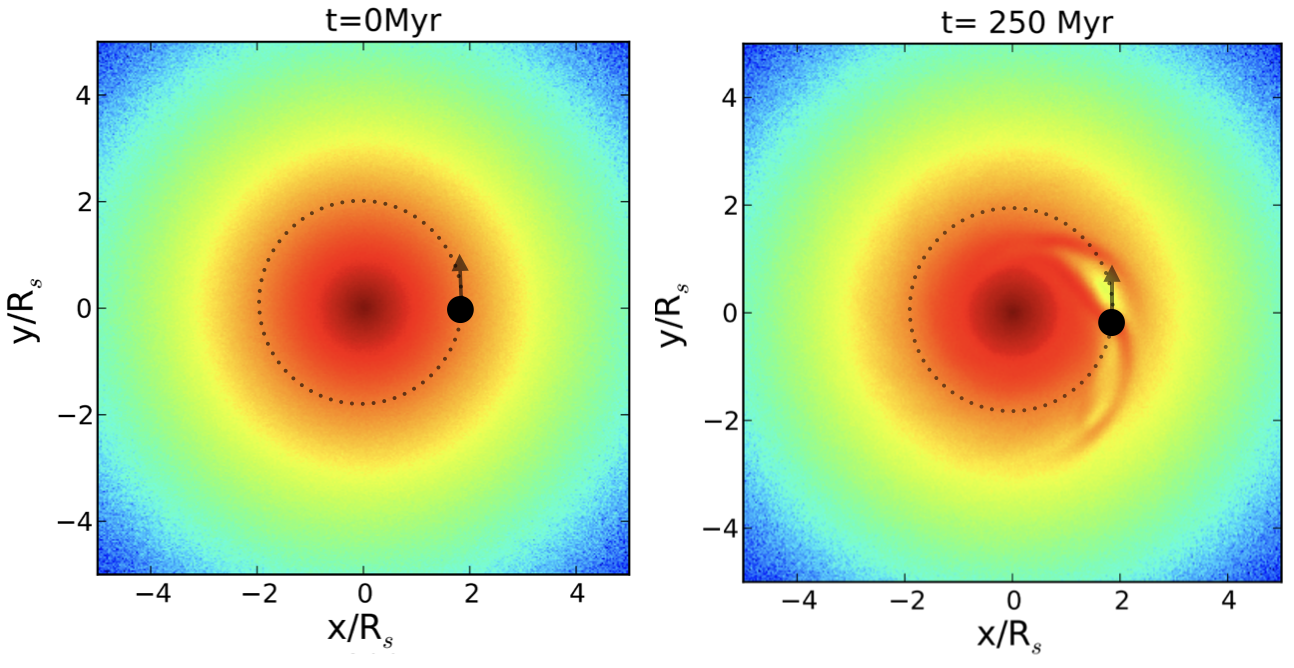}
\caption{Response of a stellar disk to a massive perturber driven on a circular orbit
(adapted from E. D’Onghia, M. Vogelsberger, and L. Hernquist, ApJ 766, 34 (2013)\cite{d2013self}).
The perturber mass is much less than $1\%$ of the mass of the system.
Still, it manages to produce a significant
response over a large swathe of the disk.}
\label{fig:DOnghia}
\end{figure}

Since a homogeneous box is not a particularly realistic model for a stellar system,
we finally present Figure~\ref{fig:DOnghia}, which is adapted from~\citet{d2013self}.
This figure shows the response density of a stellar disk to an embedded point-mass perturber, driven on a circular orbit with constant speed for $1$ azimuthal period ($250$ Myr). The disk is chosen to be fairly dynamically cold, but not so cold as to be unstable.
The response was calculated using a full $N$-body simulation in which stars interact both with the perturber and with each other.
The perturbing mass was much less than $1\%$ of the whole system's mass. Despite its relatively small mass, 
the perturber provokes a strongly amplified response in the disk surface density which extends over a large azimuthal range.
This phenomenon, which was 
first discovered by~\citet{julian1966non} using linear theory,
reflects the fact that cold stellar disks are very effective amplifiers of perturbations, as we discussed in \S\ref{sec:Where_Now}.  
As a result, it is typical of weakly stable stellar systems to develop collective responses on a global scale.

\section{Quasilinear evolution and the Balescu--Lenard equation}
\label{sec:Quasilinear_Evolution}

In many cases we do not know, nor do we really care, about the details of the microscopic DF $\micro{f}$ of a particular stellar system. What we often \textit{do} care about, however, is the structure and evolution of some suitably \textit{averaged} DF, $\langle \micro{f} \rangle$. 
If we are interested in the evolution of a system driven by random fluctuations, be they internally or externally generated, 
then the appropriate average is an \textit{ensemble average} over all possible microscopic realizations of those fluctuations.
The central idea of kinetic theory is to use the {statistics} of the realizations of the fluctuations to derive an equation for the evolution of $\langle \micro{f} \rangle$.

We expect for an ensemble average of $\micro{f}$ over microscopic realizations to be equivalent to an
average over angles and over time (on timescales $\ll t_\mathrm{relax}$).
Thus $\langle \micro{f} \rangle = f_0$ and the relevant equation describing the evolution is (c.f.\ equation~\ref{eqn:Axisymmetric_Evolution_Fourier}):
\begin{equation}
      \frac{\p f_0}{\p t} = -\frac{\p }{\p \mathbf{J}} \cdot \sum_{{\mathbf{n}}} \mi {\mathbf{n}}\,  \langle \delta \Phi^*_{{\mathbf{n}}}(\mathbf{J},t)
        \delta f_{{\mathbf{n}}}(\mathbf{J}, t) \rangle ,
        \label{eqn:Axisymmetric_Evolution_Fourier_Ensemble}    
\end{equation}
In \S\ref{sec:Linear_and_Quasilinear}, we spelled out the basic idea of \textit{quasilinear} kinetic theory, 
which is that if the `fluctuations' $\delta \Phi$ and $\delta f$ are sufficiently small, then we can predict the 
evolution of $f_0$ on the timescale $t_\mathrm{relax} \gg t_\mathrm{dyn}$ by plugging the solutions from linear response theory (\S\ref{sec:Linear_Response})  into the right hand side of~\eqref{eqn:Axisymmetric_Evolution_Fourier_Ensemble}.
Quasilinear theory works provided such a timescale separation exists, and provided we can ignore 
higher order nonlinear and nonperturbative effects like trapping, as discussed in \S\ref{sec:Nonlin_Nonpert}.

In this section we will develop quasilinear kinetic theory more quantitatively.
Our study of internally-driven evolution will culminate in the Balescu--Lenard equation which describes the slow evolution of an isolated, stable $N$-star system.
We first describe the basic physics underlying the Balescu--Lenard equation, 
then simply write it down without a proper derivation and 
discuss some of its properties (\S\ref{sec:Tale_Two_Fluxes}).
Then we offer a simple derivation of the Balescu--Lenard equation based on Rostoker's principle (\S\ref{sec:Rostoker}). 
We give several examples of this equation being used in practice (\S\ref{sec:BL_examples}), before discussing where the theory breaks down and how it might be improved (\S\ref{sec:BL_Breakdown}).

As in \S\ref{sec:Linear_Response},
we consider systems in which the fluctuations are internally generated by the finite-$N$ noise, but we note that a very similar formalism can be developed to account for externally-driven quasilinear evolution\cite{Binney1988-zy,Fouvry2018-gi}.

\subsection{A tale of two fluxes}
\label{sec:Tale_Two_Fluxes}

As a first step, we take the formal solution for the perturbation to the DF in linear theory, equation~\eqref{eqn:linear_Vlasov_formal_solution}, and plug it into the right hand side of~\eqref{eqn:Axisymmetric_Evolution_Fourier_Ensemble}.  The result is
\begin{equation}
  \frac{\p f_0}{\p t}
  = -\frac{\p}{\p \mathbf{J}}\cdot \bF,
  \label{eqn:F1_plus_F2}
\end{equation}
where the two contributions to the `action space flux' $\bF = \bF_1 + \bF_2$ on the right hand side are
\begin{equation}
    \bF_1(\mathbf{J}, t) = \sum_{{\mathbf{n}}} \mi \mathbf{n} \, \me^{- \mi {\mathbf{n}}\cdot\bOm t} \langle \delta f_{{\mathbf{n}}}(\mathbf{J}, 0) \, \delta \Phi_{{\mathbf{n}}}^{*} (\mathbf{J}, t) \rangle,
    \label{eqn:F1_time_dependent}
\end{equation}
and 
\begin{align}
    \bF_2(\mathbf{J}, t) = -\sum_{{\mathbf{n}}} {\mathbf{n}} & \, {\mathbf{n}} \cdot \frac{\p f_0}{\p \mathbf{J}} \int_0^t \md t' \me^{-\mi {\mathbf{n}}\cdot \bOm (t-t')}
 \langle \delta \Phi_{{\mathbf{n}}}^{*} (\mathbf{J}, t) \delta \Phi_{{\mathbf{n}}}(\mathbf{J},t')\rangle.
    \label{eqn:F2_time_dependent}
\end{align}
These equations are not yet written in a form that is useful for calculation, 
but their physical meaning is already apparent. 
In order to make this discussion as clear as possible,
let us deal with them in reverse order.

\begin{itemize}[itemsep=0pt, topsep=4pt]
\item The second flux $\bF_2$ is a diffusion term.
This is obvious since it is proportional to the gradient of the DF $f_0(\bJ)$
in action space.
That is to say, this term answers the following question:
``I am a star with action $\mathbf{J}$, moving along my orbit within a bath of potential fluctuations $\delta \Phi$.  If I think of myself as a massless test particle, then on average, how are those potential fluctuations going to nudge me around?"  It is natural that the answer involves the correlation of fluctuations sampled at action space location $\mathbf{J}$.
Nothing about the expression~\eqref{eqn:F2_time_dependent} would formally change had we insisted that $\delta\Phi$ was externally imposed, but here we are going to use the solution for $\delta \Phi$ derived in \S\ref{sec:Wakes} assuming the fluctuations are internally generated and that the system is stable. Thus, we say that the flux $\bF_2$ describes \textit{orbital diffusion due to dressed discreteness noise}.
\item On the other hand, the first flux $\bF_1$ provides the answer to the obvious rebuttal: ``Wait! I'm not really a test particle, so diffusion cannot be the whole story.  I have finite mass, and so I drive potential fluctuations of my own. To what extent is my motion altered by the part of $\delta \Phi$ that I have myself produced?" Physically this term represents \textit{dynamical friction}, which Chandrasekhar intuited back in the 1940s.
The key idea is that when any massive body moves through a sea of other bodies, it creates a `wake' behind it in response (just as in Figures~\ref{fig:Magorrian}-\ref{fig:DOnghia}).
This wake then backreacts, or drags, on the massive body, slowing it down. 
 It is therefore natural that the flux $\bF_1$ involves the correlation between $\delta \Phi$ at action space location $\mathbf{J}$ and the initial discreteness noise at the same location,
$\delta f_{{\mathbf{n}}}(\mathbf{J}, 0)$.
The existence of such a term is also a necessary consequence of energy (and momentum) conservation: since $\bF_2$ scatters stars to larger energies on average, in a self-consistent system there must be another term dragging them back to lower energies.
\end{itemize}

Although we have taken all stars to be of equal mass for simplicity, one can show that in a multi-mass system, the more massive a body is the larger the wake it induces, and so the stronger the drag force it feels. This leads for instance to massive bodies losing energy and spiraling in towards the center of their host galaxy.
Mathematically, for a subsystem of very massive bodies, the frictional term in~\eqref{eqn:F1_plus_F2} is typically much more important than the diffusive term. By contrast, a very light subsystem will contribute only weakly to the total gravitational potential fluctuation, and hence will experience little dynamical friction; for this subsystem, 
the diffusive term in~\eqref{eqn:F1_plus_F2} can dominate over the frictional one. This is also an astrophysically relevant scenario: for instance, a key problem of galactic evolution is to understand the diffusion of stellar orbits (Figure~\ref{fig:Frankel}) due to interactions with transient spiral waves and/or giant molecular clouds\cite{Binney1988-zy}. 
For the present single-mass system, though, $\bF_1$ and $\bF_2$ are typically comparable in magnitude.

The next step in deriving a proper quasilinear kinetic equation is to take the 
solution for the dressed potential from linear theory $\delta\Phi[\delta f(0)]$, equation~\eqref{eqn:dressed_potential_fluctuations}, and plug it into the right hand sides of~\eqref{eqn:F1_time_dependent}-\eqref{eqn:F2_time_dependent}. One then further assumes that stellar orbits are initially uncorrelated and takes $t\gg t_\mathrm{dyn}$.
The details of this procedure are rather lengthy and do not convey much physical insight, so we refer the reader to, e.g.,\ the works by~\citet{Chavanis2012-lu,Fouvry2018-gi}, and here just write down the final answer. Equation~\eqref{eqn:F1_plus_F2} takes the Fokker--Planck form
 \begin{equation}
  \frac{\p f_0}{\p t}
  = \frac{\p}{\p \mathbf{J}}\cdot \left[- \bA (\mathbf{J}) f_0 + 
    \msD (\mathbf{J})\cdot\frac{\p f_0}{\p \mathbf{J}}\right],
\end{equation}
with the friction vector
\begin{align}
\bA (\mathbf{J}) =&  \pi (2\pi)^d m \sum_{{\mathbf{n}}, {\mathbf{n}}'} {\mathbf{n}} \int \md \mathbf{J}' \delta ({\mathbf{n}} \cdot \bOm - {\mathbf{n}}' \cdot \bOm')
 \vert \psi^\md_{{\mathbf{n}}{\mathbf{n}}'}(\mathbf{J}, \mathbf{J}', {\mathbf{n}}'\cdot \bOm') \vert^2 \, {\mathbf{n}}' \cdot \frac{\p f_0}{\p \mathbf{J}'},
\label{eqn:BL_A}
\end{align}
and diffusion tensor
\begin{align}
    \msD (\mathbf{J}) =& \pi (2\pi)^d m \sum_{{\mathbf{n}}, {\mathbf{n}}'} {\mathbf{n}} \otimes {\mathbf{n}} \int \md \mathbf{J}' \delta ({\mathbf{n}} \cdot \bOm - {\mathbf{n}}' \cdot \bOm')
\vert \psi^\md_{{\mathbf{n}}{\mathbf{n}}'}(\mathbf{J}, \mathbf{J}', {\mathbf{n}}'\cdot \bOm') \vert^2 f_0(\mathbf{J}').
    \label{eqn:BL_D}
\end{align}
We can put these expression together to arrive at the \textit{Balescu--Lenard equation}
\begin{align}
&\frac{\p f_0(\mathbf{J})}{\p t} = \pi (2\pi)^d m \frac{\p}{\p \mathbf{J}} \cdot  \sum_{{\mathbf{n}}, {\mathbf{n}}'} {\mathbf{n}}  \int  \md \mathbf{J}' \delta ({\mathbf{n}} \cdot \bOm - {\mathbf{n}}' \cdot \bOm')
\vert \psi^\md_{{\mathbf{n}} {\mathbf{n}}'}(\mathbf{J}, \mathbf{J}', {\mathbf{n}}' \cdot \bOm')
\vert^2 \left( {\mathbf{n}} \cdot \frac{\p}{\p \mathbf{J}}
-
{\mathbf{n}}' \cdot \frac{\p}{\p \mathbf{J}'}
\right) f_0(\mathbf{J}) f_0(\mathbf{J}').
\label{eqn:BL}
\end{align}

\subsubsection{Properties of the Balescu--Lenard equation}
\label{sec:Physics_of_BL}

At first glance the Balescu--Lenard equation might look implausibly complicated, but the physics underlying it is simple. Indeed, with hindsight, one could argue that this is the simplest equation
one could write down with all the desired properties (particularly that it must overcome the shortcomings of Chandrasekhar theory that we listed in \S\ref{sec:Where_Now}).
For instance:
\begin{itemize}[itemsep=0pt, topsep=4pt]
\item The equation is expressed in angle-action variables and therefore naturally accounts for the inhomogeneity of the system and its associated nontrivial orbital structure.
\item Related to this, unlike in Chandrasekhar's theory, in which we had to stipulate a maximum impact parameter $b_\mathrm{max}$ to fix the Coulomb logarithm (e.g.\ equation~\ref{eqn:Sum_deltavperpsquared}), the Balescu--Lenard equation suffers from no large scale divergence and no such cutoff is needed.
 The largest scale of fluctuations in the system is accounted for naturally through the lowest-order basis elements $\Phi^{(p)}$ and wavenumbers $\bn$.
\item It takes the form of a Fokker--Planck equation in action space, with both friction and diffusion terms present, as it must following the physical arguments below equation~\eqref{eqn:F2_time_dependent}.
\item The right hand side is an antisymmetric functional of $f_0(\bJ)$ and $f_0(\bJ')$.
This tells us that the system is driven by \textit{pairwise interactions}.
In other words, the Balescu--Lenard equation describes a collection of orbits at action space location $\bJ$
interacting with orbits at all other locations $\bJ'$.
It is remarkable that in this approximation (truncating at second order in small fluctuations), the collective evolution of an amazingly complicated 
$N$-star system acts \textit{as if it were driven by uncorrelated two-body encounters} --- though they interact not through the usual bare coupling $\psi = -G/\vert \br - \br' \vert$ but through the dressed pairwise interaction potential $\psi^\mathrm{d}$.
This is a direct manifestation of Rostoker's principle, which serves as the basis for the alternative derivation shown in \S\ref{sec:Rostoker}.
\item The Dirac delta function on the right hand side encodes the fact that interactions are not only pairwise but also \textit{resonant}. In other words, the only meaningful interactions occur if some linear combination of frequencies of the star at $\bJ$ matches a linear combination of the frequencies of the star at $\bJ'$.     This makes sense, since pairs of stars that do not satisfy this condition will undergo an interaction that vanishes 
on average --- each star will feel an equal `pull' and `push' from the other as it traverses many mean field orbits, with a net result of zero.
\item The equation conserves mass, energy, and satisfies a $H$-theorem for Boltzmann entropy. In addition, the right hand side exactly vanishes for inhomogeneous Boltzmann DFs (i.e.\ thermal
distributions) of the form $f_0(\bJ) \propto \me^{-\beta H_0(\bJ)}$ (although these DFs are not realizable in 
actual galaxies since they contain infinite mass).
\item In the limit where collective amplification is ignorable ($\mathsf{M}=\mathsf{0}$, equation~\ref{Fourier_M}), we can replace $\psi^\mathrm{d}\to\psi$ and recover the \textit{Landau equation} which describes the evolution of an $N$-body system driven by a  `bare' resonant interactions\cite{Chavanis2013-sa}.
\end{itemize}

Finally, in the limit of a homogeneous system one can use the angle-action mapping~\eqref{eq:AA_homogeneous}, 
and thereby recover (up to constants) the classic Balescu--Lenard collision operator from electrostatic plasma kinetics
\begin{align}
&\frac{\p f_0(\mathbf{v})}{\p t} \propto \frac{\p}{\p \mathbf{v}} \cdot  \sum_{{\mathbf{k}}} {\mathbf{k}}  \int  \md \mathbf{v}' \delta [ \mathbf{k} \cdot (\bv -  \bv')]
\frac{1}{k^4 \vert \epsilon_{{\mathbf{k}}}({\mathbf{k}}' \cdot \bv')
\vert^2} \, \mathbf{k} \cdot \left(  \frac{\p}{\p \mathbf{v}}
- \frac{\p}{\p \mathbf{v}'}
\right) f_0(\mathbf{v}) f_0(\mathbf{v}').
\label{eqn:BL_Homogeneous}
\end{align}

In Table~\ref{table:BL_Comparison}, we provide a direct comparison between the various terms in the homogeneous equation~\eqref{eqn:BL_Homogeneous} and the corresponding inhomogeneous equation \eqref{eqn:BL}.
\begin{table}
\caption{Comparison of various pieces entering the Balescu--Lenard equation for homogeneous (equation~\ref{eqn:BL_Homogeneous}) and inhomogeneous (equation~\ref{eqn:BL}) stellar systems.\label{table:BL_Comparison}}
\begin{ruledtabular}
\begin{tabular}{ccc}

&
Homogeneous
&
Inhomogeneous
\\
\hline
Orbital distribution
& 
$\displaystyle f_0(\bv, t)$
& 
$\displaystyle f_0(\bJ, t)$
\\
Basis decomposition
& 
$\displaystyle \psi(\br, \br') \propto \sum_{\mathbf{k}} \me^{\mi \mathbf{k}\cdot(\br-\br')} / k^{2} $
&
$\displaystyle \psi(\br, \br') \propto \sum_{p} \Phi^{(p)}(\br) \, \Phi^{(p) *}(\br')  $ 
\\
Response matrix
&
$\displaystyle \propto \frac{1}{k^2}  \!\! \int \!\! \md \mathbf{v} \, \frac{{\mathbf{k}} \cdot \p f_{0} / \p \bv}{\omega - {\mathbf{k}} \cdot \bv } $
&
$\displaystyle \propto \sum_{{\mathbf{n}}} \!\! \int \!\! \md \mathbf{J} \, \frac{{\mathbf{n}} \cdot \p f_{0} / \p \mathbf{J}}{\omega - {\mathbf{n}} \cdot \bOm } \, \Phi^{(p) *}_{{\mathbf{n}}} (\mathbf{J}) \, \Phi^{(q)}_{{\mathbf{n}}} (\mathbf{J})$
\\
Dressing of perturbations
&
$\displaystyle \propto 1 / \vert \epsilon_{\mathbf{k}}(\mathbf{k}' \cdot \mathbf{v}') \vert^{2}$
&
$\displaystyle \vert \psi^\mathrm{d}_{\mathbf{n}\mathbf{n}'}(\bJ, \bJ', \mathbf{n}'\cdot\bOm')\vert^2$
\\
Resonance condition
&
$\displaystyle \mathbf{k}\cdot(\bv-\bv')=0 $
&
$\displaystyle \mathbf{n} \cdot \bOm = {\mathbf{n}}' \cdot \bOm'$
\end{tabular}
\end{ruledtabular}
\end{table}
In particular, for a homogeneous system we have $\vert \psi^\mathrm{d}_{\mathbf{k}\mathbf{k}'}\vert^2 \propto \delta_{\mathbf{k}}^{\mathbf{k}'}k^{-4}\vert \epsilon_{\mathbf{k}} \vert^{-2}$, where $\epsilon$ is the dielectric function. This is why the homogeneous equation \eqref{eqn:BL_Homogeneous} involves only a single sum over wavenumbers $\mathbf{k}$, rather than the double sum over $\bn, \bn'$ that we had in the inhomogneous equation \eqref{eqn:BL}. The factor $\propto \vert \epsilon_{\mathbf{k}}(\mathbf{k}'\cdot\mathbf{v}') \vert^{-2}$ in \eqref{eqn:BL_Homogeneous} is the classic dielectric shielding factor. In near-Maxwellian plasmas, this factor tends to suppress two-body interactions on large (bigger than Debye) scales, but in stellar systems it can strongly amplify such interactions  --- see \S\ref{sec:BL_Breakdown} for a detailed discussion.

\subsubsection{An historical aside}

The Balescu--Lenard equation was first derived in the context of homogeneous electrostatic plasmas\cite{Balescu1960-cn,Lenard1960-rm}. The main mathematical difference in the stellar dynamical case is that one must formulate the equation in angle-action variables.

The basic idea of formulating a quasilinear diffusion equation  in action space seems to have been 
first arrived at by~\citet{Kaufman1970-za,Kaufman1971-ld,Kaufman1972-vv},
in the context of inhomogeneous plasmas.
The idea of doing \textit{gravitational} kinetic theory in angle action variables was introduced around the same time by~\citet{kalnajs1971dynamics}, who was predominantly interested in the linear properties of disk galaxies and their associated spiral 
modes (\S\ref{sec:Linear_Response}).
A proper diffusion equation of the form
$\partial_t f_0 = \partial_{\bJ}\cdot [\mathsf{D}\cdot \partial_{\bJ} f_0]$ was not written down in the stellar-dynamical context until a few years later, by~\citet{dekker1976spiral}.
For a thorough overview of the history and of linear and quasilinear kinetics in angle-action variables (including their plasma precedents), 
see~\citet{chavanis2023kinetic}.

The first derivation of the full Balescu--Lenard kinetic equation for stellar systems was in a somewhat opaque 1987 treatise by~\citet{Luciani1987-do}.
The same equation was then arrived at in 2010 by~\citet{Heyvaerts2010-aq} using the BBGKY hierarchy; then in 2012 by~\citet{Chavanis2012-lu}
using the Klimontovich description; again in 2017 by~\citet{Heyvaerts2017-ts} using a Fokker--Planck description; then in 2018 by~\citet{Fouvry2018-gi} using stochastic equations and Novikov's theorem; 
and finally in 2021 by~\citet{Hamilton2021-qe} using Rostoker's principle, as we now detail.

\subsection{Alternative derivation of the Balescu--Lenard equation from Rostoker's principle}
\label{sec:Rostoker}

\textit{Rostoker's principle}\cite{Rostoker1964-qv,Rostoker1964-rt,Diamond2010-mh} tells us that we can think of a secularly evolving plasma or stellar system as a collection of $N$ uncorrelated particles undergoing two-body encounters (\`a la Chandrasekhar), provided we replace the bare Coulombic/Newtonian interaction $\psi(\br, \br')$ of each two-body encounter with its dressed counterpart $\psi^\md$.
Roughly speaking, this is because particles are weakly coupled, and because the main effect of collective interactions upon the potential fluctuation spectrum is to dress each particle with its collective wake (equation~\ref{eqn:dressed_potential_fluctuations}).
Here we use this principle to derive the Balescu--Lenard equation~\eqref{eqn:BL} in a short and simple way following~\citet{Hamilton2021-qe}.
(Note we do not actually need to deal explicitly with fluctuations here, so we will drop the `$0$' subscript on the ensemble-averaged DF).

First, consider a `test' star with coordinates $(\btheta, \mathbf{J})$ and a `field' star with coordinates $(\btheta', \mathbf{J}')$. Mathematically, Rostoker's principle tells us to forget about all the other stars, and treat the interaction of these two stars as if they were an isolated system with specific two-body Hamiltonian (units of (velocity)$^2$):
\begin{equation}
\label{eqn:twoody_Hamiltonian}
    h = H_0(\mathbf{J}) + H_0(\mathbf{J}') + m \psi^\md(\btheta,\mathbf{J}, \btheta',\mathbf{J}'),
\end{equation}
where $H_0(\mathbf{J})$ is the mean field Hamiltonian.
Here $m\psi^\md(\btheta,\mathbf{J}, \btheta',\mathbf{J}')$ is the dressed specific potential energy between a star at phase space location $(\btheta, \mathbf{J})$ and a star at $(\btheta', \mathbf{J}')$. (It consists of the usual Newtonian attraction plus collective effects; if we ignore these then $\psi^\md \to -G/\vert \br - \br'\vert$). 
Let us expand $\psi^\md$ as a Fourier series in the angle variables:
\begin{equation}
\psi^\md(\btheta,\mathbf{J}, \btheta',\mathbf{J}')= \sum_{{\mathbf{n}}, {\mathbf{n}}'} \me^{\mi({\mathbf{n}} \cdot \btheta - {\mathbf{n}}' \cdot \btheta') } \, \psi^\md_{{\mathbf{n}} {\mathbf{n}}'}(\mathbf{J}, \mathbf{J}',{\mathbf{n}}' \cdot \bOm').
\end{equation}
Here $\psi^\md_{{\mathbf{n}} {\mathbf{n}}'}(\mathbf{J}, \mathbf{J}',{\mathbf{n}}' \cdot \bOm')$ is the dressed potential interaction we derived in \S\ref{sec:Wakes}, and we have put in the correct frequency dependence ($\omega = {\mathbf{n}}' \cdot \bOm'$), since it is physically clear that these are the only frequencies available to the system. (Alternatively one can cheat: just think of $\psi^\mathrm{d}/m$ as some function of $\btheta, \btheta', \bJ, \bJ'$,  and then identify the correct functional form at the end by comparison with~\ref{eqn:BL}).

Let us treat the two-body interaction as a perturbation to each star's motion.
Using Hamilton's equation $\md \mathbf{J}/\md t = -\p h / \p \btheta$, we find that to 
zeroth order in this perturbation the test and field stars just follow their mean field trajectories $\btheta = \btheta_0 + \bOm t$ and $\btheta' = \btheta'_0 + \bOm' t$ indefinitely.  To first order, similarly to equation~\eqref{eqn:deltaJ_1}, the result of their interaction is to nudge each other to new actions $\mathbf{J}+\delta \mathbf{J}$ and $\mathbf{J}'+\delta \mathbf{J}'$ respectively, where
\begin{align}
\delta \mathbf{J}(\btheta_0, \mathbf{J}, \btheta_0', \mathbf{J}', \tau) 
= 
-  m
 \sum_{{\mathbf{n}}, {\mathbf{n}}'} \mi {\mathbf{n}} \, \psi^\md_{{\mathbf{n}} {\mathbf{n}}'}(\mathbf{J}, \mathbf{J}', {\mathbf{n}}' \cdot \bOm') \, \me^{\mi({\mathbf{n}} \cdot \btheta_0 - {\mathbf{n}}' \cdot \btheta_0')}  \frac{\me^{\mi({\mathbf{n}} \cdot \bOm - {\mathbf{n}}' \cdot \bOm')\tau}-1}{\mi ({\mathbf{n}} \cdot \bOm - {\mathbf{n}}' \cdot \bOm')} ,
 \label{eqn:delta_J} 
\end{align}
The result for $\delta \mathbf{J}'$ is the same as~\eqref{eqn:delta_J} except we replace the first factor $\mi {\mathbf{n}} \to -\mi {\mathbf{n}}'$. Note that because the two stars nudge each other, this pairwise interaction
conserves the total energy of the pair.
This is different from considering an interaction
between a test star and a background star (as in \S\ref{sec:Chandra}).

\begin{figure}[htbp!]
\centering
\includegraphics[width=0.45\linewidth]{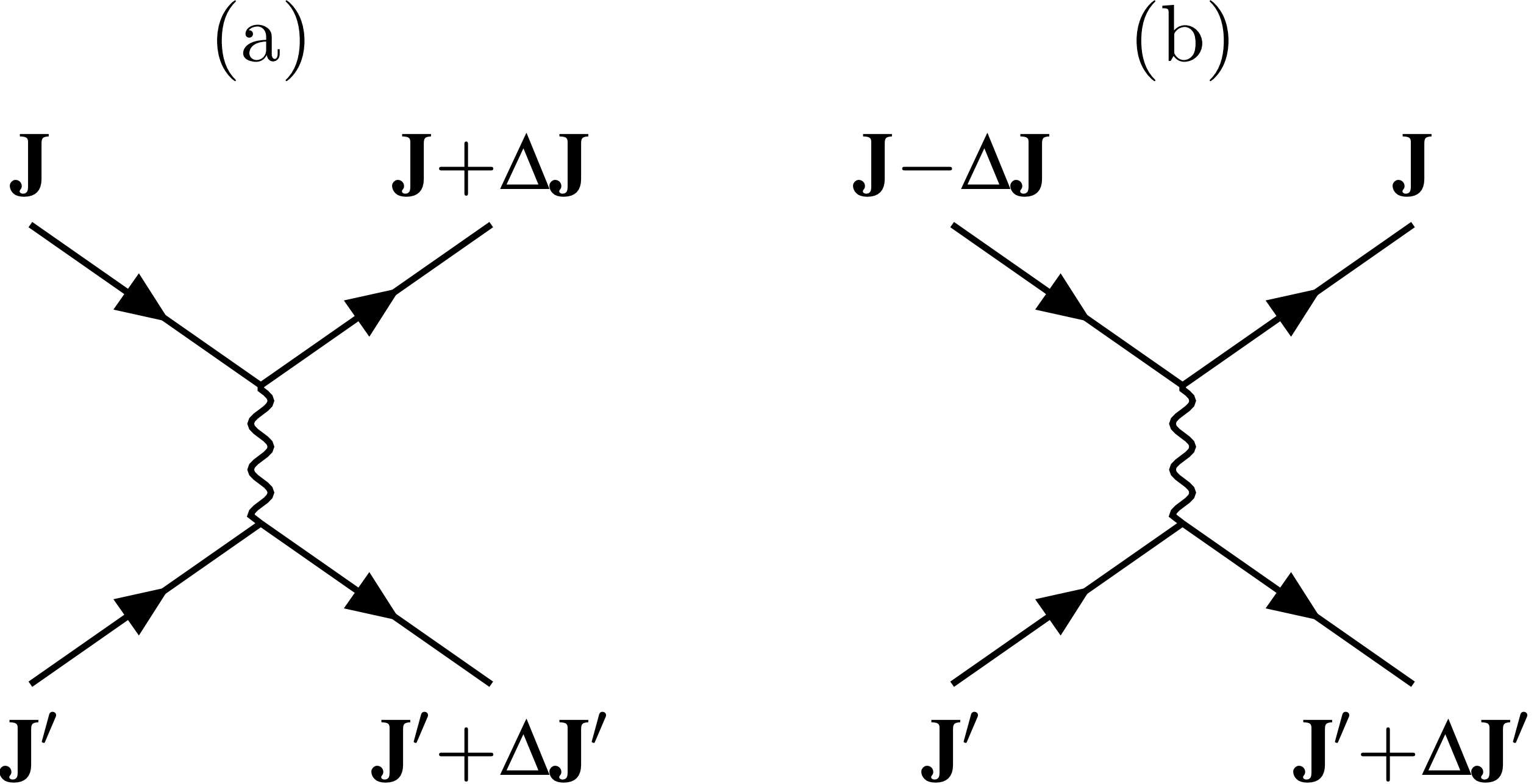}
\caption{Two possibilities for `scattering' in action coordinates. A test star's action changes by $\Delta \mathbf{J}$ during an `encounter' with a field star whose action changes from $\mathbf{J}'$ to $\mathbf{J}' + \Delta \mathbf{J}'$.
The DF $f(\mathbf{J})$ will be decremented if the test star is kicked out of state $\mathbf{J}$ (process a), but will be incremented if it is kicked in to state $\mathbf{J}$ from $\mathbf{J}-\Delta \mathbf{J}$ (process b). To write down a master equation we must also account for the inverse processes (by reversing the directions of the arrows).
\label{fig:BL_Feynman}}
\end{figure}

Next, following Rostoker, we consider the relaxation of our entire system to consist of nothing more than an uncorrelated set of dressed two-body encounters.
Then it is easy to write down a master equation for the DF $f(\mathbf{J})$.  To do so, we account for (a) test stars being nudged out of the state $\mathbf{J}$ and in to some new state $\mathbf{J}+\Delta \mathbf{J}$, as illustrated in Figure~\ref{fig:BL_Feynman}a, and (b) test stars being nudged in to the state $\mathbf{J}$ from $\mathbf{J} - \Delta \mathbf{J}$, as in Figure~\ref{fig:BL_Feynman}b.
Processes (a) and (b) are both deterministic and time-reversible,
so we must also account for their inverses, which would be represented by the same diagrams but with the arrows pointing in the opposite direction.
Let the \textit{transition rate density} be $w(\Delta \mathbf{J}, \Delta \mathbf{J}' \vert \mathbf{J}, \mathbf{J}')$. This quantity is defined such that $w(\Delta \mathbf{J}, \Delta \mathbf{J}' \vert \mathbf{J}, \mathbf{J}')\,  \md \Delta \mathbf{J} \md \Delta \mathbf{J}' \tau$ is the probability that a given test star with initial action $\mathbf{J}$ is scattered to the volume of phase space $\md \Delta \mathbf{J}$ around $\mathbf{J}+\Delta \mathbf{J}$, by a given field star with action $\mathbf{J}'$ that is itself scattered to the volume element $\md \Delta \mathbf{J}'$ around $\mathbf{J}'+\Delta \mathbf{J}'$, in a time interval $\tau$ that is much longer than an orbital period but much shorter than the relaxation time.
Assuming the system's equilibrium state
is invariant under time reversal, $f$
must satisfy the master equation
\begin{align}
 \frac{\p f(\mathbf{J})}{\p t} 
 = \frac{(2\pi)^d}{m} \int \md \mathbf{J}' \md \Delta \mathbf{J} \md \Delta \mathbf{J}' 
 \frac{1}{2} \Big[ & w(\Delta \mathbf{J}, \Delta \mathbf{J}' \vert  \mathbf{J}, \mathbf{J}') [ -f(\mathbf{J}) f(\mathbf{J}') + f(\mathbf{J} +\Delta \mathbf{J})f(\mathbf{J}' + \Delta \mathbf{J}')]
\nn
\\
&
 + w(\Delta \mathbf{J}, \Delta \mathbf{J}' \vert  \mathbf{J}-\Delta \mathbf{J},\mathbf{J}')
[ f(\mathbf{J}-\Delta \mathbf{J}) f(\mathbf{J}') - f(\mathbf{J})f(\mathbf{J}' + \Delta \mathbf{J}') ] 
 \Big].
\label{eqn:Calc_Rostoker}
\end{align}
(The factor of $1/2$ accounts for the fact that, when we integrate over all possible $\bJ'$, $\Delta \bJ$ and $\Delta \bJ'$, the contribution from each diagram in Figure~\ref{fig:BL_Feynman} accounts for all possible interactions, and both diagrams are equivalent. The `accounting trick' of using both diagrams and then halving the answer leads very simply to the Balescu--Lenard equation).

By expanding the integrand in \eqref{eqn:Calc_Rostoker} for weak interactions, i.e.\ for $\Delta \mathbf{J}, \Delta \mathbf{J}' \ll \mathbf{J}, \mathbf{J}'$, up to second order in small quantities, we can show that
\begin{equation}
 \frac{\p f(\mathbf{J})}{\p t} 
 = 
 \frac{\p}{\p \mathbf{J}} \cdot \int \md \mathbf{J}' 
\left[ \msA (\bJ , \bJ') \cdot  f(\mathbf{J}')\frac{\p f}{\p \mathbf{J}} +
\msB (\bJ, \bJ') \cdot f(\mathbf{J})\frac{\p f}{\p \mathbf{J}'}
\right].
\label{eqn:kinetic_equation}
\end{equation}
where $\msA$ is the $3\times 3$ matrix
\begin{align}
\label{eqn:A}
\msA (\mathbf{J},\mathbf{J}') &= \frac{(2\pi)^d}{2m} \int \md \Delta \mathbf{J} \md \Delta \mathbf{J}' w(\Delta \mathbf{J}, \Delta \mathbf{J}' \vert \mathbf{J}, \mathbf{J}') \, \Delta \mathbf{J} \otimes \Delta \mathbf{J} \equiv \frac{(2\pi)^d}{2m} \frac{\langle \Delta \mathbf{J} \otimes \Delta \mathbf{J} \rangle_\tau}{\tau},
\end{align}
where $\langle \Delta \mathbf{J} \otimes \Delta \mathbf{J} \rangle_\tau$ is the expectation value of $\Delta \mathbf{J} \otimes \Delta \mathbf{J}$ after a time interval $\tau$ for a given test star action $\mathbf{J}$ and field star action $\mathbf{J}'$. 
The matrix $\msB$ is identical to $\msA$ except we replace $\langle \Delta \mathbf{J} \otimes \Delta \mathbf{J} \rangle_\tau \to \langle \Delta \mathbf{J} \otimes \Delta \mathbf{J}' \rangle_\tau$. In a homogeneous system, $\mathbf{J}$ and $\mathbf{J}'$ can be taken to be the (suitably normalized) linear momenta of the test and field star; then momentum conservation implies $\Delta \mathbf{J} = - \Delta \mathbf{J}'$ so that $\msA = -\msB$. With this, one can easily recover equation (2.36) of the analogous 
calculation for a homogeneous plasma by~\citet{Diamond2010-mh}.

Now we put the above results together. Since by assumption stars at a given action are randomly distributed in angles, we can calculate the expectation value $\langle \Delta \mathbf{J} \otimes \Delta \mathbf{J} \rangle_\tau$ by averaging over the initial phases $\btheta_0, \btheta_0'$. (This is analogous to the integral over impact parameters $b$ in Chandrasekhar's theory, as in equation~\ref{eqn:Sum_deltavperpsquared}). Thus we have
\begin{align}
     \langle \Delta \mathbf{J} \otimes \Delta \mathbf{J} \rangle_\tau =
\int \frac{\md \btheta_0}{(2\pi)^d} \frac{\md \btheta_0'}{(2\pi)^d} \,\delta \mathbf{J}(\btheta_0, \mathbf{J}, \btheta_0', \mathbf{J}', \tau) \otimes \delta \mathbf{J}(\btheta_0, \mathbf{J}, \btheta_0', \mathbf{J}', \tau),
\label{eqn:expectation_value}
\end{align}
where $\delta \mathbf{J}$ is given in equation~\eqref{eqn:delta_J}.
Plugging~\eqref{eqn:expectation_value} and~\eqref{eqn:delta_J} in to~\eqref{eqn:A} and taking the limit $\tau \to \infty$, and making use of the following two identities
\begin{subequations}
\begin{align}
    &\psi^\md_{{\mathbf{n}} {\mathbf{n}}'}(\mathbf{J},\mathbf{J}' , \omegaR) = [\psi^{ \md}_{-{\mathbf{n}}, -{\mathbf{n}}'}(\mathbf{J},\mathbf{J}' , - \omegaR)]^* \quad \mathrm{for} \,\,\, \omegaR \,\,\, \mathrm{real} ,
  \label{eqn:identity_psid}  
    \\
   & \lim_{\tau \to \infty} \left[ \vert \me^{\mi x\tau}-1 \vert^2/x^2\tau\right] = 2\pi\delta (x),
    \label{eqn:identity_exp}
\end{align}
\label{eqn:identity_double}%
\end{subequations}
we find the following expression for the matrix $\msA$:
\begin{align}
\label{eqn:A_final}
\msA (\mathbf{J}, \mathbf{J}') = \pi (2\pi)^d m \sum_{{\mathbf{n}}, {\mathbf{n}}'} & \,
{\mathbf{n}} \otimes {\mathbf{n}}\,
\delta ({\mathbf{n}} \cdot \bOm - {\mathbf{n}}' \cdot \bOm')\vert \psi^\md_{{\mathbf{n}} {\mathbf{n}}'}(\mathbf{J}, \mathbf{J}', {\mathbf{n}}' \cdot \bOm') \vert^2 .
\end{align}
The result for $\msB$ is identical to~\eqref{eqn:A_final} 
except we replace the factor ${\mathbf{n}} \otimes {\mathbf{n}}$ with $- {\mathbf{n}} \otimes {\mathbf{n}}'$. 
Putting the explicit formulae for $\msA (\mathbf{J},\mathbf{J}')$ and $\msB (\mathbf{J},\mathbf{J}')$ in to the kinetic equation~\eqref{eqn:kinetic_equation}, we recover the  Balescu--Lenard equation~\eqref{eqn:BL}.

Thus, just as in plasma physics\cite{Diamond2010-mh}, the Balescu--Lenard collision operator can be interpreted as arising from the uncorrelated pairwise scattering of \textit{dressed} stars as they traverse their (otherwise unperturbed) mean field orbits.


\subsection{Examples}
\label{sec:BL_examples}

Implementing the Balescu--Lenard equation~\eqref{eqn:BL} in practice is hard, and has only been attempted a handful of times.
Here, we mention some of these implementations,
starting with an artificial one-dimensional gravitational system (\S\ref{sec:Example_BL_1D}),
then moving on to a hot sphere (\S\ref{sec:Example_BL_Sphere}) and finally a cold disk (\S\ref{sec:Example_BL_Disk}).
Other applications include the inhomogeneous Hamiltonian Mean Field model\cite{Benetti2017-gi},
as well as the relaxation of the eccentricities\cite{bar2018scalar}
and orientations\cite{Fouvry2019-sphere} of stellar  orbits
around supermassive black holes.

\subsubsection{Example: Balescu--Lenard in one-dimensional `gravity'}
\label{sec:Example_BL_1D}

Recently,
\citet{roule2022long} considered the evolution of
a one-dimensional `gravitational' $N$-body system.
Their system is gravitational in the sense that density and potential are linked through Poisson's equation.
Yet, because it is one-dimensional the
resulting bare interaction between particles is $\psi(x, x') \propto \vert x-x'\vert$, and the force between any two particles is proportional to $\mathrm{sgn}(x-x')$.
In this simple setup, there is a one-to-one mapping between action $J$ and energy  $E$, 
and the authors tend to use $E$ to characterize orbits.
The initial DF the authors choose for the particles is closely analogous to the isotropic DF that generates the 
three-dimensional Plummer sphere whose orbits we studied in \S\ref{sec:Central}. Crucially for the application of Balescu--Lenard theory, the DF is linearly stable.

\citet{roule2022long}simulated this one-dimensional $N$-body system directly many times over with random initial conditions, in order to produce an ensemble of systems over which to average.
This was to be compared with the prediction from the Balescu--Lenard equation~\eqref{eqn:BL}.
They also performed simulations `without collective effects'.
To do this they first integrated the mean field motion of random ensembles of massive but non-interacting particles in their chosen model.
They then used the fluctuating gravitational potential generated by these particles as the input to a test particle simulation.
Since test particles, by definition, do not produce potential fluctuations of their own, they experience no friction (so $\bF_1 = \mathbf{0}$ in equation~\ref{eqn:F1_plus_F2}) but they do experience diffusion ($\bF_2 \neq \mathbf{0}$). These test particle simulations were to be compared to the predictions of `Landau' theory, i.e.\ diffusion along the lines of Balescu--Lenard but without collective amplification (\S\ref{sec:Physics_of_BL}).

\begin{figure}[htbp!]
\centering
\includegraphics[width=0.55\linewidth]{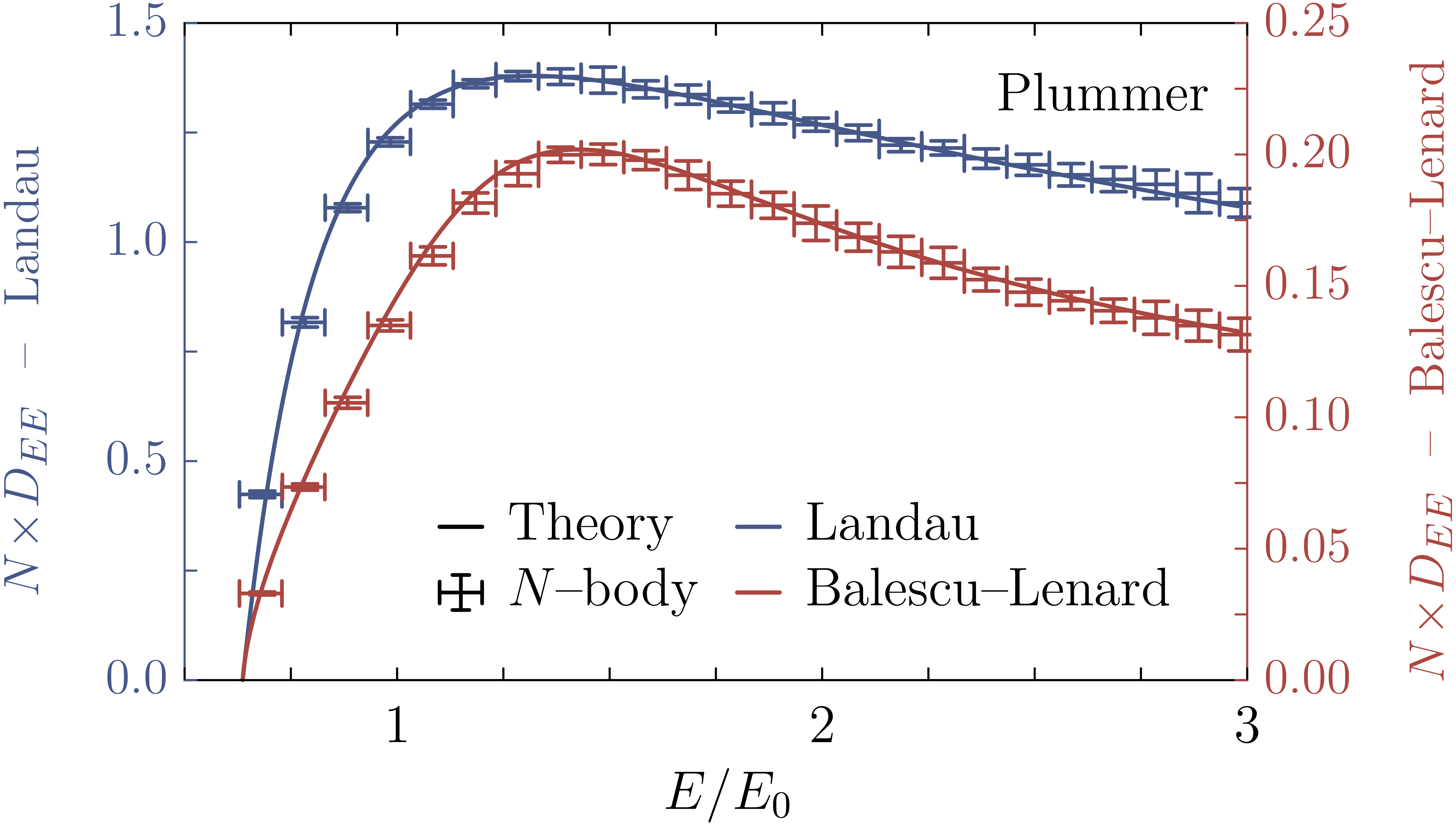}
\caption{Diffusion coefficients of particles in a one-dimensional self-gravitating system (\S\ref{sec:Example_BL_1D}).
Crosses show the results of $N$-body simulations, while
solid curves show the theoretical results based on the Balescu--Lenard equation~\eqref{eqn:BL}
(when collective interactions are switched on) or the corresponding Landau equation (when they are switched off).
Note that the vertical axis scale is different for red and blue points.
Figure taken from M. Roule, J.-B. Fouvry, C. Pichon, and P.-H. Chavanis, PRE 106, 044118 (2022)\cite{roule2022long}.
}
\label{fig:Roule_1D}
\end{figure}

Figure~\ref{fig:Roule_1D} shows~\citet{roule2022long}'s results.
The red crosses in this figure show measurements of the diffusion coefficient in energy of particles over a time long compared to the dynamical time $t_\mathrm{dyn}$ but short compared to the relaxation time $t_\mathrm{relax}$.
The red solid  line shows the theoretical `Balescu--Lenard' prediction.
It demonstrates an excellent fit, confirming that equation~\eqref{eqn:BL}
is capturing the physics of this system.
Meanwhile the blue crosses and blue solid line shows that the `Landau' theory is an excellent predictor of the diffusion coefficient in the case without collective effects.

Let us emphasize that the red and blue data in Figure~\ref{fig:Roule_1D} are plotted on different $y$-axes.
In particular, the Landau diffusion coefficient is $\approx 1$ in these units, which
 would correspond to a relaxation time $t_\mathrm{relax}\sim Nt_\mathrm{dyn}$, similar to Chandrasekhar's theory (equation~\ref{eqn:2BR}).
What is very notable is that over a large range of energies $E$,
the Balescu--Lenard diffusion coefficient is ten times smaller than the Landau one,  corresponding to a relaxation time $t_\mathrm{relax} \sim 10Nt_\mathrm{dyn}$. 
In other words, this is a self-gravitating system (to the extent that this is `gravity' at all) in which collective effects tend to quench, rather than amplify, the rate of evolution. 

While this result is unusual, it is not unique.
A similar result was obtained by Weinberg\cite{Weinberg1989-cj}, who studied the dynamical friction drag on a satellite falling into a spherical galaxy (a similar calculation to the one we will present in \S\ref{sec:Dynamical_friction}). Weinberg found that when collective effects were included, the wake that the satellite induced in the galaxy was highly symmetric and that this led to a drag which was \textit{weaker} than in the bare case without collective effects. The lesson we draw from these examples is that collective effects can have a major impact upon the evolution of inhomogeneous stellar systems, but it may be difficult to guess in advance what that impact will be.



\subsubsection{Example: Balescu--Lenard in a hot sphere}
\label{sec:Example_BL_Sphere}

On the surface, Balescu--Lenard theory is much more general than Chandrasekhar's theory of two-body relaxation (\S\ref{sec:Chandra}). One might therefore ask (i) whether it contains Chandrasekhar's theory as a limiting case, and (ii) given that Chandrasekhar's theory works very well in hot spheres (\S\ref{sec:Where_Now}), 
does a direct application of the results of Balescu--Lenard theory to those systems give an answer that differs from Chandrasekhar's theory?

The answer to question (i) is yes: one can recover Chandrasekhar's theory from the Balescu--Lenard formalism by ignoring collective effects and  confining oneself to interactions on small scales over which the curvature of mean field orbits can be neglected (replacing them with straight lines\cite{Chavanis2013-sa}).

\begin{figure}[htbp!]
\centering
\includegraphics[width=0.45\linewidth]{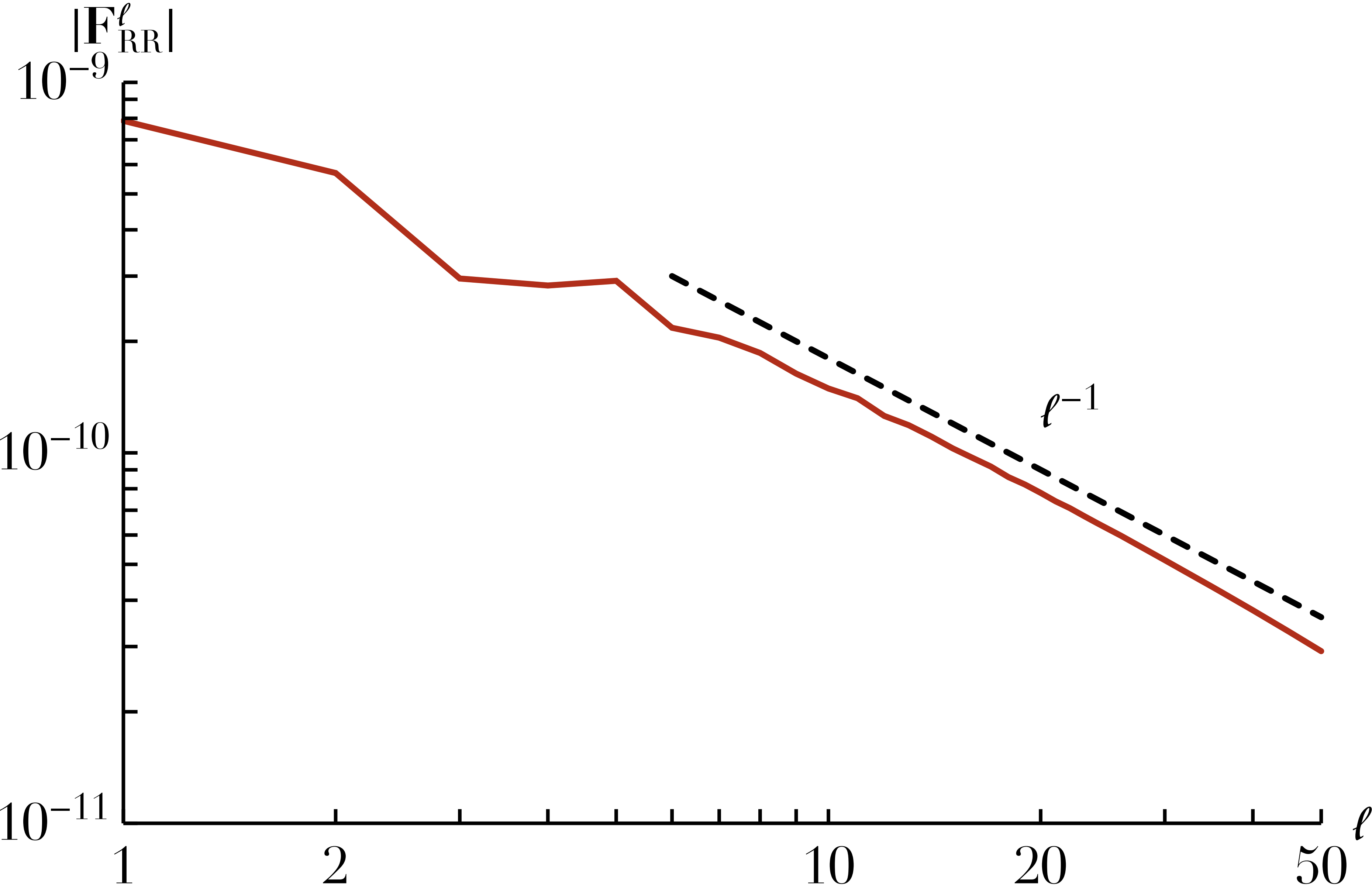}
\caption{Balescu--Lenard flux, $\bF_{\ell}$ (see equations~\ref{eqn:F1_plus_F2}-\ref{eqn:F2_time_dependent}), for fluctuations with 
spherical harmonic quantum number $\ell$ (an angular scale $\sim \ell^{-1}$).
The magnitude of this flux scales as $\ell^{-1}$ for large $\ell \gg 1$.
This confirms that the Balescu--Lenard equation recovers the Coulomb logarithm behavior from Chandrasekhar's theory in the limit of local encounters (\S\ref{sec:Chandra}).
Figure taken from J.-B. Fouvry, C. Hamilton, S. Rozier, and C. Pichon, MNRAS 508, 2210 (2021)\cite{Fouvry2021-ra}.
}
\label{fig:Coulomb_Recovery}
\end{figure}

\begin{figure}[htbp!]
\centering
\includegraphics[width=0.75\linewidth]{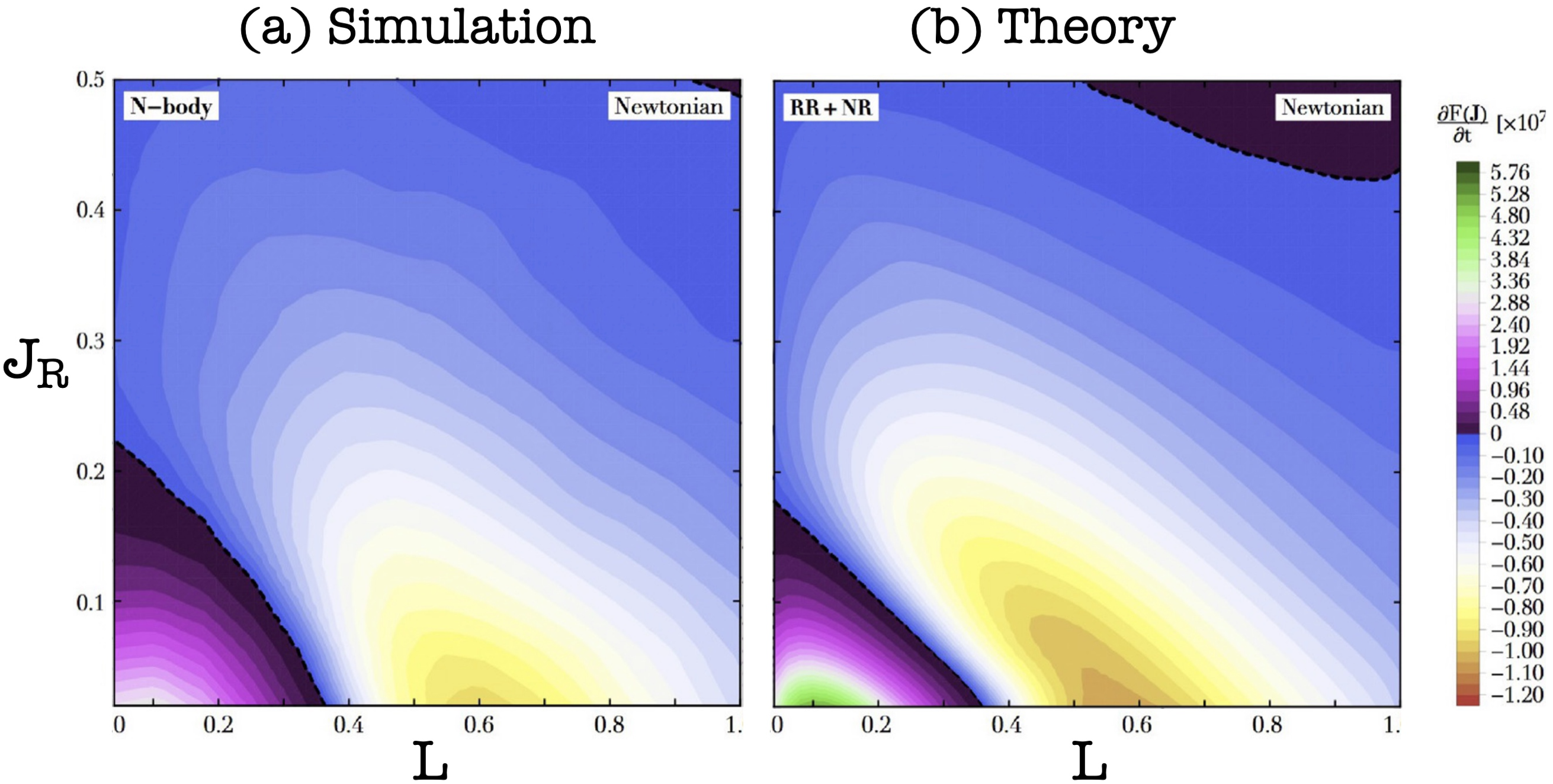}
\caption{Contour plot of the evolution rate $\partial f_0/\partial t$ in action space, 
for a simple model of a spherical cluster as measured from (a) direct $N$-body simulations and (b) a theory which combines the Balescu--Lenard equation~\eqref{eqn:BL} on large scales with the Chandrasekhar theory on small scales. 
Figure taken from J.-B. Fouvry, C. Hamilton, S. Rozier, and C. Pichon, MNRAS 508, 2210 (2021)\cite{Fouvry2021-ra}.}
\label{fig:Fouvry21b}
\end{figure}

To answer question (ii) requires a lot more work.
\citet{Fouvry2021-ra} investigated it by evaluating the right hand side of equation~\eqref{eqn:BL}
for an \textit{isochrone sphere}, which is a toy model of a spherical star cluster.
Using a known (stable) DF, they calculated the Balescu--Lenard flux $\bF$ as a sum of contributions $\bF^\ell$ from spherical harmonic quantum numbers $\ell = 0, 1,2, ...$.
As illustrated in Figure~\ref{fig:Coulomb_Recovery}, they found that on small scales $\ell\gg 1$, the modulus $\vert \bF_\ell\vert$ scales as $\propto \ell^{-1}$.
Hence, a sum over $\ell$ recovers the small scale logarithmic divergence familiar from Chandrasekhar's theory, a result already obtained by~\citet{Weinberg1986-friction} when considering dynamical friction (see Figure~{5} therein).
Such a divergence is always present in a kinetic theory based on weak $\pm1/r^2$ interactions. This strongly suggests that Chandrasekhar's theory --- which is much simpler to implement than the Balescu--Lenard theory --- is  being recovered by equation~\eqref{eqn:BL} on small scales.
This motivated~\citet{Fouvry2021-ra} to calculate the Balescu--Lenard flux as accurately as possible on large scales ($\ell \leq \ell_\mathrm{crit}$, with $\ell_\mathrm{crit}\approx 12$, say), and to calculate the 
Chandrasekhar flux on small scales ($\ell > \ell_\mathrm{crit}$),  and then add the two contributions together.
The advantage of this over just using Chandrasekhar's theory is that one
is not plagued by any large-scale divergence
(which is normally cured by imposing a maximum impact parameter $b_\mathrm{max}$), 
and in principle one accounts for any resonant and/or collective phenomena that are ignored by Chandrasekhar.
The result of this exercise, which is shown in Figure~\ref{fig:Fouvry21b}, is a slightly improved agreement between the contour map of $\partial f_0/\partial t$
in action space computed from theory as compared to $N$-body simulations.
As we discussed in \S\ref{sec:Where_Now}, the agreement based on Chandrasekhar's theory alone is already good, and as such the large scale Balescu--Lenard `correction' is small.
 
In summary, perhaps the most intellectually satisfying approach to spherical cluster kinetics is to couple a a large-scale Balescu--Lenard flux with a small-scale Chandrasekhar flux. But in practice, Chandrasekhar's theory applied to all scales (with an additional cutoff at large scales $b_\mathrm{max}$) 
is almost as good a description,
and is both much simpler to understand and easier to implement.

\subsubsection{Example: Balescu--Lenard in a cold disk}
\label{sec:Example_BL_Disk}

\begin{figure}[htbp!]
\centering
\includegraphics[width=0.85\linewidth]{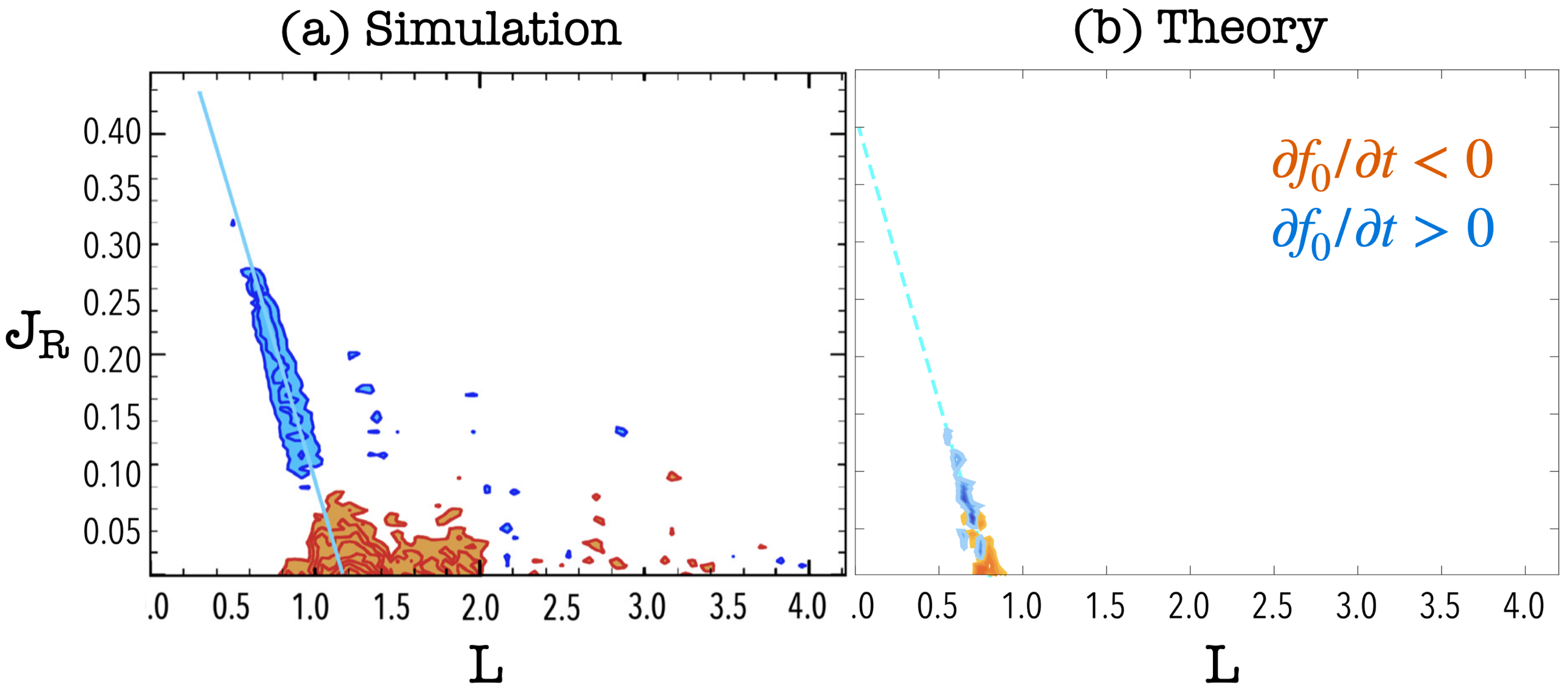}
\caption{Change in the DF $f_0(L, J_R)$ for an initially smooth, razor thin stellar disk, (a) as measured in an $N$-body simulation, (b) as predicted by the Balescu--Lenard equation~\eqref{eqn:BL}
(figure adapted from J.-B. Fouvry, C. Pichon, J. Magorrian, and P.-H. Chavanis, A\&A 584, A129 (2015)\cite{Fouvry2015-nk}).
The DF develops a ridge/groove feature which renders it unstable to spiral Landau modes --- see Figure~\ref{fig:groove}.}
\label{fig:Fouvry15}
\end{figure}

As a final example, let us discuss the dynamics of an initially stable, razor thin stellar disk, following the $N$-body simulations of~\citet{Sellwood2012-ro} and the kinetic predictions from~\citet{Fouvry2015-nk}.
The model employed by these authors is the stable Mestel disk, with a DF $f_0$ whose $J_R=0$ slice we plotted with a dotted black curve back in Figure~\ref{fig:groove}a. Notably, this DF is very smooth in action space. However,
\citet{Sellwood2012-ro}'s simulations of this disk shows that after many dynamical times $t_\mathrm{dyn}$, a grooved/ridged feature always develops in the action space DF (see Figure~\ref{fig:Fouvry15}a).
The time taken for this feature to develop increases linearly with the number of particles $N$ employed, suggesting it is driven by the finite-$N$ noise.
Remarkably, this newly-modified DF is found in the simulations to be 
unstable to the formation of exponentially growing spiral waves.

In fact, this modified DF is precisely the one studied by~\citet{De_Rijcke2019-uo} in their linear stability analysis (\S\ref{sec:Response_Example_Imhomogeneous}), the $J_R=0$ slice through which is shown with a solid black curve in Figure~\ref{fig:groove}a. The corresponding spiral Landau mode they calculated from linear theory (which is shown in Figure~\ref{fig:groove}b) matches that extracted by~\citet{Sellwood2012-ro} from his simulations, confirming that his spiral waves are true linear instabilities of the modified DF. Thus, our stable, isolated stellar disk has driven itself unstable!

We can usefully contrast this behavior with that of a plasma. If we initiated e.g.\ a finite-$N$ electrostatic plasma with a Maxwellian DF, then the plasma would exhibit thermal fluctuations $\delta \Phi,\,  \delta f \propto N^{-1/2}$, but its mean-field Maxwellian $f_0$ would simply never evolve.  That is because the Maxwellian is the ultimate maximum entropy state in a homogeneous plasma, a perfect thermodynamic equilibrium.
But for self-gravitating systems there is no such thermodynamic equilibrium (\S\ref{sec:Thermal_Equilibrium}), and so even a linearly a stable stellar disk is never at rest: instead it gradually carves a groove in itself and drives itself unstable,
even without any external influences.
It is as if a stable plasma were to spontaneously grow a bump on its own tail!

Before the system becomes unstable, its evolution can be predicted using the
Balescu--Lenard equation~\eqref{eqn:BL}, which after all is supposed to describe the long term evolution of stable systems driven by amplified finite-$N$ noise.
\citet{Fouvry2015-nk} evaluated the right hand side of equation~\eqref{eqn:BL}
for the same Mestel disk DF used by~\citet{Sellwood2012-ro}. The resulting prediction is shown in Figure~\ref{fig:Fouvry15}b.
While there are some important differences between this panel and the result in panel (a),  the key physics is quite clear: the grooved/ridged DF is formed because of dressed, resonant coupling between finite-$N$ fluctuations in the disk, which happen to be strongly enhanced along a particular direction in action space (in this case $\bn=(2,-1)$, shown with light blue lines in the Figure).
Unfortunately, the Balescu--Lenard equation is so cumbersome to implement in practice that nobody has managed to push the theory any further than this.

\subsection{Breakdown of Balescu--Lenard theory; fundamental differences between galaxies and plasmas}
\label{sec:BL_Breakdown}

As discussed at the beginning of this section and in \S\ref{sec:Nonlin_Nonpert}, any linear or quasilinear kinetic theory will break down if strongly nonlinear effects such as orbit trapping are important to the evolution. We will develop a method for dealing with such systems in \S\ref{sec:Nonlinear_Kinetics}.
However, there is another way in which the Balescu--Lenard theory can break down
even in the absence of strong nonlinearities.
Namely, the Balescu--Lenard equation~\eqref{eqn:BL}
is derived under a key assumption which is that \textit{we may ignore the direct influence of the Landau modes}.
This assumption is often a poor one when applied to real galaxies.
This important point is often neglected in the literature, and a discussion of it allows us to 
highlight some fundamental differences between stellar systems and plasmas.

First, 
Balescu--Lenard theory is designed to describe the evolution of an $N$-body system in which dressed star-star interactions alone drive the evolution.
Mathematically, this assumption enters the Balescu--Lenard theory when one uses equation~\eqref{eqn:dressed_potential_fluctuations}  for the potential fluctuations. This equation 
is valid on timescales long compared to the 
damping time of the system's most weakly damped Landau mode $\omega_\mathrm{m}$, 
which requires
\begin{equation}
    t_\mathrm{relax} \gg 1 / \vert \mathrm{Im} \, \omega_\mathrm{m} \vert.
    \label{eqn:t_relax_requirement}
\end{equation}
This requirement should already be making us worried. Recall that for real galaxies $t_\mathrm{relax}$ may be of order only a few Gyr, and many stellar systems are capable of harboring weakly damped Landau modes that live for many $t_\mathrm{dyn}$, which is $\approx 250$ Myr in the Milky Way. Moreover, the right hand side of~\eqref{eqn:t_relax_requirement} by definition \textit{diverges} during the transition from a stable to unstable DF described in \S\ref{sec:Example_BL_Disk}, so the Balescu--Lenard theory is clearly invalid in that regime.
To make matters worse, in this regime the Balescu--Lenard equation actually gives nonsensical answers. 
To see this, note that a Landau mode with Im $\omega_\mathrm{m} \to 0$ corresponds to a divergence in $\psi^\mathrm{d}$ (equation~\ref{dressed_basis}) at some real frequency, and the Balescu--Lenard equation~\eqref{eqn:BL} involves an integral of $\vert \psi^\mathrm{d}\vert^2$ over all real frequencies.
Thus, a naive application of the Balescu--Lenard theory to this marginally stable regime will necessarily return an unphysical, divergent result.

How might we do better?
Note that in plasma kinetics, the classic quasilinear (CQL) theory of wave-particle interactions\cite{drummond1961non} does just the opposite of the Balescu--Lenard theory: it includes the Landau modes while
ignoring the dressed discreteness noise.  (Throughout this discussion we use the abbreviation `CQL' to distinguish the classic theory of particles interacting with Landau modes from the generic quasilinear scheme of constructing kinetic equations by multiplying two linear quantities together, see~\S\ref{sec:Linear_and_Quasilinear}).
This assumption \textit{is} justified in certain plasma regimes.
In fact, in near-Maxwellian electrostatic plasmas,
the Balescu--Lenard and CQL regimes are physically separated 
by the Debye length $\lambda_\mD$.

Let us be more precise. 
In an electron plasma, the `dressing' of finite-$N$ noise amounts to Debye shielding (Figure~\ref{fig:Debye}).
This means that on scales larger than $\lambda_\mD$ the amplitude of `dressed noise' drops to zero and one
can ignore particle-particle interactions altogether, i.e.\ the contribution to the Balescu--Lenard equation from these scales is negligible.
Meanwhile, there exist Landau modes  on all scales: 
these are Langmuir waves, which involve longitudinal oscillations of electrons as they try to maintain 
the plasma's quasi-neutrality. Langmuir waves oscillate very rapidly --- typically their associated timescale, 
the inverse plasma frequency $\omega_\mathrm{p}^{-1}$,
is the shortest physical timescale in the system.
Now, we know from Landau's theory of wave-particle damping that
these modes will damp only if they can resonate with electrons with velocities $v \approx \omega_\mathrm{p}/k$,
where $k$ is the wavenumber of the Langmuir oscillation in question.
Thus for small $k$ (large scales), the only electrons capable of resonating with the wave are those 
moving extremely fast, and since these are few in number, small $k$ Langmuir waves tend to be only very weakly damped, 
and so wave-particle interactions must be taken into account in the kinetic theory via a CQL collision operator.
Conversely, for very large $k$ the waves are able to resonate with relatively slowly-moving particles
(the `thermal bulk'), of which there are many --- hence,  large-$k$ Langmuir oscillations are typically very strongly damped. 
Again it turns out that the separation between these two regimes occurs at the Debye scale $k_\mD = 2\pi/\lambda_\mD$.
In short, in an electron plasma the electric potential fluctuations on scales $k \gg k_\mD$ are dominated by collectively-dressed noise and therefore well-described by Balescu--Lenard theory (a `noise-only' regime), 
whereas on scales $k \ll k_\mD$ Landau modes dominate and one may use CQL theory (`Landau-modes-only').

The reader has already anticipated where this argument fails when applied to self-gravitating systems.
In this case there is no Debye shielding, so there is nothing to suppress discreteness noise on \emph{any} scale.
On small scales (say, the interparticle distance) collective effects are very weak, 
so the power spectrum just reflects
bare Poisson noise. This is why Chandrasekhar's theory works so well there (\S\ref{sec:Example_BL_Sphere}).
On large scales (comparable to the size of the system) there typically exists 
a combination of weakly-damped Landau modes and collectively-amplified finite-$N$ noise, and crucially
\textit{it is near the frequencies
of weakly-damped Landau modes that the collective amplification of the finite-$N$ noise tends to be most pronounced.}
In other words, the power spectrum of the continuum resulting from dressed noise ($\propto \vert \psi^\mathrm{d}_{\bn\bn'}\vert^2$) is sharply peaked at frequencies $\bn'\cdot\bOm'$
which are very close to the (real part of the) frequency $\omega_\mathrm{m}$ of weakly-damped Landau modes.
This argument shows the logical gap that can arise when naively applying the Balescu--Lenard equation to stellar systems.
For in many cases, 
it is only worth putting in the effort to implement the Balescu--Lenard equation if its predictions are 
significantly different from those of the much simpler Chandrasekhar two-body relaxation theory, i.e.\ if there is significant
collective dressing of finite-$N$ noise on large scales. 
But this dressing in turn is typically only significant \textit{if the system harbors weakly damped Landau modes}.
 It is therefore somewhat self-contradictory to derive Balescu--Lenard theory by arguing these modes fade away rapidly --- 
if they really did so, there might be no interesting collective behaviour to speak of, and therefore no need for the Balescu--Lenard equation! (The 1D system discussed in \S\ref{sec:Example_BL_1D} is an exception to this rule, since in that case the collective effects tended to \textit{suppress} interactions rather than amplify them, and there are no weakly damped Landau modes.)

It follows that typically in stellar systems, amplified noise and Landau modes cannot naively be decoupled.
A comprehensive kinetic theory must therefore account for both types of potential fluctuation self-consistently.
This is also a challenge that arises in plasma kinetics
since neither the Balescu--Lenard or CQL theory is strictly valid on scales
$k \sim k_\mD$, and a general theory must incorporate both effects.  
Several plasma theorists attempted to formulate such a general theory in the 1960s, notably~\citet{Rogister1968-tb,Rogister1969-gn,oberman1970advanced} and~\citet{hatori1968kinetic,hatori1969nonstationary}.
These authors considered a homogeneous electrostatic plasma
 and derived a set of self-consistent coupled equations for the particle distribution function and the energy in the electric field fluctuations, accounting for both Balescu--Lenard and CQL-type effects. 
 It turns out that the divergence of the Balescu--Lenard part of the theory when Im $\omega_\mathrm{m} \to \pm 0$
 is accompanied by a corresponding divergence in the opposite sense in the CQL contribution.
 This renders the generalized theory divergence-free for any Im $\omega_\mathrm{m}$.
 (In fact, this cancellation of divergences arises already in linear theory --- see~\citet{hamilton2023linear}).
A complete, unified kinetic theory of stellar systems would account for both types of potential fluctuations (dressed particle noise and Landau modes), while also
taking account of the inhomogeneity of the system using angle-action variables.  This is a topic of ongoing research\cite{Hamilton2020-sm,hamilton2023linear}.

\section{Nonlinear kinetic theory}
\label{sec:Nonlinear_Kinetics}

The evolution of a system's mean DF $f_0$ is always `nonlinear' in the sense that it depends on the product of two `fluctuations' $\delta \Phi$ and $\delta f$ (equation~\ref{eqn:Axisymmetric_Evolution_Fourier_Ensemble}).
In \S\ref{sec:Quasilinear_Evolution} we discussed a quasilinear approximation to this evolution
in which those fluctuations were each calculated using linear theory.
Here, we discuss nonlinearities that cannot be captured with a quasilinear approach.
We begin in \S\ref{sec:Which_Nonlinear} by discussing some of the ways in which these nonlinearities develop,
and we identify one that is typically most important for stellar systems, namely when orbits
are strongly distorted and/or trapped by a single coherent perturbation.
Then in \S\ref{sec:Resonance_Diffusion} we write down and solve a kinetic equation which describes this regime, using the pendulum machinery that 
we developed for individual orbits in \S\ref{sec:Pendulum_Approximation}.
Finally we give two concrete examples of this nonlinear kinetic theory in practice: first in the context of galactic bar-dark matter halo interactions and dynamical friction (\S\ref{sec:Dynamical_friction}), 
and then regarding the saturation of spiral instabilities (\S\ref{sec:saturation}).

\subsection{Which nonlinear regimes are relevant for stellar systems?}
\label{sec:Which_Nonlinear}

Consider a stellar system that is initially completely quiet ($\delta f = \delta \Phi = 0$), and subject it to some perturbation.
Then we know that at least for a short time, the system will be well described by linear theory (\S\ref{sec:Linear_Response}),
and so e.g.\ its potential/density response will consist of a combination of amplified wakes and Landau modes.
If the system is unstable (and in some cases even if it is weakly stable) the response will grow large enough that the system 
will eventually cease to be well described by linear theory.

Why, physically, does linear theory break down? Let us list a few of the main possibilities. 

\begin{itemize}[itemsep=0pt, topsep=4pt]
\item  One reason can be that the DF $f_0$ relaxes quasilinearly, in the manner already discussed in \S\ref{sec:Quasilinear_Evolution}.
Since linear theory is based on a constant $f_0$ (e.g.\ equation~\ref{eqn:linear_Vlasov_formal_solution}),
such an evolution will eventually break the initial linear prediction (effectively `resetting the initial value problem' with a new DF and fluctuation spectrum).
This will be the dominant source of nonlinearity if the evolution time of $f_0$ according to the relevant quasilinear theory --- which is naively on the order of $t_\mathrm{relax} \sim \varepsilon^{-2} t_\mathrm{dyn}$, where $\varepsilon \sim \vert \delta \Phi/H_0\vert$ is the typical fluctuation strength, see equation~\eqref{eqn:Axisymmetric_Evolution_Fourier_Ensemble} --- 
is shorter than the timescale over which any alternative nonlinearities can develop.
\item One such alternative, which is especially important if the system is unstable, is that nonlinear evolution can arise in the form of  direct coupling between Landau modes. That is, the (initially independent) Landau mode solutions attain such large amplitudes that they no longer respond solely to the mean Hamiltonian $H_0$ and the action space gradient $\partial f_0/\partial \bJ$ (as would be dictated by linear theory), but instead start to notice each others' gravitational fields. Landau mode coupling can lead to modifications both of the potential fluctuation spectrum and the mean stellar distribution function $f_0$. For this effect to be significant, the system usually has to harbor at least two Landau modes that reach amplitudes $\varepsilon \sim 1$
on a short enough timescale that they can interact \textit{before} being destroyed by some other process (such as the quasilinear evolution referred to above). The specific nonlinear physics involved here depends on whether (i) only a handful of Landau modes are excited to large amplitudes, 
or (ii) a broad spectrum of many such modes is excited.

Case (i), in which only a handful of modes are important, has been studied to some extent in galactic dynamics, mostly in the context of global spiral-spiral interactions\cite{Tagger1987-sf,Sygnet1988-eq} (in what plasma physicists would refer to as `three wave coupling'). However, the timescale required for even a rather vigorously-growing global spiral instability to reach a nonlinear amplitude is usually on the order of several Gyr (see~\S\ref{sec:saturation} for a quantitative estimate). This is often comparable to the timescale over which the bulk of the galaxy tends to evolve (\S\ref{sec:Where_Now}).
This leaves little time for such modes to interact nonlinearly (recall from \S\ref{sec:Introduction} that \textit{a typical disk galaxy is only $\approx50$ orbits old}!) As a result, spiral-spiral coupling and similar effects seem to be \textit{present but not dominant} in galactic disk evolution.

Case (ii) on the other hand, in which many modes interact, is usually referred to by plasma physicists as the regime of `wave turbulence'\cite{kadomtsev1965plasma}. That is, plasma physicists tend to think of such a system as consisting of `particles' (described by $f_0$) and `waves' (Landau modes, which dominate the fluctuation spectrum). 
Then, they construct kinetic theories in which wave-wave coupling plays a dominant role, and 
in which the evolution of $f_0$ is deemed secondary or just ignored altogether.
Some studies of this flavor have been conducted in galactic dynamics, mainly with reference to 
tightly wound spiral waves\cite{toomre1969group,toomre1977theories,Mark1977-yx}).  
Some authors in the 1970s\cite{norman1978non,ter1979solitons} took the plasma analogy so far as to construct 
versions of spiral structure theory governed by a nonlinear Schr\"{o}dinger equation, and attempted to describe spirals as solitons/modulational instabilities in the way that~\citet{zakharov1972collapse} described Langmuir collapse. However, no serious numerical evidence was ever offered to bolster this formalism.
Moreover, as in case (i), it is only approximately correct to freeze the stellar distribution function $f_0$ 
in order to isolate the evolution of nonlinearly interacting modes.
Even if it were possible, the dynamical youth of galaxies means that the number of wave cycles over which such modes can possibly have interacted is much less than is normally assumed in plasma theory.
\item A third possibility is for the system to be `dynamically degenerate'.
For example, with ${ \p f_0 / \p \bJ = 0 }$ and ${ \p H_0 / \p \bJ = 0 }$,
equation~\eqref{eqn:Fluctuation_Evolution_Fourier_Linearized} exactly vanishes
and the linearized dynamics of fluctuations becomes trivial.
The system's evolution is then intrinsically non-linear, and one must resort to statistical closure theory to make progress\cite{Krommes2002-ka}.
Astrophysically such regimes occur when describing the diffusion of stellar orbital orientations around supermassive black holes\cite{Kocsis2015,Fouvry2019-pf} or dynamical friction in harmonic cores in the center of galaxies\cite{Read2006,Kaur2022}. This regime bears deep connexions with (anomalous) two-dimensional diffusion in magnetized plasmas\cite{Taylor1971}.
\item A fourth possibility is that stellar orbits can be trapped, or otherwise strongly distorted, by a single coherent perturbation.
For instance, stars can be trapped at the corotation resonance of a rotating bar or spiral (Figure~\ref{fig:3x3_rotating}e).
This effect can be important if the libration time $t_\mathrm{lib}$ of trapped orbits
is shorter than or comparable to the evolution time of the distribution function $f_0$ in the relevant region of phase space,
\textit{and} shorter than the timescale over which some other effect (like incoherent `collisions' or the sweeping of resonances through phase space) is able to interrupt the libration process, as discussed in \S\ref{sec:Radial_Migration}.
\end{itemize} 

The last of these nonlinear effects, orbit trapping by a single perturbation, seems to be important in sculpting the evolution of both real and simulated galaxies\cite{Tremaine1984-wt}, 
and there is good evidence that it has occurred in our own Galaxy\cite{Chiba2020-th,Chiba2021-cc}.
Therefore, for the rest of this section we will focus only on this type of nonlinearity.
We already showed in \S\ref{sec:Pendulum_Approximation} how one can describe individual orbits in this regime
using the pendulum approximation.
Below, we show how a collection of pendulum orbits can be described in a kinetic formalism, although we will ignore self-consistency (i.e.\ Poisson's equation).
We will, however, allow for the possibility that the nonlinearity is interrupted due to stochastic `collisions', by appending an ad-hoc diffusion operator to our (otherwise collisionless) kinetic equation.

We emphasize, though, that
in certain contexts more than one of the above nonlinear effects can be important in different parts of the system and at different times.
For instance, in~\citet{sellwood2022spiral}'s simulations of grooved unstable stellar disks
a spiral Landau mode initially grows exponentially in accordance with linear theory.
Once it reaches a large enough amplitude it begins to trap stars nonlinearly at its corotation resonance, which causes its amplitude to saturate (\S\ref{sec:saturation}). At the same time, the spiral interacts quasilinearly with stars at the Lindblad resonances, draining its angular momentum and carving a new groove in the DF at the location of the ILR.
This newly-grooved disk is unstable to a new set of linear spiral eigenmodes, and so on in a recurrent cycle\cite{sellwood1991spiral,Sellwood2019-gb}.

\subsection{Kinetic theory in the pendulum approximation}
\label{sec:Resonance_Diffusion}

In \S\ref{sec:Pendulum_Approximation},
we described the nonlinear dynamics of individual stars orbiting near resonances with a long-lived potential perturbation rotating rigidly with some fixed pattern speed.
We did this by isolating a particular resonance, 
defining an associated set of slow and fast angle-action variables, and averaging out the fast angles.
We were left with dynamics in the slow angle-action space $(\varphi, I)$ governed by the pendulum Hamiltonian $h$ (equation~\ref{eqn:resonant_Hamiltonian}), all at a given (fixed) fast action $J_\mathrm{f}$.
We now study how an ensemble of stars behaves near such a resonance, following~\citet{hamilton2023galactic}.

Let us therefore consider a subset of stars that all share the same fast action 
${J}_\mathrm{f}$.
We will describe the density of these stars in the $(\varphi, I)$ plane 
using a smooth DF
which we call $f(\varphi, I, t)$, defined such that the number of stars in the phase space area element $\md \varphi \, \md I$ surrounding $(\varphi, I)$ at time $t$ is proportional to
$\md \varphi\,  \md I f(\varphi, I, t)$.
Since $(\varphi, I)$ are canonical variables, the kinetic equation governing $f$ is
\begin{equation}
    \frac{\partial f}{\partial t} + [f, h] = C[f],
    \label{eqn:kinetic_equation_general}
\end{equation}
where $[\cdot,\cdot]$ is the Poisson bracket encoding the smooth advection in the $(\varphi, I)$ plane
\begin{equation}
    [ f,h] = \frac{\partial f}{\partial \varphi} \frac{\partial h}{\partial I} - \frac{\partial f}{\partial I} \frac{\partial h}{\partial \varphi},
    \label{eqn:Poisson_bracket}
\end{equation}
and we have included a possible `collision operator' $C[f]$, which encodes the effect of any stochastic fluctuations in the potential.

For concreteness, let us consider the interaction of a collection of stars
(described by $f$) interacting with a rigidly rotating stellar `bar' (a collection of millions of stellar orbits at the center of a galaxy that rotates like a dumbbell). Then for $h$ we can use the pendulum form~\eqref{eqn:resonant_Hamiltonian} with $k=1$. Meanwhile, for $C[f]$ we choose a very simple diffusive form with constant diffusion coefficient $D$
\begin{equation}
    C[f] = D \, \frac{\partial^2 f}{\partial I^2}.
    \label{eqn:collision_operator}
\end{equation}
This corresponds to every star in our ensemble being kicked with a stochastic white noise forcing in the slow action, $\dot{I} = \eta(t)$,
with statistics $\langle \eta(t) \rangle = 0$
and $\langle \eta(t)\eta(t') \rangle =2D\delta(t-t')$.
(For simplicity we will ignore any such forcing in $\varphi$ and $J_\mathrm{f}$).
The diffusion coefficient $D$ can be {calculated from} a given theoretical model of, e.g.\@, scattering by passing stars and molecular clouds\cite{Binney1988-zy,Jenkins1990-qs,S_De_Simone2004-gr}, dark matter substructure\cite{Penarrubia2019-su, El-Zant2020-oc, Bar-Or2019-ge}, 
spurious numerical heating in simulations\cite{Weinberg2007-bv,Ludlow2021-wk}, or can potentially be calibrated from data\cite{Ting2019-la,Frankel2020-vy} (see below).
We emphasize that we make no attempt at self-consistency here. We simply impose a diffusion coefficient by hand, and we do not account at any stage for the self-gravity of the perturbed DF (the stars interact only with the bar, not with each other\cite{Weinberg1985-en,Dootson2022-cu}).

Before attempting to solve the kinetic equation~\eqref{eqn:kinetic_equation_general}, one more simplification is in order.
Naively, equation~\eqref{eqn:kinetic_equation_general} seems to depend on three parameters: $G, F$ (equation~\ref{eqn:resonant_Hamiltonian}) and $D$ (equation~\ref{eqn:collision_operator}).  However, we can reduce this to one effective parameter by introducing certain dimensionless variables.
First we note that the typical timescale for a star to diffuse all the way across the resonance (a distance $\approx 2I_\mathrm{h}$ in action space) is the \textit{diffusion time}, 
\begin{equation}
    t_\mathrm{diff} \equiv \frac{(2I_\mathrm{h})^2}{2D}= \frac{8F}{GD}.
\label{eqn:diffusion_time}
\end{equation}
Now let us define the dimensionless variables
\begin{subequations}
\begin{align}
\tau &\equiv \sqrt{GF}t = \frac{2\pi \,t}{t_\mathrm{lib}},
\label{eqn:dimensionless_time}
   \\
   j &\equiv \sqrt{\frac{G}{F}} \, I = \frac{2I}{I_\mathrm{h}},
   \label{eqn:dimensionless_action}
  \\
  \Delta &\equiv \sqrt{\frac{G}{F^3}}D = \frac{4\, t_\mathrm{lib}}{\pi\, t_\mathrm{diff}},
   \label{eqn:dimensionless_diffusion}
   \end{align}
   \label{eqn:dimensionless_pendulum}%
\end{subequations}
where the libration time $t_\mathrm{lib}$ is given in equation~\eqref{eqn:t_libration} and the island half-width $I_\mathrm{h}$ is defined in equation~\eqref{eqn:half_width}.
Clearly, $\tau$ is a dimensionless measure of time normalized by the libration time, $j$ is a dimensionless measure of the slow action variable relative to the resonant island width, and $\Delta$ is the \textit{diffusion strength}, i.e.\ the ratio of the libration timescale to the diffusion timescale.
Treating $f$ as a function of these variables, i.e.\ writing\footnote{Note that $f$ still depends parametrically on $\mathbf{N}$ and ${J}_\mathrm{f}$, though we suppress the explicit dependence here to keep the notation clean.} $f(\varphi, j,\tau)$, equation~\eqref{eqn:kinetic_equation} becomes:
\begin{equation}
    \frac{\partial f}{\partial \tau} + j\frac{\partial f}{\partial \varphi} - \sin \varphi\frac{\partial f}{\partial j}= \Delta \frac{\partial^2 f}{\partial j^2}.
    \label{eqn:kinetic_equation_dimensionless}
\end{equation}

We see that in these variables the kinetic equation depends on a single parameter, the dimensionless diffusion strength $\Delta \approx t_\mathrm{lib}/t_\mathrm{diff}$.
The regime $\Delta \gg 1$ corresponds to very strong diffusion,
whereas 
$\Delta \ll 1$ corresponds to very weak diffusion; the `collisionless limit' in which stars only interact with the bar and nothing else is $\Delta =0$. 
We can estimate $\Delta$ heuristically by noting that
$t_\mathrm{diff}$ is related to the relaxation time $t_\mathrm{relax}$, which is the time required for {any stochastic process} to change a star's action by order of itself.
Then from equation~\eqref{eqn:diffusion_time} we have
\begin{equation}
\frac{t_\mathrm{diff}}{t_\mathrm{relax}} \sim \left(\frac{I_\mathrm{h}}{J_\mathrm{s, res}}\right)^2
\sim \bigg\vert \frac{\Psi_1}{H_0} \bigg\vert \sim \varepsilon \ll 1.
\label{eqn:ratio_diff_relax}
\end{equation}  
Here $\varepsilon \sim \vert \Psi_1/H_0 \vert$ is the dimensionless strength of the bar; 
for the Milky Way at corotation $\varepsilon \approx 0.02$\cite{Chiba2020-th}.
Putting equations~\eqref{eqn:dimensionless_diffusion} and~\eqref{eqn:ratio_diff_relax} together, we estimate 
\begin{equation}
\Delta \approx \frac{t_\mathrm{lib}}{\varepsilon t_\mathrm{relax}} \approx 1 \times
\left( \frac{\varepsilon}{0.02} \right)^{-1}
\left( \frac{t_\mathrm{lib}}{2\, \mathrm{Gyr}} \right)
\left( \frac{t_\mathrm{relax}}{100\, \mathrm{Gyr}} \right)^{-1}. \label{eqn:Delta_numerical}
\end{equation}  
Typically, important Galactic bar resonances like the corotation resonance have libration times of $t_\mathrm{lib} \approx 2$ Gyr\cite{Chiba2021-cc}.
{The relaxation time $t_\mathrm{relax}$ depends on what we consider to be driving the diffusion, i.e.\ whether it is 
two-body relaxation (\S\ref{sec:Chandra}), 
transient spiral arms (\S\ref{sec:Radial_Migration}), 
finite-$N$ granularity noise (\S\ref{sec:Quasilinear_Evolution}), or whatever.
In \S\ref{sec:Where_Now}, we estimated that the relaxation time of angular momenta in the Milky Way disk is $t_\mathrm{relax} \approx 60$ Gyr, which would give $\Delta \approx 2$.
More generally, the estimate
\eqref{eqn:Delta_numerical} tells us that $\Delta$ can be of order unity \textit{even for systems with relaxation times much longer than the age of the Universe}.}
The important point is that it is much easier to diffuse across the width of the resonance than it is to diffuse across the whole of action space, and it is the diffusion across resonances that governs the dynamics\cite{johnston1971dominant,Auerbach1977-xe,Tolman2021-ly}.

\begin{figure}[htbp!]
\centering
\includegraphics[width=0.9\linewidth]{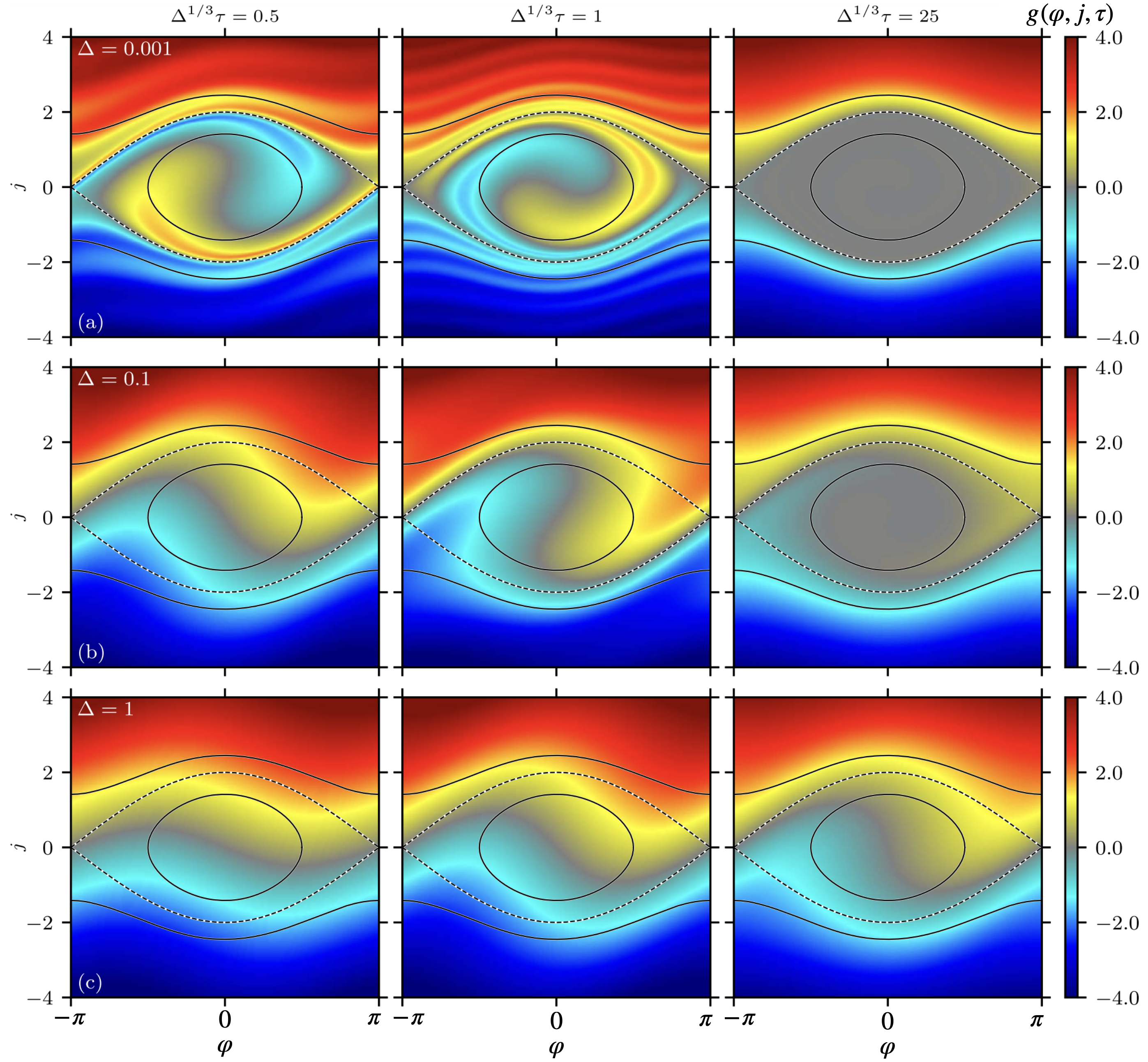}
\caption{Contours of $g(\varphi, j, \tau)$ (related to $f$ through equation~\ref{eqn:dimensionless_DF}), computed by solving the kinetic equation~\eqref{eqn:kinetic_equation_dimensionless} numerically with the initial condition $g(\tau=0) = j$. Rows (a)--(c) show the time evolution of the solution for $\Delta = 0.001,\, 0.1,\, 1$ respectively. The right column represents the steady-state solution.
Adapted from C. Hamilton, E. A. Tolman, L. Arzamasskiy, and V. N. Duarte, ApJ 954, 12 (2023)\cite{hamilton2023galactic}.
}
\label{fig:g_phij}
\end{figure}

We can solve equation~\eqref{eqn:kinetic_equation_dimensionless} analytically in the limits $\Delta = 0$ and $\Delta \gg 1$, but it is easier to just solve it numerically for a given $\Delta$.  To do this, let us take as our initial condition a simple linear DF
\begin{equation}
    f(\varphi, j, \tau=0) = f_0(j=0) + \alpha j \equiv f_\mathrm{init}(j),
    \label{eqn:linear_DF}
\end{equation}
where ${ \alpha \equiv \partial f_0 / \partial j |_{j=0} }$
is the gradient of the unperturbed DF at resonance.
In Figure~\ref{fig:g_phij}, we show with colored contours the resulting evolution of the dimensionless auxiliary DF
\begin{equation}
    g(\varphi, j, \tau) \equiv \frac{f(\varphi, j, \tau) - f_0(j=0)}{\alpha},
    \label{eqn:dimensionless_DF}
\end{equation}
whose initial value is
$g_\mathrm{init}(j) \equiv g(\varphi, j, \tau=0) =  j$. 
Each row corresponds to a different value of $\Delta$, and within each row, from left to right we plot the solution
at different `times' $\Delta^{1/3} \tau$. 

First consider row (a), which is in the limit of very weak diffusion ($\Delta =0.001$). Physically, we expect that in this limit the DF will \textit{phase mix} (c.f. Figure \ref{fig:PhaseMixing}) around and within the resonant island.
The reason for phase mixing is that in the absence of diffusion ($\Delta = 0$), the trajectories of individual stars trace contours of constant $h$ in the $(\varphi, j)$ plane. And, since adjacent contours correspond to slightly different libration/circulation periods, the initial DF gets sheared out along these contours, producing ever finer-scale structures in the phase space. If $\Delta$ were exactly zero, and in the absence of any coarse-graining, 
this process would continue indefinitely.
However, for our very small but finite value of $\Delta = 0.001$, diffusion is able to wipe out the very smallest-scale features, without needing to invoke any coarse-graining.
We see that by the third column, $g$ is approximately zero everywhere
inside the separatrix, and is smeared almost evenly
on contours of constant $h$ outside of the separatrix. The DF has reached a steady state; crucially, in this state the DF is \textit{symmetric} around $\varphi = 0$ at any fixed $j$.

Next consider {row} (c) in Figure~\ref{fig:g_phij}, which corresponds to rather strong (but not unrealistic! --- see equation \ref{eqn:Delta_numerical}) diffusion, $\Delta = 1$.  In this case, the resonance has a much less dramatic effect on $g$ at any time. This again is as expected since the initial linear DF~\eqref{eqn:linear_DF} is annihilated by the collision operator $C[f]$ (equation~\ref{eqn:collision_operator}). In other words, wherever the bar perturbation induces some curvature in the DF, strong diffusion immediately tries to remove it and hence to restore the linear initial condition.
Another way to say this is that the effect of diffusion is to counteract the phase-mixing of orbits, in particular within the resonant island.
Importantly, this means that in steady state the DF is strongly \textit{asymmetric} in $\varphi$ at a given $j$, in contrast to row (a).
Finally, the case $\Delta = 0.1$ in row (b) is intermediate between these two regimes. 

\begin{figure}[htbp!]
\centering
\includegraphics[width=0.9\linewidth]{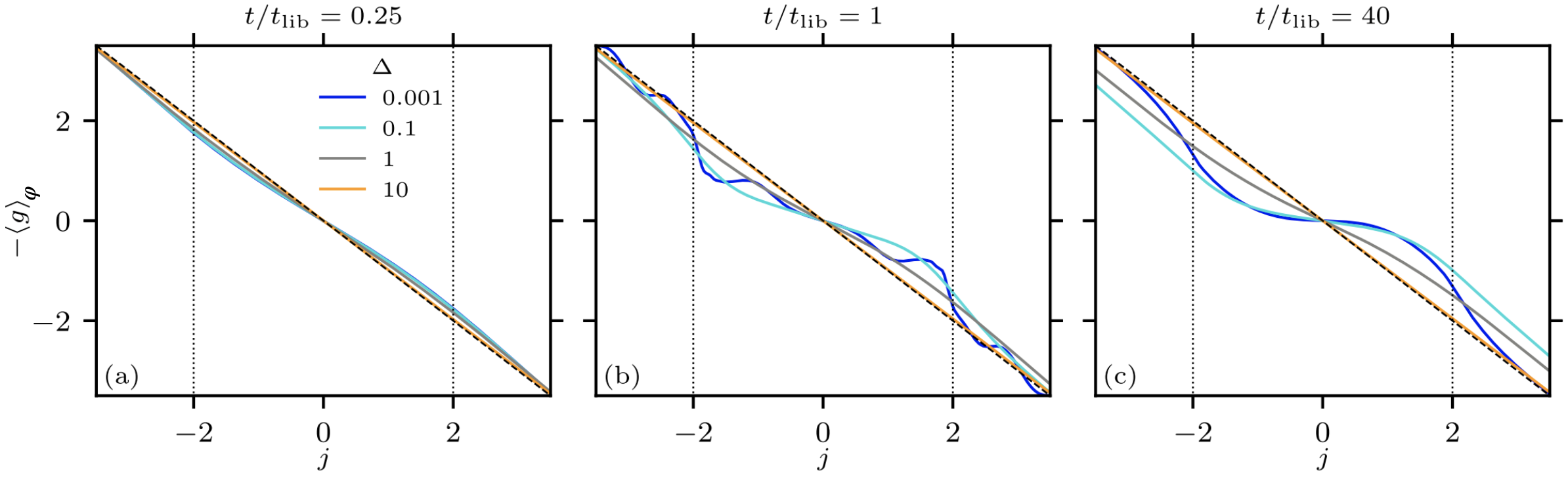}
\caption{Time evolution of the slow-angle-averaged DF, 
$\langle g \rangle_\varphi$ (see equation~\ref{eqn:phi_averaged_g}), for the same values of $\Delta$ as used in Figure~\ref{fig:g_phij}. Different panels correspond to different times, and the right hand panel is the steady-state solution. The vertical dotted lines show the maximum extent of the separatrix, $j = \pm 2$ (see equation~\ref{eqn:dimensionless_action}).
Adapted from C. Hamilton, E. A. Tolman, L. Arzamasskiy, and V. N. Duarte, ApJ 954, 12 (2023)\cite{hamilton2023galactic}.
}
\label{fig:g_average}
\end{figure}

It is also instructive to average the solution $f(\varphi, j, \tau)$ over the slow angle $\varphi$ and investigate the resulting DF 
\begin{equation}
    \langle f \rangle_\varphi \equiv \frac{1}{2\pi} \int_{-\pi}^\pi \md \varphi \, f(\varphi, j, \tau).
\end{equation}
In Figure~\ref{fig:g_average} we plot the corresponding auxiliary DF (see equation~\ref{eqn:dimensionless_DF})
\begin{equation}
    \langle g \rangle_\varphi=   \frac{\langle f \rangle_\varphi - f_0(j=0) }{\alpha},
    \label{eqn:phi_averaged_g}
\end{equation}
as a function of $j$ for various $\Delta$ values, and at different times (measured in units of the libration time $t_\mathrm{lib}=2\pi/\omega_\mathrm{lib}$).
The key point is that for $\Delta \to 0$, the
phase mixing of orbits in and around the resonant island causes the angle-averaged DF to be \textit{flattened} there (see the blue line in panel c).
Diffusion ($\Delta > 0$) suppresses this flattening effect in an attempt to preserve the initial linear profile.

We now show how these results can be applied to two real astrophysical problems.

\subsubsection{Example: Dynamical friction on a galactic bar}
\label{sec:Dynamical_friction}

As a galactic bar ages, it transfers angular momentum to its host galaxy and consequently its rotation rate slows.
The mechanism responsible is dynamical friction: the bar produces a perturbation in the DF of the stars and dark matter particles that surround it (c.f.\ Figure~\ref{fig:Magorrian}), and that perturbation back-reacts to produce a torque on the bar, draining its angular momentum.
Problems of this kind 
have been the focus of many classic studies in galactic dynamics\cite{Lynden-Bell1972-ve,Tremaine1984-wt}, and 
are strongly analogous to wave-particle interaction problems in plasma (e.g.\ Landau damping).
Here we illustrate the utility of our nonlinear pendulum kinetics, and the important role of the diffusion strength $\Delta$, 
by calculating the dynamical friction torque on a galactic bar in a simple model. 

We assume that the center of the bar is at the origin and that its long axis sits in the plane of an initially axisymmetric disk of stars. We let the bar rotate around its center in the $\phi$ direction with pattern speed $\pattern$, and consider its gravitational interaction with the stars. 
One can write down a rather general formula for the torque $T$ felt by the bar as follows.
Let the bar have potential $\delta \Phi(\br,t)$.
Then from Hamilton's equation~\eqref{eqn:Hamilton_J} the specific torque felt by an individual star due to the bar is
\begin{equation}
    \frac{\md L}{\md t} = -\frac{\partial \delta \Phi}{\partial \theta_\phi}.
\end{equation}
Let the DF of stars be $f$ (normalized such that $\int \md\theta_\phi \md \theta_R\md L \md J_R f$ is the total mass in stars).  Then by Newton's third law,
the total torque on the bar is equal to 
\begin{equation}
T(t) =  \int \md \btheta \md \mathbf{J} \, 
f(\btheta, \mathbf{J},t) \frac{\partial \delta \Phi(\btheta, \mathbf{J}, t)}{\partial \theta_\phi}.
\label{eqn:torque_general}
\end{equation}
The challenge is to compute $f$ for a given perturbation $\delta \Phi$, and then perform the integral~\eqref{eqn:torque_general}. 

The simplest way to calculate $f$ is to resort to linear theory (and ignore diffusion), and use~\eqref{eqn:linear_Vlasov_formal_solution}.
In this case, putting $\delta\Phi_{\mathbf{N}} (\mathbf{J}, t) = \delta\Phi_{\mathbf{N}} (\mathbf{J}) \, \me^{-\mi N_\phi \pattern t}$ and
performing the integral over $t'$ in equation~\eqref{eqn:linear_Vlasov_formal_solution}, one finds a `linear torque', $ T^\text{lin} = \sum_{\mathbf{N}}  T^\text{lin}_{\mathbf{N}}$, where the contribution from resonance $\mathbf{N}$ is
\begin{equation}
          T^\text{lin}_{\mathbf{N}}(t) \equiv (2\pi)^2  N_\phi \int \md \mathbf{J} \,\vert \delta \Phi_{\mathbf{N}} (\bJ) \vert^2 \mathbf{N}\cdot \frac{\partial f_0}{\partial \mathbf{J}}  \frac{\sin[(\mathbf{N}\cdot \bm{\Omega} - N_\phi \pattern)t]}{\mathbf{N}\cdot \bm{\Omega} - N_\phi \pattern}.
    \label{eqn:Torque_linear}
\end{equation}
(We emphasize that in deriving this formula we have ignored collective effects, because we have not accounted for the contribution to $\delta \Phi$ that arises self-consistently from the perturbed stellar distribution.
While collective effects are typically very important for galaxy evolution, in the present case it turns out that they do not significantly alter the character of the response\cite{Weinberg1985-en}, and we will continue to neglect them for simplicity).
Taking the
time-asymptotic limit $t\to \infty$, one arrives at the classic `LBK torque formula'\cite{Lynden-Bell1972-ve}
\begin{equation}
        T^\mathrm{LBK}_{\mathbf{N}} \equiv (2\pi)^2  N_\phi \int \md \mathbf{J}  \vert \delta \Phi_{\mathbf{N}} (\bJ) \vert^2 \mathbf{N}\cdot \frac{\partial f_0}{\partial \mathbf{J}}  \pi \delta \left(\mathbf{N}\cdot \bm{\Omega} - N_\phi \pattern \right) .
    \label{eqn:LBK}
\end{equation}
The LBK formula~\eqref{eqn:LBK} predicts that angular momentum is transferred to and from the stars exclusively at resonances.
It is also directly analogous to the classic formula for the Landau damping rate of a Langmuir wave in an electrostatic plasma\cite{Landau1946-aj,Ichimaru1965-zm} or Landau damping in more general geometry\cite{Kaufman1972-vv,Nelson1999-in}.
Importantly, for many {realistic} DFs $f_0(\mathbf{J})$, the LBK torque is finite and negative, implying a long term transfer of angular momentum away from the bar and hence a decay in its pattern speed. (As an example, at corotation the sign of the LBK torque is the sign of $\partial f_0/\partial L$, 
which is typically negative whenever the density profile of the galaxy decreases with radius.  This is similar to a plasma velocity distribution whose tail satisfies $\partial f_0/\partial v < 0$). In practice, the time-asymptotic limit may not be valid since the torque can take several dynamical timescales to converge, by which time the pattern speed may have changed significantly\cite{Weinberg2004-ss}.
Nevertheless, the LBK formula~\eqref{eqn:LBK} is a good benchmark against which we can compare the magnitude of the torque arising from more sophisticated calculations.

Since we used the linearized solution \eqref{eqn:linear_Vlasov_formal_solution} to compute $f$ we did not account for the nonlinear orbit trapping that can occur
sufficiently close to each resonance, nor did we include any finite $\Delta$.
We can include these effects by redoing the calculation using our pendulum formalism (\S\ref{sec:Pendulum_Approximation}).
Converting to slow-fast angle-action variables $(\bm{\theta}',\mathbf{J}')$ \textit{around each resonance} $\mathbf{N}$, 
Fourier expanding the potential $\delta \Phi$ in slow angle space
via $\delta \Phi(\mathbf{r}, t) = \sum_{\mathbf{k}} \Psi_{\mathbf{k}}(\mathbf{J}') \me^{\mi \mathbf{k}\cdot \btheta'}$
as in equation~\eqref{eqn:Hamiltonian_slowfastangles}, and expanding the DF as $f = \sum_{\mathbf{k}} f_{\mathbf{k}}(\mathbf{J}',t) \me^{\mi \mathbf{k}\cdot \bm{\theta}'}$, 
one can show that the total `nonlinear torque' on the bar $T^\text{non-lin}$ can be written as a sum over $\mathbf{N}$ of contributions (see~\citet{hamilton2023galactic} for details\footnote{Note, however, that~\citet{hamilton2023galactic} were working in three dimensions, since they were considering the interaction of a bar with a spherical halo. Here we are working in two dimensions, so our DF $f$ is normalized differently.}):
\begin{equation}
T^\text{non-lin}_{\mathbf{N}}(t) =
    2(2\pi)^2 N_\phi 
\sum_{k>0} k
\int 
\md \mathbf{J}' 
\, \mathrm{Im} \left[
f_{k}(\mathbf{J}', t) 
   \Psi^*_{k}(\mathbf{J}') \right].
    \label{eqn:nonlinear_torque}
\end{equation}
Here we have used the shorthand $f_{(0,k)} = f_k$, and similarly for $\Psi_k$.
In using expressions like \eqref{eqn:nonlinear_torque}, one must be careful to properly divide phase space into non-overlapping sub-volumes that are dominated by individual resonances $\mathbf{N}$ (see~\S{4.5} of~\citet{Chiba2022-qt} for a discussion).

\begin{figure}[htbp!]
\centering
\includegraphics[width=0.9\linewidth]{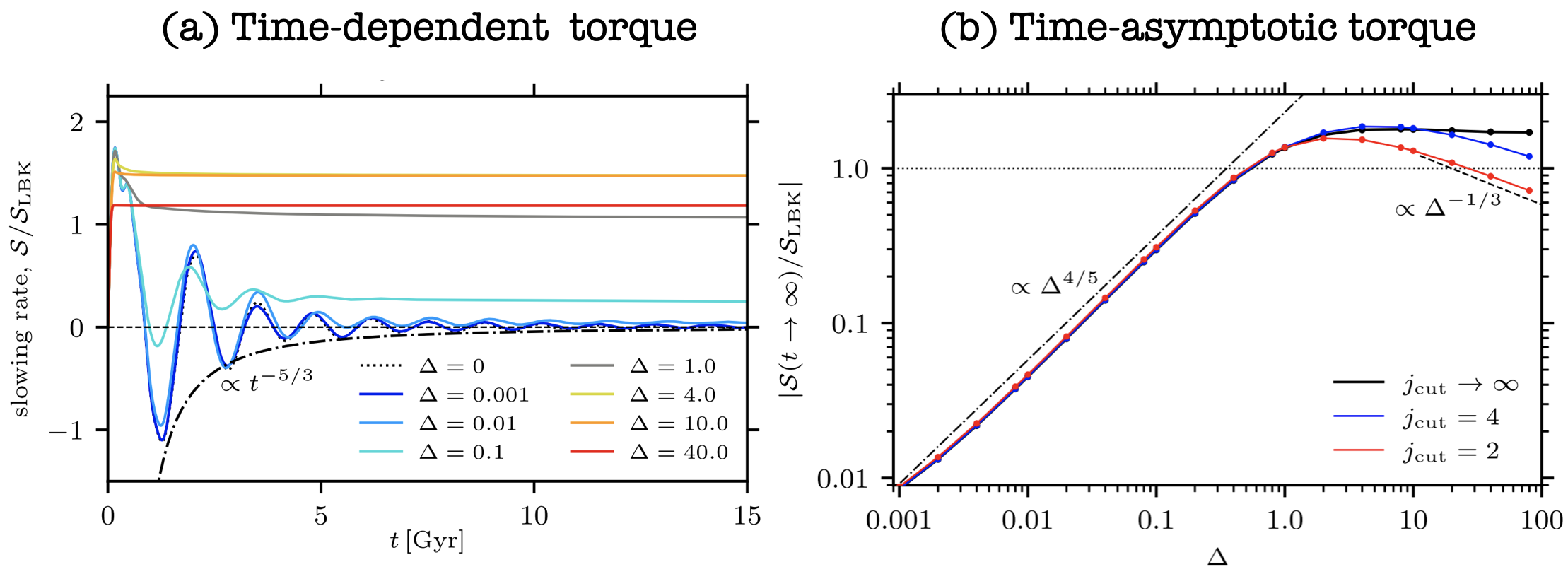}
\caption{Adapted from  C. Hamilton, E. A. Tolman, L. Arzamasskiy, and V. N. Duarte, ApJ 954, 12 (2023)\cite{hamilton2023galactic}.
(a) Dimensionless torque on the bar (i.e.\ its `slowing rate')
$S(t) / S_\mathrm{LBK}$ due to stars at the corotation resonance, with $J_R = 50$ kpc$^2$ Gyr$^{-1}$ (i.e.\ roughly circular orbits).
The libration time is $t_\mathrm{lib} \approx 2$ Gyr.
Colored lines show results for different $\Delta > 0$, while the dotted line shows the collisionless ($\Delta = 0$) result.
(b) Steady-state value of $ S(t) / S_\mathrm{LBK}$ as a function of $\Delta$ (note that both axes are on a logarithmic scale, and the different lines just correspond to different choices of integration limits).}
\label{fig:Torque}
\end{figure}

In reality, the torque on the bar involves a sum over all possible resonances $\mathbf{N}$ and an integral over all of action space, 
and  each such contribution comes with a slightly different value of  $\Delta$.
In order to isolate a representative quantity that only depends on a single value of $\Delta$,
we will focus henceforth on the \textit{corotation torque density} $S(t)$,
which is the contribution to the torque only arising from the corotation resonance $\Omega_\phi = \pattern$ (so $N_\phi=2$ and $k=1$, see \S\ref{sec:CCF})
and only considering stars with a particular $J_R$, namely
\begin{equation}
    S(t \vert J_R) = 4 (2\pi)^2 \int \md L  \Psi_1\,  \mathrm{Im} \, f_1.
    \label{eqn:corotation_torque_density}
\end{equation}
For a given $\Delta$ we can compute $\mathrm{Im}\,f_1$ using the numerical solution to the kinetic equation~\eqref{eqn:kinetic_equation_dimensionless} that we showed in Figure~\ref{fig:g_phij}.
We will compare this to the benchmark LBK value (see equation~\ref{eqn:LBK}), which is the result of computing the torque density using linear theory with $\Delta = 0$:
\begin{equation}
    S_\mathrm{LBK}(J_R) = 4 (2\pi)^2 \int \md L \, \vert \Psi_1 \vert^2 \frac{\partial f_0}{\partial L}  \pi \delta [2(\Omega_\phi - \pattern)] . 
    \label{eqn:LBK_density}
\end{equation}
In Figure~\ref{fig:Torque}a, we show the bar's dimensionless slowing rate
$S(t) / S_\mathrm{LBK}$
as a function of time
for various values of $\Delta \geq 0$, 
fixing 
$J_R = 50$ kpc$^2$ Gyr$^{-1}$ (which is much less than $L_\mathrm{CR}$, meaning the stars are initially on roughly circular orbits --- see Figure \ref{fig:Plummer}).
Note that $S_\mathrm{LBK}$ is negative, so that a positive value of $ S(t) / S_\mathrm{LBK}$ means the bar feels a negative torque (slowing it down).
In the collisionless case ($\Delta = 0$, black dotted line) we see that the slowing rate oscillates on the timescale $\approx t_\mathrm{lib} \approx 2$ Gyr,
and its envelope also decays over time.
In the time-asymptotic limit for $\Delta = 0$ we find
\begin{equation}
    S(t\to \infty) = 0, \,\,\,\,\,\,\,\,\,\,\,\,\,\,\, (\Delta =0),
    \label{eqn:S_TW84}
\end{equation}
This zero torque result is a consequence of the angle-symmetry of the stellar density distribution ($\partial_\varphi f = 0$ at fixed $j$) that arises when 
the DF completely phase mixes within and around the resonant island (Figure~\ref{fig:g_phij}a). 
Simply put, in the fully phase-mixed state there are the same number of stars `pushing' on the bar as `pulling' on it.
This is the direct analogue of~\citet{ONeil1965-uy}'s famous result in plasma theory: when nonlinear trapping  is included (and diffusion neglected), the Landau damping rate of an electrostatic wave asymptotes to zero, because of the symmetrization of the `bouncing' particle distribution in the trough of the wave.
However, as one deviates from $\Delta = 0$ to finite $\Delta > 0$, the behavior changes. 
The 
slowing rate in the steady state is manifestly positive for all $\Delta > 0$.
This is a consequence of the fact that while phase-mixing attempts to abolish the asymmetry of the slow-angle distribution, 
diffusion replenishes it (e.g.\ Figure~\ref{fig:g_phij}b) leading to a finite $\mathrm{Im} \, f_1$ in equation~\eqref{eqn:corotation_torque_density}.
Moreover, for $\Delta \gtrsim 1$,
the slowing rate no longer oscillates on the libration timescale, because no trapped star is able to complete a full libration before being de-trapped by diffusion.

To make these statements more quantitative we turn to panel (b) of Figure~\ref{fig:Torque}, in which we show the time-asymptotic slowing rate $S(t\to\infty)$ as a function of $\Delta$.
We see that $S(t\to\infty)$ grows almost linearly with $\Delta$ until $\Delta\approx 1$, after which it saturates at a level similar to that given by the naive LBK theory, even though LBK theory ignores both diffusion and nonlinearity!
This is as expected since, roughly speaking, for any $\Delta \gtrsim 1$
there is no nonlinear trapping and the effect of `diffusion' is just to broaden the delta-function in~\eqref{eqn:LBK}.
That is, \textit{diffusion relinearizes the dynamics}.
Again, this effect was also appreciated in the plasma theory of electric wave damping:
if small-scale
collisions are able to knock particles out of the wave trough before they can complete a full bounce orbit, then
O'Neil's nonlinear trapping effects are interrupted, and the standard (quasi)-linear Landau damping/growth rate is recovered\cite{johnston1971dominant,Auerbach1977-xe}.

 These results suggest that real galactic bars will always feel some non-zero torque, 
 since finite diffusion will always replenish some asymmetry in the angular distribution of the stars as viewed in the bar frame.
They also suggest that for systems with moderate to strong diffusion ($\Delta \gtrsim 1$), 
the LBK formula~\eqref{eqn:LBK} provides a robust order-of-magnitude estimate of the frictional torque. Moreover, we have treated the bar's pattern speed, $\pattern$, as constant here, 
but very similar arguments apply to the case where the `interruption' of trapping is not due to collisions (i.e.\ finite $\Delta$) but rather due to the fact that a changing $\pattern$ causes resonance locations to sweep through phase space. In that case, the LBK formula 
holds in what~\citet{Tremaine1984-wt} called the `fast regime',
namely when 
the bar slows down rapidly enough that the resonances move by their own width $I_\mathrm{h}$
on a timescale shorter than $t_\mathrm{lib}$.
In the opposite `slow regime', trapped stars are dragged to larger radii as the corotation resonance moves outwards, a process that seems to have occurred in the Milky Way~\cite{Chiba2021-cc}.
A general theory that connects smoothly the fast (LBK or Landau-like) regime and the slow (nonlinear, O'Neil-like) regime was recently completed by~\citet{chiba2023dynamical}. 

\subsubsection{Example: Saturation of spiral instabilities}
\label{sec:saturation}

Finally, let us discuss how pendulum kinetics can be used to understand the saturation mechanism of spiral instabilities and predict the resulting saturation amplitude.

\begin{figure}[htbp!]
    \centering
    \includegraphics[width=0.7\textwidth]{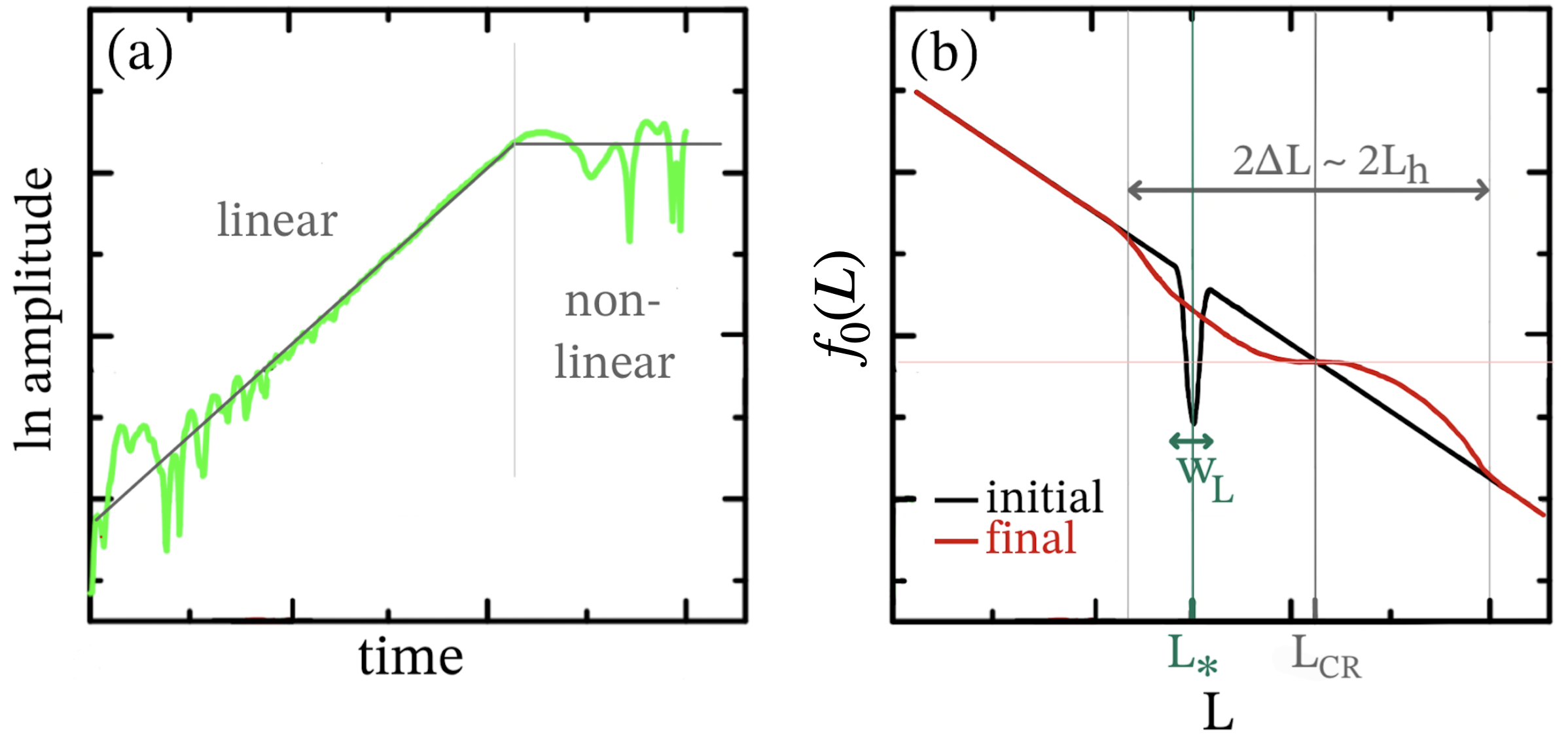}
    \caption{Taken from C. Hamilton, MNRAS 528, 5286 (2024)\cite{hamilton2024saturation}.
    (a) Growth and saturation of a spiral mode in an $N$-body simulation of a razor thin stellar disk (adapted from~\citet{sellwood2022spiral}, green line). The gray line illustrates the simplification used in this paper: we separate the spiral's life into a linear phase of perfectly exponential growth, and a nonlinearly saturated phase of constant amplitude. (b) Illustration of the change to the axisymmetric DF $f_0(L, t)$ in the vicinity of a corotation resonance.
    The initial instability is driven by some sharp feature in the DF (here a \textit{groove}, centered on $L=L_*$ with width $w_\mathrm{L}$, see~\citet{sellwood2022spiral}).
    In the saturated state, the DF is flattened around the resonance $L = L_\mathrm{CR}$.}
    \label{fig:Amplitude}
\end{figure}

We already saw (\S\ref{sec:Response_Example_Imhomogeneous}) that spiral waves can arise in stellar disks as exponentially growing, linear instabilities of the disk's underlying DF $f_0$, particularly in cases where the disk is sufficiently cold/dense (Figure~\ref{fig:Zang}) or grooved (Figure~\ref{fig:groove}).
This exponential growth phase cannot continue forever. Simulations by~\citet{sellwood2022spiral}
and theoretical work by~\citet{hamilton2024saturation}
suggest that the saturation occurs because of orbit trapping at the corotation resonance. 
To understand their argument, we present 
Figure~\ref{fig:Amplitude} (adapted from~\citet{hamilton2024saturation}).

The green line in panel (a)
shows the logarithmic amplitude of a spiral density wave with constant pattern speed $\pattern$, from an $N$-body simulation of an initially grooved, razor thin disk.
Clearly, once some noisy transients have faded away the mode undergoes very well-defined exponential growth, confirming that it is a linear mode of the initial DF.
Eventually, though, the exponential growth stops and the mode  settles at a roughly constant amplitude.
The gray line illustrates schematically the simplification we will use here, which is that the transition from the linear phase (exponentially growing amplitude) to the nonlinear phase (constant amplitude) happens instantaneously.
In panel (b), we sketch the corresponding DF of angular momenta in the disk, $f_0(L)$, for circular orbits ($J_R = 0$).
The initial unstable distribution is shown with a black line.
The idea is that once the mode has grown to large enough amplitude that it starts to resonantly trap stellar orbits at corotation,
the DF in the vicinity of that resonance will be flattened  (as shown with the red line), 
just as we saw in \S\ref{sec:Dynamical_friction}.
This flattening erases the sharp feature in $f_0$
that originally drove the instability\footnote{In linear theory, the angular momentum transfer at each resonance $\mathbf{N}$ is governed by equation \eqref{eqn:LBK}.  The right hand side of this is clearly zero if one such resonance dominates and the gradient of the DF vanishes there. The reason that corotation is the dominant resonance is that the disk is initially rather cold. The corotation resonance acts most strongly upon perfectly circular orbits, whereas Lindblad resonances require significant radial epicyclic motion to have any effect.}.

\citet{hamilton2024saturation} showed that by calculating the angular momentum content of the spiral in the linear (i.e.\ exponentially growing) and nonlinear (i.e.\ fully saturated) regimes, and equating the two expressions, one can read off an approximate formula for the saturation amplitude of the spiral potential. This is
\begin{equation}
    \vert \delta \Phi_m \vert =  \left( \frac{128}{9\pi^2} \right)^2 \frac{2}{m}\, \frac{\beta^2}{\vert \partial \Omega_\phi/\partial L \vert_{L_\mathrm{CR}}},
\label{eqn:Saturation_Amplitude}
\end{equation}
where $\beta = \mathrm{Im}\,\omega_\mathrm{m}$ is the mode's growth rate.
This result can be made more physically enlightening if we use~\eqref{eqn:F_and_G} and~\eqref{eqn:t_libration} to express it as 
\begin{equation}
   \omega_\mathrm{lib} =  \frac{256}{9\pi^2} \, m^{1/2} \beta \approx 2.88\, m^{1/2} \beta.
   \label{eqn:Libration_Saturation}
\end{equation}
In other words, saturation occurs when the nonlinear libration frequency $\omega_\mathrm{lib}$ is a few times larger than the 
linear growth rate $\beta$ (so nonlinear orbit distortions occur before the mode amplitude can grow significantly). 
 Note that for one-armed spirals, equation~\eqref{eqn:Libration_Saturation} is identical to equation (12) of~\citet{Dewar1973-rf}, who studied the saturation of \textit{linear} momentum transfer between electrons and a single longitudinal plasma wave.

One can also relate~\eqref{eqn:Saturation_Amplitude} to the surface density perturbation of the spiral  $\delta\Sigma_m$ relative to the background stellar disk $\Sigma_0$:
\begin{align}
    \bigg \vert \frac{\delta \Sigma_m}{\Sigma_0} \bigg \vert &= \frac{32768}{81\pi^4} \, \xi^{-1} \cot p \left( \frac{\beta}{\Omega_\mathrm{p}} \right)^2
    \label{eqn:Rel_Surf}
    \\
    &\approx 0.5 \times \left(\frac{\xi}{0.5}\right)^{-1} \left(\frac{\cot p}{1.5}\right) \left( \frac{\beta/\Omega_\mathrm{p}}{0.2} \right)^{2},
        \label{eqn:Rel_Surf_numerical}
\end{align}
where $\xi$ is the active fraction of the disk (\S\ref{sec:Response_Example_Imhomogeneous}) and $p$ is the spiral's pitch angle.
Thus, spiral instabilities tend to saturate once their amplitudes reach several tens of percent of the background density. 

Last of all, it is enlightening to estimate the timescale $t_\mathrm{sat}$ over which saturation of realistic spirals might be achieved. To do this, first recall (equation \ref{eqn:Libration_Saturation}) that saturation occurs when $\omega_\mathrm{lib}(t_\mathrm{sat}) \sim \beta$.  Second, a basic scaling of equation~\eqref{eqn:t_libration} tells us that at corotation, $\omega_\mathrm{lib}(t) \sim  \varepsilon(t)^{1/2}\pattern$, where $\varepsilon(t) \sim \vert  \Phi_m(t) /H_0\vert$ is the dimensionless amplitude of the mode.  Third, in the linear regime we have  $\varepsilon(t) =\varepsilon(0)\me^{\beta t}$.
Putting these three ingredients together, we get a saturation time
\begin{equation}
     t_\mathrm{sat} \sim  \frac{2}{\beta} \left[ \ln \left( \frac{\beta}{\pattern}\right)  + \frac{1}{2} \ln \left( \frac{1}{\varepsilon(0)}\right)\right].
\end{equation}
In particular, if the initial fluctuation level is set by Poisson noise alone then $\varepsilon(0) \sim N^{-1/2}$, and we have (very roughly)
\begin{align}
     t_\mathrm{sat} 
     &\sim  \frac{1}{2\beta}  \ln N \\
     & \sim 2.5\, \mathrm{Gyr} \times \left( \frac{T_\mathrm{p}}{250 \, \mathrm{Myr}} \right)
    \left( \frac{\beta/\Omega_\mathrm{p}}{0.2} \right)^{-1}\left(\frac{\ln N}{25}\right),
\end{align}
where $T_\mathrm{p} = 2\pi/\pattern$ is the orbital period at corotation.
Thus, if it has to exponentiate out of Poisson noise, a rather vigorously growing spiral with $\beta/\pattern \gtrsim 0.1$ will saturate after a few Gyr.
More realistic galactic noise tends to be of higher amplitude than is implied by Poisson statistics (\S\ref{sec:Example_BL_Disk}), though this will only change the timescale $t_\mathrm{sat}$ by a logarithmic factor.


\section{Further reading}
\label{sec:Discussion}

There are many more connections between plasma theory and stellar dynamics that we have not touched on in this tutorial article.
Let us mention just a few of them here, alongside references for further reading.

First, the most basic connection between the two fields is the theory of orbits in spatially smooth and constant or time-varying potentials, the regularity/chaoticity of those orbits, and the associated theories of canonical transformations, Hamiltonian perturbation theory, and guiding center motion. 
These are classic problems in the general theory of dynamics\cite{Lichtenberg2013-pw}, 
and Hamiltonian perturbation theory in angle-action variables has long been a tool of theorists (see Chapter 3 of BT08),
but it is only relatively recently that angle-action variables have been adopted across mainstream galactic astronomy\cite{Binney2013-cf}. They are now ubiquitous tools with which astronomers study phase space data of stellar systems, particularly from the Gaia satellite\cite{Trick2019-qp}.
Chaotic orbits are known to exist in triaxial (i.e.\ non-axisymmetric) galaxies and star clusters (see, e.g.\@, the books by~\citet{contopoulos2002order,Merritt2013-le}), but are mostly ignored in kinetic theories, which usually focus on the evolution of near-axisymmetric or near-spherical systems.  Guiding center and gyrokinetic theory have close analogues in stellar dynamics, particularly for describing stellar motion near supermassive black holes, where the orbits are closed Keplerian ellipses, which can be averaged over to form Keplerian `rings' or `wires'\cite{sridhar1999stellar,bar2018scalar}.
Despite this, the Hamiltonian theories of guiding center motion, averaged Lagrangians, ponderomotive effects and so on developed by the likes of~\citet{cary2009hamiltonian,dodin2008positive} in plasma theory have not (yet) received much attention in galactic dynamics.

Second, another common thread between plasma kinetics and stellar dynamics lies in the construction of collisionless equilibria and 
their stability properties. Though we have barely touched this subject here, developing stable candidate distribution functions is a crucial part of stellar-dynamical modeling, 
completely analogous to the construction of Vlasov--Poisson
or Vlasov--Maxwell\cite{channell1976exact}
equilibria in plasma.
Similarly, Nyquist diagrams\cite{nyquist1932regeneration}
and Penrose's criterion\cite{penrose1960electrostatic} are just as applicable to stellar systems as they are to plasmas.
On the gravitational side, perhaps the best introductions to this subject (written in a language plasma physicists can appreciate) are to be found in \S\S4-5 of BT08 and the 1994 book by~\citet{palmer1994stability}.
The theory of
\textit{nonlinear} stability, formation of BGK-like modes\cite{bernstein1957exact}, and so on have been developed to some extent in stellar dynamics\cite{fridman2012physics,goodman1988instability,vandervoort2003stationary}, usually with direct reference to previous plasma results.

Third, we mention a major area of stellar-dynamical kinetics that we have only touched on here,
the theory of spiral structure.
This was one of the original areas in which the galactic-plasma overlap was found to be most fruitful\cite{kalnajs1971dynamics,Lynden-Bell1972-ve,Dewar1972-ar,Kalnajs1976-gg,dekker1976spiral}, since it concerns problems familiar to plasma practitioners such as wave propagation, WKB theory, wave-particle interaction, phase mixing, and so on.
Despite this, the theory is far from complete:
opinions still differ over the mechanisms that provoke spiral responses, the lifetimes of individual spiral patterns, the importance of gas flows and star formation,
and so on --- for an array of perspectives, see e.g.\ reviews by~\citet{athanassoula1984spiral,dobbs2014dawes,Sellwood2021-pb}.

\begin{figure}[htbp!]
\centering
    \includegraphics[width=0.9\textwidth]{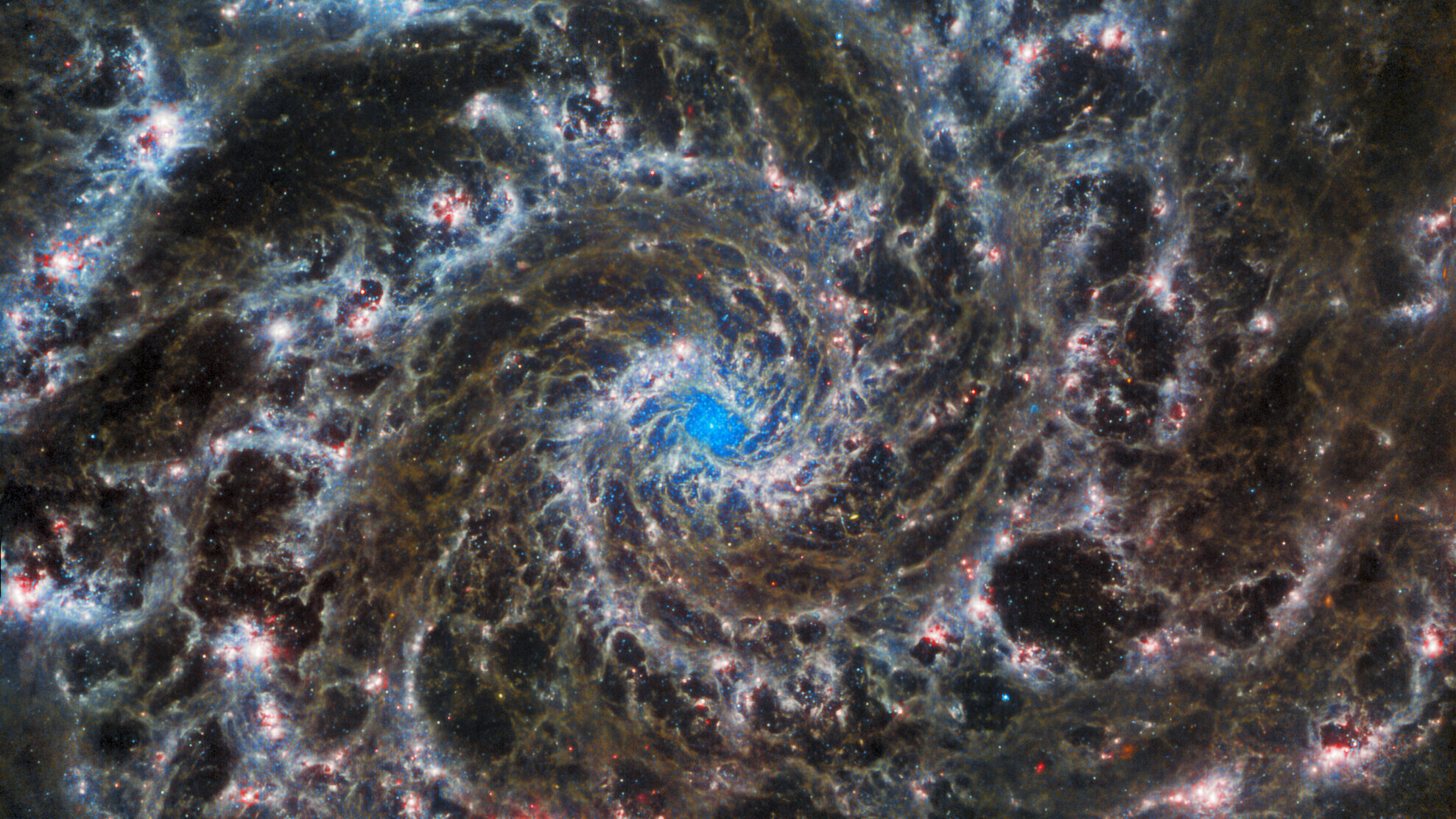}
    \caption{The Phantom Galaxy. Credit: ESA/Webb, NASA \& CSA, J. Lee and the PHANGS-JWST Team.}
    \label{fig:Phantom}
\end{figure}

Finally, while in this article we have focused on isolated and rather quiescent stellar systems, long gone are the days when astronomers thought of galaxies as `island universes'.
In the modern view galaxies are continually bombarded by infalling dark matter subhaloes and dwarf galaxies\cite{bird2012radial} 
and are connected to other galaxies via the cosmic web\cite{libeskind2018tracing}.
Nor can one always think of `stellar' (or even stellar + dark matter) components of galaxies in isolation without also considering their gaseous components. These give rise to star formation, and provide significant dynamical feedback via active galactic nuclei\cite{hopkins2016stellar} and via turbulence, shocks, supernova bubbles, etc. in the interstellar gaseous medium (see Figure \ref{fig:Phantom} for illustration).
Galaxies are therefore inherently multi-scale (both spatially and temporally), multi-component, open, dissipative systems that are invariably found out of --- and often far from --- equilibrium.
Kinetic theories that are able to deal with such complexity are in their infancy\cite{PichonAubert2006}.
\\
\\
We hope that our plasma colleagues consider this tutorial article an open invitation to participate in a subject that is 
still in its theoretical adolescence; in which simulations with realistic numbers of particles are routine\cite{Wang2015-nbody};
in which observations are astonishing and improving all the time; and, most of all, in which the Universe constantly impresses upon 
us its complex, kaleidoscopic beauty.

\newpage

\begin{acknowledgments}
This article grew out of the notes that we each compiled as lecturers of the 
Kinetic Theory of Stellar Systems MMathPhys course at the University of Oxford. 
We thank Prof.\ James Binney, who designed the course and was its first lecturer, 
as well as Prof.\ Paul Dellar and Prof.\ Alexander Schekochihin who 
taught the sister courses on kinetic theory of gases and plasmas\cite{schekochihin2022lectures} respectively, and we gratefully acknowledge the hospitality of Merton College, Oxford.
We also thank Uddipan Banik, James Binney, Thomas Foster, Shaunak Modak, Michael Nastac, Christophe Pichon, Frank van den Bosch, Martin Weinberg, George Wong and Tomer Yavetz for feedback on the manuscript.
C.~H.\ is supported by the John N. Bahcall Fellowship Fund at the Institute for Advanced Study.
This research was supported in part by grant NSF PHY-2309135 to the Kavli Institute for Theoretical Physics (KITP).
\end{acknowledgments}

\section*{Data Availability Statement}

Data sharing is not applicable to this article as no new data were created or analyzed in this study.

\appendix

\section{The epicyclic approximation}
 \label{sec:epicyclic_approximation}

In this Appendix, we derive the angle-action variables for near-circular orbits in a central potential $\Phi(R)$, using the so-called \textit{epicyclic approximation}. 
To do this, let us first write down the Hamiltonian for radial motion in this potential
\begin{equation}
    H (R, v_{R}) = \frac{1}{2} v_{R}^2 + \Phi_\mathrm{eff}(R),
\end{equation}
 where $\Phi_\mathrm{eff}(R) \equiv \Phi(R) + L^2/(2R^2)$ is the effective potential, and 
 $L = R^2 \dot{ \phi}$ is a constant.
 The Hamiltonian equations of motion then give
 \begin{equation}
     \frac{\md^2 R}{\md t^2} = - \frac{\partial \Phi_\mathrm{eff}(R)}{\partial R}.
 \end{equation}
Perfectly circular motion corresponds to the right hand side of this equation being set to zero, i.e.\ a minimum of the effective potential.
To work out the lowest-order correction to this circular motion, 
 we can expand $\Phi_\mathrm{eff}(R)$ around $R_\mathrm{g}$
 and find
 \begin{equation}
     \frac{\md^2 (R-R_\mathrm{g})}{\md t^2} = - \Omega_R^2 (R-R_\mathrm{g}) + ...
 \end{equation}
 where $\Omega_R^2 = (\partial^2 \Phi_\mathrm{eff}/\partial R^2)\vert_{R_\mathrm{g}}$.
 This just corresponds to a harmonic oscillator with angular frequency $\Omega_R$.  
This radial motion may be alternatively described using a pair of radial angle-action coordinates $(\theta_R, J_R)$ satisfying
 \begin{subequations}
 \begin{align}
     J_R &= \frac{\dot{R}^2 + \Omega_R^2 (R-R_\mathrm{g})^2}{2\Omega_R} \equiv \frac{E_R}{\Omega_R}, 
          \label{eqn:def_JR_epi}
\\
R &= R_\mathrm{g} - a_R \cos \theta_R,
\label{eqn:def_R_epi}
\end{align}
\label{eqn:def_JR_R_epi}%
 \end{subequations}
 where the amplitude of the radial epicycle is
 \begin{equation}
     a_R \equiv \frac{\sqrt{2E_R}}{\Omega_R}.
      \label{eqn:amplitudes_epicycle}
 \end{equation}
(Note we have chosen our angle coordinate $\theta_R$ such that $\theta_R=0$ corresponds to minimum $R$, i.e. the pericentre $R=R_\mathrm{a}$.  Other authors often choose different conventions).
Similarly, the
azimuthal part of the problem can also be described by angle-action variables:
 \begin{subequations}
 \begin{align}
   L &= R_\mathrm{g}^2 \, \Omega_\phi,
   \label{eqn:def_L_epi}
   \\
   \phi &= \theta_\phi + \frac{2\Omega_\phi a_R}{\Omega_R R_\mathrm{g}}\sin \theta_R,
           \label{eqn:def_phi_epi}
    \end{align}
    \label{eqn:def_L_phi_epi}%
 \end{subequations}
 and the frequencies can be written in the following form in terms of $\Phi$, all evaluated at $R=R_\mathrm{g}$:
 \begin{subequations}
 \begin{align}
 \Omega_\phi^2 &= \frac{1}{R} \frac{\partial \Phi}{\partial R} \equiv \Omega_\mathrm{c}^2,
\label{eqn:Circular_Frequency}
\\
     \Omega_R^2 &= \frac{2\Omega_\phi}{R} \frac{\partial(\Omega_\phi R^2)}{ \partial R} \equiv \kappa^2.
     \label{eqn:Radial_Epicyclic_Frequency}
\end{align}
\label{eqn:def_Frequency_epi}%
\end{subequations}
One can also extend the epicyclic approximation to include vertical harmonic oscillations or higher order corrections, but this will not concern us here.
See~\S{3.2.3} of BT08 and~\citet{Dehnen1999} for more details.

\section{The basis method}
\label{sec:BasisMethod}

To solve equation~\eqref{self_psid},
we assume
that we have at our disposal a complete set of \textit{biorthogonal basis elements}
${ (\Phi^{(p)} (\br) , \rho^{(p)} (\br)) }$, which satisfy
\begin{subequations}
\begin{align}
& \Phi^{(p)} (\br) = \!\! \int \!\! \md \br' \, \psi (\br , \br') \, \rho^{(p)} (\br') ,
\label{def_basis_1}
\\
& \!\! \int \!\! \md \br \, \Phi^{(p) *}(\br) \, \rho^{(q)} (\br) = - \mcE \delta_{p}^{q} ,
\label{def_basis_2}
\end{align}
\label{def_basis}%
\end{subequations}
where
${\psi (\br , \br') = - G / |\br - \br'| }$ and $\mcE$ is an arbitrary constant with units of energy. The `completeness' requirement is that we can expand an arbitrary function $\Psi(\br) = \sum_p A_p \Phi^{(p)}(\br)$ for some set of constant coefficients $A_p$.  
Various such basis sets do exist, and the particular set you should choose depends on the geometry of the system you are interested in.  For instance, in a spherical geometry one might take the basis to consist of 
products of radial Bessel functions and angular spherical harmonics
(see, e.g.\@,~\citet{Kalnajs1976-gg,petersen2023predicting} for more details). The idea is to project~\eqref{self_psid} onto this basis and then read off the coefficients to get an explicit expression for $\psi^\md$.

Thus, let us begin by expanding the pairwise interaction potential ${ \psi (\br , \br') } = -G/\vert \br - \br' \vert$ for a fixed value of $\br'$ as
\begin{equation}
\psi (\br , \br') = \sum_{p} u_{p} (\br') \, \Phi^{(p)} (\br),
\label{expansion_U_temp}
\end{equation}
where the coefficient ${ u_{p} (\br') }$ has units of $(\mathrm{mass})^{-1}$.
To find this coefficient, we multiply both sides of this equation by 
$\rho^{(q) *}(\br)$, integrate over $\br$, and use both properties~\eqref{def_basis}.
The result is
\begin{align}
u_{p} (\br') & \, = -\frac{1}{\mcE} \!\! \int \!\! \md \br \, \psi (\br , \br') \, \rho^{(p) *} (\br)
\nonumber
\\
& \, = -\frac{1}{\mcE} \Phi^{(p) *} (\br').
\label{calc_up}
\end{align}
Plugging this back in to equation~\eqref{expansion_U_temp} we get
\begin{equation}
\psi (\br , \br') = - \frac{1}{\mcE}\sum_{p} \Phi^{(p)} (\br) \, \Phi^{(p) *} (\br') .
\label{eq:quasi_sep}
\end{equation}
The fact that the bare interaction can be 
expanded in this `quasi-separable' form is one of the main advantages of the basis method.

What we really need is the Fourier transform of this function:
\begin{equation}
\psi_{{\mathbf{n}}{\mathbf{n}}'} (\mathbf{J} , \mathbf{J}') = - \frac{1}{\mcE}\sum_{p} \Phi^{(p)}_{{\mathbf{n}}} (\mathbf{J}) \, \Phi^{(p) *}_{{\mathbf{n}}'} (\mathbf{J}') ,
\label{bare_basis_app}
\end{equation}
where ${ \Phi^{(p)}_{{\mathbf{n}}} (\mathbf{J}) }$ stands for the Fourier transformed basis elements
naturally defined following equation~\eqref{Fourier_expansion}.
Thus we now have an explicit expression for the functions $\psi_{{\mathbf{n}}{\mathbf{n}}'} (\mathbf{J} , \mathbf{J}')$ which enter the right hand side of equation~\eqref{eqn:delta_Phi_integral_equation}.

We now perform a very similar expansion of the dressed interaction $\psi^\md$.
By analogy with~\eqref{bare_basis_app}, 
we  assume that, at fixed $\omega$, this function can be expanded as
\begin{equation}
\psi^\md_{{\mathbf{n}}{\mathbf{n}}'} (\mathbf{J} , \mathbf{J}' , \omega) = - \frac{1}{\mcE} \sum_{p , q} \Phi^{(p)}_{{\mathbf{n}}} (\mathbf{J}) \, \msE_{pq}^{-1} (\omega) \, \Phi^{(q) *}_{{\mathbf{n}}'} (\mathbf{J}') ,
\label{assumption_dressed_basis}
\end{equation}
where ${ \msE (\omega) }$ is a dimensionless, as-yet-unknown,
matrix which depends only on $\omega$ (and we have assumed its inverse exists).
Injecting this decomposition into equation~\eqref{self_psid}, we obtain
\begin{align}
\sum_{p , q} \Phi^{(p)}_{{\mathbf{n}}} (\mathbf{J}) & \, \msE_{pq}^{-1} (\omega) \, \Phi^{(q) *}_{{\mathbf{n}}'} (\mathbf{J}') = \sum_{p ,q} \Phi^{(p)}_{{\mathbf{n}}} (\mathbf{J}) \, \big[ \msM (\omega) \, \msE^{-1} (\omega) \big]_{pq} \, \Phi^{(q) *}_{{\mathbf{n}}'} (\mathbf{J}') 
+ \sum_{p , q} \Phi^{(p)}_{{\mathbf{n}}} (\mathbf{J}) \, \msI_{pq} \, \Phi^{(q) *}_{{\mathbf{n}}'} (\mathbf{J}') ,
\label{relation_psid}
\end{align}
where we introduced the identity matrix $\msI$,
as well as the system's dimensionless \textit{response matrix} ${ \msM (\omega) }$ as
in equation~\eqref{Fourier_M}.
Since the basis is biorthogonal, we can identify the elements in equation~\eqref{relation_psid},
and we obtain the matrix relation
\begin{equation}
\msE^{-1} (\omega) = \msM (\omega) \, \msE^{-1} (\omega) + \msI ,
\label{relation_E_M}
\end{equation}
which immediately tells us that
\begin{equation}
\msE (\omega) = \msI - \msM (\omega) .
\label{expression_E}
\end{equation}
To conclude, from~\eqref{assumption_dressed_basis} we have
\begin{equation}
\psi^\md_{{\mathbf{n}}{\mathbf{n}}'} (\mathbf{J} , \mathbf{J}' , \omega) = - \frac{1}{\mcE} \sum_{p , q} \Phi^{(p)}_{{\mathbf{n}}} (\mathbf{J}) \, [\msI - \msM (\omega)]_{pq}^{-1} \, \Phi^{(q) *}_{{\mathbf{n}}'} (\mathbf{J}') .
\label{eq:psid_generic}
\end{equation}
Thus we have done what we anticipated in \S\ref{sec:Linear_response_Fourier_Laplace}, which is essentially to invert an operator $\msI -\msM$, with $\msM$ encoding collective effects.
We can easily recover the bare interaction, equation~\eqref{bare_basis_app}, by ignoring collective effects, i.e.\ by setting $\msM (\omega) = \mszero$ in~\eqref{eq:psid_generic}.

\bibliography{main}

\end{document}